\documentclass[amsmath,amssymb,11pt]{article}
\usepackage{jheppub2}
\pdfoutput=1
\usepackage{graphicx,epsfig}
\usepackage{float}
\usepackage{subfig}
\usepackage{subfloat}
\usepackage[utf8]{inputenc}
\usepackage{hyperref}
\usepackage{caption}

\usepackage{color}

\newcommand*{\affmark}[1][*]{\textsuperscript{#1}}

\newcommand{\beq}{\begin{equation}}

\newcommand{\eeq}{\end{equation}}

\newcommand{\be}{\begin{equation}}

\newcommand{\ee}{\end{equation}}

\def	\eps	{\epsilon}

\def	\del	{\nabla}
\def	\lf	{\left (}
\def	\rt	{\right )}
\def	\ket	{\rangle}

\def	\comma	{\quad , \quad}

\title{Emergence of Spacetime: From Entanglement to Einstein}

\author{Andrew Svesko\affmark[a]}
\affiliation{\affmark[a]Department of Physics, Arizona State University, 
Tempe, Arizona 85287, USA}
\emailAdd{asvesko@asu.edu}

\abstract{Here I develop the connection between thermodynamics, entanglement, and gravity. I begin by showing that the classical null energy condition (NEC) can arise as a consequence of the second law of thermodynamics applied to local holographic screens. This is accomplished by essentially reversing the steps of Hawking's area theorem, leading to the Ricci convergence condition as an input, from which an application of Einstein's equations yields the NEC. Using the same argument, I show  logarithmic quantum corrections to the Bekenstein-Hawking entropy formula do not alter the form of the Ricci convergence condition, but obscure its connection to the NEC. Then, by attributing thermodynamics to the stretched horizon of future lightcones -- a timelike hypersurface generated by a collection of radially accelerating observers with constant and uniform proper acceleration -- I derive Einstein's equations from the Clausius relation $T\Delta S_{\text{rev}}=Q$, where $\Delta S_{\text{rev}}$ is the \emph{reversible} entropy change. Based on this derivation I uncover a local first law of gravity, $\Delta E=T\Delta S-W$, connecting gravitational entropy $S$ to matter energy $E$ and work $W$. I then provide an entanglement interpretation of stretched lightcone thermodynamics by extending the entanglement equilibrium proposal.
 Specifically I show that the condition of fixed volume can be understood as subtracting the irreversible contribution to the thermodynamic entropy. Using the $\text{AdS}_{3}/\text{CFT}_{2}$ correspondence, I then provide a microscopic explanation of the `thermodynamic volume' -- the conjugate variable to the pressure in extended black hole thermodynamics -- and reveal the super-entropicity of $\text{AdS}_{3}$ black holes is due to the gravitational entropy overcounting the number of available dual $\text{CFT}_{2}$ states. Finally, I conclude by providing a recent generlization of the extended first law of entanglement, and study its non-trivial $2+1$- and $1+1$-dimensional limits. This thesis is self-contained and pedagogical by including useful background content relevant to emergent gravity.}

\begin{document}

\maketitle

\newpage

\section*{Acknowledgements and Dedication}

{ \itshape  Physics is a practice that cannot and should not be done by oneself. I have had the good fortune of learning from and working with some inspiring individuals who have my gratitude. I would like to therefore briefly acknowledge all of those who have aided me along the way in reaching this goal, starting with my thesis committee.

 I would like to thank Professor Cindy Keeler for taking me on for multiple projects not even discussed in this dissertation. You have pushed me and given me a unique perspective of the field. Professor Damien Easson, who was my first research rotation advisor, thank you for giving me a first glimpse at what research is like and giving me the freedom to pursue my own project. Professor Tanmay Vachaspati, thank you for asking me honest questions forcing me to think about the `big picture'. And lastly my thesis advisor, Professor Maulik Parikh, who I was told normally does not take on first year students. You were one of the big reasons why I decided to come to ASU. Not only have you provided great insight into science, you taught me how to be self-reliant and given me tremendous freedom. I couldn't have asked for a better advisor. 

There are other professors of physics, both at ASU and other institutions, who deserve my thanks. In particular Professor Ted Jacobson at University of Maryland, who has greatly influenced the ideas that are in this thesis and given me encouragement, and Professor Clifford Johnson of University of Southern California, who has influenced my mode of thinking and lent me support and hospitality -- each of you inspire me to be a better physicist. I must also thank Dr. Darya Dolenko of ASU for showing me how to be a better instructor and granting me the opportunity to try different styles of teaching. 

Individuals who are not (yet) professors of physics have my gratitude as well. Felipe Rosso and Batoul Banihashemi, I thank them for their insight, kindness, and hospitality. I must also express my thanks to the entire cosmology graduate student office. Each of you have made my time at ASU a joy. Even though I might get more work done at home, I look forward to coming into the office everyday for the lively conversation. I must also thank the group of postdoctoral researchers in the cosmology office, two of whom I should single out: George Zahariade and Victoria Martin. Both of you have inspired me to be a better physicist, but, most of all, I am fortunate to say that you are both colleagues and dear friends. 

Lastly, there are numerous people outside of the world of physics who have my deepest appreciation. These include the friends/family I didn't expect to find in Arizona, as well as my friends and family back home in Oregon. Listing you all would take up too much time, but I must highlight a few individuals, starting with my older sister Kacie Svesko. She showed me what it takes to be a good student, and has harbored in me a healthy spirit of competitiveness. I also thank my parents, Mike and Lorie Svesko, who have encouraged me every step of the way, in more ways than a few, including buying my very first textbook on physics. And, of course, my wonderful wife Abby Mccoy, who has shown me love and tremendous patience from the time we were in high school through now as a PhD graduate.}


\newpage

\section{OVERVIEW} \label{sec:overview}

\noindent The discovery that black holes carry entropy \cite{Bekenstein72-1,Hawking74-1}, 
\beq S_{BH}=\frac{A_{\mathcal{H}}}{4G}\; , \label{BHent}\eeq
provides the two following realizations: (i) A world with gravity is holographic \cite{Susskind:1994vu}, and (ii) spacetime is emergent \cite{Jacobson:1995ab}. The former of these comes from the observation that the thermodynamic entropy of a black hole (\ref{BHent}) goes as the area of its horizon $A_{\mathcal{H}}$, and the latter from noting that black holes are spacetime solutions to Einstein's equations. In fact, black holes are not the only spacetime solutions which carry entropy; any solution which has a horizon, e.g., Rindler space and the de Sitter universe, also possess a thermodynamic entropy proportional to the area of their respective horizons. The fact that Rindler space carries an entropy is particularly striking as there the notion of horizon is observer dependent. This leads to the proposal that an arbitrary spacetime -- which may appear locally as Rindler space -- is equipped with an entropy proportional to the area of a local Rindler horizon, and that thermodynamic relationships, e.g., the Clausius relation $T\Delta S=Q$, have geometric meaning. Specifically, 
\beq T\Delta S=Q\Rightarrow G_{\mu\nu}+\Lambda g_{\mu\nu}=8\pi GT_{\mu\nu}\; . \label{ClausiusGrav}\eeq
That is, Einstein gravity arises from the thermodynamics of spacetime \cite{Jacobson:1995ab}. 

Recently it was shown how to generalize (\ref{ClausiusGrav}) to higher derivative theories of gravity \cite{Parikh:2017aas}. By attributing a temperature and entropy to a stretched future lightcone -- a timelike hypersurface composed of the worldlines of constant and uniformly radially accelerating observers -- the equations of motion for a broad class of higher derivative theories of gravity are a consequence of the Clausius relation $T\Delta S_{\text{rev}} =Q$, where $\Delta S_{\text{rev}}$ is the \emph{reversible} entropy, i.e., the entropy growth solely due to a flux of matter crossing the horizon of the stretched lightcone. This result shows that arbitrary theories of gravity arise from the thermodynamics of some underlying microscopic theory of spacetime. We will review the geometric set-up of stretched lightcones and the derivation of Einstein's equations in Chapter \ref{sec:gravfromthermo}, as well as uncover a local first law of gravity, connecting matter thermodynamics with spacetime thermodynamics. Moreover, while stretched lightcones are interesting surfaces to consider, they are not the only geometric construction for which the spirit of \cite{Jacobson:1995ab} can applied. In Chapter \ref{sec:gravfromthermo}, we will also show how the Clausius relation applies equally to causal diamonds -- the set of all events that lie in both the past and future of some point on a causal curve. Specifically, we will show causal diamonds can be understood as systems in thermal equilibrium, for which the Clausius relation gives rise to gravitational field equations for a broad class of gravity theories.

There are other aspects of general relativity, which, from the perspective of classical gravity, have an obscure orgin. Such is the case for the null energy condition (NEC) -- an \emph{ad hoc} covariant constraint on the type of matter allowed in a spacetime. While the condition depends on the energy-momentum tensor of matter, the NEC itself does not seem to have a consistency requirement coming from standard quantum field theory. This suggests the NEC arises from a combined theory of matter and gravity, such as string theory \cite{Parikh14-1}, where Einstein's equations rewrite the NEC as a geometric inequality, namely, the Ricci convergence condition. Following the spirit of \cite{Jacobson:1995ab}, it can be shown that the NEC is a consequence of the second law of thermodynamics applied to local holographic screens \cite{Parikh:2015ret,Parikh:2016lys}. That is to say, as reviewed in Chapter \ref{sec:NECfromthermo}, by assuming the second law of (spacetime) thermodynamics can be applied to local screens, reminiscent of local Rindler horizons, we will obtain the NEC as a geometric consequence.

Despite some successes in deriving (\ref{BHent}) in specific cases \cite{Strominger96-1,Rovelli96-1}, it is still unclear what the physical degrees of freedom encoded in $S_{BH}$ correspond to microscopically. Similarly, the underlying microscopics of spacetime giving rise to Einstein's equations is obscure. A potential explanation comes from studying entanglement entropy (EE) of quantum fields outside of the horizon. For a generic $(d+1)$ quantum field theory (QFT) with $d>1$, the EE of a region $A$ admits an area law \cite{Bombelli:1986rw,Srednicki:1993im}
\beq S^{EE}_{A}=c_{0}\frac{\mathcal{A}(\partial A)}{\epsilon^{d-1}}+\text{subleading divergences}+S_{\text{finite}}\; , \label{SQFT}\eeq 
where $\epsilon$ is a cutoff for the theory, illustrating that the EE is in general UV divergent, and $\mathcal{A}$ is the area of the $(d-1)$ boundary region $\partial A$ separating region $A$ from it's complement.  Identifying $\frac{c_{0}}{\epsilon^{d-1}}\to\frac{1}{4G}$ suggests $S_{BH}$ to be interpreted as the leading UV divergence in the EE for quantum fields outside of a horizon.

Further progress can be made when we consider quantum field theories with holographic duals. Specifically, in the context of $\text{AdS}_{d+2}/\text{CFT}_{d+1}$ duality \cite{Maldacena98-1}, one is led to the Ryu-Takayanagi (RT) conjecture \cite{Ryu06-1}:
\beq S^{EE}_{A}=\frac{\mathcal{A}(\gamma_{A})}{4G^{(d+2)}}\; , \label{RT}\eeq
which relates the EE of holographic CFTs (HEE) to the area of a $d$-dimensional (static) minimal surface $\gamma_{A}$  in $\text{AdS}_{d+2}$ whose boundary is homologous to $\partial A$.\footnote{The RT conjecture has a covariant generalization, in which the static minimal surface $\gamma_{A}$ is replaced by an extremal surface $\Sigma_{A}$, \cite{Hubeny:2007xt}.} The RT formula (\ref{RT}) is specific to CFTs dual to general relativity, and does not include quantum corrections. The proposal was proved in \cite{Lewkowycz:2013nqa}, and has been extended to include quantum corrections \cite{Faulkner13-1}, and for CFTs dual to higher derivative theories of gravity \cite{Dong:2013qoa}. When the minimal surface $\gamma_{A}$ is the horizon of a black hole, one observes that black hole entropy is equivalent to HEE, $S_{HEE}|_{\gamma_{A}=\mathcal{H}}=S_{BH}$ \cite{Casini:2011kv}. 

Similar to the situation with black hole thermodynamics, this observation suggests that gravity emerges from quantum entanglement, i.e., spacetime is built from entanglement \cite{VanRaamsdonk10-1,Bianchi12-1}. To take on this proposal, one can study the properties of HEE and look for the resulting geometric consequences. Indeed, the EE of a QFT generically satisfies a first law reminiscient of the first law of thermodynamics \cite{Blanco:2013joa,Wong:2013gua}
\beq \delta S^{EE}_{A}=\delta \langle H_{A}\rangle\; .\label{firstlawEE}\eeq
Here $\delta S^{EE}_{A}$ is the variation of the EE of region $A$, while $\delta \langle H_{A}\rangle$ is the variation of the modular Hamiltonian $H_{A}$ defined by $\rho_{A}\equiv e^{-H_{A}}$. When one specializes to the case where the region $A$ is a ball of radius $R$, the modular Hamiltonian can be identified with the thermal energy of the region. 

For holographic CFTs the first law of entanglement entropy (\ref{firstlawEE}) can be understood as a geometric constraint on the dual gravity side. By substituting (\ref{RT}) into the left hand side (LHS) of (\ref{firstlawEE}), and relating the energy-momentum tensor of the CFT to a metric perturbation in AdS, one arrives at the \emph{linearized} Einstein equations \cite{Lashkari13-1}:
\beq \delta S^{EE}_{A}=\delta \langle H_{A}\rangle \Rightarrow G_{\mu\nu}+\Lambda g_{\mu\nu}=8\pi GT_{\mu\nu}\; .\eeq
By considering the higher derivative gravity generalization of (\ref{RT}), similar arguments lead to the linearized equations of motion for higher derivative theories of gravity \cite{Faulkner13-2}. The non-linear behavior of gravitational equations of motion is encoded in a generalized form of (\ref{firstlawEE}), where one must take into account the relative entropy of excited CFT states \cite{Faulkner:2017tkh,Haehl:2017sot}. In this way, gravity emerges from spacetime entanglement.

Recently it has been shown how to derive gravitational equations of motion from entanglement considerations without explicit reference to AdS/CFT duality, and is therefore slightly more general than the derivation in \cite{Lashkari13-1,Faulkner13-2}. This approach, first proposed by Jacobson, is the \emph{entanglement equilibrium} conjecture \cite{Jacobson16-1}, which can be stated as follows: \emph{In a theory of quantum gravity, the entanglement entropy of a spherical region with a fixed volume is maximal in the vacuum}. This hypothesis relies on assuming that the quantum theory of gravity is UV finite (as is the case in string theory) and therefore yields a finite EE, where the cutoff $\epsilon$ introduced in (\ref{SQFT}) is near the Planck scale, $\epsilon\sim\ell_{P}$, and being able to identify the entanglement entropy $S_{EE}^{A}$ with the generalized entropy $S_{\text{gen}}$, which is independent of $\epsilon$ \cite{Susskind:1994sm,Solodukhin:2011gn}:
\beq S_{EE}^{A}=S_{\text{gen}}=S^{(\epsilon)}_{BH}+S^{(\epsilon)}_{\text{mat}}\;.\label{Sgen}\eeq
Here $S^{(\epsilon)}_{BH}$ is the Bekenstein-Hawking entropy (\ref{BHent}) expressed in terms of renormalized gravitational couplings, and $S^{(\epsilon)}_{\text{mat}}$ is the renormalized EE of matter fields. The generalized entropy $S_{\text{gen}}$ is independent of $\epsilon$ as the renormalization of gravitational couplings is achieved via the matter loop divergences. 

When one interprets the EE as the generalized entropy, one may therefore assign EE to surfaces other than cross sections of black hole horizons, or the minimal surfaces identified in the RT formula (\ref{RT}). In this way, without assuming holographic duality, one discovers a connection between geometry and entanglement entropy. Furthermore, taking into consideration the underlying thermodynamics of spacetime \cite{Jacobson:1995ab}, this link provides a route to derive dynamical equations of gravity -- not from thermodynamics, but from entanglement. 

With these consderations in mind, the variation of the EE of a spherical region at fixed volume is given by 
\beq \delta S^{A}_{EE}|_{V}=\frac{\delta A|_{V}}{4G}+\delta S_{\text{mat}}=0\;,\eeq
i.e., the vacuum is in a maximal entropy state. In the case of small spheres, this entanglement equilibrium condition is equivalent to imposing the full non-linear Einstein equations at the center of the ball \cite{Jacobson16-1}. Recently this maximal entropy condition has been generalized to include higher derivative theories of gravity, where $S^{(\epsilon)}_{BH}$ in (\ref{Sgen}) is replaced by the higher derivative extension of gravitational entropy, the Wald entropy $S^{(\epsilon)}_{\text{Wald}}$, in which case the maximal entropy condition becomes
\beq \delta S^{A}_{EE}|_{W}=\delta S_{\text{Wald}}|_{W}+\delta S_{\text{mat}}=0\;, \label{maxentcondhigher}\eeq
where the volume $V$ must be replaced with a new local geometrical quantity called the \emph{generalized volume} $W$. This condition, when applied to small spheres, is equivalent to imposing the linearized equations of motion for a higher derivative theory of gravity \cite{Bueno16-1}. 

In Chapter \ref{sec:gravfroment} we extend the work of \cite{Parikh:2017aas} and \cite{Bueno16-1} and provide an entanglement interpretation to stretched lightcone thermodynamics. We accomplish this by first deriving a ``first law of stretched lightcones", and show that it is geometrically equivalent to an entanglement equilibrium condition. By comparing the entanglement equilibrium and (reversible) equilbrium thermodynamic pictures of deriving Einstein's equations, we will show how the two are related by showing that the leading contribution to the generalized volume $\bar{W}$ is precisely the entropy change due to the natural increase of the stretched lightcone. This not only sheds light on the microscopic origins of the thermodynamics of stretched lightcones, but also provides another derivation of the non-linear (semi-classical) Einstein equations and (linearized) equations of motion of higher derivative theories of gravity from spacetime entanglement. 

As already eluded to, progress in understanding the nature of  black hole entropy can be made if we utilize the AdS/CFT correspondence. In fact, there are a number of ways AdS/CFT duality can be used to provide a microscopic explanation of the Bekenstein-Hawking entropy formula (\ref{BHent}). One of the first ways this was done was accomplished by Strominger \cite{Strominger:1997eq}. Specifically, for black holes whose near horizon geometry is locally $\text{AdS}_{3}$, the Bekenstein-Hawking formula is equal to the logarithm of the asymptotic density of $\text{CFT}_{2}$ states, i.e., the Cardy entropy is equal to the Bekenstein-Hawking entropy\footnote{His derivation relied on the well-known result by Brown and Henneaux \cite{Brown:1986nw}, that any consistent theory of quantum gravity on $\text{AdS}_{3}$ is equivalent to a $\text{CFT}_{2}$, by showing that the generators defining the asymptotic symmetry  group of $\text{AdS}_{3}$ satisfies an algebra equal to two copies of a Virasoro algebra with central charges $c_{R}=c_{L}=3L/2G$.}. 

Strominger's observation can be used to understand the mircoscopics not just of black hole thermodynamics, but also \emph{extended} black hole thermodynamics\footnote{Otherwise known as \emph{black hole chemistry}. For a recent review of EBHT, see \cite{Kubiznak:2016qmn}.} (EBHT) \cite{Caldarelli:1999xj,Sekiwa:2006qj,Kastor:2009wy}, where black holes in (A)dS spacetimes have a dynamical pressure $p=-\Lambda/8\pi G$, thermodynamic volume $V$, and where the black hole mass becomes the enthalpy. Just as black hole thermodynamics is expected to have a microscopic interpretation, so too should EBHT. 

 In Chapter \ref{sec:microvol}, we present a microscopic explanation for the thermodynamic volume $V$ for specific $\text{AdS}_{3}$ black holes; revealing in certain cases $V$ will restrict the number of allowed $\text{CFT}_{2}$ states such that the Bekenstein-Hawking formula (\ref{BHent}), given by the Cardy entropy, overcounts the number of microstates. This leads to a microscopic interpretation of black hole ``super-entropicity" -- a characteristic of AdS black holes whose entropy is larger than Schwarzschild-AdS black holes (spacetimes which were once thought to carry a maximal entropy) \cite{Cvetic:2010jb}. 

 EBHT also leads to new insights into the entanglement of conformal field theories. Specifically, using EBHT as motivation, the first law of entanglement entropy (\ref{firstlawEE}) may be extended to include variations of the central charge \cite{Kastor:2014dra}. In this way, the extended first law of entanglement considers not just state variations but also variations of the CFT itself\footnote{The variations of the central charge also has an interpretation in the EBHT picture: a flow along the isotherms in a $p-V$ plane. Such flows represent the cycles of black hole ``heat engines" \cite{Johnson:2014yja}, and are thought to be equivalent to RG flows of the dual CFT \cite{Johnson:2018amj}.}. We provide a novel generalized derivation of the extended first law in Chapter \ref{sec:extfirstlawhighlow}, such that it holds for an arbitrary theory of gravity and variety of entangling surfaces, and study its $2+1$- and $1+1$-dimensional limits. The $2+1$-dimensional limit leads to a general expression of the thermodynamic volume in terms of the horizon entanglement entropy and central charge of the dual CFT, matching the specific microscopic expressions recently found in \cite{Johnson:2019wcq}.


To summarize, let us now provide a road map of this thesis. Based on \cite{Parikh:2015ret,Parikh:2016lys},  in Chapter \ref{sec:NECfromthermo} we derive the null energy condition by way of the Ricci convergence condition via the second law of thermodynamics applied to local holographic screens. From this derivation we show that the Ricci convergence condition is stable under logarithmic (1-loop) quantum corrections to horizon entropy. In Chapter \ref{sec:gravfromthermo}, following \cite{Parikh:2017aas}, we present a complete derivation of the field equations for a broad class of theories of gravity using the Clausius relation. A particularly novel aspect of this derivation is that it holds for the (timelike) stretched horizons of future lightcones, not the null horizons of lightsheets. We then use our Clausius relation and uncover a local first law of gravity, as shown in \cite{Parikh:2018anm}, which combines elements of both matter and spacetime thermodynamics. We conclude this chapter by deriving gravitational equations of motion using a similar approach to \cite{Parikh:2017aas}, applied to the past (conformal Killing) horizons of causal diamonds, first demonstrated in \cite{Svesko:2018qim}. 

Starting in Chapter \ref{sec:gravfroment}, the thesis transitions from spacetime thermodynamics to spacetime entanglement and the microphysics of black hole thermodynamics. We begin by applying the entanglement equilibrium conjecture to stretched future lightcones, and explicitly connect to the equivalent derivation using equilibrium thermodynamics presented in Chapter \ref{sec:gravfromthermo}. From there, in Chapter \ref{sec:microvol} and based on \cite{Johnson:2019wcq}, we uncover the microscopic origins of the thermodynamic volume in extended black hole thermodynamics and the ``super-entropicity"  of AdS black holes using $\text{AdS}_{3}/\text{CFT}_{2}$ duality. Motivated by extended black hole thermodynamics, in Chapter \ref{sec:extfirstlawhighlow} we then present a new derivation of the extended first law of entanglement, generalizing previous versions and evaluating its lower dimensional limits where we find an intriguing new expression for the thermodynamic volume. This final chapter is based on the recent work \cite{Rosso:2020zkk}. Some final thoughts and remarks are given in the conclusion, Chapter \ref{sec:conclusion}. 

To keep this thesis self-contained, we include Chapter \ref{sec:history} outlining the history and philosophy of emergent gravity. Multiple appendices are also included to present background on the fundamentals of spacetime thermodynamics  in Appendix \ref{app:thermofund} and spacetime entanglement in Appendix \ref{app:entfund}, as well as detailed calculations left out of the body of the thesis for the sake of pedagogy.


\newpage

\section{FOUR ROADS TO  EMERGENT GRAVITY} \label{sec:history}

\hspace{2mm}

\emph{...the fundamental laws of physics, when discovered, can appear in so many different forms that are not apparently identical at first, but, with a little mathematical fiddling you can show the relationship...there is always another way to say the same thing that doesn't look at all like the way you said it before...}

$\quad\quad\quad\quad\quad\quad\quad\quad\quad\quad\quad\quad\quad\quad$ -- \emph{Richard Feynman, on the simplicity of Nature}

\vspace{12mm}

\emph{Emergent gravity} rests on the notion that gravity is not fundamental, at least in the sense of the standard model of particle physics.  Rather, gravity arises from some underlying  microscopic theory of spacetime or by some other means altogether. Here we provide a broad, non-exhaustive historical and philosophical review of the emergent gravity paradigm.


\subsection{Induced Gravity}
\noindent

Perhaps the earliest description of emergent gravity comes from Sakharov's \emph{induced gravity} in 1967 \cite{Sakharov:1967pk}. He observed that many condensed matter or fluid systems give rise to collective phenomena, such as the fluid mechanics approximations of Bose-Einstein condensation. As such, Sakharov found that spacetime curvature can be induced from quantum field theory on an arbitrary background, with dynamics emerging as a mean field approximation of some underlying microscopic degrees of freedom. 

The basic proposal of Sakharov's induced gravity rests on the following three elementary assumptions \cite{Visser:2002ew}: (i) Assume that spacetime is described by an arbitrary Lorentzian manifold with metric $g_{\mu\nu}$, for which matter lives on described by quantum field theory; (ii) Quantize matter and nothing else -- make no further assumptions about the dynamics of the \emph{classical} background spacetime; (iii) Consider the quantum field theory to at least 1-loop. Combined, these three assumptions lead to the following 1-loop effective action:
\beq I_{\text{1-loop}}=\int d^{4}x\sqrt{-g}\left[c_{0}+c_{1}R+c_{2}(\text{curvature squared terms})\right]\;.\label{1loopeffact}\eeq
When we compare this effective action to the standard action for Einstein gravity (plus curvature corrections), 
\beq I_{\text{EH+higher}}=\int d^{4}x\sqrt{-g}\left[\frac{1}{16\pi G}R-2\Lambda+\alpha(\text{curvature squared terms})+\mathcal{L}_{\text{matter}}\right]\;,\eeq
we find that the 1-loop effective action (\ref{1loopeffact}) automatically contains terms proportional to the cosmological constant, the Einstein-Hilbert Lagrangian, and curvature squared terms. In other words, 1-loop quantum effects of a quantum field theory living on an arbitrary background give rise to what we would interpret as classical Einstein gravity (plus corrections)\footnote{Sakharov's mechanism of induced gravity isn't the only way to `induce' gravity from quantum excitations on a background. In string theory, for example, the low-energy effective action includes the Einstein-Hilbert action (along with a dilaton and Kalb-Ramond field), and can be understood as arising from the quantum excitations of strings living on an arbitrary curved background. Curvature squared contributions appear upon including $\alpha'$ corrections to the string effective action. While these stringy arguments lead to an induced gravity, we find Sakharov's technique heuristically compelling and therefore only review this approach and its off-shoots here.}.

More explicitly, for a scalar field of mass $m$ coupled (potentially non-minimally) to spacetime curvature, the effective action at 1-loop can be computed using heat kernel techniques and the Seeley-DeWitt expansion (see, \emph{e.g.}, \cite{Vassilevich:2003xt} for a review), leading to \cite{Visser:2002ew}
\beq I_{\text{1-loop}}=\int d^{4}x\sqrt{-g}\left[\frac{1}{16\pi G}R-2\Lambda+\alpha_{1}R^{2}+\alpha_{2}C_{abcd}^{2}+\mathcal{L}_{\text{matter}}\right]\;,\label{1-loopaction2}\eeq
where $C_{abcd}^{2}$ is the Weyl tensor squared. Here, moreover, $\Lambda, G$, $\alpha_{1}$, and $\alpha_{2}$ are regularized couplings. For example, the regulated Newton's constant $G$ is related to tree-level constant $G_{0}$ via
\beq \frac{1}{G}=\frac{1}{G_{0}}-\frac{k_{1}}{2\pi}\text{str}\left[\kappa^{2}-m^{2}\log\left(\frac{\kappa^{2}}{m^{2}}\right)\right]+\text{UV finite}\;,\label{regG}\eeq
with $k_{1}$ related to the Seeley-Dewitt coefficient $a_{1}$, $\text{str}$ is the supertrace summing over all particle species, and $\kappa^{2}$ is some regularization scale introduced to identify the UV divergences. Similar expressions hold for the other gravitational couplings of the theory. 

The overall point is that vacuum fluctuations due to the matter sector influence the gravitational couplings. Sakharov's interpretation is then to assume that the 1-loop physics is dominant, where he further imposes the regulator be near the Planck scale, $\kappa\approx M_{\text{PL}}$, and to set all tree-level constants\footnote{He also seems to ignore the 1-loop corrected couplings $\Lambda,\alpha_{1}$ and $\alpha_{2}$.} to zero. The effect of setting the bare coupling $G_{0}$ to zero is that Newton's constant $G$ is \emph{induced} at 1-loop. Therefore, a suggestive conclusion that can drawn from this line of thinking is that classical Einstein gravity (and potentially higher curvature theories if we insist on including the additional couplings) is not fundamental at all -- rather it behaves as an emergent phenomena, like that of the critical behavior of the Ising model, arising from 1-loop matter effects on a Lorentzian background. 

There are, of course, criticisms to Sakharov's picture. First and foremost is that this version of induced gravity lacks predictive power. Indeed, when the remaining gravitational couplings are included, the smallness of the cosmological constant, for example, must be put in by hand. Each of the couplings, in fact, seem to require an amount of fine tuning. As such, induced gravity offers an interpretation of classical gravity, but seemingly, for the moment, nothing more. 

Briefly, we should note that the observations (\ref{1-loopaction2}) and (\ref{regG}) can lead to proposals other than Sakharov's original interpretation. One is that, rather than demanding for 1-loop dominance, impose 1-loop finiteness. Then, for example, the 1-loop contribution to Newton's constant (\ref{regG}) is finite, with
\beq \frac{1}{G}=\frac{1}{G_{0}}-\frac{1}{2\pi}\text{str}\left[k_{1}m^{2}\log\left(\frac{m^{2}}{\mu^{2}}\right)\right]+\text{two loops}\;,\eeq
where $\mu$ is some mass scale conveniently chosen to keep the argument of the logarithm dimensionless and where we assumed $\text{str}(k_{1})=\text{str}(k_{1}m^{2})=0$. This type of quantum field theory compensation can be traced back to Pauli, and is in the spirit of supersymmetry. The effect of these finiteness constraints requires strong constraints on the particle content of the theory -- strong enough to require physics beyond the standard model \cite{Visser:2002ew}. These types of constraints are so extensive, however, that it seems improbable for such a compensation to occur in nature. This interpretation also goes counter to Sakharov's original proposal in that $1/G_{0}$ is still present.

Another possibility is to assume both 1-loop dominance and 1-loop finiteness. That is, assume the finiteness constraints mentioned above, and that all tree-level coefficients vanish. Such an idea has been proposed by Frolov and Fursaev \cite{Frolov:1996aj,Frolov:1996qh,Frolov:1997xd,Frolov:1997up}. In this case Newton's constant at 1-loop becomes 
\beq \frac{1}{G}=-\frac{1}{2\pi}\text{str}\left[k_{1}m^{2}\log\left(\frac{m^{2}}{\mu^{2}}\right)\right]+\text{two loops}\;.\eeq
This approach is appealing in that it is a modification to Sakharov's original interpretation, where the gravitational constant is induced solely by the loop corrections coming from the quantum field theory of matter living on the background. The modification, however, is not so slight, as it requires a tight restriction of the allowed particle spectrum of the theory due to the finiteness constraints. Nonetheless, Frolov's and Fursaev's version of induced gravity continues to maintain an appeal as it provides a possible microscopic explanation of the Bekenstein-Hawking entropy for black holes (more on this later). 

A fourth proposal is to relax both 1-loop finiteness and dominance, and instead opt for (at least) 1-loop renormalizability. That is to say, have all of the bare coupling constants, such as $G_{0}$, to absorb any of the undesired UV divergences, just as one does when renormalizing QED in Minkowski space. The consequence now, however, is that we find new renormalizability constraints requiring modifications to the couplings of the standard model particle spectrum, again requiring new physics. Unfortunately, we might have an even worse problem: not only would the cosmological constant remain radiatively unstable (the true meaning of the cosmological constant problem), but we also expect similar radiative instabilities to appear in the other gravitational couplings, including $G$ \cite{Visser:2002ew}.

Despite the drawbacks of the various aforementioned proposals, induced gravity remains to be an intriguing viewpoint. This is in part due to its elegance: any quantum field theory in an arbitrary curved background with Lorentzian signature will, by 1-loop, generate \emph{classical} Einstein gravity (plus corrections). Gravity need not even exist at tree-level! This line of thinking tells us that the geometry of the background behaves as an external classical field, automatically leading to semiclassical gravity. Crucially, gravity was never put into a quantum theory or quantized -- it just appeared from a quantized field theory living on some classical background. This, combined with the observation that, thus far, the only experiments we can currently perform with gravity only lie in the semiclassical regime, suggests that perhaps we need not quantize gravity at all. If this is too difficult to accept, however, Sakharov's philosophy tells us something else: deriving classical aspects of gravity from any candidate theory of quantum gravity, such as the inverse square law, is not the hard part. Classical aspects of gravity arise virtually for free once we find ourselves in the limit that we have a Lorentzian manifold and a quantum field theory that can be defined, at least perturbatively, on this classical spacetime.


\subsection{Spacetime Thermodynamics}
\noindent

\begin{figure}[t]
\centering
 \includegraphics[width=9.3cm]{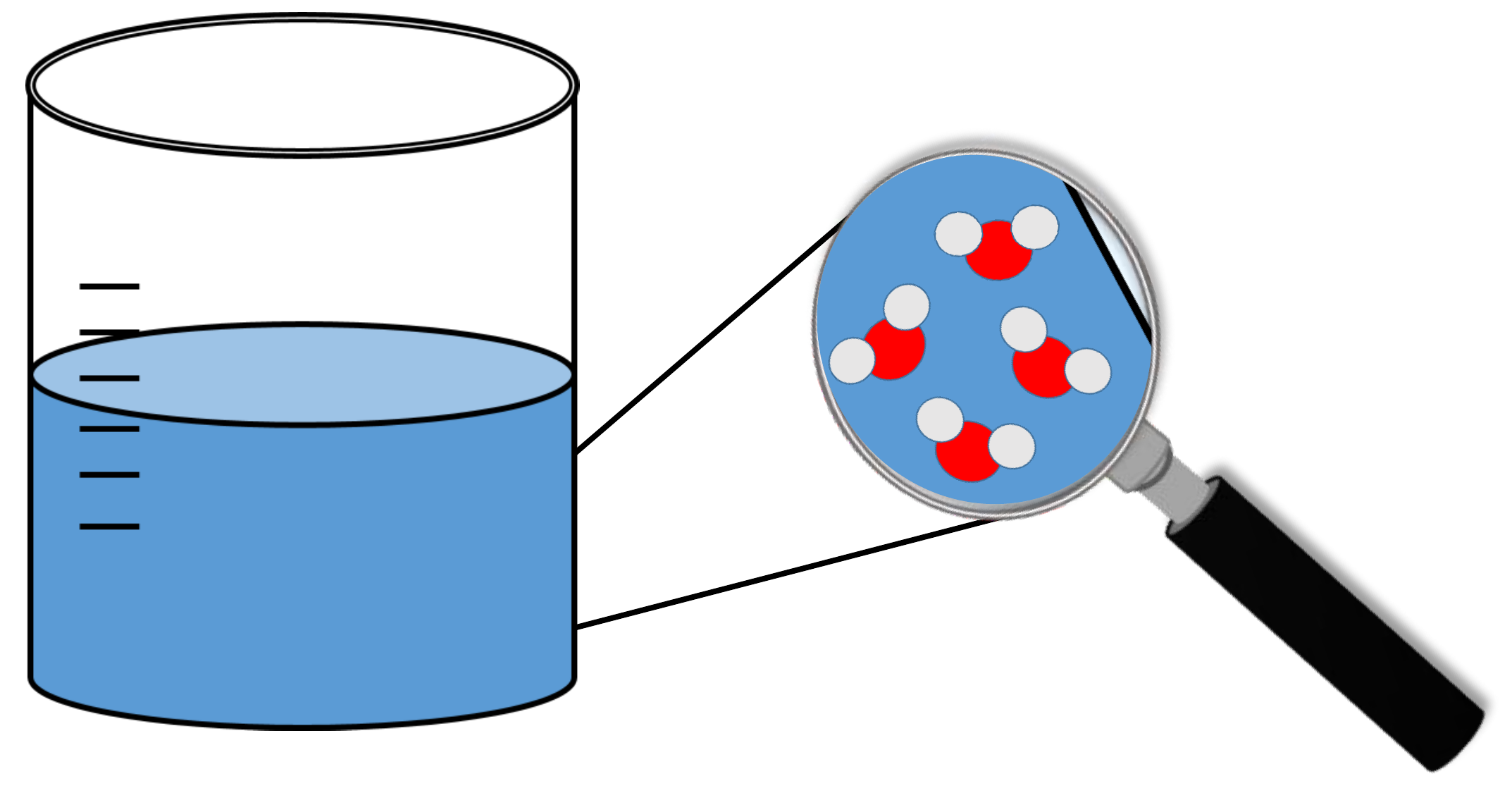}
 \caption[A Macro- and Microscopic View of Water.]{A macro- and microscopic view of water. The fact that water can be heated tells us it has an atomic structure.
}
 \label{macrowater}\end{figure}

Imagine you have a beaker of water. Macroscopically we can measure the temperature of the water, study its heat exchange with the container it rests in, and, with the laws of thermodynamics, study how the energy and entropy of the system change. Of course, we know that if we were to use a powerful enough microscope we could study the thermodynamic properties of water from a (quantum) statistical point of view. Thermodynamics -- from which macroscopic properties of a system are obtained -- is therefore a phenomenological placeholder for a more fundamental, underlying microscopic description.  In the case of water, moreover, we could have figured out that an atomic structure of water exists, even without probing those scales. This is due to, as noted by Boltzmann, the fact that we can heat water, and therefore there is an inherent internal mechanism allowing us to store the associated energy into the water's microscopic degrees of freedom. Heat and temperature are simply macroscopic measures of this underlying microscopics. More generally, because matter can get hot, we know there exists a microscopic description of matter.

Spacetime, in many ways, behaves like the thermodynamic limit of water, where it is possible to associate a temperature and entropy to local patches of an arbitrary spacetime. This realization comes from observing that black holes, the de Sitter universe and Rindler frames -- \emph{spacetime} solutions to Einstein's field equations -- carry with them a set of thermodynamic principles via semi-classical gravity. The leap, then, is to assume \emph{any} spacetime has a set of well-defined thermodynamics -- spacetime can be ``hot". By the aforementioned Boltzmann's principle, this suggests spacetime should have an atomic structure, for which the thermodynamic entropy is counting these ``atoms of spacetime". If, moreover, we treat spacetime like a fluid, the gravitational field equations acquire the interpretation as an equation of state; the conventional geometric language may be recast into some thermodynamic relation. This viewpoint, known as \emph{thermodynamical gravity} or \emph{spacetime thermodynamics}, first taken seriously by Ted Jacobson in 1995 \cite{Jacobson:1995ab}, tells us classical gravity is an emergent phenomena, arising from the thermodynamics of some more fundamental, microscopic theory of spacetime. 

\begin{figure}[t]
\centering
 \includegraphics[width=9.3cm]{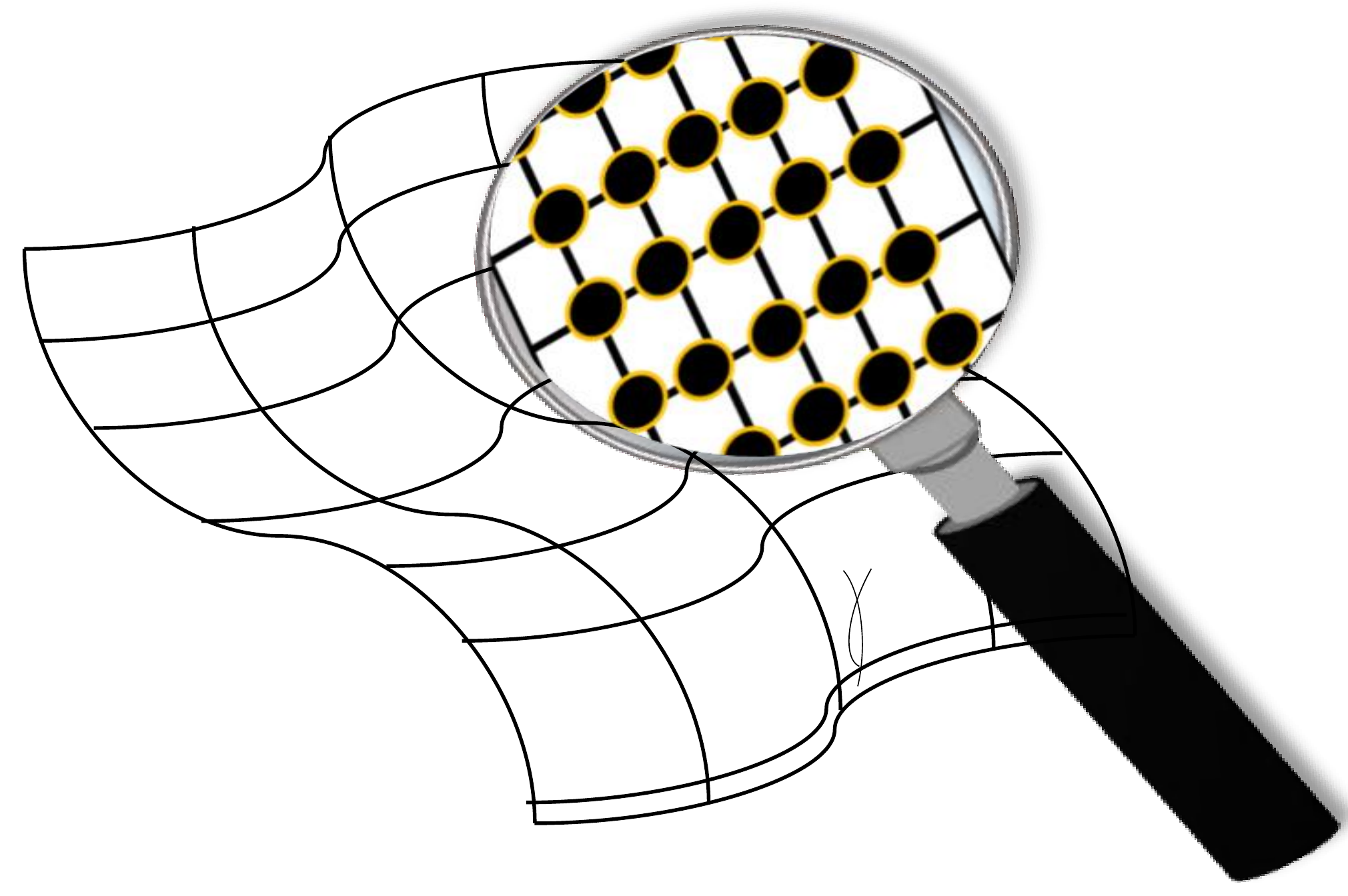}
 \caption[A Heuristic Depiction of the Microscopic Structure of Spacetime.]{A heuristic depiction of the microscopic structure of spacetime. Classical gravity, \emph{i.e.,} the spacetime continuum is viewed as the hydrodynamic limit of some more fundamental microscopic spacetime.
}
 \label{macrowater}\end{figure}

We provide a more detailed review of Jacobson's construction in Appendix \ref{app:thermofund}, however, let's highlight the essence of his argument here. The idea is to pick an arbitrary point $p$ in an arbitrary spacetime $g_{\mu\nu}$, and restrict to a sufficiently small region such that a spacelike foliation can be defined about $p$ with respect to some time coordinate $t$. The point $p$ will be contained in a codimension-2 spacelike patch for which a null congruence (called a ``lightsheet" $\mathcal{H}$) generated by a tangent vector $k^{\mu}$ will emanate from. The lightsheet will have a cross-sectional area $A$ defined as the integral of the expansion of the null congruence. As we follow the lightsheet forward in time $t$, the area is subject to change if matter, characterized by an energy-momentum tensor $T_{ab}$, enters or leaves the lightsheet.  The lightsheet, moreover, serves as a \emph{local} Rindler horizon for an appropriate set of accelerating observers with a constant and uniform acceleration $a$. 

Everything we have described thus far is done so using only geometric reasoning. Now we can rephrase this set-up using thermodynamic language. Specifically, motivated by Unruh, we assume our local Rindler observers will detect a thermal bath with an Unruh-Davies temperature proportional to their acceleration
\beq T=\frac{\hbar a}{2\pi}\;.\eeq
Since the acceleration is uniform and constant, we have that our local Rindler horizon is described by a system in thermal equilibrium with its surroundings. Matter entering or leaving the system, as measured with respect to the locally accelerating observers, is interpreted as heat\footnote{Why heat and not some other energy flux? This goes back to a standard interpretation of heat from ordinary (matter) thermodynamics: heat measures the flow of energy into macroscopic unobservable degrees of freedom. Since the Rindler observers have access only to the exterior of the local horizon, the energy flux is being carried into unobservable degrees of freedom, and therefore attains the interpretation of heat $Q$.} $Q$. 

If our local horizon is to be treated as a thermal system, we expect it to have an associated thermodynamic entropy. The only meaningful geometric quantity at hand is the area of the lightsheet. We are therefore led to make a critical assumption: \emph{the entropy $S$ of the lightsheet is proportional to the area $A$ with some universal constant $\eta$}. Then, the entropy change is proportional to change in area:
\beq \Delta S=\eta \Delta A\;.\eeq
This assumption states that local horizons, like their global black hole horizon counterparts, exhibit holography. 

Our geometric construction has now been reinterpreted in thermodynamic terms. What remains is how the heat $Q$ relates to the entropy change $\Delta S$. This leads to a second assumption: \emph{the entropy change $\Delta S$ is associated with the flow of heat across the lightsheet, which, when in thermal equilibrium, is given by the Clausius relation:}
\beq Q=T\Delta S\;.\label{clasrel}\eeq

Putting everything together, and using Raychaudhuri's equation, we find that the Clausius relation (\ref{clasrel}) is geometrically equivalent to Einstein's field equations being held about the point $p$:
\beq Q=T\Delta S\Rightarrow G_{\mu\nu}(p)+\Lambda g_{\mu\nu}(p)=\frac{2\pi}{\hbar \eta}T_{\mu\nu}(p)=8\pi GT_{\mu\nu}(p)\;,\eeq
where the cosmological constant $\Lambda$ arises as an integration constant, and $\eta=1/4G\hbar$ is required for consistency with the Bekenstein-Hawking formula. Since the point $p$ is completely arbitrary, the construction is valid at any (non-singular) point in the spacetime, and so we have Einstein's equations holding about every point in the spacetime. In this way, holographic thermodynamics applied to local horizons gives rise to Einstein's field equations, interpreted now as an equation of state. 

Since Jacobson's original derivation \cite{Jacobson:1995ab}, there has been much work in spacetime thermodynamics. Notably, this includes studying non-equilibrium effects (\emph{e.g.} \cite{Eling:2006aw}), where the gravitational entropy is corrected by an $f(R)$ term such that field equations arise from a detailed balance equation; deriving higher curvature equations (\emph{e.g.} \cite{Padmanabhan:2009ry,Parikh:2009qs,Brustein:2009hy,Guedens:2011dy,Parikh:2017aas}), where the entropy-area relation is replaced by Wald's entropy functional, and surfaces other than local Rindler horizons \cite{Guedens:2011dy,Parikh:2017aas}. 

Like models of induced gravity, spacetime thermodynamics is not without its criticisms. First and foremost are the input assumptions, that the change in entropy is described by the Clausius relation, and that this entropy variation is proportional to the area change of the local holographic screen. The former of these assumptions is largely acceptable for a thermal system in near equilibrium. There is a subtley, however: the Clausius relation is in fact given by $Q\leq T\Delta S$, where $\Delta S$ here includes reversible and irreversible changes to the entropy. Equality occurs when there is no irreversible contribution to the entropy change. Therefore, a hidden assumption in spacetime thermodynamics is that the $\Delta S$ due to a heat flux is really a reversible entropy change. To our knowledge this was treated properly for the first time in \cite{Parikh:2017aas} and is reviewed in detail in Chapter \ref{sec:gravfromthermo}. 

The second assumption, that $\Delta S\propto\eta\Delta A$, is well motivated by black hole thermodynamics, such that consistency requires $\eta=\frac{1}{4G\hbar}$. This assumption has recently been called into question, however, where it has been shown on rather general grounds the constant of proportionality must satisfy $\eta\leq\frac{1}{8G\hbar}\neq\frac{1}{4G\hbar}$ \cite{Carroll16-1}. It would seem that spacetime thermodynamics is then inconsistent. There are potential loopholes and alternatives to this problem, however, including changing the form of the entropy $S$ \cite{Carroll16-1}, or modifying Jacobson's argument by considering compact local horizons \cite{Jacobson16-1,Parikh:2017aas,Svesko:2018qim}.

Whether the original formulation of spacetime thermodynamics carries on remains to be seen. Its core philosophy, however, continues to heavily  influence the field, and has branched off into other versions of emergent gravity. Like Sakharov's proposal, the overarching lesson of Jacobson's derivation is that classical gravity is not fundamental, but instead arises as a collective phenomena, akin to the hydrodynamic limit of water.


\subsection{Entropic Gravity}
\noindent

A related cousin to Jacobson's version of spacetime thermodynamics is Erik Verlinde's \emph{entropic gravity} \cite{Verlinde:2010hp}, arriving nearly 15 years later.  In this proposal the gravitational force is interpreted as an \emph{entropic force}: an effective macroscopic force describing the statistical tendency for entropy to increase in a system composed of several degrees of freedom\footnote{There are many known examples of entropic forces in bio- and polymer physics.}. Entropic forces are not fundamental in the particle physics sense in that there is no mediator boson associated with an entropic force, and, moreover, is independent of the microscopic details of the system. As such, the gravitational force, as understood by Newton or Einstein, is the result of a collective phenomenon, and thus emergent. 

Verlinde's argument, similar to Jacobson's thermodynamic derivation, relies on the holographic principle. Specifically, the description of a volume of space is encoded in $N$ bits of information living on the boundary of this space, where the total number of bits is proportional to the area $A$ of this holographic screen:
\beq N=\frac{Ac^{3}}{G\hbar}\;.\label{Nbits}\eeq
Here we have introduced a suggestive set of physical constants for dimensional purposes. We then assume the energy $E$ of the boundary system is distributed evenly among the $N$ bits, such that the average energy per bit is given by the equipartition theorem $E=\frac{1}{2}Nk_{B}T$, where $T$ is the temperature associated with the $N$ bits. We may interpret this energy as the rest energy of a particle with an effective mass $M$, such that $T=\frac{2Mc^{2}}{Nk_{B}}$. 

Now imagine placing a particle of mass $m$ a Compton wavelength away from the screen, $\Delta x=\frac{\hbar}{mc}$. The particle will experience an entropic force $F$ because there is a tendency for the entropy of the $N$ bits living on the boundary to increase. The entropy increase occurs because, just as with Bekenstein's original thought experiment, when a particle is one Compton wavelength from the horizon, it is considered to be a part of the screen, increasing the screen's entropy by an amount of a single bit, $\Delta S=2\pi k_{B}$, in order to satisfy the second law of thermodynamics. The entropic force the particle experiences is given by 
\beq F=T\frac{\Delta S}{\Delta x}\;.\label{entforce1}\eeq
Finally, asserting that the boundary is spherical such that $A=4\pi R^{2}$, we find combining (\ref{Nbits}) with (\ref{entforce1}) yields
\beq F=T\frac{\Delta S}{\Delta x}=\frac{GMm}{R^{2}}\;.\label{entforce2}\eeq
For consistency, we interpret $G$ as Newton's constant, and we find we have derived Newton's law of gravitation using the thermodynamics of holographic screens, from which gravity is interpreted as an entropic force. From here Verlinde goes on to show the particle will experience an acceleration $a$ proportional to the temperature $T$ -- just like local Rindler observers -- that is equal to an entropy gradient characterized by a Newtonian potential, $a=-\nabla\Phi$. 

Entropic gravity rests on four assumptions: (i) space itself has at least a single emergent holographic direction\footnote{That is to say, the holographic screens storing information act like stretched horizons of a black hole, where on one side space is defined, and on the other space has not yet emerged.}; (ii) there exists a change in entropy in the emergent dimension as a particle is lowered toward the screen; (iii) the information is encoded in $N$ bits living on the screen, where the maximum number of bits is proportional to the area of the screen, and (iv) the energy of the system is divided evenly among each of the $N$ bits. These assumptions, moreover, are all one needs to derive Einstein's equations using local thermodynamic principles \cite{Verlinde:2010hp}. Unlike Jacobson's argument, however, one need not use Raychaudhuri's equation of expanding null congruences and the local holographic screens are time-like. 

While Jacobson's and Verlinde's derivation of Einstein's equations have a thermodynamic origin, Verlinde's model offers potentially observable consequences. In particular, in 2011 it was argued that late time cosmic acceleration -- often described using dark energy -- can be interpreted as a gravitational-entropic force \cite{Easson:2010av}. Moreover, Verlinde argued that entropic gravity contains an additional ``dark" gravitational force which can account for the profiles of particular galactic rotation curves, doing away with the need for dark matter \cite{Verlinde:2016toy}. 

Due to what entropic gravity offers, both fundamentally and observationally, Verlinde's theory has endured much criticism, from theoretical and experimental viewpoints alike. For example, \cite{Visser:2011jp} demonstrated that since Newton's gravitational force is conservative, heavy constraints are placed on the form of the entropy and temperature functions. Moreover, while the gravitational fields for a large set of galactic rotation curves are consistent with entropic gravity \cite{Brouwer:2016dvq}, Verlinde's proposal is inconsistent with the rotation curves of dwarf galaxies \cite{Pardo:2017jun}. Despite its controversy, entropic gravity remains the best model of emergent gravity that can be tested experimentally, and for that reason should be taken seriously.


\subsection{Spacetime Entanglement}
\noindent

The most recent incarnation of emergent gravity comes from a current popular area of interest commonly referred to as \emph{spacetime entanglement}. Due to the rapid development of the area, and how it has become an interdisciplinary study of quantum gravity, the subject is a genuine paradigm shift in scientific thinking. In many ways, spacetime entanglement is the culmination of several ideas starting with black hole physics, and includes many of the themes of emergent gravity reviewed above. The elementary statement of spacetime entanglement is that classical aspects of spacetime, including connectivity, are all encoded in entangling degrees of freedom of some underlying theory of quantum gravity. Classical gravitational dynamics, moreover, arises from basic relations non-gravitational microscopic degrees of freedom obey. Since the viewpoint makes use of quantum entanglement, many of the statements made have an information theoretic/computational interpretation; colloquially, spacetime entanglement exemplifies the aphroism ``it from qubit".  

Here we provide only a cursory review of the history and philosophy of spacetime entanglement, as some of the details are the subject of this thesis.  The viewpoint is most sharply defined for systems which exhibit gauge/gravity duality, specifically AdS/CFT duality, though it is believed spacetime entanglement is thought to apply more generally \cite{Bianchi:2012ev,Jacobson16-1}. This is in part because the entropy associated with a horizon can be interpreted as a type of entanglement entropy. Indeed, the entropy of a black hole not necessarily confined to a maximally symmetric background behaves as the leading UV divergent contribution to the entanglement entropy due to field fluctuations across either side of the horizon (\emph{e.g.,} \cite{Callan:1994py,Jacobson:1994iw,Frolov:1996aj,Frolov:1996qh,Frolov:1997xd}). The connection between entanglement and horizon entropy is deepened when one realizes that the entanglement entropy of $d+1$-dimensional QFTs generically satisfies an area law \cite{Bombelli:1986rw,Srednicki:1993im}. Therefore, the expectation is that 
\beq S_{\text{EE}}\approx S_{\text{BH}}\;,\label{EEBHrel}\eeq
where $S_{\text{EE}}$ is the entanglement entropy with respect to field degrees of freedom divided between (at least) two subregions, and $S_{\text{BH}}$ is the Bekenstein-Hawking area formula, which computes the entropy associated with Killing horizons in Einstein gravity. 

It is reassuring that the expectation (\ref{EEBHrel}) holds for explicit microscopic models. Such is the case when we consider CFTs in Minkowski space dual to a gravity theory in AdS in one dimension higher. In particular, in its most concise form, the entanglement entropy of a CFT in vacuum reduced to a ball, upon invoking AdS/CFT, is equal to the Bekenstein-Hawking entropy of  a massless Schwarzschild-AdS black hole with a hyperbolically sliced horizon \cite{Blanco:2013joa}:
\beq S^{\text{CFT}}_{\text{EE}}=S_{\text{BH}}^{(M=0)}\;.\label{SEESBHball}\eeq
 This derivation will be explored in more detail in Appendix \ref{app:entfund}. 

\begin{figure}[t]
\centering
 \includegraphics[width=9.3cm]{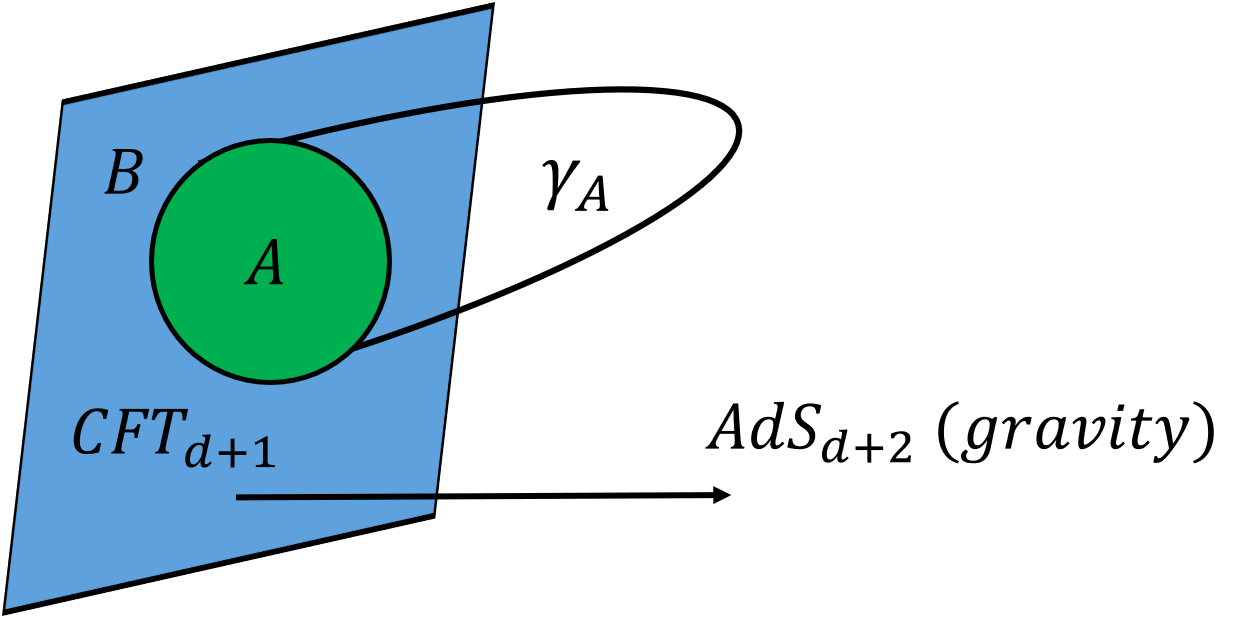}
 \caption[An Illustration of Holographic Entanglement Entropy.]{An illustration of holographic entanglement entropy. The entanglement entropy of a $d+1$-dimensional CFT reduced to a region $A$ is equal to the area of minimal surface $\gamma_{A}$ extending into $d+2$-dimensional AdS, whose boundary is homologous to $\partial A$.
}
 \label{adscftRTex}
\end{figure}

Therefore, entanglement entropy of a holographic CFT is deeply related to spacetime geometry. In fact, the relation (\ref{SEESBHball}) is a special case of the Ryu-Takayanagi proposal \cite{Ryu06-1,Ryu06-2}, which states that the entanglement entropy of $d+1$- holographic CFTs in a boundary region $A$ is equal to the area $\mathcal{A}$ of a $d$-dimensional minimal surface $\gamma_{A}$ protruding in $d+2$-AdS, where the edge of $\mathcal{A}$ is equal to the boundary of $A$ (\ref{RT})
\beq S^{EE}_{A}=\frac{\mathcal{A}(\gamma_{A})}{4G^{(d+2)}}\;.\eeq
When the minimal surface is that of a horizon of a black hole, and the boundary region $A$ is a ball, the Ryu-Takayanagi formula reduces to (\ref{SEESBHball}).

Now we see that any statements about CFT entanglement translate into statements about spacetime geometry, through (\ref{RT}). This observation was used to derive Einstein's equations from entanglement considerations. Loosely, the argument is as follows. The perturbation to any state $\rho_{A}$ of a generic quantum subsystem $A$  will obey the \emph{first law of entanglement entropy}
\beq \delta S_{A}=\delta\langle H_{A}\rangle\;,\label{firstlawEErev}\eeq
where $H_{A}$ is the modular Hamiltonian defined by expressing $\rho_{A}=e^{-H_{A}}/\text{tr}e^{-H_{A}}$. In the event $A$ is a ball shaped region of radius $R$ and $\rho_{A}$ describes the vacuum state of a CFT in Minkowski space reduced to the ball, the modular Hamiltonian can be explicitly written down as \cite{Blanco:2013joa}
\beq H_{A}=2\pi\int_{A}d^{d-1}x\frac{R^{2}-|\vec{x}-\vec{x}_{0}|^{2}}{2R}T_{tt}^{\text{CFT}}\;.\eeq
The first law of entanglement then becomes
\beq \delta S_{A}=2\pi\int_{A}d^{d-1}x\frac{R^{2}-|\vec{x}-\vec{x}_{0}|^{2}}{2R}\delta \langle T_{tt}^{\text{CFT}}\rangle\;.\eeq

We now assume our CFT has a holographic dual and employ the Ryu-Takayangi prescription (\ref{RT}), replacing the variation of the entanglement entropy with a variation of the geometric entropy, for which vacuum state variations are interpreted as linear perturbations to pure AdS. The variation of the CFT modular Hamiltonian is understood to be a variation of the gravitational energy-momentum tensor associated with linearized metric perturbations. In a fashion similar to the derivation of Einstein's equations from the Clausius relation, the first law of (holographic) entanglement is equivalent to the geometric constraint that the linearized Einstein's equations hold locally in a perturbed asymptotically AdS spacetime \cite{Lashkari13-1,Faulkner13-2}:
\beq \delta S_{A}=\delta\langle H_{A}\rangle \Rightarrow G_{\mu\nu}+\Lambda g_{\mu\nu}=8\pi G T_{\mu\nu} \;(\text{linearized})\;.\eeq

The arguments mentioned above apply equally well to higher curvature theories of gravity, where the area of the spherical entangling surface is replaced by a Wald entropy functional evaluated on the spherical entangling surface (which is a Killing horizon). Recently, moreover, this derivation was generalized to derive non-linear gravitational equations of motion, where the first law of entanglement is modified\footnote{There is another way of generalizing the first law of entanglement. Motivated by extended black hole thermodynamics, where one introduces a dynamical cosmological constant, the first law may be extended so as to include not only perturbations to the CFT state, but also the CFT itself by including variations of the central charge \cite{Kastor:2014dra}.} so as to include the effects of excited CFT states \cite{Faulkner:2017tkh,Haehl:2017sot}. All in all, in the context of AdS/CFT, classical gravity emerges from CFT entanglement living on the boundary of the bulk spacetime; gravitational dynamics is governed by entangled CFT degrees of freedom.

As eluded to before, classical spacetime and its dynamics is thought to be obtained from entanglement on more general grounds. This was partially realized by Jacobson in 2015 via the \emph{entanglement equilibrium} conjecture \cite{Jacobson16-1}: in any theory of quantum gravity the entanglement entropy of ball regions of fixed volume is maximal in vacuum, formally given by, 
\beq \delta_{g,\rho}S^{B}_{\text{EE}}=\delta_{g,\rho}S_{\text{BH}}+\delta_{g,\rho}S_{\text{mat}}\;.\label{enteqheur}\eeq
Here $S^{B}_{\text{EE}}$ is the entanglement entropy of a quantum state reduced to a ball $B$, where the causal diamond $D(B)$ is the union of the past and future domains of dependence of $B$; $\delta_{g,\rho}$ is symbolic for allowing both the background geometry $g$ and state $\rho$ change; $S_{\text{BH}}$ is the Bekenstein-Hawking gravitational entropy, representing the UV (quantum gravitational) entanglement entropy, and $S_{\text{mat}}$ is the matter entanglement entropy representing the correlations of IR (quantum field theoretic) degrees of freedom. Based on the right hand side of (\ref{enteqheur}), it is clear one assumes that the Hilbert space of states reduced to the ball $\mathcal{H}_{B}$ may be factorized into UV and IR contributions, $\mathcal{H}_{B}=\mathcal{H}_{\text{UV}}\otimes\mathcal{H}_{\text{IR}}$. 

 The (non-linear) Einstein equations arise by showing an off-shell geometric identity known as the `first law of causal diamond mechanics' -- a statement analogous  to the equilibrium version of the first law of black hole mechanics but applied to causal diamonds in a perturbed maximally symmetric background -- is equivalent to Einstein's equations holding locally, upon an application of the entanglement equilibrium condition (\ref{enteqheur}). The first law of entanglement also makes an appearance here, specifically applied to the state variation of the matter entanglement entropy, $\delta_{g,\rho}S_{\text{mat}}=\delta_{\rho} S_{\text{mat}}=\langle H_{B}\rangle$.  Importantly, we emphasize that Jacobson's derivation does not rely on AdS/CFT as the background need not be AdS. The linearized equations of motion for higher curvature theories of gravity were computed using entanglement equilibrium in \cite{Bueno16-1}. Additional details of entanglement equilibrium are given in Appendix \ref{app:cdmechanicsandEE}. 

The spacetime thermodynamics and spacetime entanglement programs are deeply related. This is because the first law of entanglement can be interpreted as the first law of thermodynamics for equilibrium systems. Therefore, when the first law of entanglement is applied to regions of spacetime, it naturally leads to statements about equilibrium thermodynamics applied to spacetime. This is made particular clear via entanglement equilibrium. In fact, an assumption baked into the entanglement equilibrium condition (\ref{enteqheur}) is that the causal diamond is in thermodynamic equilibrium with its surroundings. This is because the condition $\delta S_{\text{EE}}^{B}=0$ is equivalent to demanding that for a fixed energy, a small region  should be well described by a thermal Gibbs state, such that the causal diamond represents a canonical ensemble with fixed degrees of freedom and volume. The thermodynamics of causal diamonds in maximally symmetric backgrounds was further analyzed in \cite{Jacobson:2018ahi}, and used to derive non-linear gravitational field equations via the Clausius relation in \cite{Svesko:2018qim}. Due to the geometric similarities of causal diamonds and stretched future lightcones -- a timelike stretched horizon of the future of a lightcone -- \cite{Svesko:2018qim} also demonstrated the condition entanglement equilibrium holds for geometric regions other than causal diamonds.

Classicality of spacetime emerges from entanglement in other ways too. For example, one observation is spacetime connectivity can be interpreted as entangled regions of spacetime (namely, a pair of maximally entangled black holes) connected via an Einstein-Rosen bridge. This proposal has been aptly named ``ER = EPR" \cite{Maldacena:2013xja}. One of its claimed successes is that it resolves the Almheiri, Marolf, Polchinski, Sully (AMPS) firewall paradox, though this is still up for debate. The field of \emph{tensor networks} -- a representation of many body quantum systems based on their entanglement structure -- has also been applied to spacetime entanglement, where spacetime is literally built up qubit by qubit (see, \emph{e.g.}, \cite{Hayden:2016cfa}). Using tensor networks to model bulk/boundary duality, one is naturally led to reinterpret aspects of bulk locality by rewriting the usual dual CFT statements in the language of \emph{quantum error correction} \cite{Almheiri:2014lwa,Pastawski:2015qua}. These discretized methods also lead to reinterpretations of holographic entanglement, where, for example, the Ryu-Takayanagi relation arises from an error correcting code \cite{Harlow:2016vwg}, and that it may be recast in terms of \emph{bit threads}\footnote{Bit threads are divergenceless vector fields with Planck thickness, where the entanglement entropy of a boundary region is given by maximum number of bit threads that emanate from it, rather than the minimal surface whose boundary is homologous to the boundary region.} \cite{Freedman:2016zud}, doing away with minimal surfaces altogether.

So far it is not entirely clear how the framework of spacetime entanglement will shape up in the end. The field is teeming with new ideas, some of which might not lead to anything fruitful. What is clear, however, is that spacetime entanglement has dramatically altered our way of approaching questions about quantum gravity. Perhaps most of all, spacetime entanglement lends further evidence that classical gravity, \emph{i.e.}, spacetime geometry, is not fundamental: it is emergent.



\newpage

\section{THERMODYNAMIC ORIGIN OF THE NULL ENERGY CONDITION}  \label{sec:NECfromthermo}


 The null energy condition (NEC) plays a critical role in classical general relativity. It is used in proving a host of gravitational theorems, from the area theorem that states that classical black holes cannot shrink \cite{Bardeen73-1}, to singularity theorems that guarantee the existence of the Big Bang \cite{Hawking70-1}. The NEC is also invoked in excluding bouncing cosmologies and exotic spacetimes containing traversable wormholes and time machines, which might otherwise be exact solutions of Einstein's equations \cite{molinaparis99-1,Parikh15-2,Hawking92-1,Farhi87-1,Morris88-2}. And in asymptotically AdS spaces, the validity of the NEC is equivalent to a c-theorem in the holographic dual theory \cite{Freedman99-1}. The NEC is usually expressed as the condition
\be
T_{\mu\nu} v^\mu v^\nu \geq 0 \; ,	\label{TNEC}
\ee
where $v^\mu$ is any light-like vector. Here $T_{\mu\nu}$ is the energy-momentum tensor of matter, suggesting that the NEC should be a property of matter. However, our best framework for describing matter -- quantum field theory -- does not appear to have a consistency requirement of the form of (\ref{TNEC}), even as a classical limit. Moreover, several explicit examples of effective theories that violate (\ref{TNEC}) but that are nevertheless not in manifest conflict with the principles of quantum field theory are now known. Thus the origin of a vitally important aspect of general relativity has been mysterious. With no apparent fundamental principle from which the NEC flows, the validity of the NEC has been called into question \cite{Barcelo02-1,Rubakov14-1}.

Motivated by this failure to derive the NEC in some classical limit of quantum field theory, it has been proposed that the NEC should be regarded as a property not purely of matter but of a combined theory of matter and gravity \cite{Parikh14-1}. In such a theory, Einstein's equations imply that the NEC can be reformulated in a quite different, though equivalent, form as
\be
R_{\mu\nu} v^\mu v^\nu \geq 0 \; ,	\label{RNEC}
\ee
where $R_{\mu\nu}$ is the Ricci tensor. This is now a constraint on spacetime geometry, rather than on energy densities; indeed, it is this geometric form of the null energy condition, known as the Ricci or null convergence condition, that is ultimately invoked in gravitational theorems. Despite its importance, the NEC is invoked \emph{ad hoc}, lacking a clear origin\footnote{Recently it has been shown that precisely this condition can be derived from string theory \cite{Parikh14-1}, which of course is a theory of both matter and gravity. For a closed bosonic string propagating in an arbitrary graviton-dilaton background, the Virasoro constraints of the effective action lead precisely to (\ref{RNEC}) in Einstein frame, including even the contractions with null vectors. This is a very satisfying derivation of the null energy condition for a number of reasons: It is another example of the beautiful interplay between the worldsheet and spacetime, the Virasoro constraints are none other than Einstein's equations in two dimensions, and there is a physical principle -- worldsheet diffeomorphism invariance -- that is associated with the null energy condition.}.

Our goal here is to derive the NEC using the principles of emergent gravity. Our premise is that gravity arises from the coarse-graining of some underlying microscopic theory. 
As we will see, the derivation has its appeal because it relies on a universal theory, namely thermodynamics. In fact, a relation between thermodynamics and the null energy condition is already present in black hole physics. Recall that the NEC is used in deriving the second law of thermodynamics for black holes \cite{Bardeen73-1}. The logic runs as follows:
\be
\begin{split}
T_{\mu\nu} v^\mu v^\nu \geq 0& \Rightarrow R_{\mu\nu} v^\mu v^\nu \geq 0 \Rightarrow \dot{
\theta} \leq 0 \Rightarrow \theta \geq 0\Rightarrow \dot{A} \geq 0 \Rightarrow \dot{S} \geq 0 \; .
\end{split}
\ee
Here $\theta$ is the expansion of a pencil of null generators of a black hole event horizon and the dot stands for a derivative with respect to an affine parameter, which can be thought of as time. The first arrow follows from Einstein's equations, the second from the Raychaudhuri equation, the third from avoidance of horizon caustics, the fourth from the definition of $\theta$, and the last from the definition of Bekenstein-Hawking entropy. Ideally, we would like to able to reverse all these arrows so that the NEC flows from the second law of thermodynamics, rather than the other way around \cite{Chatterjee:2012zh}. However, although the first and last arrows can readily be reversed, provided we assume Einstein gravity and the validity of the gravitational equations, the remaining arrows do not appear to be reversible. In particular, a serious problem with reversing the arrows is that the second law is a global statement, whereas the NEC is a local condition. 

 However, recall Jacobson's now famous observation \cite{Jacobson:1995ab}, where he obtained Einstein's equations -- which are also local -- from the Clausius relation (essentially the first law of horizon thermodynamics) applied to local Rindler horizons. Thus a global law was ``gauged," which was a pre-requisite for obtaining the local gravitational equations of motion. In the same vein, we will show that the null energy condition too, in the form of the Ricci or null convergence condition, (\ref{RNEC}), comes out of thermodynamics applied to a local holographic screen. In a nutshell, just as Jacobson regarded the {\em first} law as an input and obtained Einstein's equations as an output (reversing the laws of black hole mechanics, as it were), we shall regard the {\em second} law as an input and obtain the null energy condition as an output.

Note that we will consider only the classical null energy condition. Much effort in the literature \cite{Wall:2009wi,Lashkari:2014kda,Kontou:2015yha,Bousso:2015wca,Faulkner:2016mzt,Hartman:2016lgu} has been directed at proving a quantum null energy condition, $\langle T_{\mu \nu} \rangle k^\mu k^\nu \geq 0$, or generalizing the concept to some kind of averaged null energy condition. Indeed, the standard null energy condition is known to be violated even by Casimir energy. So why focus on the classical NEC? First, the properties of the classical stress tensor are of independent interest. Typically, whenever exotic matter is proposed  in the literature e.g. phantom fields, galileons, ghost condensates, etc., the gravitational consequences are worked out by coupling Einstein gravity to the classical stress tensor of such matter. So it is important to prove the generic properties of this tensor. Second, in attempts to prove the quantum null energy condition, the validity of the classical NEC is often assumed -- yet this needs to be proven. Third, it is not obvious that the expectation value of the quantum stress tensor, as computed, has any gravitational consequences. A quantum null energy condition $\langle T_{\mu \nu} \rangle k^\mu k^\nu \geq 0$ would certainly be meaningful if there were a semi-classical Einstein equation of the form $G_{\mu \nu} = 8 \pi G \langle T_{\mu \nu} \rangle$. However, such an equation is not known to have any rigorous derivation. By contrast, whatever be the ultimate theory of quantum matter coupled to quantum gravity, it surely admits a well-defined $\hbar = 0$ limit of classical gravity coupled to classical matter, which is the situation considered here.


\subsection{From the Second Law to the NEC}

Before entering into the details, let us summarize the logic of the derivation. First we will quote a statistical-mechanical result about the non-positivity of the second time-derivative of entropy. This is a very general result which holds for virtually all near-equilibrium thermodynamic systems. Next we will propose a prescription for associating thermodynamic systems to patches of null congruences in spacetime. We will then show that, in the vicinity of any point in spacetime, null congruences corresponding to near-equilibrium thermodynamic systems can always be found. By the quoted result, these then necessarily have non-positive second time-derivative of entropy. Finally, substituting this into the Raychaudhuri equation will imply the Ricci convergence condition, (\ref{RNEC}), which is the geometric form of the null energy condition.

\subsection{Time Derivatives of Entropy}

Consider then a finite thermodynamic system and let $S_{\rm max}$ be its maximum coarse-grained entropy. For systems already at equilibrium, $S = S_{\rm max}$, and $\dot{S}, \ddot{S} = 0$. For systems approaching equilibrium, $S < S_{\rm max}$ and the second law says that $\dot{S} \geq 0$. Now, since the entropy tends to a finite maximum value as it approaches thermal equilibrium, and since $\dot{S}\geq0$, it seems intuitively reasonable that the first time derivative of entropy will be a decreasing function of time: $\ddot{S}\leq0$. This inequality, which will be crucial below, indeed holds for a great many systems of interest. For such systems, the coarse-grained entropy satisfies
\be
S \geq0,\quad\dot{S}\geq0,\quad\ddot{S}\leq0 \; . \label{Sconditions}
\ee
For example, consider a clump of particles, with some initial Gaussian density distribution, $\rho \sim \exp(-r^2/2)$, diffusing outwards with diffusion constant $D$. The diffusion equation implies that $\rho(r, t) = (2\pi(1+2Dt))^{-3/2} \exp \left ( - \frac{r^2}{2(1+2Dt)} \right )$.
It is then easy to check that the entropy, $S = - \int d V \rho \ln \rho$, obeys $\ddot{S} = - \frac{2}{3}\dot{S}^2$ at all times, so that (\ref{Sconditions}) holds.

In fact, this is a very general property. As reviewed below, it can be shown quite generally  that $\ddot{S} \leq 0$ for virtually all near-equilibrium systems approaching internal equilibrium. That is, finite, closed systems at late times inevitably obey (\ref{Sconditions}). By near-equilibrium, we mean systems that are characterized by $(\dot{S}/S)^2 \ll |\ddot{S}/S|$, which follows from $S \sim S_{\rm max}$ in this context. For systems that are not near equilibrium, $\ddot{S}$ can generically have either sign and hence (\ref{Sconditions}) may or may not hold; the diffusing gas is an example of a system in which (\ref{Sconditions}) does hold even though the system is never near equilibrium unless the gas is placed in a finite volume.

Let us now be more precise and show (\ref{Sconditions}) is guaranteed to hold for near-equilibrium systems, following a proof by \cite{Falkovich:2004} showing that typical near-equilibrium thermodynamic systems relaxing to equilibrium must have $\ddot{S} \leq 0$. Consider a phase space density $\rho$ associated with a reduced description of the system (due to coarse-graining). Suppose the system is close to thermodynamic equilibrium. Then the phase space density is near  the value $\rho_{0}$ that maximizes the entropy:

\be
\rho=\rho_{0}+\delta \rho \; .
\ee
Then,
\beq
\begin{split}
S(\rho_{0}+\delta \rho)&=-\int (\rho_{0}+\delta \rho)\ln(\rho_{0}+\delta \rho) \nonumber \\
&\approx S_{\rm max}-\int \left ( \rho_{0}^{-1}\frac{(\delta \rho)^{2}}{2}\right ) \; ,
\end{split}
\eeq
where $S_{\rm max}=-\int  \rho_{0}\ln \rho_{0} $
and we have used the fact that $\delta S|_{\rho_{0}}=0$. Near equilibrium, the time-derivative of
the density fluctuation satisfies a linear Onsager relation:
\be
\delta\dot{\rho}=\hat{L}\delta \rho \; ,
\ee
where the Onsager $\hat{L}$ matrix is taken to be symmetric. As Onsager showed \cite{Onsager:1931}, the symmetry of $\hat{L}$ follows from the principle of microscopic reversibility, so long as the macroscopic thermodynamic state variables are themselves time-invariant; this is the case for all but a few ``exceptional" systems of interest (usually involving magnetic fields). It seems quite likely that the thermodynamics of the microscopic theory of gravity satisfies these time-invariance properties; here we assume that this is the case. ($\hat{L}$ is presumably also invariant under time-translations.) When $\hat{L}$ is symmetric, we can expand $\delta \rho$ into orthonormal eigenfunctions of $\hat{L}$:
\be
\delta \rho=\sum_{k}\sqrt{\rho_{0}}a_{k}\psi_{k} \; ,
\ee
where $\hat{L}\psi_{k}=\lambda_{k}\psi_{k}$. 
Now
\be
\dot{S}=-\int \rho_{0}^{-1}\delta \rho (\hat{L}\delta \rho)  \; .
\ee
Then the second law implies
\be
-\sum_{j,k}\int \left(a_{j}a_{k}\lambda_{k}\psi_{j}\psi_{k}\right) \geq0\Rightarrow\lambda_{k}\leq 0 \; , \label{lambdak}
\ee
for all $k$. That is, the second law indicates that the eigenvalues of the operator $\hat{L}$ are real (and non-positive). 
Now consider the second derivative:
\begin{eqnarray}
\ddot{S} & =&-\int \rho_{0}^{-1}\left[\delta\dot{\rho}(\hat{L}\delta \rho)+\delta \rho (\hat{L}\delta \dot{\rho})\right] \nonumber\\
&=&-\int  \rho_{0}^{-1}\left[\left(\hat{L}\delta \rho \right)^{\! 2}+\delta \rho \left(\hat{L}^{2}\delta \rho \right)\right] \; .
\end{eqnarray}
Inserting the eigenfunction expansion, we find
\be
\ddot{S} 
=-2\sum_{k}a_{k}^{2}\lambda_{k}^{2} \; ,
\ee
so that
\be
\ddot{S} \leq0 \; .
\ee
Note from (\ref{lambdak}) that if $\dot{S} = 0$ then $\ddot{S} = 0$ while if $\dot{S} > 0$ then $\ddot{S} < 0$.


\subsection{Thermodynamics of Spacetime}

Next, let us attempt to connect thermodynamics to local regions of spacetime. The motivation is as follows. The Bekenstein-Hawking entropy formula associates entropy to the area of black hole horizons. The formula is universal, applying to the horizons of all kinds of black holes in any number of dimensions. It even applies to de Sitter horizons. But most strikingly, the formula is also considered to hold (as an entropy density) for acceleration horizons. Since such horizons could be anywhere, this suggests that there might be a local entropy associated with the areas of patches of certain null surfaces. The idea of emergent gravity is to assume that this local entropy is similar to entropy in statistical-mechanical systems. That is, we assume that gravitational entropy arises as the coarse-grained entropy of some microscopic system of Planckian degrees of freedom associated with patches of certain null surfaces. What these degrees of freedom are is unknown and also largely irrelevant. It is not even clear whether these degrees of freedom live in spacetime or, because they have to account for an entropy that scales as an area, in some dual space in one lower dimension. We do know that for stationary horizons (including de Sitter and Rindler horizons), there is also an associated temperature. It therefore seems natural to assume that the underlying microscopic system is in fact a thermodynamic system. These two points are the basis for the idea that gravity might be described locally by some dual thermodynamic system. Despite little being known about the underlying system, the emergent gravity paradigm has met with great success due to Jacobson's remarkable result \cite{Jacobson:1995ab} that Einstein's equations follow from what is essentially the first law of thermodynamics. Here, the only feature we will need to assume is that the underlying system either is already at, or is approaching, internal equilibrium via the second law of thermodynamics. 

Since the second law of thermodynamics is perhaps the most universal law in physics, this is not much of an assumption; we merely need to assume that the system is closed over the time-scales of interest. Moreover, since the idea is that the system is dual to an infinitesimal region of spacetime, the requirement that it be closed over infinitesimal times also seems natural.

Next, we would like to have a prescription for how to choose our null congruences. In Jacobson's paper, the thermodynamic system was taken to be instantaneously at equilibrium, and hence the corresponding null congruence was chosen to be a local Rindler horizon, with vanishing expansion and shear at the point of interest. Here we are interested in the second law, so we allow for non-equilibrium systems with increasing entropy. Correspondingly, we allow our congruences to have positive, or at least non-negative, local expansion. Our prescription then is very simple: we postulate that every non-contracting infinitesimal open patch of the integral curves of every null geodesic congruence is associated with a thermodynamic system obeying the second law; the
restriction to non-contracting patches enforces the second law of thermodynamics, which is the basic premise from which we will derive the null energy condition. Through a given spacetime point $p$ with a given future-directed null vector $v^\mu$ in the tangent space at $p$, there are infinitely many non-contracting geodesic congruences with tangent $v^\mu$ at $p$. We associate thermodynamic systems to {\em all} such infinitesimal patches. A particular class of expanding congruences consists of future light cones of earlier spacetime points. Among these, a special limiting case consists of the integral curves emanating from the future light cone of a point in the infinite past of $p$. Near $p$, the patch of such a stationary congruence is a local planar Rindler horizon, corresponding to an equilibrium system. Thus our prescription covers both equilibrium and non-equilibrium systems; it generalizes Jacobson's local Rindler horizons to patches whose local expansion can be not only zero, but also positive.

With this background, we identify the gravitational entropy of our infinitesimal patch with the coarse-grained entropy of a thermodynamic system. Then
\be
S = \frac{A}{4} \; .
\ee
It is implicit in this formula that classical physics is described by Einstein gravity minimally coupled to matter; for higher-curvature theories of gravity, or for non-minimally coupled gravity \cite{Chatterjee:2012zh}, the Bekenstein-Hawking entropy would have to be replaced by its appropriate generalization, such as the Wald entropy \cite{Wald:1993nt}. Next, we identify the affine parameter of the null congruence with the time parameter in our thermodynamic system. Then
\be
\dot{S} = \frac{A}{4} \theta \; ,
\ee
and
\be
\ddot{S} = \frac{A}{4} \lf \theta^2 + \dot{\theta} \rt \; . \label{ddot}
\ee
Here we are assuming that $\theta$ is roughly constant over the surface; this is valid because the surface is infinitesimal. Notice that the near-equilibrium condition, $(\dot{S}/S)^2 \ll |\ddot{S}/S|$, translates to $\theta^2 \ll |\dot{\theta}|$.

Now because the congruence is null, its generators obey the optical Raychaudhuri equation:
\be
\dot{\theta}= - \frac{1}{2} \theta^2 - \sigma^2 + \omega^2 - R_{\mu\nu} v^\mu v^\nu \; .
\ee
By hypersurface-orthogonality, $\omega^2 = 0$. The shear, $\sigma$, can always be chosen to vanish at a point. Choose an initial surface near or enclosing this point. In this region the shear will be small compared to $\theta$. Moreover, for small enough affine parameter $\lambda$ the shear will remain small compared to $\theta$. Then, for small times, $\sigma^2$ is negligible. 
We therefore drop the $\sigma$ and $\omega$ terms from Raychaudhuri's equation. Then we have
\begin{eqnarray}
R_{\mu \nu} v^\mu v^\nu & = & - ( \dot{\theta} + \theta^2 ) + \frac{1}{2} \theta^2 \nonumber \\
& = & - \frac{\ddot{S}}{S} + \frac{1}{2} \left (\frac{\dot{S}}{S} \right)^{\! \! 2} \label{RandS}
\end{eqnarray}
Now, for systems that are already at equilibrium, $\dot{S}$ and $\ddot{S}$ are both zero. Hence
\be
R_{\mu \nu} v^\mu v^\nu = 0 \; . \label{Rzero}
\ee
Next, consider systems approaching equilibrium. Then $\dot{S} > 0$. For systems that are far from equilibrium, $\ddot{S}$ can have either sign. Therefore, for expanding patches that correspond to far-from-equilibrium thermodynamic systems, the two terms on the right of (\ref{RandS}) could have different signs so that nothing can be inferred about the sign of $R_{\mu \nu} v^\mu v^\nu$ without knowing the precise values of $\dot{S}$ and $\ddot{S}$; no general statement can be made for such systems. However, for patches that correspond to near-equilibrium systems, we are guaranteed that  $\ddot{S} \leq 0$. The existence of such systems would guarantee that $R_{\mu \nu} v^\mu v^\nu \geq 0$. 

To complete the proof, we show existence of such congruences by construction. In the vicinity of the point $p$, $R_{\mu \nu} v^\mu v^\nu$ is a constant, namely $R_{\mu \nu}(p) v^\mu v^\nu$. Call this constant $C$. We will shortly determine the sign of $C$ from thermodynamics. Solving the Raychaudhuri equation for a shear-free congruence, we find
\be
\theta = \sqrt{2C} \tan \left (-\sqrt{\frac{C}{2}} \lambda + b \right ) \; ,
\ee
where $b$ is a constant of integration; different choices of $b$ correspond to different congruences. Choosing $b = 0$, we see that $\theta$ vanishes for $\lambda = 0$. Suppose we consider some open patch for very small $\lambda$ (but not including the point $\lambda = 0$, where the sign of $\theta$ changes). Then
\be
\theta \approx - C \lambda \comma \dot{\theta} \approx -C \; .
\ee
If $\theta = \dot{\theta} = 0$ then $C = 0$; stationary (equilibrium) congruences require (\ref{Rzero}). Otherwise, since $\lambda$ is chosen to be small, we see that $\theta^2 \ll |\dot{\theta}|$. This translates to $(\dot{S}/S)^2 \ll |\ddot{S}/S|$, which means that the system is indeed near equilibrium. We have thus shown, by explicit solution of the Raychaudhuri equation, that congruences corresponding to stationary (equilibrium) or near-equilibrium systems exist everywhere.

But if the system is near equilibrium, then we know from statistical mechanics that $\ddot{S} < 0$. By (\ref{ddot}), this in turn means $\dot{\theta} < 0$, so that $C > 0$, which is to say
\be
R_{\mu \nu} v^\mu v^\nu > 0 \; .
\ee
Therefore, for both equilibrium and non-equilibrium thermodynamic systems, we find $R_{\mu \nu} v^\mu v^\nu \geq 0$. This is precisely the geometric form of the null energy condition, (\ref{RNEC}). Since $v^\mu$ is any arbitrary future-directed null vector, this establishes the null energy condition.



\subsection{ Quantum Corrections to Entropy and the NEC}

Above we showed that the NEC, in the form (\ref{RNEC}), arises from the second law of thermodynamics, applied locally, in the same spirit as \cite{Jacobson:1995ab}. However, our derivation only considered \emph{classical} matter and gravity. The natural next question is to ask whether quantum effects lead to violations of the NEC. Indeed, it is known that the matter form of the NEC is violated when first order quantum effects are taken into account, e.g., by Casimir energy. Nevertheless, it is not clear that this indicates a violation in the Ricci convergence condition  (\ref{RNEC}). To understand this, consider the semi-classical Einstein equations,
\be
G_{\mu\nu}=8\pi G\langle T_{\mu\nu}\rangle\;, \label{semi} 
\ee
which describe the backreaction of quantum fields on a classical background. The effect of the fluctuating quantum fields is captured by the renormalized expectation value of the energy-momentum tensor $\langle T_{\mu\nu}\rangle$ over a particular background. The relevance of  $\langle T_{\mu \nu} \rangle$ to spacetime geometry relies on the validity of an equation of the form of (\ref{semi}), but we are not aware of any rigorous derivation of this equation as the semi-classical limit of a theory of both quantum matter and quantum geometry. Indeed, an equation which treats gravity classically but matter quantum-mechanically appears to be in some tension with the spirit of string theory in which matter and gravity are treated in a unified manner. In principle $\langle T_{\mu\nu}\rangle$ can be derived from an effective action $S_{\rm eff}(g_{\mu\nu})$ describing the quantum matter fields propagating on the background metric $g_{\mu\nu}$. In that case, generally one finds that $\langle T_{\mu\nu}\rangle$ will depend on higher-curvature terms (see, e.g., \cite{Birrell82-1}). The field equations, therefore, will in general include higher-curvature corrections to Einstein's equations, severing the link between the NEC as a constraint on matter  (\ref{TNEC}) and the NEC as a constraint on geometry  (\ref{RNEC}). Thus a violation in  (\ref{TNEC}) does not imply a violation in  (\ref{RNEC}), and vice versa. 

Here we take a different approach. Rather than calculating $\langle T_{\mu \nu} \rangle$, and then trying to determine its gravitational implications, the novel idea here is to directly determine $R_{\mu \nu} v^\mu v^\nu$ in the semi-classical theory. Specifically, we use the known form of the quantum-corrected version of the Bekenstein-Hawking entropy \cite{Kaul00-1} to obtain the Ricci convergence condition. We find that, if we replace the  Bekenstein-Hawking entropy of a horizon with its one-loop generalization and apply the second law of thermodynamics, we again arrive at exactly the Ricci convergence condition (\ref{RNEC}). Quantum corrections, at least of the type that contribute to the entropy, do not appear to alter the condition; if these were the only quantum corrections, then, for example, singularity theorems would continue to hold even in the semi-classical theory.



 Much effort has been put into calculating quantum corrections to the NEC on the matter side \cite{Graham07-1,Kontou:2015yha,Bousso:2015wca}. Fortunately, there is an easier way to address this question. The key point is that the Raychaudhuri equation depends only on the geometry of spacetime and not on the theory in which the geometry arises. In particular, it should hold also for the geometry that arises in an effective theory of gravity that includes one-loop corrections. Furthermore, the Raychaudhuri equation contains the actual geometric object of interest, namely $R_{\mu \nu} v^\mu v^\nu$. It is the positivity of this term that controls the possible existence of singularities, say. By contrast, the gravitational implications of $\langle T_{\mu \nu} \rangle$ rely on the unclear question of how quantum matter couples to gravity. If, for example, the left-hand side of Einstein's equations are modified by the inclusion of geometric counter-terms, then the sign of $\langle T_{\mu \nu} \rangle v^\mu v^\nu$ does not have any obvious bearing on the sign of $R_{\mu \nu} v^\mu v^\nu$.

The sign of $R_{\mu \nu} v^\mu v^\nu$ is determined by the Raychaudhuri equation once we know $\theta, \dot{\theta}$. Our underlying (and non-trivial) assumption is that the semi-classical theory can continue to be described by thermodynamics. Under that assumption, we need to express geometric quantities like $\theta, \dot{\theta}$ in terms of thermodynamic quantities, specifically time derivatives of the coarse-grained entropy. The one-loop quantum-corrected formula for the gravitational entropy is
\be 
S=\frac{A}{4}+c\ln A+\mathcal{O}(1)\;,
\ee
where $c$ is a constant. As before we have suppressed Newton's constant here, so that $A$ is measured in Planck units. Such a logarithmic correction \cite{Kaul00-1} to the Bekenstein-Hawking entropy arises in a great variety of contexts. These include Carlip's derivation using the Virasoro algebra associated with two-dimensional conformal symmetry at the horizon \cite{Carlip:2000nv}, the partition function of the BTZ black hole \cite{Govindarajan:2001ee}, one-loop effects  \cite{Fursaev95-1,Mann:1997hm}, type-A (Euler density) contribution to the trace-anomaly induced effective action \cite{Cai:2009ua,Aros10-1}, along with many others; see, e.g. \cite{Medved:2004eh,Page05-1} for a review. The technical reason for this evident universality of the leading correction to the Bekenstein-Hawking entropy is that all the microscopic derivations ultimately invoke the Cardy formula. 

Note that the positivity of $S$ implies that
\be
c \geq -\frac{A}{4 \ln A} \; . \label{boundonc}
\ee
Typically, $c$ is of order unity. In fact, the majority of calculations agree that
\be
c = -\frac{3}{2} \; , \label{c=-3/2}
\ee
with some other approaches giving a result that differs by a factor of order unity. As $A$ is measured in Planck units, the validity of an approximately classical regime requires that $A \gg 1$. Then we have from (\ref{boundonc}) that $\left(\frac{A}{4} + c \right) > 0$.

The time derivative of the entropy is given by
\be
\dot{S}= \theta \left ( \frac{A\theta}{4} + c \right ) \; .
\ee
Hence $\dot{S} \geq 0 \Rightarrow \theta \geq 0$: increasing entropy corresponds to expanding congruences, unsurprisingly. The second derivative of the entropy is
\be
\ddot{S}=\frac{A}{4}\left[\left(\frac{A}{4} + c\right)\dot{\theta}+\theta^{2}\right]\;.
\ee
Next, as we are regarding the gravitational entropy to be the coarse-grained entropy of some dual thermodynamic system, we invert the geometric quantities $A$, $\theta$, and $\dot{\theta}$ in terms of the thermodynamic quantities $S$, $\dot{S}$, and $\ddot{S}$. We find
\be
A(S) = 4c \, W \! \left ( \frac{e^{S/c}}{4c} \right ) \; ,
\ee
where $W$ is the Lambert W-function, and
\be
\theta = \frac{\dot{S}}{\frac{A(S)}{4} + c} \comma \dot{\theta} = \frac{\ddot{S}}{\frac{A(S)}{4} + c}  -\frac{\dot{S}^2 A(S)}{4} \left ( \frac{1}{\frac{A(S)}{4} + c} \right )^{\! 3} \; .
\ee
We can now again consider the two types of thermal systems. For systems at equilibrium we  have $\dot{S}=\ddot{S}=0$, so that $\theta=\dot{\theta}=0$, leading to $R_{\mu\nu}v^{\mu}v^{\nu}$ via the Raychaudhuri equation. For systems approaching equilibrium we find, using $\dot{S}\geq0$ and $\ddot{S}\leq0$, that
\begin{eqnarray}
R_{\mu \nu} v^\mu v^\nu 
& = & \frac{1}{\frac{A(S)}{4} + c} \left [ - \ddot{S} + \frac{1}{2} \left ( \frac{\dot{S}}{\frac{A(S)}{4} + c} \right )^{\! 2} \left( \frac{A(S)}{4} - c \right ) \right ] > 0 \; ,
\end{eqnarray}
provided $c< A/4$. This is indeed the case since $A \gg 1$ and explicit calculations indicate that $c$ is of order unity, (\ref{c=-3/2}). Therefore, even in the context of semi-classical gravity, we again recover the geometric form of the null energy condition from the second law of thermodynamics.


\section*{Summary and Future Work}

 The null energy condition was initially proposed as a plausible but ad hoc requirement on matter. This condition, which does not seem to follow from any first principles, has sweeping consequences when matter is coupled to gravity. Here we have taken a different view: we regard the null energy condition not as an ad hoc characteristic of matter, but as a fundamental property of gravity. Moreover, we have shown that this property, in the form of the Ricci convergence condition, follows directly from an assumption that some underlying conventional non-gravitational microphysics accounts for the Bekenstein-Hawking entropy and obeys the second law of thermodynamics. It is remarkable that the point-wise classical null energy condition, which in its matter form has so far been impossible to derive from quantum field theory, follows in its geometric form so readily from the thermodynamics of emergent gravity. It is a satisfying result because the universality of the null energy condition -- which is supposed to hold for all physical spacetimes -- is traced to another universal condition, namely the second law of thermodynamics. 

Here, the underlying premise has been that all non-contracting infinitesimal open patches of the integral curves of null geodesic congruences can be associated with thermodynamic systems. How then, should we interpret geodesic congruences that are locally contracting? One can imagine several alternatives. First, it may well be that the existence of congruences with $\theta < 0$ (or in which $\theta$ changes sign) merely indicates that our premise is wrong. This is certainly a logical possibility. But the same critique could be applied to Jacobson's original paper, which restricts discussion to patches of null congruences with vanishing $\theta$ (``local Rindler horizons"), an even more restrictive set of congruences than the one we consider. In both cases, however, accepting the premise leads to a non-trivial result (Einstein's equations, null energy condition). Perhaps one could regard this as evidence for the assumption. Second, it may be that the correct way to associate thermodynamics with geometry is to start from the microscopic system. In this case, not every geometric surface or congruence need correspond to something that has a meaningful microscopic interpretation. In this approach, if we start with microscopic thermodynamic systems that obey the second law, we should necessarily consider only null congruences with $\theta \geq 0$, and we need not inquire about the interpretation of other congruences. Third, it may be that all congruences, even those with $\theta < 0$, do in fact correspond to thermodynamic systems. For suppose we have a contracting patch. We could simply identify thermodynamic time with negative affine parameter, $\lambda$. Then $\theta < 0$ would still correspond to $\dot{S} > 0$. The Raychaudhuri equation is invariant under $\lambda \leftrightarrow - \lambda$, and so we would still obtain the null energy condition as a consequence of thermodynamics; in this way, patches in which $\theta < 0$ can be accommodated as well. That leaves only patches for which $\theta$ changes sign. But these are rare events of measure zero; one can speculate that these may correspond to rare violations of the second law. 

We have also taken a novel approach to studying quantum effects in semi-classical gravity. In particular, we have shown that the Ricci convergence condition remains stable under one-loop quantum corrections to the Bekenstein-Hawking entropy. If this were the entirety of the effect (which we do not claim), it would mean that quantum effects at one-loop do not, for example, prevent the occurrence of cosmological or black hole singularities. 

There are at least two clear instances of quantum effects violating the matter form of the null energy condition: Hawking radiation and Casimir energy. Hawking radiation, however, is really a non-perturbative effect; this is easiest to understand by noting that Hawking radiation can be expressed as a tunneling process \cite{Parikh00-1,Parikh04-1}. But Casimir energy certainly violates the matter NEC at one-loop. How is our result to be reconciled with the general expectation that the matter null energy condition should be violated by one-loop effects? Here it is important to recognize that it is not definitively known how Casimir energy actually gravitates. One can imagine several possibilities. Since quantum corrections inevitably induce gravitational counter-terms, these would generically sever the link between the matter and the geometry form of the NEC. Thus it could be that the matter NEC is indeed violated by one-loop quantum effects, but the geometric one is not. Alternatively, it could be that vacuum expectation values of $T_{\mu \nu}$ do not gravitate for unknown reasons related to the resolution of the cosmological constant problem. Or it could be that there are additional quantum gravity effects  that are not captured by the logarithmic correction to the entropy considered here. Finally, it could be that only classical spacetime physics corresponds to thermodynamics in the dual theory, and that the approach here is invalid.


\newpage

\section{GRAVITY FROM EQUILIBRIUM THERMODYNAMICS} \label{sec:gravfromthermo} 

The fact that black holes -- example spacetimes -- come equipped with a temperature and thermodynamic entropy, 
\beq S_{\text{BH}}=\frac{A}{4}\left(\frac{c^{3}k_{B}}{G\hbar}\right)\;,\eeq
suggests a deep interplay between gravity, quantum mechanics, and thermodynamics. Moreover, the fact that de Sitter and Rindler horizons -- which are observer-dependent and therefore could be anywhere -- also have thermodynamic properties suggests that holographic entropy and temperature are actually more generally applicable concepts in spacetime, i.e., black holes are not required. Taking this idea significantly further, Jacobson \cite{Jacobson:1995ab} attributed thermodynamic properties even to \emph{local Rindler horizons}: planar patches of certain null congruences passing through arbitrary points in spacetime, and are not event horizons in any global sense. The locality of local Rindler ``horizons" has the effect that local equations follow from thermodynamic equations. Specifically, Einstein's equations follow from the Clausius theorem, $Q = T \Delta S$. Other classical properties of spacetimes, e.g., the null energy condition, can be obtained from the second law of thermodynamics \cite{Parikh:2015ret,Parikh:2016lys}. 

Here we present a new formulation: we attribute thermodynamic properties to the future light cone of any point, $p$, in an arbitrary spacetime. A future light cone can be regarded as a kind of spherical Rindler horizon because the worldlines of observers with constant outward radial acceleration asymptote to it. In fact, it will be more convenient to consider the stretched future light cone, a timelike codimension-one hypersurface. Indeed, we will define our stretched future light cone as a timelike congruence of worldlines with approximately constant and uniform radial acceleration. By constant, we mean that the proper acceleration of any single worldline does not change along the worldline; by uniform, we mean that all worldlines share the same proper acceleration. 

Given the relation between temperature and acceleration, it then seems natural to attribute a constant and uniform temperature to this surface. In fact, entropy is also a somewhat better-motivated property of our surface than of local Rindler horizons. This is because a future light cone separates its interior from the exterior spacetime; the interior is causally disconnected from the exterior, in the same sense that the interior of a black hole is. It seems therefore plausible that we might associate entropy to spacelike sections of the light cone, for example as the entanglement entropy between the interior and exterior regions. By contrast, a finite strip of Rindler horizon (unlike an infinite global Rindler horizon) does not separate space into two disconnected regions, and it is not obvious that it should possess an entropy. Another appealing feature of our formulation is that the interior of a future light cone resembles that of black holes or de Sitter space in that it admits compact spatial sections. 

These geometric aspects motivate the premise of this section, which is that holographic thermodynamic properties can be associated locally with the stretched future light cone emanating from an arbitrary point $p$ in an arbitrary spacetime. We will then show that the Clausius theorem, properly understood, yields Einstein's equation at $p$,
\be
Q = T \Delta S \Rightarrow R_{ab} - \frac{1}{2} R g_{ab} + \Lambda g_{ab} = 8 \pi G T_{ab} \; ,
\ee
much as the association of thermodynamics with local Rindler horizons leads to Einstein's equation emerging as an equation of state \cite{Jacobson:1995ab}. 


Besides its conceptual appeal, the stretched future light cone formulation of local holographic thermodynamics also offers a significant new result: it permits the extension of Jacobson's result to a wide class of theories of gravity. It has been a longstanding challenge to obtain the gravitational equations of motion for general, higher-curvature theories of gravity from  thermodynamics. Broadly, we can divide earlier attempts into two categories: (i) those that aim to derive the equations of motion for $f(R)$ theories of gravity via a nonequilibrium modification of the Clausius theorem to account for internal entropy production terms \cite{Eling06-1}, and (ii) those that aim to derive the gravitational equations for general theories of gravity \cite{Parikh:2009qs,Brustein:2009hy,Padmanabhan:2009ry,Guedens:2011dy,Dey:2016zka}. The approaches that fall into category (i) have been critically reviewed in \cite{Guedens:2011dy}, which points out that this nonequilibrium approach can never lead to theories beyond $f(R)$ gravity. The attempts that fall into category (ii) mainly use a ``Noetheresque" approach, in which the local entropy is expressed as an integral of a Noether current \cite{Parikh:2009qs,Brustein:2009hy,Guedens:2011dy,Dey:2016zka} over spacelike sections of a local Rindler plane. Unfortunately, all the early papers using the Noetheresque approach contained technical errors, as reviewed by Guedens \emph{et al.} \cite{Guedens:2011dy}. Although the authors of \cite{Guedens:2011dy} fixed the technical problems, the derivation nonetheless appears quite unphysical, with the entropy not always proportional to the area even for Einstein gravity. The present work applies the Noetheresque approach of Parikh and Sarkar \cite{Parikh:2009qs} to the setting of a stretched future light cone, rather than to local Rindler planes. As we shall see, the geometry of the new setup allows the technical problems in earlier derivations to be overcome while still preserving an entropy proportional to the area for Einstein gravity. We will describe the earlier literature of the Noetheresque approach, as well as its technical challenges, in more detail later.

Here we consider those gravitational theories whose Lagrangian consists of a polynomial in the Riemann tensor (with no derivatives of the Riemann tensor, for simplicity). For all such theories, after replacing the Bekenstein-Hawking entropy with the Wald entropy, we find that Clausius' theorem again implies the field equations of classical gravity:
\be
Q = T \Delta S \Rightarrow P_{a}^{\;\;cde}R_{bcde}-2\nabla^{c}\nabla^{d}P_{acdb}-\frac{1}{2}Lg_{ab}=8\pi G T_{ab} \; ,
\ee
where the equation on the right is, as we shall describe, the generalization of Einstein's equations for these higher-curvature gravitational theories, up to an undetermined cosmological constant term.

In summary, the main goals of this section are, first to formulate a definition of the stretched future light cone and, second, to derive the (generalized) Einstein equations from the premise that local holographic thermodynamic properties can be attributed to stretched future light cones.


\subsection{Einstein's Equations from the Stretched Future Lightcone} \label{subsec:Einfromstretch}
\noindent


\subsubsection{Geometry of Stretched Lightcones}
\noindent

We begin with a review of the construction of the stretched lightcone (for more details see \cite{Parikh:2017aas}). For concreteness, let us first restrict to pure $D$-dimensional Minkowski space. In Minkowski space there are $\binom{D+1}{2}$ independent Killing vectors $\chi^{a}$ corresponding to spacetime translations and Lorentz transformations. The flow lines of Cartesian boost vectors, e.g., $x\partial^{a}_{t}+t\partial^{a}_{x}$, trace the worldlines of Rindler observers, i.e., observers traveling with constant acceleration in some Cartesian direction. 

The stretched future lightcone can be viewed as a spherical Rindler horizon generated by the radial boost vector:
\beq \xi^{a}\equiv r\partial^{a}_{\;t}+t\partial^{a}_{\;r}=\sqrt{x^{i}x_{i}}\partial^{a}_{t}+\frac{tx^{j}}{\sqrt{x^{i}x_{i}}}\partial^{a}_{\;j}\;, \label{xi}\eeq
where $r$ is the radial coordinate and $x^{i}$ are spatial Cartesian coordinates. We define the stretched future lightcone as a congruence of worldlines generated by these radial boosts. Unlike their Cartesian boost counter-parts, which preserve local Lorentz symmetry, the radial boost vector is not a Killing vector in Minkowski space; this is because radial boosts are not isometries in Minkowski space. 

The flow lines of $\xi^{a}$ trace out hyperbolae in Minkowski space. Let us define a codimension-1 timelike hyperboloid via the set of curves which obey
\beq r^{2}_{\text{Mink}}-t^{2}=\alpha^{2}\;,\label{hyperboloid}\eeq
where $t\geq0$ and $\alpha$ is some length scale with dimensions of length. This hyperboloid can be understood as a stretched future lightcone emanating from a point $p$ at the origin. The constant-$t$ sections of the hyperboloid are $(D-2)$-spheres with an area given by 
\beq A_{\text{Mink}}(t)=\Omega_{D-2}(\alpha^{2}+t^{2})^{(D-2)/2}\;.\label{areamink}\eeq
Here we have that $\xi^{2}=-\alpha^{2}$, and is therefore an unnormalized tangent vector to the worldlines of the spherical Rindler observers. The normalized velocity vector is defined as $u^{a}=\xi^{a}/\alpha$, with $u^{2}=-1$, and has a proper acceleration with magnitude
\beq a_{\text{Mink}}=\frac{1}{\alpha}\;. \label{xiacc}\eeq
The stretched future lightcone, in Minkowski space, can therefore be understood as a congruence of worldlines of a set of constant radially accelerating observers, all with the same uniform acceleration of $1/\alpha$. 

Let us now consider what happens in an arbitrary spacetime. In the vicinity of any point $p$, spacetime is locally flat. The components of a generic metric tensor can always be expanded using Riemann normal coordinates (RNC):
\beq g_{ab}(x)=\eta_{ab}-\frac{1}{3}R_{acbd}(p)x^{c}x^{d}+...\;, \label{RNC}\eeq
where the Riemann tensor is evaluated at the point $p$, the origin of the RNC system. Here $x^{a}$ are Cartesian coordinates and $\eta_{ab}$ is the Minkowski metric in Cartesian coordinates. Since a generic spacetime is locally flat, there still exist the $\binom{D+1}{2}$ vectors $\chi^{a}$ which preserve the isometries of Minkowski space, locally, however, they are no longer exact Killing vectors; the presence of quadratic terms $\mathcal{O}(x^{2})$ in the RNC expansion (\ref{RNC}) indicates that these vectors will not satisfy Killing's equation and Killing's identity at some order in $x$. The specific order depends on the nature of the vector $\chi^{a}$, e.g., for Lorentz boosts the components are of order $\mathcal{O}(x)$. Therefore, for the generators of local Lorentz transformations, Killing's equation and Killing's identity will fail as
\beq \nabla_{a}\chi_{b}+\nabla_{b}\chi_{a}\approx\mathcal{O}(x^{2})\;,\quad \nabla_{a}\nabla_{b}\chi_{c}-R^{d}_{\;abc}\chi_{d}\approx\mathcal{O}(x)\;.\eeq
We call these local Cartesian boost vectors $\chi^{a}$ approximate Killing vectors. 

The radial boost vector (\ref{xi}) is therefore not a Killing vector in an arbitrary spacetime for two reasons: (i) It is not a Killing vector in Minkowski space, and (ii) the addition of curvature via the RNC expansion leads to a further failure of Killing's equation and Killing's identity. Specifically, 
\beq
\begin{split}
&\nabla_{t}\xi_{t}=0+\mathcal{O}(x^{2})\;,\quad \nabla_{t}\xi_{i}+\nabla_{i}\xi_{t}=0+\mathcal{O}(x^{2})\;,\\
&\nabla_{i}\xi_{j}+\nabla_{j}\xi_{i}=\frac{2t}{r}\left(\delta_{ij}-\frac{x_{i}x_{j}}{r^{2}}\right)+\mathcal{O}(x^{2})\;.
\end{split}
\label{Killingfailure}\eeq
Observe that the $t-t$ and $t-i$ components satisfy Killing's equation at $\mathcal{O}(1)$, while the $i-j$ components fail to obey Killing's equations even at leading order. This means that Killing's identity will also fail; in fact it fails to order $\mathcal{O}(x^{-1})$. We also note that on the $t=0$ surface our radial boost vector is an instantaneous Killing vector. 

\begin{figure}[t]
\centering
 \includegraphics[width=10cm]{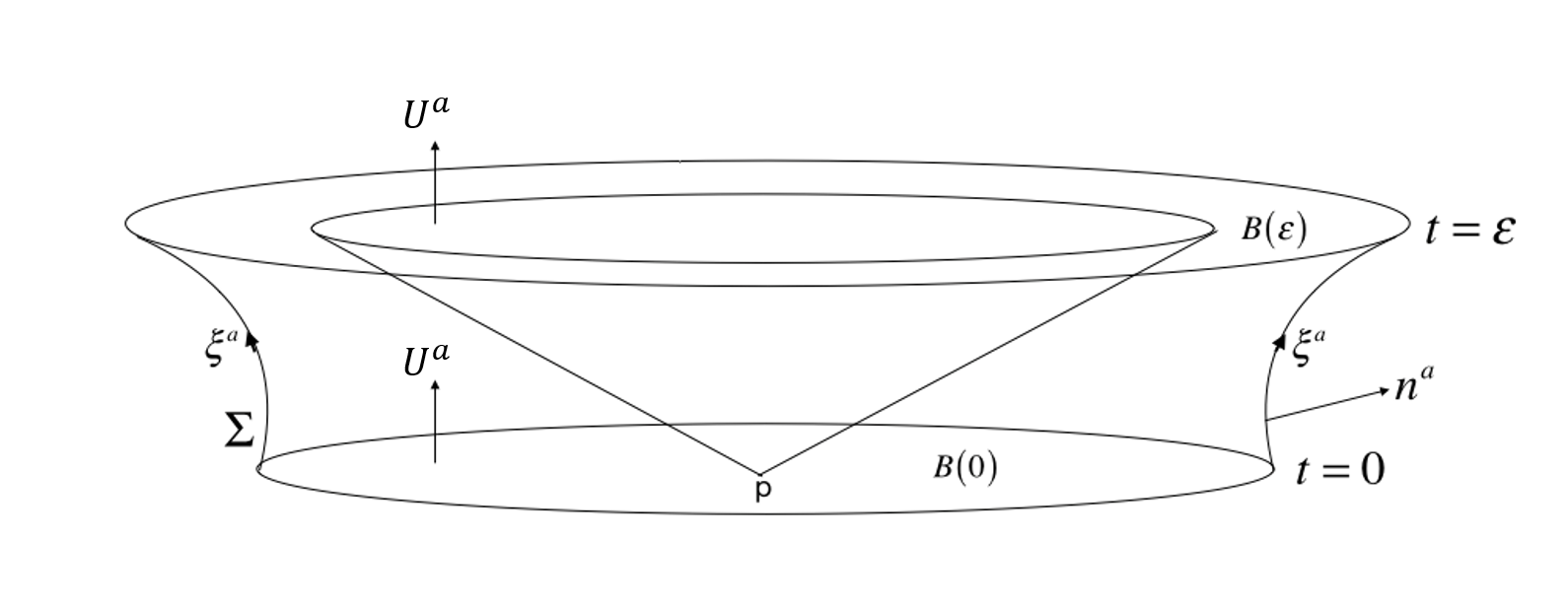}
 \caption[The Stretched Future Lightcone.]{A congruence of radially accelerating worldlines $\xi^{a}$ with the same uniform proper acceleration $1/\alpha$ generates the stretched future light cone of point $p$, and describes a timelike hypersurface, $\Sigma$, with unit outward-pointing normal $n^{a}$. The boundary of $\Sigma$ consists of the two codimension-two surfaces $\partial \Sigma(0)$ and $\partial \Sigma(\epsilon)$ given by the constant-time slices of $\Sigma$ at $t=0$ and $t=\epsilon$, respectively. The co-dimension-1 spatial ball $B$ is the filled in co-dimension-2 surface $\partial \Sigma$.
}
 \label{sigma}\end{figure}

In an arbitrary spacetime our notion of stretched future lightcone must be modified. In a curved spacetime it is straightforward to show that 
\beq \xi^{2}=-\alpha^{2}+\mathcal{O}(x^{4})\quad a=\frac{1}{\alpha}\left(1+\mathcal{O}(x^{4})\right)\;.\eeq
Motivated by the stretched horizon defined in the black hole membrane paradigm \cite{Price86-1}, we define the stretched future lightcone $\Sigma$ as follows: Pick a small length scale\footnote{``Small" here means $\alpha$ is much smaller than the smallest curvature scale at the point $p$, i.e., the metric is taken to be roughly flat to a coordinate distance $\alpha$ from the origin.}. Then select a subset of observers who at time $t=0$ have a proper acceleration $1/\alpha$. If we follow the worldlines of these observers we would find that generically they would not have the same proper acceleration at a later generic time. This problem can be remedied by choosing a timescale $\epsilon\ll\alpha$. Over this timescale the initially accelerating observers have an approximate constant proper acceleration, and the stretched future lightcone $\Sigma$ can be regarded as a worldtube of a congruence of observers with the same nearly-constant approximately outward radial acceleration $1/\alpha$, as can be seen in figure (\ref{sigma}).


 Let us remark on the similarities between the radial boost vector $\xi_{a}$ (\ref{xi}) generating the stretched future lightcone, and  conformal Killing vectors, which we denote by $\zeta_{a}$. Conformal Killing vectors are those which satisfy conformal Killing's equation 
\beq \nabla_{a}\zeta_{b}+\nabla_{b}\zeta_{a}=2\Omega g_{ab}\;,\label{confKeqn}\eeq
where $\Omega$ satisfies
\beq \Omega=\frac{1}{D}\nabla_{c}\zeta^{c}\;,\eeq
and is related to the conformal factor $\omega^{2}$ of $\bar{g}_{ab}=\omega^{2}g_{ab}$ via $2\Omega=\zeta^{c}\nabla_{c}\ln\omega^{2}$. 

Conformal Killing vectors also satisfy the conformal Killing identity
\beq \nabla_{b}\nabla_{c}\zeta_{d}=R^{e}_{\;bcd}\zeta_{e}+(\nabla_{c}\Omega)g_{bd}+(\nabla_{b}\Omega)g_{cd}-(\nabla_{d}\Omega)g_{bc}\;.\label{confKid}\eeq
Following the discussion above, in an arbitrary spacetime the conformal Killing vectors will become approximate conformal Killing vectors, failing to satisfy the conformal Killing equation to order $\mathcal{O}(x^{2})$ in a RNC expansion about some point $p$, and the conformal Killing identity to $\mathcal{O}(x)$.

Now, notice that the radial boost vector $\xi_{a}$ satisfies
\beq \nabla_{a}\xi_{b}+\nabla_{b}\xi_{a}=2\left(\frac{t}{ r}\right)\left(\eta_{ij}-\frac{x_{i}x_{j}}{r^{2}}\right)\delta^{i}_{a}\delta^{j}_{b}\;, \label{Keqforu}\eeq
where the $\delta^{i}_{a}\delta^{j}_{b}$ are present to project the non-zero contributions. We see that $\xi^{a}$ is a vector which satisfies Killing's equation in specific metric components, and one which fails as a modified CKV in other components. This comparison leads us to define a conformal factor associated with $\xi$:
\beq \Omega_{\xi}\equiv\frac{1}{(D-2)}\nabla_{c}\xi^{c}=\frac{t}{ r}\;,\eeq
for which one finds 
\beq \nabla_{d}\Omega_{\xi}=-\frac{1}{r^{2}}\xi_{d}\;,\quad N_{\xi}^{-1}\equiv||\nabla_{a}\Omega_{\xi}||=\frac{\alpha}{r^{2}}\;,\eeq
and
\beq u_{a}=N_{\xi}\nabla_{a}\Omega_{\xi}\;.\eeq
It is also straightforward to work out
\beq \nabla_{d}(\mathcal{L}_{\xi}g_{ab})|_{t=0}=\frac{2}{N_{\xi}}u_{d}\delta^{i}_{a}\delta^{j}_{b}\left(\eta_{ij}-\frac{x_{i}x_{j}}{r^{2}}\right)\;, \label{nabLg}\eeq
and 
\beq K_{\partial\Sigma}=\frac{1}{\alpha}(D-2)\;,\eeq
where $\mathcal{L}_{\xi}$ is the Lie derivative along $\xi_{a}$, and the extrinsic curvature of the spherical boundary $\partial\Sigma$ is $K=h^{ab}K_{ab}=g^{ab}\nabla_{b}n_{a}$, since $h_{ab}=g_{ab}-n_{a}n_{b}$. We will make use of these properties in Chapter \ref{sec:gravfroment} where we discuss the relationship between the Clausius relation and the entanglement equilibrium proposal associated with stretched lightcones. 

\subsection{Stretched Lightcone Thermodynamics}
\noindent

The reason for choosing $\Sigma$ to be a hypersurface composed of constant acceleration worldlines is that, by the relation between temperature and acceleration, $\Sigma$ then becomes an isothermal surface. However, a rigorous identification of temperature with acceleration applies only to eternally accelerating observers in Minkowski space with a Poincar\'e-invariant vacuum, whereas here we have transient acceleration in an only approximately locally flat patch of spacetime. We therefore need to justify first, why the existence of an approximately Poincar\'e-invariant vacuum state can be assumed and second, why even granted the existence of such a state, it is possible to associate a temperature with transient acceleration. 

The existence of an approximately Poincar\'e-invariant vacuum state is a consequence of the strong principle of equivalence. If we assume that free-falling observers should see the same physics locally as inertial observers in Minkowski space, then we are naturally led to assume that the quantum state responsible for local physics should be approximately the Poincar\'e-invariant state of Minkowski space; any other coherent state would have a stress tensor whose vacuum expectation value would be singular somewhere. The same prescription is used to select the Unruh state in the black hole case, ensuring that an observer falling along a geodesic sees no Hawking radiation. The validity of using the Poincar\'e-invariant state locally even has experimental support in that high-energy physics at accelerators is perfectly captured by quantum field theory in Minkowski space, even though on larger scales our spacetime is not well described by Minkowski space. 

Having justified our choice of the Poincar\'e-invariant vacuum state, we automatically find that eternally accelerating Rindler observers will detect particles with a thermal spectrum. Transient acceleration in Minkowski space was studied by Barbado and Visser \cite{Barbado:2012fy} who found that a thermal spectrum is still obtained provided the duration of acceleration is sufficiently long compared with the inverse acceleration. This condition is easy to arrange in our construction. We need to extend the worldlines of the accelerating observers over a longer time, $\tau$, much greater than the inverse acceleration, $\alpha$ (but still short enough that curvature effects are negligible). Since there is no limit to how small $\alpha$ can be, we can always do this. Our surface $\Sigma$ is then a brief segment, $0 < t < \epsilon \ll \alpha \ll \tau$ of a more extended surface traced by a congruence of such observers.  Temperature and acceleration can now be rigorously identified on the extended surface, and therefore also on $\Sigma$, so that both are isothermal surfaces. In general, the worldlines of the observers will not be integral curves of our approximate Killing vector $\xi_a$ before $t = 0$ or after $t = \epsilon$. We therefore restrict our calculation to $\Sigma$ because we need a congruence generated by the flow lines of $\xi_a$. 

With this rationale, $\Sigma$ is an isothermal surface with Davies-Unruh temperature
\be
T \equiv \frac{\hbar a}{2 \pi} = \frac{\hbar}{2 \pi \alpha} \; . \label{temp}
\ee
In particular, this means that in any integration over $\Sigma$, we can move the temperature outside the integral.

Having defined our stretched future light cone, $\Sigma$, and having associated a uniform temperature with it, we next need to define the entropy. The underlying premise of the ``thermodynamics of spacetime" is that gravitational entropy can be attributed not just to global event horizons, but also to local Rindler horizons. In the same vein, we attribute a local entropy to spacelike sections of the future light cone \cite{DeLorenzo:2017tgx}. We also attribute entropy to sections of our timelike stretched horizon, $\Sigma$. This is consistent with the black hole membrane paradigm in which the timelike stretched horizon can also be thought of as having thermodynamic properties \cite{Parikh:1997ma}. 

The form of the entropy depends on the gravitational theory under consideration. For Einstein gravity, the entropy is the Bekenstein-Hawking entropy, one quarter of the area measured in Planck units:
\be
S = \frac{A}{4 G \hbar} \; . \label{Bek}
\ee
We will first rewrite this in a useful form using the vectors $n_{a}$ and $\xi_{a}$ on $\Sigma$. Let $\omega(t)$ be the codimension-two section of $\Sigma$ at time $t$.
Its area is
\be 
A(t) \equiv \int_{\omega(t)} dA = \alpha \int_{\omega(t)} dA \, n_b \frac{1}{\alpha} n^b =  \alpha \int dA \, n_b u^a \nabla_a u^b = \int dA \, n_b u_a \nabla^a \xi^b \; .
\ee
Here we have used $a^{b} = u^{a} \nabla_{a} u^{b} = \frac{1}{\alpha} n^{b}$ and $u_{a} \equiv \frac{\xi_{a}}{\sqrt{-\xi^{a} \xi_{a}}} \approx \frac{\xi_{a}}{\alpha}$.
Next we make use of the fact that $\nabla_{a}\xi_{b}=-\nabla_{b}\xi_{a}$ for the projection of $\nabla_{a}\xi_{b}$ in the $n-\xi$ plane, as we see from the first line of (\ref{Killingfailure}). Then defining
\be
dS_{ab} \equiv \frac{1}{2}(n_{a}u_{b}-n_{b}u_{a})dA \; ,
\ee
we see that the Bekenstein-Hawking entropy at time $t$ can be expressed as 
\be
S(t) = - \frac{1}{4G \hbar} \int_{\omega(t)} dS_{ab} \del^a \xi^b = - \frac{1}{4G \hbar} \int_{\omega(t)} dS_{ab} \frac{1}{2} (g^{ac} g^{bd} - g^{ad} g^{bc} ) \del_c \xi_d \; . \label{areaxi}
\ee
Here we have written the entropy in the form $\int dS_{ab} M^{ab}$, where $M^{ab}$ is an antisymmetric tensor; this form will be helpful in deriving Einstein's equations and will generalize readily to other theories of gravity.

Now let us calculate the total change in the Bekenstein-Hawking entropy $\Delta S_{\rm tot} = S(\epsilon) - S(0)$, between $t = 0$ and $t = \epsilon$. To that end, note that the codimension-two surfaces $\omega(\epsilon)$ and $\omega(0)$ are the boundaries of the stretched future light cone, $\Sigma$. We can therefore make use of Stokes' theorem for an antisymmetric tensor field $M^{ab}$,
\be
\int_{\Sigma}d\Sigma_{a}\nabla_{b} M^{ab}=-\int_{\omega(\epsilon)}dS_{ab} M^{ab} + \int_{\omega(0)}dS_{ab} M^{ab} \; , \label{stokesthm}
\ee
where the overall minus sign arises because $\Sigma$ is a timelike surface. From (\ref{areaxi}), we find
\be
\Delta S_{\rm tot}=\frac{1}{4G \hbar}\int d\Sigma_{a}\frac{1}{2}(g^{ac}g^{bd}-g^{ad} g^{bc})(R^{e}_{\;bcd}(p) \xi_{e}+f_{bcd}) \label{DStot}
\ee
where we have approximated the Riemann tensor by its value at the point $p$, which we can do to leading order in $x$. To obtain (\ref{DStot}), we have written the Killing identity for our approximate Killing vector $\xi_a$ as
\be
\nabla_{b}\nabla_{c}\xi_{d}=R^{e}_{\;bcd}\xi_{e}+f_{bcd} \; . \label{fbcd}
\ee

The term $f_{bcd}$ accounts for the failure of Killing's identity to hold; for a true Killing vector, $f_{bcd}$ would be zero. As we see from (\ref{Killingfailure}), $\xi_a$ fails to be a Killing vector in two ways. First, because of spacetime curvature, Killing's equation generically fails at quadratic order in Riemann normal coordinates. These quadratic terms contribute terms of order $x$ to $f_{bcd}$. But second, even if spacetime were exactly Minkowski space, our $\xi_a$ generates not planar boosts, but radial boosts; these are not true isometries, as indicated by the leading-order failure of Killing's equation to hold for the $i-j$ components. This contributes terms of order ${\cal O}(x^{-1})$ to $f_{bcd}$. (In addition to these, there will also be terms ${\cal O}(1)$ in $f_{bcd}$ coming from modifications to $\xi_a$, as detailed in Appendix \ref{app:failKILC}.) We cannot discard either of these pieces of $f_{bcd}$ because they are not higher order than the $R^{e}_{\;bcd}(p)\xi_{e}$ term we would like to keep, which is of order $x$. Fortunately, we do not need $f_{bcd}$ to vanish: we only need its integral to vanish. This distinction makes a tremendous difference. We note that because the constant-$t$ sections of $\Sigma$ are spheres (to leading approximation), any odd power of a spatial Cartesian coordinate $x^i$ integrates to zero over $\Sigma$. As shown in Appendix \ref{app:failKILC} this results in the vast majority of terms of order $x$ (and ${\cal O}(1)$) in $f_{bcd}$ integrating to zero. The handful of surviving terms can be canceled by including quadratic and cubic terms in the expansion of $\xi_a$. The same is not true for the term of order $1/x$ in $f_{bcd}$, which neither vanishes upon integration, nor can be canceled by redefinitions. To leading order, we can evaluate it in $D$-dimensional Minkowski space, where we find
\be
\frac{1}{4G \hbar}\int d\Sigma_{a}\frac{1}{2}(\eta^{ac}\eta^{bd}-\eta^{ad} \eta^{bc}) f^{{\cal O}(x^{-1})}_{bcd} =  \frac{\Omega_{D-2}}{4 G \hbar} \alpha^{D-4} \epsilon^2 \; . \label{nPf-1Einstein}
\ee
Remarkably, this term actually has a physical interpretation. 

Recall that we would like to equate our entropy change to the heat flux. However, as we have defined it, $\Delta S_{\rm tot}$ is the total change in the area of our stretched future light cone. Not all of this change in area can be attributed to the influx of heat. This is because $\Sigma$ is generated by a congruence of outwardly accelerating worldlines whose area would increase even in the absence of heat. Indeed, even in  Minkowski space with no heat flux whatsoever, the area of the hyperboloid of outwardly accelerating observers increases in time, Eq. (\ref{areamink}). Therefore, before identifying the change in entropy with $T^{-1} Q$, we should first subtract this background expansion of the hyperboloid, $\Delta S_{\rm hyp}$, from $\Delta S_{\rm tot}$:
\be
\Delta S_{\rm rev} \equiv \Delta S_{\rm tot} - \Delta S_{\rm hyp}	\label{revS}
\ee
We call the difference $\Delta S_{\rm rev}$, the reversible change in entropy, in analogue with ordinary thermodynamics for which we have $Q = T \Delta S_{\rm rev}$ (the general formula in the presence of irreversible processes is $\Delta S \geq Q/T$, with saturation only for the reversible component of $\Delta S$).

Now the change in the Bekenstein-Hawking entropy from the natural expansion of the stretched future light cone can be read off from (\ref{areamink}). It is
\be
\Delta S_{\rm hyp} = \frac{\Omega_{D-2}}{4 G \hbar} \left (r^{D-2}_{\rm Mink} (\epsilon) - r^{D-2}_{\rm Mink} (0) \right ) \approx  \frac{\Omega_{D-2}}{4 G \hbar} \alpha^{D-4} \epsilon^2 \; ,
\ee
which is precisely equal to (\ref{nPf-1Einstein}). Evidently we can interpret (\ref{nPf-1Einstein}) as the natural increase in the entropy of the hyperboloid in the absence of heat flux, an increase that is eliminated by considering only the reversible part of the entropy change, Eq. (\ref{revS}).

We therefore have
\be 
\Delta{S}_{\rm rev}=\frac{1}{4 G \hbar}\int_{\Sigma} d \Sigma^{a}R_{ab}(p)\xi^{b}
\ee
Now we use the fact that $\Sigma$ was constructed to be a surface of constant and uniform acceleration. We can therefore associate with it a constant and uniform temperature, Eq. (\ref{temp}). Then
we have
\be 
T\Delta{S}_{\rm rev}=\frac{1}{8\pi\alpha G}\int_{\Sigma} d \Sigma^{a}R_{ab}(p)\xi^{b} \label{TDelS}
\ee
Meanwhile, the integrated energy flux into $\Sigma$ as measured by our accelerating observers is
\be
Q = \int_\Sigma d\Sigma^a T_{ab} u^b \approx \frac{1}{\alpha} \int_\Sigma d\Sigma^a T_{ab} (p) \xi^b \; . \label{DelQ}
\ee
where the energy-momentum tensor can again be approximated to leading order by its value at $p$. Now, in thermodynamics, heat is the energy that goes into macroscopically unobservable degrees of freedom. Since the interior of the future light of $p$ is fundamentally unobservable (being causally disconnected from the exterior), we identify the integrated energy flux, Eq. (\ref{DelQ}), as heat \cite{Jacobson:1995ab}.

Clausius' theorem, $Q = T \Delta S_{\rm rev}$, then tells us to equate the integrals in (\ref{DelQ}) and (\ref{TDelS}). But note that this equality holds for all choices of $\Sigma$. For example, we could have chosen a different surface $\Sigma$ by having a different choice of $\alpha$ or by varying $\epsilon$. In particular, since the surface $\Sigma$ is capped off by constant-time slices, we can also obtain a different $\Sigma$ by performing a Lorentz boost on our Riemann normal coordinate system.
It is shown in Appendix \ref{app:eqlintegrands}, that this implies that the tensors contracted with $n^a$ and $\xi^b$ in the integrands of (\ref{TDelS}) and (\ref{DelQ}) must match, up to a term that always vanishes when contracted with $n^a$ and $\xi^b$. Since $n^a \xi_a = 0$, the unknown term must be proportional to the metric. We therefore have
\be 
R_{ab}+\varphi g_{ab} =8\pi G T_{ab} \; ,
\ee
where $\varphi$ is some scalar function of spacetime. We may determine this function by demanding that the Bianchi identity hold, leading finally to Einstein's equations:
\be
R_{ab}-\frac{1}{2}Rg_{ab}+\Lambda g_{ab}=8\pi G T_{ab} \; .
\ee
Thus, gravitational equations emerge out of Clausius' theorem, $Q = \Delta S_{\rm rev}/T$, when we attribute thermodynamic properties to stretched future light cones. The cosmological constant appears as an integration constant. We have reproduced Jacobson's famous result, but using a construction based on the stretched future light cone.

It is instructive to ask why $\Delta S_{\rm rev}$ had to be positive. In fact, this follows intuitively from the way we have defined $\Sigma$ as a surface of constant acceleration, a setup that is motivated by black hole physics. Consider a sphere of observers at some radius $r$, outside some spherically symmetric body, such as a black hole. The observers stay at $r$, firing their rockets to not fall in, and are therefore all subject to the same, constant acceleration. Now suppose more matter accretes on to the source, increasing its gravitational pull. Heuristically, the observers have to move outwards in order to maintain their original acceleration. Therefore a surface of constant accelerating observers increases its area when matter falls in; this is why $\Delta S_{\rm rev}$ is positive when $Q > 0$. More precisely, explicit evaluation of $Q$ from its definition, Eq. (\ref{DelQ}), yields:
\be
Q = \frac{\Omega_{D-2}}{2} \alpha^{D-3} \epsilon^2 \left (\rho + \frac{1}{D-1} \sum_i P_i \right ) \; ,
\ee
where $\rho = -T_{tt}(p)$ and $P_i = T_{ii}(p)$. We see that $Q$ is positive when the null energy condition is obeyed. Thus our stretched future light cone has $\Delta S_{\rm rev} \geq 0$ when the null energy condition holds, analogous to the area theorem for black holes. Our stretched future light cone evidently also obeys the second law of thermodynamics.


\subsubsection{Generalized Equations of Gravity}
\noindent

One significant achievement of the stretched lightcone formulation is that the derivation of the Einstein equations can be extended to more general theories of gravity.  Extending the thermodynamic derivation of the gravitational equations to other theories of gravity has been a long-standing challenge. Many previous attempts have been made, both for specific theories of gravity such as $f(R)$ theories, and for more general diffeomorphism-invariant theories. However, all previous attempts at general derivations have been marred by errors, or appear unphysical (or both). Four early papers, which come close, deserve special mention.

Padmanabhan \cite{Padmanabhan:2009ry} attempts to rewrite the field equations in terms of thermodynamics (rather than obtaining them from thermodynamics). The author claims, without showing any calculations, that the steps can be reversed to obtain the equations from the thermodynamics. However, he uses Killing's identity for approximate Killing vectors, without apparently realizing that it fails at the same order as the equations he would be trying to derive. Moreover, his expression for the entropy appears to depend on volume, rather than area. Parikh and Sarkar \cite{Parikh:2009qs} attempt a derivation from thermodynamics, using the Noether charge. The authors recognize that Killing's identity is invalid for approximate Killing vectors, but have no convincing justification for their use of it. They consider a rectangular spacelike patch of a (stretched) local Rindler horizon and equate the difference in area between two such patches using Stokes' theorem on a timelike surface joining them. However, that timelike surface has additional boundaries that connect the edges of the rectangles (which is easiest to visualize in (2+1)-dimensional spacetime); this contribution was missed. Brustein and Hadad \cite{Brustein:2009hy} also attempt a Noether-charge derivation from thermodynamics. The authors write some equations that do not appear correct, expressing the entropy as a volume, for example. They also appear to have used Killing's identity without realizing that it fails. In their use of Stokes' theorem, they also appear to have missed the existence of extra boundary terms. Finally, Guedens \emph{et al.} \cite{Guedens:2011dy} recognize both the issues (failure of Killing's identity, existence of extra boundary terms) that have tripped up previous attempts at derivations. The authors deal with the Killing's identity problem by restricting integration to a very narrow strip of the Rindler horizon plane using the observation \cite{Guedens:2012sz} that Killing's identity can be made to hold approximately near a single null generator. However, they deal with the boundary term by choosing the second surface to have the same edges as the first one, while dipping down in a nearly null test-tube shape. Although they formally succeed in obtaining the gravitational equations from the variation of a Noether charge, their derivation appears unphysical, as they themselves note. For example, even for Einstein gravity, the entropy on the looping part of the test-tube shape is no longer proportional to its area.

The success of the approach in the present work, which is based on the paper by Parikh and Sarkar \cite{Parikh:2009qs}, is directly related to our use of a stretched future light cone. Because a stretched future light cone has closed spacelike sections (spheres, which, unlike the rectangular sections of Rindler planes, have no edges), there are no extra boundary terms in Stokes' theorem. And the failure of Killing's identity is not fatal because the vast majority of problematic terms integrate to zero over a sphere; the few remaining terms can be dealt with, as shown in detail in Appendix \ref{app:failKILC}.

Consider then the action, $I$, of a diffeomorphism-invariant theory of gravity in $D$ dimensions of the form
\be
I=\frac{1}{16\pi G}\int d^{D}x\sqrt{-g}L\left(g^{ab},R_{abcd}\right) + I_{\rm matter} \; .
\ee
Here we have written the gravitational Lagrangian, $L$, as a function of the inverse metric $g^{ab}$ and the curvature tensor $R_{abcd}$ separately. Cast in this way, the action encompasses a wide class consisting of all diffeomorphism-invariant Lagrangian-based theories of gravity that do not involve derivatives of the Riemann tensor. We then define \cite{Padmanabhan:2007en}
\be 
P^{abcd} \equiv \frac{\partial L}{\partial R_{abcd}} \; ,
\ee
where the tensor $P^{abcd}$ can be shown to have all of the algebraic symmetries of the Riemann tensor. The gravitational equation of motion of such theories is
\be
P_{a}^{\;\;cde}R_{bcde}-2\nabla^{c}\nabla^{d}P_{acdb}-\frac{1}{2}Lg_{ab}=8\pi G T_{ab} \; . \label{eom}
\ee
In particular, for Einstein gravity, we have $L = R$, and therefore
\be 
P^{abcd}_{\rm E}=\frac{1}{2}(g^{ac}g^{bd}-g^{ad}g^{bc}) \; . \label{PEinstein}
\ee
Substituting this in (\ref{eom}), we recover Einstein's equation.
 
Our goal is to derive (\ref{eom}) from local holographic thermodynamics. Here we will see that our stretched future light cone derivation of Einstein's equations extends naturally to higher-curvature theories of gravity. Our Noetheresque approach will be based on an earlier paper by one of us \cite{Parikh:2009qs}. In that work, $\Sigma$ was a planar strip of a Rindler horizon, rather than a spherical Rindler horizon. As already mentioned, this resulted in two technical problems: (i) in Stokes' theorem, $\Delta S$ did not account for all contributions from the surface $\Sigma$ because there were also extra contributions from the edges of the strip, and (ii) the failure of Killing's identity, which does not hold for approximate symmetries, led to unwanted terms that could not be eliminated over the strip. As we have already seen, choosing a spherical Rindler horizon for $\Sigma$ resolves both these issues: since a sphere has no boundaries, the problem of extra contributions in Stokes' theorem does not arise. In addition, most of the unwanted terms arising from the failure of Killing's identity integrate to zero on a sphere. Of the remaining terms, as shown in Appendix \ref{app:failKILC}, the leading one precisely cancels the natural expansion of the hyperboloid, and the few remaining ones can be dealt with by re	defining $\xi_a$, as in the case of Einstein gravity.

Now, information about the underlying gravitational theory is encoded within the thermodynamic formula for entropy. For Einstein gravity, the entropy is one quarter of the horizon area, but for more general theories of gravity we have to generalize the Bekenstein-Hawking entropy to something else. We will take that generalization to be the Wald entropy \cite{Wald:1993nt}. To obtain the Wald entropy, one first defines the antisymmetric Noether potential $J^{ab}$, associated with the diffeomorphism $x^{a}\to x^{a}+\xi^{a}$. For theories, that do not contain derivatives of the Riemann tensor, the Noether potential is
\be
J^{ab}=-2P^{abcd}\nabla_{c}\xi_{d}+4\xi_{d}\nabla_{c}P^{abcd} \; . \label{NoetherJ}
\ee
Then, when $\xi_{a}$ is a timelike Killing vector, the Wald entropy, $S$, associated with a stationary black hole event horizon 
is proportional to the Noether charge \cite{Wald:1993nt}:
\be
S =\frac{1}{8G \hbar}\int dS_{ab}J^{ab} \; .
\ee
Substituting (\ref{NoetherJ}) and (\ref{PEinstein}), we indeed recover the Bekenstein-Hawking entropy, Eq. (\ref{Bek}), for the case of Einstein gravity.
 
Wald's construction was designed to yield an expression for the entropy of a stationary black hole in an asymptotically flat spacetime in generalized theories of gravity. As before, we will make the nontrivial assumption of local holography, meaning that this gravitational entropy
can also be attributed locally to the future light cones of arbitrary points, and even to their timelike stretched horizons, $\Sigma$. Consider then a stretched future light cone generated by $\xi_{a}$. Analogous to (\ref{areaxi}), the Wald entropy at time $t$ is
\be
S(t) =-\frac{1}{4G \hbar}\int_{\omega(t)}dS_{ab}\left(P^{abcd}\nabla_{c}\xi_{d}-2\xi_{d}\nabla_{c}P^{abcd}\right) \; .
\ee
The total change in entropy between $t = 0$ and $t = \epsilon$ is $\Delta S_{\rm tot} = S(\epsilon) - S(0)$, or
\be
\Delta S_{\rm tot}
 =\frac{1}{4G \hbar}\int_{\Sigma}d\Sigma_{a}\nabla_{b}\left(P^{abcd}\nabla_{c}\xi_{d}-2\xi_{d}\nabla_{c}P^{abcd}\right) \; ,
\ee
where we have again invoked Stokes' theorem, Eq. (\ref{stokesthm}), for an antisymmetric tensor field. Then
\be
\Delta S_{\rm tot}=\frac{1}{4G \hbar}\int_{\Sigma}d\Sigma_{a}\left[-\nabla_{b}\left(P^{adbc}+P^{acbd}\right)\nabla_{c}\xi_{d}+P^{abcd}\nabla_{b}\nabla_{c}\xi_{d}-2\xi_{d}\nabla_{b}\nabla_{c}P^{abcd}\right] \; .
\ee
For Lovelock theories of gravity, which include Einstein gravity and Gauss-Bonnet gravity, it can be shown that $\nabla_b P^{abcd} = 0$ identically and so the first two terms vanish. For other theories of gravity, however, these terms do not generically vanish. By symmetry, only the contraction with the symmetric part of $\nabla_c \xi_d$ survives. As seen from (\ref{Killingfailure}), $\xi_{a}$ satisfies Killing's equation to $\mathcal{O}(x^{2})$, except for the $i,j$ indices, which means that the term cannot generically be discarded. Define
\be
q^a \equiv \nabla_{b}\left(P^{adbc}+P^{acbd}\right)\nabla_{c}\xi_{d}
\ee
We therefore have
\be
\Delta S_{\rm tot}=\frac{1}{4 G \hbar}\int_{\Sigma} d \Sigma_{a}\left(-q^a + P^{abcd}(R_{dcbe}\xi^{e}+f_{bcd})-2\xi_{d}\nabla_{b}\nabla_{c}P^{abcd}\right) \; , \label{TdStot}
\ee
where we have again taken into account the fact that $\xi_{a}$ does not satisfy Killing's identity, Eq. (\ref{fbcd}). This generalizes (\ref{DStot}). As shown in Appendix \ref{app:failKILC}, just as for the case of Einstein gravity, the unwanted term $\int_{\Sigma}d\Sigma_a P^{abcd}f_{bcd}$ can be dropped by redefining $\xi_a$ and subtracting the natural entropy increase of the hyperboloid, Eq. (\ref{revS}). In Appendix \ref{app:failKILC}, we show that the same redefinition of $\xi_a$ can also be used to eliminate $q^a$ for the non-Lovelock theories for which it does not identically vanish.

Defining the locally measured energy as before, Eq. (\ref{DelQ}),
\be 
Q=\int_{\Sigma}d\Sigma_{a}T^{a}_{\;e}u^{e}=\frac{1}{\alpha}\int_{\Sigma}d\Sigma_{a}T^{a}_{\;e}\xi^{e} \; ,
\ee
we see that $T\Delta S_{\rm rev}= Q$ can be written as
\be
\frac{1}{8\pi\alpha G}\int_{\Sigma} d\Sigma_{a}\left(P^{abcd}R_{dcbe}-2\nabla_{b}\nabla_{c}P^{abc}_{\; \; \; \; \; e}\right) \xi^e = \frac{1}{\alpha}\int_{\Sigma}d\Sigma_{a}T^{a}_{\;e}\xi^{e} \; . \label{Clausiusgeneral}
\ee
As shown in Appendix \ref{app:eqlintegrands}, the equality of these integrals under variations of $\Sigma$ implies a stronger equality of the integrands,
\be
P_{a}^{\;cde}R_{bcde}-2\nabla^{c}\nabla^{d}P_{acdb}+ \varphi g_{ab}=8\pi G T_{ab} \; ,
\ee
where $\varphi$ is an undetermined scalar function. The requirement that the energy-momentum tensor be conserved then implies that $\varphi= - \frac{1}{2}L+\Lambda'$, where $L$ is the Lagrangian and $\Lambda'$ is an integration constant. Altogether, 
\be
P_{a}^{\;cde}R_{bcde}-2\nabla^{c}\nabla^{d}P_{acdb}-\frac{1}{2}g_{ab}L+\Lambda' g_{ab}=8\pi G T_{ab} \; ,
\ee
which we recognize as having the form of the generalized Einstein's equation for our theory of gravity, Eq. (\ref{eom}). Note, however, that the cosmological constant term does not match that in (\ref{eom}), unless the integration constant $\Lambda'$ is zero. For example, if the Lagrangian $L$ already includes a cosmological term $-2 \Lambda$, then the equation of motion derived from the action will have a term $\Lambda g_{ab}$ whereas the equation we derived from thermodynamics has a term $(\Lambda + \Lambda') g_{ab}$. This discrepancy can be traced to the fact that the Wald entropy is unaffected by the cosmological constant which does not contribute to $P_{abcd}$.

To summarize, we have shown that gravitational field equations for a broad class of diffeomorphism invariant theories -- not just general relativity -- arise from spacetime thermodynamics, namely, the Clausius relation $Q=T\Delta S_{\text{rev}}$. The Clausius relation is only one of many statements in thermodynamics, but makes an appearance in the first law of thermodynamics. A natural question to ask is what do the remaining contributions to the first law of thermodynamics correspond to in our picture of local holography. We turn to this question in the next section.


\subsection{A Local First Law of Gravity} \label{subsec:firstlaw}
\noindent

The fact that black holes carry a thermodynamic entropy (\ref{BHent}) suggests to us that that laws of black hole mechanics \cite{Bardeen73-1}, should really be interpreted as the laws of black hole thermodynamics \cite{Bekenstein73-1}. The first law, for a Schwarzschild black hole, is given by 
\be
\Delta M = T \Delta S \; , \label{bhfirstlaw}
\ee
where $M$ is the Arnowitt-Deser-Misner (ADM) mass of the black hole, $T$ is its Hawking temperature, and $S$ the Bekenstein-Hawking entropy. The first law of black holes should be compared to the first law of thermodynamics for macroscopic matter systems:
\be 
\Delta E = T \Delta S_{\rm rev} - W \; , \label{firstlawthermo}
\ee
where, by the Clausius theorem, $\Delta S_{\rm rev} = Q/T$ is the reversible component of the change in entropy.

Despite the superficial similarities between (\ref{firstlawthermo}) and (\ref{bhfirstlaw}), these expressions are rather different in character. First of all, the black hole law only applies, obviously, in the presence of a black hole. Also, unlike (\ref{firstlawthermo}), the black hole law is not local: the definition of an event horizon in general relativity involves the global causal structure of spacetime. Moreover, a formal definition of the mass term calls for special asymptotic boundary conditions, in particular asymptotic flatness; generically, energy density cannot simply be integrated over finite regions of space to obtain the total energy. Hence the left-hand side of (\ref{bhfirstlaw}) has no exact definition for the realistic case of, say, an astrophysical, uncharged black hole in an expanding universe. Another distinction is that, whereas in equation (\ref{firstlawthermo}) the system can exchange energy with a thermal reservoir, there is no physical process \cite{Gao:2001ut,Jacobson:2003wv} by which the ADM mass can change because the total energy at spacelike infinity in an asymptotically flat spacetime is a conserved quantity. Instead, the $\Delta M$ in (\ref{bhfirstlaw}) refers to differences in the ADM mass under a variation in the space of static uncharged black hole solutions. Finally, the work term is notably absent in (\ref{bhfirstlaw}); indeed, neither pressure nor spatial volume admits a straightforward definition for black holes \cite{Parikh:2005qs,Dolan:2011xt,Dolan:2012jh,Kubiznak:2016qmn}.

The observation that gravitational field equations arise from the Clausius relation allows us to derive a local first law of thermodynamics that also includes gravitational entropy, i.e., the  hybrid equation,
\be
\Delta E=T \, \Delta \! \left ( \frac{A_{\rm rev}}{4G \hbar} \right )- W \; , \label{localfirst}
\ee
combining attributes of (\ref{firstlawthermo}) and (\ref{bhfirstlaw}). We find that such an equation applies, within a suitably defined region, to all matter-gravity systems that are significantly smaller than the local curvature scale of spacetime. Amusingly (and somewhat mysteriously), we can express (\ref{localfirst}) in terms of fluid properties as
\be
\rho \Delta V = T \, \Delta \! \left ( \frac{A_{\rm rev}}{4G \hbar} \right ) - p \Delta V \; , \label{fluidlaw}
\ee
where $\rho$ and $p$ are the energy density and pressure measured by inertial observers, and $V$ is the volume of a ball in Euclidean space, namely $\frac{4}{3} \pi r^3$. 

Arriving to (\ref{localfirst}) relies on three uncommon elements: (i) energy $E$ is measured with respect to accelerating observers, rather than with respect to inertial observers; (ii) the geometry of the stretched future lightcone, i.e., a co-dimension 2-sphere of constant and uniformly outward radially accelerating observers, and (iii) the use of Einstein's equation to convert the heat flux through the hypersurface into the change in gravitational entropy (essentially the reverse steps of our thermodynamic derivation of Einstein's equations). 

We begin the derivation of (\ref{localfirst}) by studying the first law of thermodynamics for matter, as would be measured with respect to radially accelerating observers. The radially-accelerating observers have a normalized four-velocity vector $u_a \equiv \xi_a/(-\xi^2)^{1/2} \approx \xi_a/\alpha$, to leading order. Let the energy-momentum tensor be $T_{ab}$. Then the energy current measured by the accelerating observers is
\be
J^a =-T^{ab}u_{b} = -\frac{1}{\alpha} T^{ab} \xi_b \; . \label{energycurrent}
\ee
If $\xi_a$ were a Killing vector, this current would be conserved by Killing's equation. However, since $\xi_a$ is not a Killing vector, we have
\be 
\int_{M}d^{4}x\nabla_{a}J^{a} = -\frac{1}{\alpha}  \int_{M}d^{4}x T^{ab} \nabla_a \xi_b \; .
\ee
Applying the divergence theorem to the left-hand side and rearranging, we find
\be
\begin{split}
& \frac{1}{\alpha}\int_{B(\epsilon)} dS_a T^{ab}\xi_{b} 
- \frac{1}{\alpha}\int_{B(0)} dS_a T^{ab}\xi_{b}= \frac{1}{\alpha}\int_{\Sigma}d\Sigma_{a}T^{ab}\xi_{b} 
 - \frac{1}{\alpha}\int_{M}d^{4}xT^{ab}\nabla_{a}\xi_{b} \; , \label{stokes}
\end{split} 
\ee
where, in accordance with Stokes' theorem, the signs depend on whether a boundary is timelike or spacelike. Here $dS^{a}=N^{a}d^3x = \partial_t^a r^2 dr d \Omega$ and $d\Sigma^{a}=n^{a}d^3x \approx n^a dt (\alpha/r) r^2(t) d \Omega$, where $dt (\alpha/r)$ is the differential of proper time on the hyperboloid. We now argue that these terms can be interpreted as the change in energy, the heat flow, and the work done, so that (\ref{stokes}) is the first law of thermodynamics for matter. 

It is evident that $E(t)$, the energy of the system at time $t$, is given by $\frac{1}{\alpha}\int_{B(t)}dS_a T^{ab}\xi_{b}$, where $B(t)$ is the three-ball section of ${\cal M}$ at constant $t$. Not only does this expression have the correct dimension of energy, but $E(t)$ is simply the Noether charge associated with the energy current density, (\ref{energycurrent}). We then find that the difference between the energy at $t = \epsilon$ and $t = 0$ is
\be 
\Delta E = \frac{1}{\alpha}\int_{B(\epsilon)}dS_a T^{ab}\xi_{b}  - \frac{1}{\alpha}\int_{B(0)}dS_a T^{ab}\xi_{b} \; ,     \label{energy}
\ee
which is indeed the expression on the left-hand side of (\ref{stokes}). It is interesting to evaluate $\Delta E$ explicitly. We first note that, to leading order in Riemann normal coordinates, the energy-momentum tensor $T^{ab}(x) = T^{ab}(P) + {\cal O}(x)$ can be replaced within the integral by its value at $P$. Referring to the components of our Killing vetor $\xi$, we then see that the off-diagonal pieces of $T^{ab}$ integrate to zero because the integral of a Cartesian spatial coordinate over a ball centered at the origin vanishes. We are therefore left with $E(t) = \frac{4 \pi}{\alpha} T^{tt}(P) \int_0^{r(t)} dr r^2 N_t \xi_t$. We can approximate the radius of the ball by the radius of the hyperboloid. Hence $\Delta E = 2 \pi T^{tt}(P) \alpha \epsilon^2$, using also $\epsilon \ll \alpha$. Similarly, the volume of $B(t)$ is $V(t) = \frac{4}{3} \pi (\alpha^2 + t^2)^{3/2}$. Then the difference between the volume of $B(\epsilon)$ and of $B(0)$ is
\be
\Delta V = 2 \pi \alpha \eps^2 \; . \label{volume}
\ee
Labeling the energy density $\rho \equiv T^{tt}(P)$, we obtain 
\be
\Delta E = \rho \Delta V \; . \label{rhodV}
\ee
It is amusing that, even though $\Delta E$ is the difference in energies as measured by accelerating observers, it can nevertheless be written in terms of $\rho$ and $\Delta V$, the energy density and volume change measured by inertial observers; it is not the case, though, that $E(t) = \rho V(t)$.

Next, consider the first term on the right in (\ref{stokes}). This is clearly the integrated energy flux into the timelike surface $\Sigma$. The sign matches too: the normal to $\Sigma$ is outward-pointing, while the energy current, $J^a$, is defined with a minus sign, (\ref{energycurrent}). Now, in thermodynamics, heat is the energy flowing into macroscopically unobservable degrees of freedom. For our observers on the stretched future light cone, the interior of the system is fundamentally unobservable, being causally disconnected. We can therefore interpret the integrated energy flux into the system as heat \cite{Jacobson:1995ab}:
\be 
Q=\frac{1}{\alpha}\int d\Sigma_{a}T^{ab}\xi_{b} \; .   \label{deltaQ}
\ee
This interpretation will be confirmed when we incorporate gravity. 

Finally, consider the last term in (\ref{stokes}). At first sight, this term does not appear to be a work term because it is an integral over a four-volume. To see that it is, consider first for simplicity a diagonal energy-momentum tensor with isotropic pressure, $T^{ij}(P)=p\delta^{ij}$. Then, working as always at leading order, we find
\be 
\frac{1}{\alpha}\int_{M}d^{4}xT^{ab}\nabla_{a}\xi_{b}\approx\frac{1}{\alpha}\int_{M}d^{4}x\frac{2pt}{r} \approx 2\pi p\alpha\epsilon^{2} \; ,
\ee
where, in the last step, we have evaluated the integral at leading order in $\epsilon$. From (\ref{volume}), we see that this is exactly equal to $p \Delta V$, the pressure-volume work done by a system, motivating the identification of the last term in (\ref{stokes}) as work.

More generally, consider an arbitrary energy-momentum tensor, for which $T^{ii}(P)=p_{i}$, and $T^{ij}\neq0$ for $i\neq j$. Now from (\ref{Killingfailure}), we have $\partial_i \xi_j \sim \frac{t}{r^{3}}x_i x_j$ for $i \neq j$. This is an odd function of the coordinates and therefore $T^{ij} \partial_i \xi_j$ vanishes under integration over the three-ball for $i \neq j$. 
Moreover, $T^{xx}\partial_{x}\xi_{x}=p_{x}\frac{t}{r^{3}}\left(y^{2}+z^{2}\right)$, and similarly for $T^{yy}$ and $T^{zz}$. Then we find
\be
W = \frac{1}{\alpha}\int_{M}d^{4}xT^{ab}\nabla_{a}\xi_{b} = \left (\frac{1}{3} \sum_{i=1}^{3}p_{i} \right ) \Delta V \; , \label{work}
\ee
which is precisely the pressure-volume work for anisotropic pressures, and is now valid for arbitrary energy-momentum tensors. 

Consulting (\ref{energy}), (\ref{deltaQ}), and (\ref{work}), we indeed find that (\ref{stokes}) can be interpreted as a first law of thermodynamics for accelerating observers moving along $\Sigma$. Our first law is local in that it is valid near an arbitrary point in a generic spacetime. As it stands though, this equation does not yet involve gravity: there is no Newton's constant and all the terms involve the energy-momentum tensor of matter, $T^{ab}$. To turn it into a local first law with gravity, we now invoke Einstein's equation.

Using Einstein's equation, $R_{ab} - \frac{1}{2} R g_{ab} + \Lambda g_{ab} = 8 \pi G T_{ab}$, in (\ref{deltaQ}) we find $Q = \frac{1}{8 \pi G \alpha}\int_{\Sigma} d\Sigma_{a} R^{ae} \xi_{e}$. The terms proportional to the metric vanish when contracted with $d \Sigma_a$ and $\xi_b$ because $\xi_a$ lies along $\Sigma$ while $n_a$ is normal to it.

Now if $\xi_a$ were a Killing vector, it would obey Killing's identity: $\nabla_b \nabla_c \xi_d = R^{e}_{\;bcd} \xi_e$. However, we already know that $\xi_a$ is not exactly a Killing vector. We therefore have $\nabla_b \nabla_c \xi_d - R^{e}_{\;bcd} \xi_e = f_{bcd}$ where $f_{bcd}$ encodes the failure of Killing's identity to hold. Then
\be
Q =\frac{1}{8\pi G \alpha}\int_{\Sigma}d\Sigma_{a} \frac{1}{2} (g^{ac} g^{bd} - g^{ad}g^{bc})  ( \nabla_{b}\nabla_{c}\xi_{d} - f_{bcd} ) \; . \label{Qint}
\ee 
We now show that the integral of the $\nabla_{b}\nabla_{c}\xi_{d}$ term evaluates to $T \Delta S$, by essentially reversing the thermodynamic derivation of Einstein's equations in the Noether charge approach \cite{Parikh:2017aas}. First, we use Stokes' theorem for an antisymmetric tensor field $A^{ab}$,
namely $\int_{\Sigma}d\Sigma_{a}\nabla_{b}A^{ab}=-\oint_{\partial\Sigma}dS_{ab}A^{ab}$, to express that integral as the difference of terms $-\frac{1}{8 \pi G \alpha} \int dS_{ab} \frac{1}{2} (g^{ac} g^{bd} - g^{ad}g^{bc}) \nabla_{c}\xi_{d}$ evaluated over the two-spheres at time $t = 0$ and $t=\epsilon$. Here $dS_{ab} = dA \frac{1}{2} (n_a u_b - u_a n_b)$. Then, since $u_a \approx \xi_a/\alpha$, we have
\be
-\frac{1}{16 \pi G \alpha^2} \int dA (n^c \xi^d - n^d \xi^c) \nabla_c \xi_d = +\frac{A}{8 \pi G \alpha} = T \frac{A}{4 G \hbar} \; .
\ee
Here we used the fact, (\ref{Killingfailure}), that the projection of $\nabla_c \xi_d$ in the $n-\xi$ plane is antisymmetric. We then made use of our judicious choice of $\Sigma$ as a surface of constant acceleration and thus temperature in writing $\xi^c \nabla_c \xi_d = \alpha n_d$ and in using $T=\frac{\hbar}{2 \pi \alpha}$. Hence the integral of the $\nabla_{b}\nabla_{c}\xi_{d}$ term can be written as $T \Delta S$, where $S$ is precisely the Bekenstein-Hawking entropy, suggesting that gravitational entropy can be associated with sections of $\Sigma$. 

We can manage the failure of Killing's identity, encoded in the $f_{bcd}$ term in the $Q$ integral, (\ref{Qint}), following the prescription described above (and further detailed in \cite{Parikh:2017aas}). This leads us to 
\be 
Q = T \Delta S- T\Delta S_{\rm hyp} \equiv T \Delta S_{\rm rev} \; .
\label{delQrev}\ee
where $\Delta S_{\rm rev}$ is the reversible part of the change in gravitational entropy, having subtracted the irreversible background expansion of the hyperboloid. A direct calculation using (\ref{deltaQ}) shows that $Q = (\rho + \frac{1}{3}\sum_i p_i) \Delta V$. Hence we have that $\Delta S_{\rm rev} \geq 0$ if the null energy condition holds.

Putting everything together, we arrive at our result:
\be
\Delta E=T \, \Delta \! \left ( \frac{A_{\rm rev}}{4G \hbar} \right )- W \; . \label{localfirstlaw}
\ee
We have found a hybrid first law that resembles both the ordinary first law of thermodynamics for matter (in that it is valid locally and has a work term) as well as the first law for black holes (in that it involves gravitational entropy). Using (\ref{rhodV}) and (\ref{work}), we can also put this in the form (\ref{fluidlaw}). In (\ref{localfirstlaw}), $\Delta E$ and $W$ refer to the energy of and work done by matter, while the middle term refers to the entropy of gravity. The result suggests that (stretched) future light cones possess thermodynamic entropy, which is perhaps not unreasonable as their interiors are causally disconnected from the outside. Note the absence of a term corresponding to the entropy of matter. This property is reminiscent of black holes: if one empties a cup of hot coffee into a black hole, the black hole's entropy increases solely due to the mass-energy of the coffee, with no extra contribution from the coffee's own thermal entropy. It is also notable that, because all terms vanish when $T_{ab}$ is zero, there is no contribution of gravitational energy in our local first law; indeed, inclusion of such energy would require a quasi-local conservation law \cite{McGrath:2012db}.

Our local first law can be extended to higher-dimensional spacetime; in particular, (\ref{delQrev}) always corresponds to subtracting the inherent area increase of the hyperboloid. More significantly, the derivation can also be extended to a broad class of higher-curvature theories of gravity: replacing the Einstein-Hilbert Lagrangian with a more general diffeomorphism-invariant theory of gravity $L=L(g^{ab},R^{abcd})$, and the Bekenstein-Hawking entropy with the Wald entropy, we arrive to
\beq \Delta E = T \Delta S^{\rm Wald}_{\rm rev} - W\;.\eeq

Historically, the laws of black hole mechanics supported, as an analogy, Bekenstein's idea that a black hole could be attributed thermodynamic entropy proportional to the horizon area; this was found to be literally true with the discovery that black holes have temperature. Here we have shown that the first law holds locally on stretched future light cones generated by families of accelerating observers, thereby supporting an analogy between entropy and (in Einstein gravity) the area of such surfaces. But since it is already known that accelerating observers perceive a temperature, our result suggests that stretched future light cones can indeed be regarded as having thermodynamic entropy. 


\subsection{Gravity from Causal Diamond Thermodynamics} \label{sec:gravfromCD}

Earlier we extended Jacobson's original derivation of Einstein's fields equations to more general theories of gravity. Motivated by the local Rindler horizon construction, we considered the stretched future lightcone. The surface is a timelike stretched horizon of the future of a lightcone generated by radial boost vectors, and, in this sense, the stretched future lightcone can be interpreted as a local spherical Rindler horizon. Since the stretched future lightcone defines a surface of constant acceleration $a$, we understand it as a system in thermal equilibrium with temperature $T\propto a$. From here we applied an elementary statement in equilibrium thermodynamics, namely, the Clausius relation $T\Delta S_{\text{rev}}=Q$, and found it was geometrically equivalent to the non-linear field equations for arbitrary theories of gravity.

Critical to our derivation was that the stretched lightcone has compact spherical sections. Moreover, while the radial boost vector $\xi$ was not an exact Killing vector, we saw that it could be treated as an approximate Killing vector, and, in fact, the way it failed to be a Killing vector, it succeeded in being a conformal Killing vector. In this sense, the stretched future lightcone, within a certain limit, can be understood to be a conformal Killing horizon. An obvious question then is whether the stretched future lightcone is the only kind of local screen one could consider to derive gravitational field equations. 

There is, in fact, another kind of holographic screen which shares many of the same features of the stretched lightcone: a causal diamond. As we will study below, causal diamonds have spherical subregions, and are generated by a true conformal Killing vector (in pure Minkowski space), and whose boundary defines a conformal Killing horizon with constant surface gravity. The constant surface gravity allows for one to interpret the causal diamond as a system in thermal equilibrium for which the standard principles of equilibrium thermodynamics\footnote{For a more thorough review of causal diamond thermodynamics in maximally symmetric spaces, see \cite{Jacobson:2018ahi}.} may be applied. 

It is then natural to use the techniques developed above but applied to causal diamonds in Minkowski space. This is the central goal of this section: Derive gravitational field equations via the Clausius relation, substituting the stretched lightcone for the causal diamond. We should remark that causal diamonds make an appearance in another context: spacetime entanglement \cite{Jacobson16-1,Bueno16-1}. In Appendix \ref{app:cdmechanicsandEE} we show precisely how a constant volume variation of the entanglement  entropy attributed to a causal diamond yields gravitational field equations. Due to the similarities between causal diamonds and stretched lightcones, this further motivates us to look for an entanglement interpretation for stretched lightcone thermodynamics, the subject of Chapter \ref{sec:gravfroment}. 


\subsubsection{Geometry of Causal Diamonds}

\begin{figure}[t]
\includegraphics[width=6.5cm]{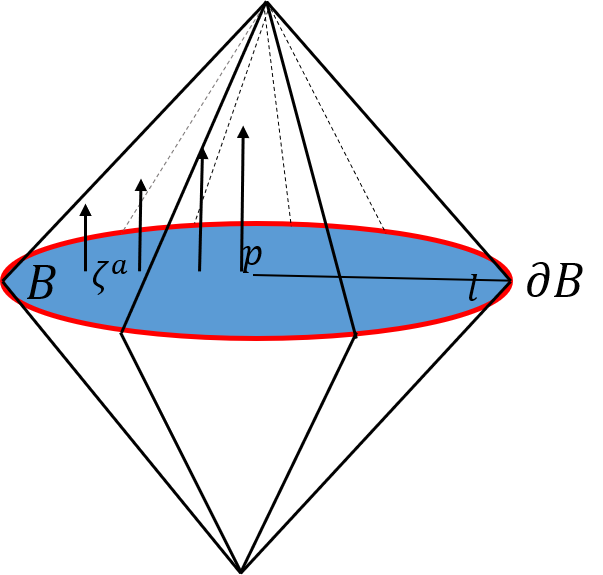}
\centering
\caption[The Causal Diamond in Minkowski Space.]{The causal diamond as the union of future and past domains of dependence of the spatial balls $B$ of size $\ell$ with boundary $\partial B$. The diamond admits a conformal Killing vector $\zeta^{a}$ whose flow preserves the diamond, and vanishes at the boundary $r=\pm\ell$.}
\label{causaldiamond2}\end{figure}

In a maximally symmetric background, a causal diamond can be defined as the union of future and past domains of dependence of its spatial slices, balls $B$ of size $\ell$ with boundary $\partial B$. The diamond admits a conformal Killing vector (CKV) $\zeta^{a}$ whose flow preserves the diamond (see figure (\ref{causaldiamond2})). 

Conformal Killing vectors are those which satisfy conformal Killing's equation 
\beq \nabla_{a}\zeta_{b}+\nabla_{b}\zeta_{a}=2\Omega g_{ab}\;,\label{confKeqn}\eeq
where $\Omega$ satisfies
\beq \Omega=\frac{1}{D}\nabla_{c}\zeta^{c}\;,\eeq
and is related to the conformal factor $\omega^{2}$ of $\bar{g}_{ab}=\omega^{2}g_{ab}$ via $2\Omega=\zeta^{c}\nabla_{c}\ln\omega^{2}$. 

Conformal Killing vectors also satisfy the conformal Killing identity
\beq \nabla_{b}\nabla_{c}\zeta_{d}=R^{e}_{\;bcd}\zeta_{e}+(\nabla_{c}\Omega)g_{bd}+(\nabla_{b}\Omega)g_{cd}-(\nabla_{d}\Omega)g_{bc}\;.\label{confKid}\eeq
Following the discussion above, in an arbitrary spacetime the conformal Killing vectors will become approximate conformal Killing vectors, failing to satisfy the conformal Killing equation to order $\mathcal{O}(x^{2})$ in a RNC expansion about some point $p$, and the conformal Killing identity to $\mathcal{O}(x)$. 

We can define a timelike normal $U^{a}$ to $B$ via
\beq U_{a}=N\nabla_{a}\Omega\;,\eeq
with
\beq N=||\nabla_{a}\Omega||^{-1}\;,\eeq
being some normalization such that $U^{2}=-1$. In fact, it can be shown in general that
\beq N=\frac{D-2}{\kappa K}\;, \label{NKk}\eeq
where $\kappa$ is the surface gravity and $K$ is the trace of the extrinsic curvature.

One also has
\beq \nabla_{d}(\mathcal{L}_{\zeta}g_{ab})|_{B}=\frac{2}{N}U_{d}g_{ab}\quad \nabla_{a}\zeta_{b}|_{\partial B}=\kappa N_{ab}\;,\eeq
where we have the binormal $N_{ab}=2U_{[a}N_{b]}$, where $N_{a}$ is the spacelike unit normal to $U_{b}$. The spatial slice $B$ is taken to be the $t=0$ slice. 

For concreteness, in $D$-dimensional Minkowski space, the CKV which preserves the causal diamond is \cite{Bueno16-1}
\beq
\begin{split}
 \zeta^{a}&=\left(\frac{\ell^{2}-r^{2}-t^{2}}{\ell^{2}}\right)\partial^{a}_{t}-\frac{2rt}{\ell^{2}}\partial^{a}_{r}\\
& = \left(\frac{\ell^{2}-r^{2}-t^{2}}{\ell^{2}}\right)\partial^{a}_{t}-\frac{2x^{i}t}{\ell^{2}}\partial^{a}_{i}\;.\label{zeta}
\end{split}
\eeq
We point out that $\zeta^{a}$ goes null on the boundary, $t=\ell\pm r$, and $\zeta^{2}=-1$ when $r=t=0$. We also have
\beq U^{a}=\partial^{a}_{t}\quad N^{a}=\partial^{a}_{r}\Rightarrow N_{ab}=2\nabla_{[a}r\nabla_{b]}t\;,\eeq
\beq \Omega=-\frac{2t}{\ell^{2}}\quad \nabla_{a}\Omega=-\frac{2\nabla_{a}t}{\ell^{2}}=2\frac{U_{a}}{\ell^{2}}\;,\eeq
and, 
\beq N=\frac{\ell^{2}}{2}\quad K_{\partial B}=\frac{(D-2)}{\ell}\;. \label{NandK}\eeq
We see that the causal diamond has constant extrinsic curvature, constant surface gravity $\kappa=2/\ell$, and $\zeta^{a}$ is an exact Killing vector on the $t=0$ surface $B$. 

Let us remark on the similarities between the radial boost vector $\xi_{a}$ (\ref{xi}) generating the stretched future lightcone, and the conformal Killing vector $\zeta_{a}$ (\ref{zeta}) preserving the causal diamond. Specifically, we find that $\xi_{a}$ satisfies
\beq \nabla_{a}\xi_{b}+\nabla_{b}\xi_{a}=2\left(\frac{t}{ r}\right)\left(\eta_{ij}-\frac{x_{i}x_{j}}{r^{2}}\right)\delta^{i}_{a}\delta^{j}_{b}\;, \label{Keqforu}\eeq
where the $\delta^{i}_{a}\delta^{j}_{b}$ are present to project the non-zero contributions. We see that $\xi^{a}$ is a vector which satisfies Killing's equation in specific metric components, and one which fails as a modified CKV in other components. This comparison leads us to define a conformal factor associated with $\xi$:
\beq \Omega_{\xi}\equiv\frac{1}{(D-2)}\nabla_{c}\xi^{c}=\frac{t}{ r}\;,\eeq
for which one finds 
\beq \nabla_{d}\Omega_{\xi}=-\frac{1}{r^{2}}\xi_{d}\;,\quad N_{\xi}^{-1}\equiv||\nabla_{a}\Omega_{\xi}||=\frac{\alpha}{r^{2}}\;,\eeq
and
\beq u_{a}=N_{\xi}\nabla_{a}\Omega_{\xi}\;.\eeq
It is also straightforward to work out
\beq \nabla_{d}(\mathcal{L}_{\xi}g_{ab})|_{t=0}=\frac{2}{N_{\xi}}u_{d}\delta^{i}_{a}\delta^{j}_{b}\left(\eta_{ij}-\frac{x_{i}x_{j}}{r^{2}}\right)\;, \label{nabLg}\eeq
and 
\beq K_{\partial\Sigma}=\frac{1}{\alpha}(D-2)\;,\eeq
where $\mathcal{L}_{\xi}$ is the Lie derivative along $\xi_{a}$, and the extrinsic curvature of the spherical boundary $\partial\Sigma$ is $K=h^{ab}K_{ab}=g^{ab}\nabla_{b}n_{a}$, since $h_{ab}=g_{ab}-n_{a}n_{b}$.


\subsubsection{Causal Diamond Thermodynamics}
\indent

Consider the past of the causal diamond, i.e., the bottom half below the $t=0$ co-dimension-2 spherical slice $\partial B$ of Fig. 2\footnote{We focus on the past of the causal diamond for reasons which we will discuss later.}. Our picture for a physical process will be comparing the entropy between a time slice at $t=-\epsilon$ for positive $\epsilon$ and $t=0$ after some energy flux has entered the past of the diamond. At the boundary $t=\ell\pm r$, $\zeta^{2}=0$, and therefore, in Minkowski space, the boundary of the causal diamond represents a conformal Killing horizon of constant surface gravity $\kappa$, and therefore an isothermal surface with Hawking temperature $T=\kappa/2\pi$. An arbitrary spacetime will include curvature corrections, however, to leading order in a RNC expansion about a point $p$, $\zeta^{2}\approx0$, and $\kappa$ remains approximately constant. If we followed the worldline of $\zeta$ from time $t=-\epsilon$ to $t=0$, we would find that $\kappa$ would be different at each of these time slices. Motivated by the set-up of the stretched lightcone, we choose a timescale $\epsilon\ll\ell$ over which the surface gravity $\kappa$ is approximately constant. Therefore, in an arbitrary spacetime $\partial B$ of the causal diamond represents a local conformal Killing horizon, which may be interpreted as an isothermal surface with constant Hawking temperature $T=\kappa/2\pi$. 

We associate with this conformal Killing horizon a gravitational entropy \cite{Nielsen:2017hxt}, i.e., time-slices $\partial B$ of the causal diamond have an attributed entropy. The form of the entropy depends on the theory of gravity under consideration, e.g., for Einstein gravity, the correct form is the Bekenstein-Hawking entropy (\ref{BHent}). Here we consider a diffeomorphism invariant theory of gravity in $D$ spacetime dimensions defined by the action $I$:
\beq I=\frac{1}{16\pi G}\int d^{D}x\sqrt{-g}L\left(g^{ab},R_{abcd}\right)+I_{\text{matter}}\;.\eeq
whose equations of motion we repeat for ease of the reader
\beq P_{a}^{\;cde}R_{bcde}-2\nabla^{c}\nabla^{d}P_{acdb}-\frac{1}{2}Lg_{ab}=8\pi G T_{ab}\;.\label{eqnsofmotion}\eeq
It is straightforward to verify that in the case of Einstein gravity, $L=R$, this reduces to Einstein's field equations. 

For a general theory of gravity of this type we must generalize the Bekenstein-Hawking entropy formula. We take this generalization to be the Wald entropy \cite{Wald:1993nt}:
\beq S_{\text{Wald}}=\frac{1}{8G\kappa}\int dS_{ab}J^{ab}\;,\eeq
where we have introduced the Noether potential associated with a diffeomorphism $x^{a}\to x^{a}+\zeta^{a}$, where we will take $\zeta^{a}$ to be a timelike (conformal) Killing vector, 
\beq J^{ab}=-2P^{abcd}\nabla_{c}\zeta_{d}+4\zeta_{d}\nabla_{c}P^{abcd}\;,\quad P^{abcd}\equiv\frac{\partial L}{\partial R_{abcd}}\;,\eeq
and have infinitesimal binormal element of $\partial B$:
\beq dS_{ab}\equiv\frac{1}{2}(N_{a}U_{b}-N_{b}U_{a})dA=\frac{1}{2}N_{ba}dA\;.\eeq
Wald's Noether charge construction of gravitational entropy was originally developed to yield an expression for the entropy of a stationary black hole in more general theories of gravity. Here we make the non-trivial assumption of local holography that this gravitational entropy can also be attributed locally to the spatial sections of causal diamonds whose structure is preserved by $\zeta_{a}$. 

For computational convenience, we will first not work directly on the horizon, but instead work on the timelike stretched horizon of the causal diamond -- a co-dimension-1 timelike surface we call $\Sigma$. At the end of the calculation we will take the limit where our stretched horizon coincides with the conformal Killing horizon. The fact that we have to take the step in which we move to the conformal  Killing horizon -- a null hypersurface -- is a marked difference with the analogous calculation using stretched future lightcones \cite{Parikh:2017aas}. 

The Wald entropy at time $t$ is
\beq S_{\text{Wald}}=-\frac{1}{4G\kappa}\int_{\partial B(t)}dS_{ab}(P^{abcd}\nabla_{c}\zeta_{d}-2\zeta_{d}\nabla_{c}P^{abcd})\;.\eeq
The total change in entropy between $t=0$ and $t=-\epsilon$ is $\Delta S_{\text{Wald}}=S_{\text{Wald}}(0)-S_{\text{Wald}}(-\epsilon)$, or, 
\beq \Delta S_{\text{Wald}}=\pm\frac{1}{4G\kappa}\int_{\Sigma}d\Sigma_{a}\nabla_{b}(P^{abcd}\nabla_{c}\zeta_{d}-2\zeta_{d}\nabla_{c}P^{abcd})\;, \label{DeltaS1}\eeq
where we have invoked Stokes' theorem for an antisymmetric tensor field $M^{ab}$:
\beq \int_{\Sigma}d\Sigma_{a}\nabla_{b}M^{ab}=\pm\left[\int_{\partial B(0)}dS_{ab}M^{ab}-\int_{\partial B(-\epsilon)}dS_{ab}M^{ab}\right]\;,\eeq
where the overall sign depends on whether $\Sigma$ is timelike ($-$), or spacelike ($+$). For our discussion of causal diamond thermodynamics we are interested in the timelike version, however, it will be illustrative for future discussion if we do not specify, for now, the signature of co-dimension-1 surface $\Sigma$. 

Moving on, we have
\beq 
\begin{split}
\Delta S_{\text{Wald}}&=\pm\frac{1}{4G\kappa}\int_{\Sigma}d\Sigma_{a}\{-\nabla_{b}(P^{adbc}+P^{acbd})\nabla_{c}\zeta_{d}+P^{abcd}\nabla_{b}\nabla_{c}\zeta_{d}-2\zeta_{d}\nabla_{b}\nabla_{c}P^{abcd}\}\;.
\end{split}
\eeq
We have yet to use any properties of $\zeta_{d}$, which to leading order is a conformal Killing vector, satisfying (\ref{confKeqn}) and  (\ref{confKid}). We have then:
\beq 
\begin{split}
\nabla_{b}(P^{adbc}+P^{acbd})\nabla_{c}\zeta_{d}&=\nabla_{b}P^{adbc}(\nabla_{c}\zeta_{d}+\nabla_{d}\zeta_{c})\\
&=2\Omega g_{cd}\nabla_{b}P^{adbc}\;,
\end{split}
\label{nabPnabz}\eeq
and
\beq
\begin{split}
 P^{abcd}\nabla_{b}\nabla_{c}\zeta_{d}&=P^{abcd}[R_{ebcd}\zeta^{e}+(\nabla_{c}\Omega)g_{bd}-(\nabla_{d}\Omega)g_{bc}]\\
&=P^{abcd}R_{ebcd}\zeta^{e}+2P^{abcd}(\nabla_{c}\Omega)g_{bd}\;,
\end{split}
\label{Pnabnabz}
\eeq
where we used that $P^{abcd}$ shares the same algebraic symmetries of the Riemann tensor. Substituting (\ref{nabPnabz}) and (\ref{Pnabnabz}) into (\ref{DeltaS1}) yields
\beq 
\begin{split}
\Delta S_{Wald}&=\pm\frac{1}{4G\kappa}\int_{\Sigma}d\Sigma_{a}\{P^{abcd}R_{ebcd}\zeta^{e}-2\zeta_{d}\nabla_{b}\nabla_{c}P^{abcd}+2P^{abcd}(\nabla_{c}\Omega)g_{bd}-2\Omega g_{cd}\nabla_{b}P^{adbc}\}\;,
\end{split}
\label{DeltaS2}\eeq
where the overall $+$ ($-$) sign indicates that $\Sigma$ is a timelike (spacelike) surface. In Appendix \ref{app:cdmechanicsandEE}, we consider the spacelike surface and provide an alternative derivation to the first law of causal diamond mechanics for higher derivative theories of gravity as presented in \cite{Bueno16-1}. 

Using that $d\Sigma_{a}=N_{a}dAd\tau=\partial_{a}^{r}dAd\tau=x_{i}/r\partial_{a}^{i}dAd\tau$, and that we are integrating over a spherically symmetric  region, we find that to leading order in the RNC expansion, that the final two terms integrate to zero since we are integrating over a timelike surface with spherical compact sections. Thus, to leading order, 
\beq \Delta S_{\text{Wald}}\approx\frac{1}{4G\kappa}\int_{\Sigma}d\Sigma_{a}(P^{abcd}R_{ebcd}\zeta^{e}-2\zeta_{d}\nabla_{b}\nabla_{c}P^{abcd})\;.\label{DeltaStotlead}\eeq
The two terms we neglect here, of course, have higher order contributions due to the RNC expansion, and in order to derive the non-linear equations of motion we must deal with these higher order contributions. We follow the technique developed in \cite{Parikh:2017aas}, in which we modify the conformal Killing vector $\zeta_{a}$ by adding $\mathcal{O}(x^{3})$ corrections and higher such that they remove the undesired higher order effects of the two terms we neglect. The details may be found in the Appendix \ref{sec:appendB}. 

The above expression (\ref{DeltaStotlead}) represents the leading order contribution to the total entropy variation, including the effect due to the natural increase of the spatial sections of the (past) causal diamond -- an irreversible thermodynamic process. Presently we are interested in the change in entropy due to a flux of matter crossing the conformal horizon -- a reversible thermodynamic process\footnote{We can consider the following analogy to help describe this process and our use of the terms `irreversible' and reversible': Imagine we have a box a gas sitting on a burner. When the box opens the gas will leave the box simply due to a free expansion, which has an associated irreversible entropy increase. The heating of the box will also lead to a reversible entropy increase. The natural increase of our diamond -- to the past of $t=0$ -- is analogous to the free expansion of the gas and we therefore identify this process as having an associated irreversible entropy increase.}. We therefore remove the entropy due to the natural increase of the diamond $\bar{S}$:
\beq
\begin{split} 
\bar{S}&=-\frac{1}{4G\kappa}\int_{\partial B}dAN_{i}U_{t}\left[P^{ittj}2\partial_{t}\zeta_{j}+P^{tijk}\partial_{j}\zeta_{k}\right]\\
&=\frac{1}{4G\kappa}\int_{\partial B}dA\frac{4}{r\ell^{2}}x_{i}x_{j}P^{ittj}\\
&=\frac{1}{4G\kappa}\frac{2\kappa K}{(D-2)}\frac{1}{(D-1)}\left(\sum_{i}P^{itti}\right)\Omega_{D-2}r^{D-1}\;,
\end{split}
\eeq
where to get to the second line we used that $\partial_{i}\zeta_{j}\propto\delta_{ij}$, which cancels with its contraction with $P^{tijk}$, and $\partial_{t}\zeta_{j}=-2x_{j}/\ell^{2}$, and in the third line we used that $2/\ell^{2}=\kappa K/(D-2)$, and again the fact we are integrating over a spherical subregion. To this order $P^{abcd}$ is constant, allowing us to pull it through the integral.

We may arrange the above suggestively as\footnote{As written, $\bar{S}$ is a bit misleading. It would appear that $\bar{S}$ goes like the volume rather than the area. However, this is in fact not the case. Indeed, in the case of general relativity, using $K=(D-2)/\ell$, and that on the $t=0$ slice $\partial B$, $r=\ell$, it is straightforward to show that $\bar{S}=A/4G$, where $A$ is the area of the spherical subregion $\partial B$.}
\beq \bar{S}=\frac{1}{2G}\frac{K}{(D-2)}\left(\sum_{i}P^{itti}\right)\int_{B}dV\;.\eeq
This expression\footnote{In the context of general relativity, we note that the this expression is nothing more than the Smarr formula for a maximally symmetric ball in flat space -- the ``thermodynamic volume" is notably absent \cite{Jacobson:2018ahi}. This is because we are considering perturbations about Minkowski spacetime. Even if we considered perturbations about a more general MSS, the thermodynamic volume would be subdominant.} is recognized to be the leading contribution  of the \emph{generalized volume} $\bar{W}$ (\ref{Wbar})
\beq
\begin{split}
\frac{K}{2G(D-2)}\int_{B}dVP^{abcd}U_{a}U_{d}h_{bc}\equiv\frac{K}{2G}\bar{W}\;,
\end{split}
\eeq
that is, 
\beq\Delta\bar{S}=\frac{K}{2G}\Delta\bar{W}\;,\label{natentincreasediamond}\eeq
where $\Delta \bar{S}=\bar{S}(0)-\bar{S}(-\epsilon)$, and $\Delta \bar{W}=\bar{W}(0)-\bar{W}(-\epsilon)$. The generalized volume $W$ (\ref{Wvol}),
\beq W=\frac{1}{(D-2)P_{0}}\int_{B}dV(P^{abcd}U_{a}U_{d}h_{bc}-P_{0})\;,\eeq
was introduced in \cite{Bueno16-1} as the higher derivative analog of the spatial volume $V$ of the causal diamond and is kept fixed in the higher derivative extension of maximal entropy condition (\ref{maxentcondhigher}). The theory dependent constant $P_{0}$ defined by $P^{abcd}$ in a maximally symmetric solution to the field equations, $P^{abcd}_{MSS}=P_{0}(g^{ac}g^{bd}-g^{ad}g^{bc})$. In the case of Einstein gravity it is straightforward to show that $W$ reduces to $V$. The construction of $W$ is reviewed in more detail in Appendix \ref{app:cdmechanicsandEE}.

We see from (\ref{natentincreasediamond}) that the entropy change $\Delta\bar{S}$ due to the natural increase of the diamond is proportional to the change of the generalized volume $\Delta\bar{W}$.  Since the area on a future time slice $\partial B(0)$ is smaller than the that of $\partial B(-\epsilon)$, one has $\Delta\bar{S}>0$. Note that this is not the case for time-slices to the future of $t=0$, and therefore the thermodynamics of causal diamonds is peculiar; we will have more to say about this in the discussion. 

We thus define the reversible entropy variation as
\beq
\begin{split}
 \Delta S_{\text{rev}}&\equiv\Delta S_{\text{Wald}}-(\Delta\bar{S})=\Delta S_{\text{Wald}}-\frac{K}{2G}\Delta\bar{W}\\
&=\frac{1}{4G\kappa}\int_{\Sigma}d\Sigma_{a}\left(P^{abcd}R_{ebcd}\zeta^{e}-2\zeta_{d}\nabla_{b}\nabla_{c}P^{abcd}\right)\;.
\end{split}
\label{DSrev}
\eeq
Calling this variation the reversible change in entropy is analogous to the Clausius relation in ordinary thermodynamics $Q=T\Delta S_{\text{rev}}$. 


\subsubsection{Gravity from Thermodynamics}
\indent

Next, following \cite{Jacobson:1995ab,Parikh:2017aas}, define the integrated energy flux across $\Sigma$ as 
\beq Q=\int_{\Sigma}d\Sigma_{a}T^{ab}\zeta_{b}\;,\eeq
where the energy momentum tensor can be approximated to leading order by its value at $p$. As we make the transition to the conformal Killing horizon, the interior of $\Sigma$ becomes causally disconnected from its exterior, allowing us to identify $Q$ as heat -- energy which flows into macroscopically unobservable degrees of freedom. 

The Clausius relation $T\Delta S_{\text{rev}}=Q$ for our set-up results in the geometric constraint:
\beq
\begin{split}
&\int_{\Sigma}d\Sigma_{a}\left(P^{abcd}R_{ebcd}\zeta^{e}-2\zeta_{d}\nabla_{b}\nabla_{c}P^{abcd}\right)=8\pi G\int_{\Sigma}d\Sigma_{a}T^{ab}\zeta_{b}\;.
\end{split}\eeq
Since this holds for all causal diamonds $\Sigma$, we may equate the integrands leading to
\beq (P^{aecd}R_{becd}-2\nabla^{d}\nabla^{c}P_{abcb})N^{a}\zeta^{b}=8\pi GT_{ab}N^{a}\zeta^{b}\;.\eeq
At the boundary, $t=\ell+r$, i.e., when the timelike stretched surface moves to the conformal Killing horizon, one has $g_{ab}N^{a}\zeta^{b}=0$. Therefore, at the conformal Killing horizon, the above is valid up to a term of the form $f g_{ab}$, where $f$ is some yet to be determined scalar function. The form of $f$ can be determined by demanding covariant conservation of $T_{ab}$. Specifically, we are led to 
\beq P^{aecd}R_{becd}-2\nabla^{d}\nabla^{c}P_{abcb}-\frac{1}{2}Lg_{ab}+\Lambda g_{ab}=8\pi GT_{ab}\;,\eeq
where $L(g^{ab},R_{abcd})$, and $\Lambda$ is some integration constant. We recognize the above as the equations of motion for a general theory of gravity. In this way we see that the equations of motion for a theory of gravity arise from the thermodynamics of causal diamonds. We have reproduced the results of \cite{Parikh:2017aas}, however, using the geometric construction of causal diamonds. 

This approach to deriving the equations of motion offers a thermodynamic perspective to the derivation of linearized equations of motion from the entanglement equilibrium proposal as presented in \cite{Bueno16-1}. In particular, we found that the generalized volume $\bar{W}$ can be interpreted as the natural increase of the causal diamond. To apply the Clausius relation for a reversible thermodynamic process, we removed this increase and, therefore, $\bar{W}$ is the contribution which generates irreversible thermodynamic processes in the causal diamond construction. We note that removing $\bar{W}$ also appears in the first law of causal diamond mechanics (\ref{1stlawcdoff}), and consequently the entanglement equilibrium condition (\ref{1stlawEEmod}).

It is interesting to compare the above construction with that of the stretched future lightcone. As shown in \cite{Parikh:2017aas}, the non-linear equations of motion for the same class of theories of gravity arise as a consequence of the Clausius relation applied to the stretched future lightcone -- a co-dimension-1 timelike hyperboloid. Unlike the above derivation, one need not take the limit that the stretched horizon goes to a null surface. This is because the stretched horizon of the future lightcone acts as a causal barrier between observers living on the exterior of the cone from its interior, allowing for a well-defined notion of heat even in the absence of a Killing horizon. In the causal diamond set-up we had to take the limit that the stretched horizon moves to the conformal Killing horizon for technical reasons; it is unclear what the physical reason for this may be as the energy passing through the past causal diamond seemingly has a well-defined notion of heat. 

Moreover, in the future stretched light cone set-up, one similarly removes the entropy change due to the natural expansion of the hyperboloid. In light of the result above, that the entropy change due to the natural increase in the diamond may be interpreted as the generalized volume, naively we guess that the natural entropy change of the hyperoloid might have a similar interpretation. This suggests that we can think about the derivation of the gravitational equations of motion using the stretched future lightcone construction from an entanglement entropy perspective, i.e., perhaps the gravitational equatons of motion arise from an entanglement equilibrium condition, analogous to that given in \cite{Jacobson16-1,Bueno16-1}. We explore this idea in the next chapter.


\section*{Summary and Future Work}
\noindent

The work detailed above has extended and provided new insights into the `thermodynamical gravity' paradigm. Specifically, in (\ref{subsec:Einfromstretch}) we defined the stretched future light cone, argued that it is natural to associate temperature and holographic entropy with it, and shown that the reversible thermodynamic equation -- the Clausius relation $Q = T\Delta S_{\rm rev}$ -- directly leads to the generalized Einstein equations for all diffeomorphism-invariant theories of gravity whose Lagrangian contains no derivatives of the Riemann tensor. Then, as summarized in (\ref{subsec:firstlaw}), we used the Clausius theorem to derive a local first law of gravity -- a hybrid equation connecting matter and spacetime thermodynamics. A comparable derivation, where we replaced stretched lightcones with causal diamonds was given in Appendix \ref{sec:gravfromCD}. Combined, these results further strengthen the relation between thermodynamics and geometry. There are several extensions to the work described above, some of which are currently underway. Let's outline a few of these now. 


\noindent


\textbf{Horizon Thermodynamics without Horizons}

First we emphasize that the presented derivation of Einstein's equations not only extended Jacobson's original argument \cite{Jacobson:1995ab} to include general theories of gravity, but is valid without needing to work directly on a horizon, i.e., a null hypersurface. This observation is interesting as it suggests we can consider scenarios where spacetime thermodynamics was thought not to apply: stars. A thought-experiment can be imagined thusly: a collection of contant and uniformly radial accelerating observers sit above a star, with a temperature proportional to their acceleration. As the star deforms in some way, e.g., the star increases in mass $M$ via an accretion process, for the observers to maintain their same acceleration, i.e., to remain in thermal equilibrium, they must move outward, increasing the radius of co-dimension-2 spherical slice of the stretched future lightcone. Therefore, the change in geometry of a star $\Delta M$ -- which does not have an event horizon -- results in a change in thermodynamic entropy $\Delta S$ as measured by non-inertial observers\footnote{A necessary ingredient to accurately describe the physics of the thermodynamic response measure by accelerating observers outside a star is the existence of a quantum vacuum state. With some work it can be shown that such a vacuum state can be constructed and that a set of local Rindler observers will measure to populated with thermal radiation. }. Therefore, we arrive to a relation $\Delta M\propto\Delta S$, similar to the first law of black hole thermodynamics.

This observation suggests that the classic Smarr relation \cite{Smarr:1972kt} relating thermodynamic variables of a black hole, e.g., a static black hole, 
\beq M=2T_{H}S\;,\label{Smarrstatic1}\eeq
where $T_{H}$ is the Hawking temperature, can be extended to account for the thermodynamics of timelike stretched horizons, and, moreover, that the proportionality constant relating $\Delta M$ and $\Delta S$ depends on the location of timelike hypersurface $\Sigma$. In particular, we may write the extended Smarr formula in terms of the physical (Unruh) temperature\footnote{Here we term the Unruh temperature $T_{p}$ the physical temperature, as it is the physical temperature measured by accelerating observers. There are two other `temperatures' we can relate the Unruh temperature to, namely the Hawking temperature  $T_{H}=T_{p}\alpha$ -- the temperature of a black hole -- and the Tolman temperature $T_{T}=T_{H}/\alpha$, which is the blue shifted Tolman temperature. It would appear as though the Tolman temperature and Unruh temperature are equivalent, however, this is only the case of the near horizon limit, then $\lim_{r\to r_{H}}T_{T}/T_{p}=1$. Otherwise, the Unruh temperature and Tolman temperature are generally different measures of temperature.} $T_{p}=a/2\pi$, where $a$ is the proper acceleration of the non-inertial observers, and a redshift factor $\alpha$
\beq M=2T_{p}\alpha S\;,\eeq
where now $T_{p}$ is taken to be constant. Depositing matter onto the star causes an increase in radius via $r\to r+dr$  leading to
\beq dM=\frac{2T_{p}\alpha}{\left[1+\frac{M}{\alpha^{2}r}\right]}\left(1+\frac{M}{2\alpha^{2}r}\right)dS\;.\eeq
In the far field limit, $r\to\infty$ we have that $\alpha\to1$ yielding
\beq \lim_{r\to\infty}dM=2T_{p}dS\;,\eeq
while in the limit we approach the horizon $r\to r_{H}$ ($\alpha\to0$), 
\beq \lim_{r\to r_{H}}dM=T_{H}dS\;,\eeq
which is simply the first law of static black holes. The above argument can be extended to systems which include rotation, charge, and even a cosmological constant. Another interesting feature of this model is that the system has a positive heat capacity, unlike the traditional static black hole scenario.



\noindent

\textbf{The Four Laws of Stretched Future Lightcones}

Above we showed that stretched future lightcones obey the second law of thermodynamics -- $\Delta S_{\text{rev}}>0$ in order for observers to maintain their same acceleration for a positive heat flux -- and was used to derive a first law of thermodynamics. Due to the similarities between black hole thermodynamics and stretched lightcones it is natural to hypothesize that stretched lightcones possess four laws of thermodynamics. Specifically, analogous to black hole thermodynamics, the four laws for stretched lightcones would be: (0) the proper acceleration $a$ is constant on the stretched horizon $\Sigma$; (1) perturbations to the stretched horizon leads to the (local) first law of gravity; (2) the area of the stretched horizon $A$, assuming the weak energy condition is a non-decreasing function of time, and (3) it is not possible to form a stretched horizon with vanishing proper acceleration. 

Making each of these statements precise involves work currently underway. For example, it is natural to attempt to extend the first law to include charge and rotation. Including rotation is straightforward in fact (requiring that we only add an angular contribution to the radial boost vector, i.e., $u_{b}=1/\alpha \xi_{b}+\Omega\xi^{\phi}_{b}$, wher $\Omega$ is the rotation parameter). We can also include charge by introducing a electromagnetic contribution to the energy-momentum tensor, such that $T^{ab}=T^{ab}_{fluid}+T^{ab}_{EM}$, and adding a electromagnetic current $J^{a}_{EM}=-1/\alpha j^{a}A^{b}\xi_{b}$ to the current, $J^{a}=J^{a}_{matter}+J^{a}_{EM}$. Then, using the arguments described in \ref{subsec:firstlaw} and \cite{Gao:2001ut}, we can, at least in principle, extend the first law of gravity to
\beq \Delta E=T\Delta S_{\text{rev}}^{Wald}+\Phi\Delta Q+\Omega\Delta J-P\Delta V\;.\eeq

We have already established that $\Delta S_{\text{rev}}>0$ via the weak energy condition, leading to $\Delta A/\Delta t\geq0$. It would be interesting, however, to establish this relation, at least in the case of Einstein gravity, via the geometry of congruences. Along the lines of \cite{Piazza:2010hz}, we can formally construct a congruence of radially accelerating observers, work out its expansion $\theta$ and, via an application of the Gauss-Codazzi equations, establish the second law. 

Collectively then we may formally write a set of four laws of stretched lightcone thermodynamics relating geometric relations to thermodynamic principles. Recently the thermodynamics of lightcones (\emph{not} stretched lightcones) was established \cite{DeLorenzo:2017tgx} and further related to black hole thermodynamics \cite{DeLorenzo:2018ghq}. It would be interesting to understand how the thermodynamics of ordinary lightcones relates to the thermodynamics of stretched lightcones. 

\noindent

\textbf{Microscopics from Entanglement}

Another potential explanation of the microscopic origins of thermodynamical gravity is entanglement. Indeed, as summarized in Chapter \ref{sec:overview}, in certain regimes black hole entropy can be understood as entanglement entropy, e.g., the correlations of quantum fields above and below the event horizon of a black hole. Applying this logic to local Rindler horizons would then suggest that the local holographic thermodynamics used to derive classical gravitational equations of motion are a consequence of some underlying principle of quantum entanglement. Making this observation precise is the subject of the following the chapter, which we move to now.


\section{GRAVITY FROM ENTANGLEMENT EQUILIBRIUM}  \label{sec:gravfroment}

\subsection{Vacuum Entropy and Gravity}
\noindent

There are many `definitions' of entropy. So far we have been focusing on thermodynamic entropy, a measure of energy which cannot be used as a useful work. From the statistical point of view, entropy can be understood as a counting of the microstates of a quantum system and the thermodynamic entropy is simply the macroscopic limit of the microscopic statistical entropy. In information theory the Shannon entropy is a measure of the uncertainty of knowledge one has about a classical message before said message is received, i.e., it measures the correlation of degrees of freedom between a message and a receiving device. Quantum (information) entropy (more formally known as the von Neumann entropy) is the measure of quantum correlations, i.e., entanglement, between two regions of space separated by a boundary. In the context of information theory, the statistical entropy can be understood as the amount of information needed to specify a microstate of the system. In this way, quantum entanglement gives rise to the microscopic accounting of entropy in a thermodynamic system\footnote{To see how Shannon entropy gives rise to statistical entropy, recall that the Shannon entropy $S$ of a probability distribution $X$ with a discrete set of probabilities $p(x_{i})$ is given by 
$$
 S(X)=-\sum_{i=1}^{n}p(x_{i})\log p(x_{i})\;.$$
Assuming that each $p(x_{i})$ is equiprobable $p=1/W$, we find $S=\log W$, which we recognize as the Boltzmann (microcanonical ensemble) entropy for $k_{B}=1$, where $W$ is the number of microstates which corresponds to a macroscopic thermodynamic state. In similar fashion we may derive the statistical entropy for other ensembles starting from, in fact, the von Neumann entropy $S_{EE}=-\text{tr}\rho\log\rho$, where $\rho$ is the thermal density matrix $\rho=e^{-\beta H}/Z$, with $H$ a Hamiltonian, $\beta$ the inverse temperature, and $Z$ the partition function.}.

A natural setting for understanding entropy as missing information comes to us from black hole physics: event horizons are locations which causally disconnect two regions of spacetime. This suggests that black hole entropy, or at least a contribution to it can be interpreted as entanglement entropy. More precisely, we can consider quantum fields living in a black hole background. While the Hawking radiation from a black hole appears thermal according to an observer sitting outside of the horizon, the global state of the radiated quantum fields is pure -- the state appears mixed when the outside remains ignorant to the degrees of freedom behind the horizon. Therefore at least a contribution to black hole entropy is entanglement entropy.

That black hole entropy might be understood as entanglement entropy of quantum fields outside and behind the horizon leads to the following puzzle: continuum $(d+1)$ quantum field theory tells us that the entanglement entropy computed via correlations in vacuum fluctuations on either side of the horizon is infinite, leading one to impose a short distance cutoff $\epsilon$
\beq S_{EE}= c_{0}\frac{A}{\epsilon^{d-1}}+...\;,\eeq
where $A$ is the area of boundary region separating a region of spacetime from its complement. Yet classical general relativity tells us that the entropy of a black hole is given by the Bekenstein-Hawking entropy, 
\beq S_{BH}=\frac{A}{4G\hbar}\;.\eeq
Comparing the two entropic relations tells us black hole thermodynamics, and, by extension, spacetime thermodynamics, demands a fundamental cutoff at the level of the Planck scale, $c_{0}/\epsilon^{d-1}\equiv1/4G\hbar$. 

This observation leads to a further puzzle: If the short distance cutoff is fixed to be at the Planck length, then the entanglement entropy depends on the number of independent quantum fields -- the `species' -- however it would appear that the Bekenstein-Hawking entropy does not. Reconciling this tension suggests that the gravitational constant $G$ appearing in the Bekenstein-Hawking entropy formula is renormalized by the same zero point fluctuations giving rise to the entanglement entropy. Therefore, black hole entropy can be understood as entanglement entropy between quantum vacuum fluctuations inside and outside of the horizon, and classical gravity knows about this entropy because the gravitational dynamics describing the theory is governed by an action `induced' from the same quantum vacuum fluctuations -- an idea first considered by Sakharov \cite{Sakharov:1967pk}.

Let's be a bit more explicit here. Assume we have a generic quantum field $\phi$ living in its ground state $|0\rangle$ on an eternal static black hole; the ground state is the Hartle-Hawking vacuum\footnote{While we are considering an eternal static black hole for simplicity, it is expected that the conclusions here should hold for a black hole which forms from collapse \cite{Jacobson:1994iw}. In this case, the quantum field is presumed to be in the Unruh vacuum state.}. The degrees of freedom encoded in $\phi$ are subdivided into the region inside the horizon, denoted `IN', and outside of the horizon, `OUT', such that the two regions are entangled. Having access only to those degrees of freedom living outside of the horizon, observers in OUT would describe the state of the OUT subsystem via the reduced density matrix:
\beq \rho_{\text{OUT}}=\text{tr}_{\text{IN}}|0\rangle\langle0|\;,\eeq
where we have (partially) traced over the degrees of freedom living inside of the horizon. $\rho_{\text{OUT}}$ describes an entangled state, since its von Neumann entropy is non-vanishing, $S_{\text{VN}}(\rho_{\text{OUT}})=-\text{tr}(\rho_{\text{OUT}}\log\rho_{\text{OUT}})\neq0$. 

It turns out that the density matrix $\rho_{\text{OUT}}$ can be also be expressed as a thermal state in the canonical ensemble \cite{Birrell82-1}:
\beq \rho_{\text{OUT}}=\frac{1}{Z}e^{-\beta_{H}\hat{H}}\;,\quad Z[\beta]\equiv\text{tr}(e^{-\beta_{H}\hat{H}})\;,\eeq
where $\hat{H}$ is the Hamiltonian of static observers outside of the black hole horizon responsible for generating time translations, $\beta_{H}$ is the inverse Hawking temperature $\beta=\frac{2\pi}{\kappa}$, and $Z$ is the canonical ensemble partition function.  In this setting, the entanglement entropy $S_{\text{EE}}$ is precisely the same as the thermal entropy
\beq S_{\text{EE}}=\left(1-\beta\partial_{\beta}\right)\log Z[\beta]\;.\label{EEthermBHstat}\eeq

The connection between the entanglement entropy (\ref{EEthermBHstat}) and the Bekenstein-Hawking entropy can be born out using the low-energy effective action $\mathcal{W}[g]$ 
\beq e^{- \mathcal{W}[g]/\hbar}=\int\mathcal{D}\phi e^{-I[\phi,g]}\;,\eeq
where in this case $g$ is understood to be the (Euclideanized) metric describing the static black hole geometry. The path integral over fields on $g$ on the right hand side is interpreted as the black hole partition function such that $W=-\hbar\log Z$. The effective action can be written down, and generically takes the form \cite{Jacobson:1994iw,Frolov:1997up}
\beq \mathcal{W}=\hbar\int_{\mathcal{M}} d^{4}x\sqrt{g}\left[a_{0}+a_{1}R+a_{2}R^{2}+a'_{2}R_{ab}^{2}+...\right]+\hbar\int_{\partial\mathcal{M}}d^{3}y K+...\;,\label{effact}\eeq
where $K$ is the Gibbons-York-Hawking boundary term, and the $+...$ refers to additional higher curvature corrections, with their corresponding boundary terms. Here $a_{0},a_{1},a_{2}$, etc. are generically induced UV divergent couplings, determined by the masses of the constituent fields $\phi$; $a_{0}$ represents the cosmological constant, while $a_{1}$  Newton's gravitational constant induced from the vacuum fluctuations across the horizon\footnote{Specifically, for an induced model of gravity consisting of non-minimally coupled Dirac fermions of mass $m_{d}$ and scalar fields with masses $m_{s}$, the induced gravitational constant becomes \cite{Frolov:1996aj,Frolov:1997xd}: $G^{-1}=\frac{1}{12\pi}\left(\sum_{s}(1-6\xi_{s})m_{s}^{2}\log m_{s}^{2}+2\sum_{d}m^{2}_{d}\log m^{2}_{d}\right)$}. 

The entanglement entropy, given by the thermal entropy (\ref{EEthermBHstat}), will receive contributions from the entire effective action (\ref{effact}), many of which are dependent on the quantum state. The most singular contribution, however, turns out to be universal for all states with the same UV structure. Since we are integrating over a metric that we have assumed is time independent, the integral over the spacetime volume $\int d^{4}x\sqrt{g}$ will only provide us with a term proportional to the inverse temperature $\beta$, and will therefore not contribute to the entropy. This takes care of the cosmological constant $a_{0}$ and Einstein-Hilbert contributions. The remaining leading contribution is then the Gibbons-York-Hawking boundary term, which is proportional to the area of the black hole horizon $A_{\mathcal{H}}$ \cite{Jacobson:1994iw,Jacobson:2012yt}. Therefore, by this heuristic model, the entanglement entropy  arising from correlated vacuum fluctuations across a horizon is, to leading order, given by the Bekenstein-Hawking entropy of the black hole
\beq S_{\text{EE}}(\rho_{\text{OUT}})=\frac{A_{\mathcal{H}}}{4G\hbar}+...\;,\eeq
where $G$ is understood to be Newton's constant induced from the same vacuum fluctuations. In this way, the Bekenstein-Hawking entropy captures the leading UV divergence of the entanglement entropy, while the subleading UV divergent contributions, denoted here by $+...$, are accounted for by the higher curvature corrections of the induced effective action (\ref{effact}). The entire UV divergent structure of the entanglement entropy, then, can be combined into a single UV cutoff dependent Wald entropy.

Now, if we assume that the quantum theory of gravity from which the classical theory of gravity is induced is UV finite (as often claimed in string theory), then the entanglement entropy must be finite. This suggests an additional mystery: the entanglement entropy is UV finite but the gravitational entropy largely depends on renormalized gravitational couplings, and is therefore not expected to be UV finite by itself. There is mounting evidence (see, e.g., \cite{Susskind:1994sm}), however, that the generalized entropy 
\beq S_{\text{gen}}=S^{(\epsilon)}_{BH}+S^{(\epsilon)}_{\text{mat}}\;,\eeq
is independent of the UV cutoff $\epsilon$. Here $S^{(\epsilon)}_{BH}$ is the Bekenstein-Hawking entropy dependent on the renormalized gravitational coupling (where we momentarily neglect higher curvature contributions), and $S^{(\epsilon)}_{\text{mat}}$ is a renormalized entanglement entropy of matter fields. The two contributions to $S_{\text{gen}}$ therefore conspire to make $S_{\text{gen}}$ finite, suggesting that it be identified with the entanglement entropy. 

When one assumes $S_{EE}=S_{\text{gen}}$, we may assign entanglement entropy to surfaces other than cross sections of black hole horizons or minimal surfaces in AdS spaces -- this is indeed very natural from the perspective of entanglement entropy. The above observation led Jacobson to propose the entanglement equilibrium conjecture \cite{Jacobson16-1}
\beq \delta S^{A}_{EE}|_{V}=\frac{\delta A|_{V}}{4G}+\delta S_{\text{mat}}=0\;,\eeq
i.e., the vacuum is in a maximal entropy state -- any perturbation and matter fields and geometry inside the ball leads to a decrease in entanglement -- where it was shown this condition is equivalent to imposing the non-linear Einstein equations at the center of small balls. Thus, gravity emerges from entanglement, not thermodynamics. 

Jacobson's set-up relied on studying the geometry of causal diamonds\footnote{Particularly, spherical spatial subregions in geometries that are a perturbation of a maximally symmetric background. Each such subregion defines a causal diamond, which admits a conformal Killing vector $\zeta^{a}$ whose flow preserves the diamond} and working out a geometric identity termed the first law of causal diamonds. Moreover, the entanglement equilibrium conjecture was extended to incorporate higher derivative theories of gravity \cite{Bueno16-1} by including the subleading UV divergent contributions captured by curvature squared terms present in the effective action (\ref{effact}). In this case the maximal entropy condition becomes 
\beq \delta S^{A}_{EE}|_{W}=\delta S_{\text{Wald}}|_{W}+\delta S_{\text{mat}}=0\;, \eeq
where the volume $V$ is replaced with a new local geometrical quantity called the generalized volume $W$. This condition, when applied to small spheres, is equivalent to imposing the linearized equations of motion for a higher derivative theory of gravity. 

In what follows we briefly describe how to extend the work of \cite{Parikh:2017aas} and \cite{Bueno16-1} by deriving a first law of stretched lightcones, analogous to the first law of causal diamonds (FLCD), and showing that it is equivalent to an entanglement equilibrium condition, and that this is equivalent to a derivation of the non-linear Einstein's equations, and linearized equations for higher derivative theories of gravity. Moreover, we will show that the condition of fixed (generalized) `volume' can be understood as subtracting the entropy due to the natural increase of the stretched lightcone -- the irreversible contribution to the thermodynamic entropy -- thereby connecting entanglement equilibrium to (reversible) equilibrium thermodynamics. 


\subsection{Entanglement of Stretched Lightcones}
\indent

Our procedure is as follows. First we compute $\delta S_{\text{Wald}}$ and derive an off-shell geometric identity analogous to the first law of causal diamonds, which we call the first law of stretched lightcones. We will use the Noetheresque approach illustrated in described in \cite{Parikh:2017aas}. Next we will show how this off-shell identity is equivalent to the variation of the entanglement entropy, following arguments presented in \cite{Bueno16-1}. Finally, we will find that the linearized form of the gravitational equations emerge from an entanglement equilibrium condition. In essence, we are simply considering Jacobson's entanglement equilibrium proposal \cite{Jacobson16-1} for the geometry of stretched lightcones in an arbitrary background (where we explicitly consider perturbations to Minkowski space). One expects to find a similar result as established in \cite{Bueno16-1}, simply by noting that the stretched lightcone shares enough geometric similarities to the causal diamond. 

Begin by recalling that $\xi_{a}$ satisfies (\ref{Keqforu})
\beq \nabla_{a}\xi_{b}+\nabla_{b}\xi_{a}=2\Omega_{\xi}\tilde{g}_{ab}\;,\eeq
where $\Omega_{\xi}=t/ r$, and we have defined
\beq \tilde{g}_{ab}=\left(\delta_{ij}-\frac{x_{i}x_{j}}{r^{2}}\right)\delta^{i}_{a}\delta^{j}_{b}\;.\eeq
The derivation of the (FLCD) presented in \cite{Bueno16-1} (and further reviewed in \cite{Svesko:2018qim}) relies on the fact that $\zeta_{a}$ is an exact conformal Killing vector in flat space; specifically the fact that $\zeta_{a}$ satisfies the conformal Killing identity.  Here the vector $\xi_{a}$ is not a conformal Killing vector, and therefore, it will not satisfy the conformal Killing identity. The issue is that $\tilde{g}_{ab}$ defined above is not the metric, and therefore this object will have a non-vanishing covariant derivative. However, since we are considering the time $t=0$ surface, the fact that $\xi_{a}$ does not satisfy the conformal Killing identity is not a problem for us because $\Omega_{\xi}$ will vanish at $t=0$. Therefore, all terms $\Omega_{\xi}\nabla\tilde{g}$ which would appear can be neglected. 

Following the steps described in \cite{Parikh:2017aas}, we can show that for our approximate conformal Killing vector\footnote{Here we have chosen to set $\hbar=1$.} $\xi_{a}$
\beq 
 S_{Wald}=-\frac{1}{4G}\int_{B}dB_{a}\{P^{abcd}R_{ebcd}\xi^{e}-2\xi_{d}\nabla_{b}\nabla_{c}P^{abcd}+2P^{abcd}(\nabla_{c}\Omega_{\xi})\tilde{g}_{bd}\}\;,
\label{deltaSLC}\eeq
where we have the volume element $dB_{a}=U_{a}dV$ of the $D-1$-ball cross section of the stretched lightcone, and $P^{abcd}=\partial L/\partial R_{abcd}$. 

Let us study the bottom line. Using $(\nabla_{c}\Omega_{\xi})|_{t=0}=-1/r^{2}\xi_{c}$, we find to leading order we have
\beq
\begin{split}
-\frac{1}{4G}\int_{B}dB_{a}2P^{abcd}(\nabla_{c}\Omega_{\xi})\tilde{g}_{bd}&=-\frac{1}{2G}\int_{B}dV\frac{P^{tijt}}{r}\left(\delta_{ij}-\frac{x_{i}x_{j}}{r^{2}}\right)\\
&=-\frac{1}{2G}\frac{1}{(D-1)}\left(\sum_{i}P^{tiit}\right)\Omega_{D-2}r^{D-2}\;.
\end{split}
\eeq
Note that this object is proportional to the surface area of the spherical subregions; in fact in the case of Einstein gravity, $P_{GR}^{abcd}=\frac{1}{2}(g^{ac}g^{bd}-g^{ad}g^{bc})$, the above simply becomes $-\frac{A_{\partial B}}{4G}$, the Bekenstein-Hawking entropy. Motivated by the derivation of the first law of causal diamonds in \cite{Bueno16-1} we might be inclined to refer to this object as the \emph{generalized area}\footnote{In fact, we could also interpret this quantity as being proportional to the generalized volume. Using $K_{\partial\Sigma}=(D-2)/\alpha$, and that we are integrating a ball of radius $\alpha$, we find that this term may be expressed as $K/2G\bar{W}$.}, however, this object appears in \cite{Parikh:2017aas} (see equations (67)-(68) of their paper), and is identified as the entropy due to the natural background expansion of the hyperboloid, $\bar{S}$. Specifically, 
\beq  \bar{S}=-\frac{1}{4G}\int_{B}dB_{a}2P^{abcd}(\nabla_{c}\Omega_{\xi})\tilde{g}_{bd}\;,\label{genarea}\eeq
and therefore, 
\beq 
\begin{split}
 S_{\text{Wald}}- \bar{S}&=-\frac{1}{4G}\int_{B}dB_{a}\{P^{abcd}R_{ebcd}\xi^{e}-2\xi_{d}\nabla_{b}\nabla_{c}P^{abcd}\}\;.
\end{split}
\eeq

Next, introduce the matter energy $H^{m}_{u}$ associated with spherical Rindler observers with proper velocity $u$, 
\beq H^{m}_{u}=\int_{B}dB_{a}T^{ab}u_{b}\;.\eeq
Then, following the same arguments given in \cite{Jacobson16-1,Bueno16-1}, we find
\beq \frac{1}{2\pi\alpha}(\delta S_{\text{Wald}}-\delta \bar{S})=-\delta H_{u}^{m}\; \label{firstlawstretchedLCon}\eeq
is equivalent to the linearized gravitational  equations of motion about flat spacetime for $L(g^{ab},R_{abcd})$ theories of gravity:
\beq \delta G^{ad}-2\partial_{b}\partial_{c}(\delta P^{abcd}_{\text{higher}})=8\pi G\delta T^{ad}\;.\eeq

The off-shell identity is simply 
\beq \frac{1}{2\pi\alpha}(\delta S_{\text{Wald}}-\delta \bar{S})+\delta H^{m}_{u}=\int_{B}\delta C_{\xi}\;,\label{firstlawstretchedLCoff}\eeq
where $\delta C_{\xi}$ represents the linearized constraint that the gravitational field equations hold. 

We can actually understand this first law of stretched lightcones as the Iyer-Wald identity \cite{Iyer:1994ys} in the case of the stretched horizon of spherical Rindler observers, rather than the dynamical horizon of a black hole. As illustrated in Appendix \ref{app:IyerWaldformLC}, we may actually interpret the generalized area as the variation of the gravitational Hamiltonian. 

Moreover, the first two terms on the LHS of (\ref{firstlawstretchedLCoff}) can be combined into a single object  \cite{Bueno16-1}, namely, the variation of the Wald entropy while keeping the generalized area constant, i.e., 
\beq \frac{1}{2\pi\alpha}(\delta S_{\text{Wald}}-\delta \bar{S})=\frac{1}{2\pi\alpha}\delta S_{\text{Wald}}|_{\bar{S}}\;,\eeq
leading to
\beq \frac{1}{2\pi\alpha}\delta S_{\text{Wald}}|_{\bar{S}}+\delta H^{m}_{u}=\int_{B}\delta C_{\xi}\;.\eeq

The Wald formalism contains the so-called JKM ambiguities \cite{Jacobson:1993vj}; one may add an exact form $dY$ linear in the field variations and their derivatives to the Noether current, and $Y$ to the Noether charge. This would lead to a modification of $S_{\text{Wald}}$ and $\bar{S}$. However, it is clear the combined modification will cancel, allowing us to write
\beq \frac{1}{2\pi\alpha}\delta S_{\text{Wald}}|_{\bar{S}}=\frac{1}{2\pi\alpha}\delta(S_{\text{Wald}}+S_{JKM})|_{\bar{S}'}\;,\eeq
where $\bar{S}'=\bar{S}+\bar{S}_{JKM}$. For more details on this calculation one need only follow the calculation presented in \cite{Bueno16-1} as it is identical in the stretched lightcone geometry. 


\subsection{Gravity from Entanglement of Stretched Lightcones}\label{subsec:gravfromentlc}
\noindent

Our aim here is to show how the first law of stretched lightcones -- an off-shell geometric identity -- can be understood as a condition on entanglement entropy. Before we consider the scenario with stretched lightcones, let us recall what happens in the case of a causal diamond. The entanglement equilibrium conjecture makes four central assumptions which we outline here. These assumptions include\footnote{Reviewed in further detail in Appendix \ref{app:cdmechanicsandEE}.} \cite{Carroll:2016lku}: (i) Entanglement separability, i.e., $S_{EE}=S_{UV}+S_{IR}$; (ii) equilibrium condition, i.e., a simultaneous variation of the quantum state and geometry of the entanglement entropy of the causal diamond is extremal, and the geometry of the causal diamond is that of a MSS; (iii) Wald entropy as UV entropy, i.e., the variation of the UV entropy is proportional to the Wald entropy at fixed generalized volume, and (iv) CFT form of modular energy, i.e., the modular energy is defined to be the variation of the expectation value of the modular Hamiltonian -- which for spherical regions may be identified with the Hamiltonian generating the flow along the CKV which preserves the causal diamond -- plus some scalar operator $X$. 

Reference \cite{Jacobson16-1} showed that the above postulates can be used to derive the full non-linear Einstein equations, while \cite{Bueno16-1} showed these postulates lead to the linearized gravitational equations for higher derivative theories of gravity. Here we will discuss how to justify the above assumptions (for a more pedagogical review, see \cite{Carroll:2016lku}) and attempt to apply a similar set of assumptions for the case of stretched lightcones. 

Assumption (i), where we require minimal entanglement between IR and UV degrees of freedom, is in fact a fundamental feature of renormalization group (RG) flows. More precisely, an RG flow requires a decoupling between high and low momentum states. Thus, in a Wilsonian effective action we would expect minimal entanglement between UV and IR modes. We also would assume that this basic feature of effective field theory to continue to hold in the theory's UV completion. This assumption is reasonably justified in both the causal diamond and stretched lightcone set-ups. 

The second assumption (ii) asserts that the vacuum state in a small region of spacetime may be described by a Gibb's energy state, and that for a fixed energy, this state will have a maximum entropy, i.e., $\delta S_{EE}=0$. Moreover, the requirement that the causal diamond is described in a MSS is simply there to prevent curvature fluctuations from producing a large backreaction which spoil the equilibrium condition. In other words, the semiclassical (linearized) equations hold if and only if the causal diamond is in thermodynamic equilibrium. Likewise, we may safely make this same assumption about the stretched lightcone: when the stretched lightcone is in thermal equilibrium, the gravitational equations hold (via the Clausius relation), and vice versa. 

Assumption (iii), like assumption (i), is also not very controversial. All that is being said is that one should identify the area $\partial B$ of the causal diamond, and, similarly, the cross-sectional area of the stretched lightcone $\partial \Sigma$, as the area of the planar Rindler horizons existing at the edge of the causal diamond, and the area of the timelike spherical Rindler horizon, respectively. Motivated by the Ryu-Takayanagi proposal, we then simply identify these areas with the entanglement entropy of each region. We should point out a difference between the two pictures, however. It is known that the entanglement entropy of the causal diamond $D[B]$, i.e., the causal domain of a spherical ball region $B$, is equivalent to the entanglement entropy of $B$ itself. Meanwhile, we are saying that the entanglement entropy of the stretched horizon, $\Sigma$, is equivalent to the ball $B$ whose boundary is $\partial\Sigma$. This has been established in the context of spherical Rindler space, which we may interpret our stretched lightcone as being: The entanglement entropy of spherical Rindler space is equal to the area of the horizon $\partial\Sigma$ \cite{Balasubramanian13-1}. 

Unlike the first three assumptions, which all rely on the underlying UV physics, assumption (iv) makes an assertion about the form of the modular Hamiltonian for IR degrees of freedom. In the case of causal diamonds one makes two observations. First, a causal diamond in Minkowski space may be conformally transformed to a (planar) Rindler wedge. Then, via an application of the Bisognano-Wichmann theorem \cite{Bisognano:1976za}, for CFTs the modular Hamiltonian $H_{\text{mod}}$, defined via the thermal state $\rho_{IR}=Z^{-1}e^{-H_{\text{mod}}}$, is proportional to the Hamiltonian generating the flow along the CKV $\zeta$, i.e., $H_{\text{mod}}=2\pi/\kappa H^{m}_{\zeta}$ \cite{Casini:2011kv}. This implies then that the variation of the modular Hamiltonian is equal to the variation of of $H^{m}_{\zeta}$, plus some additional spacetime scalar $X$, i.e., 
\beq \delta\langle H_{\text{mod}}\rangle=\frac{2\pi}{\kappa}\delta\int_{B}dB_{a}(T^{ab}\zeta_{b}+Xg^{ab}\zeta_{b})\;.\eeq
This specific assumption is interesting in that it may be explicitly checked, and has been justified \cite{Casini:2016rwj,Carroll:2016lku}, though with the stipulation that $X$ may depend on $\ell$. 

In the case of stretched lightcones, our assumption is then that the modular Hamiltonian $H^{\text{mod}}_{u}$, defined by $\rho_{\Sigma}=Z^{-1}e^{-H_{\text{mod}}}$, is proportional to the radial boost Hamiltonian,
\beq H_{\text{mod}}=2\pi\alpha\int_{B}dB_{a}T^{ab}u_{b}\;,\eeq
and that we may also include a spacetime scalar $X$. We would like to be able to similarly justify this assumption, as was accomplished in the causal diamond case. While currently this assumption is non-trivial and has not been computationally justified, we find that it is reasonable, as we now describe. 

The stretched lightcone $\Sigma$, like spherical Rindler space, can be understood as the union of Rindler planes; indeed, if we constrain ourselves to the $y=z=0$ plane, the radial boost vector $\xi^{a}=r\delta^{a}_{t}+t\partial^{a}_{r}$ reduces to a Cartesian boost vector. Each Rindler plane may be associated with a single causal diamond. The union of these causal diamonds yields a single ``radial causal diamond" \cite{Balasubramanian13-1}\footnote{This is precisely the construction of spherical Rindler space. If we were to embed spherical Rindler space into AdS, i.e., spherical Rindler-AdS space, the radial causal diamond was found to be holographically dual to a finite time strip in a boundary field theory \cite{Balasubramanian:2013lsa}.}. Therefore, the congruence of uniformly and constantly, radially accelerating observers comprising the stretched lightcone have an associated radial causal diamond. Moreover, the radial boost $\xi^{a}$ preserves the flow of the hyperboloid $\Sigma$. Our assumption is that the entanglement entropy of the stretched lightcone is that of the radial causal diamond which is also that of spherical region $B$. Thus we define the modular Hamiltonian as above and assume that it is proportional to the Hamiltonian generating the flow of $\Sigma$. For similar arguments given in \cite{Casini:2016rwj,Carroll:2016lku}, we expect -- but have not proved -- that for CFTs we may also modify the modular Hamiltonian by a spacetime scalar.

Let us now briefly show how the first law of stretched lightcones -- an off-shell geometric identity -- can be understood as a condition on entanglement entropy. In particular, we can follow the discussion given in \cite{Bueno16-1}. We perform a simultaneous (infinitesimal) variation of the entanglement entropy on a stretched lightcone of $S_{EE}$ with respect to the geometry and quantum state. By entanglement separability, $\delta S_{EE}$ takes the form
\beq \delta S_{EE}=\delta S_{UV}+\delta S_{IR}\;,\eeq
where the UV contribution is state independent and is assumed to be given by $\delta S_{UV}=\delta (S_{\text{Wald}}+S_{JKM})_{\bar{S}'}$, while the IR contribution comes from the modular Hamiltonian via the first law of EE, $\delta S_{IR}=\delta\langle H_{\text{mod}}\rangle=2\pi\alpha\delta\langle H^{m}_{u}\rangle$. 
Then, using the first law of entanglement entropy for a system in which the background geometry is also varied
\beq \delta S_{EE}=\delta(S_{\text{Wald}}+S_{JKM})+\delta\langle H_{\text{mod}}\rangle\;,\eeq
 we arrive to
\beq \frac{1}{2\pi\alpha}\delta S_{EE}|_{\bar{S}'}=\int_{B}\delta C_{\xi}\;,\eeq
valid for minimally coupled, conformally invariant matter fields. 

Thus, there is an equivalence between the following statements: (i) $S_{EE}$ is maximal in vacuum for all balls in all frames, and (ii) the linearized higher derivative equations hold everywhere. In other words, the entanglement equilibrium condition is equivalent to the linearized higher derivative equations of motion to be satisfied, and vice versa. This equivalence may be verified via a simple modification of the calculations presented in \cite{Bueno16-1}. We also note that here we considered perturbations about Minkowski space, however, one could, in principle, generalize this to a maximally symmetric spacetime, and while the above discussion was particular to theories of gravity described by $L(g^{ab},R_{abcd})$, i.e., those which do not depend on the derivatives of the Riemann tensor, we could have included those derivatives as well.


\section*{Summary and Future Work}
\indent

Motivated by \cite{Jacobson16-1,Bueno16-1}, we showed how to derive the linearized gravitational equations of motion from the entanglement equilibrium proposal, i.e., that the entanglement entropy for spherical entangling regions is maximal in the vacuum. We did this by first deriving an off-shell geometric identity, the first law of stretched lightcones, and showed that it was equivalent to the first law of entanglement entropy in the case of spherical subregions and conformally invariant matter. In the derivation of the first law of stretched lightcones we found an expression for the generalized area, which is nothing more than the entropy due to the natural expansion of the stretched lightcone. To complete this derivation, however, we to had make the non-trivial assumption that the entanglement entropy of the spherical entangling region $\partial\Sigma$ is the entanglement entropy of $\Sigma$, and the  modular Hamiltonian $H_{\text{mod}}$ is proportional to the radial boost Hamiltonian $H^{m}_{u}$.  

The entanglement equilibrium condition associated with causal diamonds can be related to the Clausius relation by assigning thermodynamic propertis to the conformal Killing horizon \cite{Svesko:2018qim}. In this way we can show that the entanglement of causal diamonds considered in \cite{Jacobson16-1,Bueno16-1} can be interpreted via local holographic thermodynamics, and that the full non-linear equations arise from $\Delta Q=T\Delta S_{\text{rev}}$. Moreover, as eluded to above with the stretched lightcone geometry,  $\Delta S_{\text{rev}}$ is defined as the entropy solely due to a matter flux crossing the conformal horizon. We found that the quantity $\frac{K}{2G}\bar{W}$, where $\bar{W}$ is the generalized volume, can be understood as the entropy of the natural increase of the causal diamond. 

We can summarize our findings of \ref{subsec:gravfromentlc} and the equivalent statement for causal diamonds \cite{Jacobson16-1,Bueno16-1} as
\beq T\delta S_{EE}|_{\bar{S}'}=\int_{B}\delta C\;.\eeq
Here $\bar{S}'$ is the irreversible entropy due to the natural change of the background geometry -- identified as the generalized volume in the case of causal diamonds, or the generalized area in the case of stretched lightcones -- and where $T$ is the temperature associated with the horizon of the surface, namely, the Hawking temperature  $T_{H}=\kappa/2\pi$ in the case of causal diamonds, or the Unruh-Davies temperature $T=1/2\pi \alpha$ in the case of stretched lightcones. Entropy being maximal in the vacuum implies that the linearized constraint is satisfied, leading to the linearized form of the equations of motion of higher derivative theories of gravity, or, in the special case of Einstein gravity, the full non-linear equations. 


\textbf{Comparison to Other Approaches of `Emergent Gravity'}
\noindent

Before we examine potential future avenues of research, let us briefly compare three approaches of `emergent gravity': (i) the method of local causal horizons via spacetime thermodynamics, e.g., \cite{Jacobson:1995ab,Parikh:2017aas}, (ii) the entanglement equilibrium approach described here \cite{Svesko:2018qim} and \cite{Jacobson16-1,Bueno16-1}, and the approach taken using holographic entanglement entropy (HEE) and AdS/CFT \cite{Lashkari13-1,Faulkner13-2,Faulkner:2017tkh,Haehl:2017sot,Lewkowycz:2018sgn}. 

In the (i), one assigns thermodynamic/entropic properties to local causal horizons, such as the (null) local planar Rindler horizons/causal diamonds, or the (timelike) stretched horizons of future lightcones. The derivation of the equation of state makes use of  `physical process' analysis such that set-up is inherently \emph{dynamical}: a dynamical entropy change leads to a dynamical change of local geometry of a single background spacetime. An advantage of this approach is that we can readily attain the full non-linear gravitational field equations since we are studying local horizons about a point, which, by construction, the resulting equations of motion satisfy the Riemann normal coordinate expansion at each order. As noted earlier, a disadvantage of the thermodynamic method is that the entropy functional lacks a precise physical interpretation, i.e., what is the entropy in spacetime thermodynamics? In \cite{Svesko:2018qim} and this chapter, we strengthened the bridge between the thermodynamical and entanglement equilibrium approaches (i) and (ii), where we connected the physical process derivations using local causal diamonds or stretched lightcones to their equilibrium state counterparts. This does not prove, but deeply suggests the entropy appearing in spacetime thermodynamics is an entanglement entropy due to fluctations near local causal horizons.

To contrast, the entanglement equilibrium approach uses an equilibrium state form of the first law, not the physical process version. This is consistent with the fact we obtain the constraint for linearized equations of motion. That is, we do not expect to attain evolution equations to arise from an equilibrium condition. Yet, we are able to get (local) dynamical equations using the equilibrium condition. This is because the dynamics of a diffeomorphism invariant theory of gravity is entirely determined by evaluating the constraints in all possible Lorentz frames, and locality because we are focusing on small balls in a perturbed MSS. It turns out, moreover, that the linearized first order variation of the Einstein tensor evaluated in a Riemann normal coordinate expansion about the center of the small ball $p$, is equivalent to the full non-linear Einstein tensor about $p$, $\delta G_{ab}|_{\text{RNC}}= G_{ab}(p)$. This allows us to recover the full (local) non-linear Einstein equations. In contrast, the non-linear equations for higher derivative theories are not consistent with the RNC, i.e., $\partial_{c}\partial_{d}\delta P^{acdb}_{\text{high}}|_{\text{RNC}}\neq \nabla_{c}\nabla_{b}P^{acdb}_{\text{high}}(p)$. This is because $P^{abcd}$ is quadratic in the Riemann tensor, and the linearization of the higher order contributions using the RNC expansion come in at the same level. In other words, the non-linear equations of higher curvature theories of gravity at a point cannot be derived by only imposing linearized equations -- we require information beyond the first order perturbations. 

The third method (iii) requires one use AdS/CFT and the Ryu-Takayanagi formula, specifically the Casini-Huerta-Myers map \cite{Casini:2011kv}. This map says that the vacuum entanglement entropy of a $d$-dimensional CFT reduced to ball-shaped regions in flat space can be reinterpreted as the thermal entropy of a CFT on a hyperbolic cylinder at a temperature inversely proportional to the radius of the cylinder. In the event the CFT is holographic, the thermal entropy is shown  to be \emph{dual} to the horizon entropy of a massless $(d+1)$-dimensional AdS black hole with a hyperbolically sliced (or, the AdS-Rindler patch of pure AdS). This is simply the RT prescription applied to spherical entangling surfaces. Then, the first law of entanglement entropy of the CFT on the boundary is dual to the first law of (\emph{global}) horizon thermodynamics with respect to Killing horizons in pure AdS. Consequently, perturbations to the CFT vacuum are dual to perturbations in the AdS geometry, which must satisfy the \emph{linearized} gravitational field equations. Put another way, gravity emerges as a dual description of the entanglement entropy degrees of freedom. Similar to the entanglement equilibrium approach, this HEE method uses an equilibrium state first law. Unlike the previously described methods, HEE considers global horizons in the bulk, not local horizons on some dynamical spacetime, which is why this method can only get linearized equations, even when the bulk is described by Einstein gravity.

\hspace{2mm}

Let us now discuss potential directions for future work.


\textbf{Local First Laws}
\noindent

We now have two derivations of the gravitational equations of motion via a thermodynamic process, and an application of the Clausius relation $T\Delta S_{\text{rev}}=Q$. As reviewed in \ref{subsec:firstlaw}, it was shown that one may write down a hybrid first law of gravity and thermodynamics 
\beq \Delta E=T\Delta S_{\text{rev}}-\mathcal{W}\;,\label{firstlawlocgrav}\eeq
connecting matter energy $E$ and work $\mathcal{W}$ with the gravitational entropy $S$ evaluated on the stretched future lightcone of any point in an arbitrary spacetime. It would be interesting to see if we can find a similar first law of causal diamonds. In fact, recently, Jacobson and Visser have established a first law for a causal diamond in a maximally symmetric space, analogous to the first law of black hole mechanics \cite{Jacobson:2018ahi}. In this set-up, the causal diamond is equipped with a cosmological constant, and one discovers that a local gravitational first law of causal diamonds is reminiscent of the Smarr formula for a ball in a maximally symmetric space. Moreover, if one wishes to interpret this first law as a Clausius relation, then the causal diamond, classically, is a thermodynamic system with a negative temperature. It would be interesting to study the thermodynamic behavior of the causal diamond, as well as look for a similar local first law for stretched lightcones, and verify that the stretched lightcone is a thermodynamic system with positive temperature. 



\textbf{Non-Linear Equations of Motion}
\noindent

It is interesting that we were able to derive the full non-linear gravitational equations of motion via a reversible process, while we only found the linearized equations of motion via the entanglement equilibrium condition. This is because we restricted ourselves to first order perturbations of the entanglement entropy and background geometry.  Higher order perturbations to the entanglement entropy lead to a modified form of the first law of entanglement entropy, e.g., the second order change in entanglement entropy is no longer proportional to the expectation value of the modular Hamiltonian (\ref{firstlawEE}), but rather one must include the relative entropy. Moreover, as pointed out in \cite{Bueno16-1}, using higher order terms in the RNC expansion and higher order perturbations to the entanglement entropy could make it possible to derive the fully nonlinear equations of an arbitrary theory of gravity. Indeed, these ideas were recently incorporated in the context of holographic entanglement entropy to derive the non-linear contributions to gravitational equations \cite{Faulkner:2017tkh,Haehl:2017sot,Lewkowycz:2018sgn}. Due to the simlarity between the holographic and entanglement equlibrium approaches, developments in one is likely to inform the other. 

We should also point out that the way we derived the non-linear gravitational equations via a physical process was by modifying $\zeta_{a}$ and $\xi_{a}$ to deal with the fact that $\zeta_{a}$ and $\xi_{a}$ are both approximate Killing vectors. It would be interesting to see whether these modifications have a microscopic interpretation and could be employed in the context of entanglement equilibrium  such that the non-linear equations of motion arise without needing to consider second order perturbations to the entanglement entropy. 



\textbf{Entanglement of Spherical Rindler Horizons}
\noindent

This is not the first time spherical Rindler horizons have appeared in the literature on holography or entanglement entropy. In particular, spherical Rindler horizons make an appearance in the mathematical construction of minimal entangling surfaces necessary to derive the (static) version of the Ryu-Takayanagi formula \cite{Fursaev:2013fta}. It would be interesting to better understand how the spherical Rindler horizon can be understood as an entangling surface, and how it relates to the minimal surfaces used in \cite{Lewkowycz:2013nqa,Dong:2013qoa} -- we note that the stretched future lightcone $\Sigma$, despite having properties reminiscent of a black hole horizon, is not a minimal surface. Since $\Sigma$ is timelike, it would be interesting to see if it plays a role in the covariant formulation of holographic entanglement entropy \cite{Hubeny:2007xt}. 

Spherical Rindler horizons also make an appearance in another version of the `spacetime from entanglement paradigm'. Specifically, in \cite{Balasubramanian13-1,Balasubramanian:2013lsa} it was shown that when spherical Rindler space is embedded in AdS, it has an entropy proportional to the area of the spherical Rindler horizon. Moreover, spherical-Rindler-AdS space was shown to be dual to a UV sector of the boundary field theory, and that one can define a \emph{differential entropy} -- a UV divergenceless quantifying the collective ignorance a family of local observers in a CFT who make measurements over a finite time -- which reproduces the entropy of circular holes in $AdS_{3}$, and, more generally, reconstructs bulk curves on a spatial slice of $AdS_{3}$. These ideas have since been generalized to use the differential entropy to reconstruct bulk surfaces in any dimension \cite{Czech:2014wka}, bulk surfaces which vary in time, and that for a broad class of holographic backgrounds possessing generalized planar symmetry, the differential entropy and gravitational entropy are equivalent \cite{Headrick:2014eia}. It would be interesting to see if it is possible to interpret the first law of stretched lightcones as a condition on the differential entropy.


\newpage

\section{MICROSCOPIC HERALDS OF THERMODYNAMIC  VOLUME} \label{sec:microvol}

The laws of black hole mechanics \cite{Bardeen73-1} were originally observations about black hole geometry, and were only recognized as laws of thermodynamics after Hawking provided a convincing quantum mechanical argument that black holes have a temperature \cite{Hawking74-1,Hawking75-1}. Recognizing the laws of black hole mechanics as thermodynamic statements, in part, comes about by comparing to the analogous statements for ordinary matter systems. The first law of (static) black hole thermodynamics, however, 
\beq \Delta M=T\Delta S\;\eeq
is clearly missing a $p\Delta V$ term. Since the $p\Delta V$ quantity in the first law of ordinary thermodynamics\footnote{Even the hybrid first law of gravity derived in \cite{Parikh:2018anm}.}  is typically associated with work done by a system or substance, e.g., an ideal gas in a piston, where $p$ is the pressure and $\Delta V$ is the change in volume of the gas, it is initially unclear how to interpret such a quantity in the context of black hole physics. Indeed it is not obvious what is meant by the `volume' of a black hole, as, for example, in a Schwarzschild black hole at $r<r_{H}$ the $r$ coordinate is timelike, and therefore a volume of $V=4\pi r_{H}^{3}/3 $ does not make much sense.

We can, however, make progress by embedding black holes into spacetimes with a cosmological constant. In such spacetimes, pressure makes a natural appearance in gravity when we include in a cosmological constant $\Lambda$ as the dynamical pressure of a fluid
\beq p=-\frac{\Lambda}{8\pi}\;,\label{ccasp}\eeq
Cosmological observations strongly suggest that $\Lambda>0$, which would translate to a negative pressure system, signaling thermodynamic instability. Alternatively, when $\Lambda<0$, as is the case for AdS spacetimes, $p>0$ leading to a well defined thermodynamic system. Embedding black holes in backgrounds with $\Lambda\neq0$ also suggest the ADM mass $M$ of the black hole should be interpreted as the enthalpy $H$, rather than the internal energy $U$ \cite{Kastor:2009wy}, leading to the extended first law of thermodynamics\footnote{Assuming for the moment that no other dynamical quantities, like charge and angular momentum, are in play.}
\beq \Delta M=\Delta H=T\Delta S+V\Delta p\;,\label{eq:first-law}\eeq
where the volume $V$ is simply the thermodynamic conjugate variable to pressure $p$
\beq V\equiv\left(\frac{\partial H}{\partial p}\right)_{S}\;.\label{thermovolgen}\eeq
In this context the appearance of enthalpy is natural: forming a black hole of volume $V$ requires removing a region of spacetime of size $V$ at a cost of $pV$. Enthalpy $H$ is the energy which captures the creation of such a thermodynamic system.

Since the beginning of the program of black hole extended thermodynamics, the thermodynamic volume (\ref{thermovolgen}) is a bit mysterious. In the simple case of static black holes (with no additional non--trivial scalar sector) it has the geometric interpretation as the naive spherical volume occupied by the black hole. For example, the case of static black holes, the thermodynamic volume is simply the geometric volume constructed by the naive use of the horizon radius, e.g., in $D=d+1=4$ spacetime dimensions, 
\beq V=\frac{4}{3}\pi r^{3}_{H}\;.\eeq
 In general, however,  the thermodynamic volume is non--geometrical~\cite{Cvetic:2010jb,Johnson:2014xza}. For example, when the rotation of a black hole is included, the thermodynamic volume and naive geometric volume expressions no longer coincide, instead the volume will depend on this rotation (for a review see, \emph{e.g.}, \cite{Dolan:2012jh}). In such general settings, it becomes a truly independent variable from the entropy, and the physics associated with it becomes richer. 

Despite its mysterious nature, the thermodynamic volume can nonetheless be used to classify different types of black holes using the so-called \emph{reverse isoperimetric inequality} \cite{Cvetic:2010jb}:
\begin{equation}
 \mathcal{R}\equiv\left(\frac{(d-1)V}{\omega_{d-2}}\right)^{\frac{1}{d-1}}\left(\frac{\omega_{d-2}}{4S}\right)^{\frac{1}{d-2}}\geq1\;,\label{revisogend}
\end{equation}
where $V$ is the thermodynamic volume, and~$S$ is the gravitational entropy. Also, the quantity $\omega_{n}{=}2\pi^{(n+1)/2}/\Gamma[(n+1)/2]$ is the standard volume of the round unit sphere. It was conjectured in \cite{Cvetic:2010jb} the inequality (\ref{revisogend}) is saturated by Schwarzschild--AdS black holes (including the Banados, Teitelboim and Zanelli (BTZ) black hole \cite{Banados:1992wn}) in $d{=}3$),  with 
$\mathcal{R}{=}1$. Black holes where $\mathcal{R}{>}1$ are said to be sub--entropic, such as Kerr--AdS \cite{Cvetic:2010jb} and STU black holes \cite{Caceres:2015vsa}. Systems with $\mathcal{R}{<}1$, such as the ultra--spinning limit of Kerr--AdS black holes \cite{Hennigar:2014cfa,Hennigar:2015cja}, are super--entropic. Unlike their higher-dimensional counterparts, in $d{=}3$, the rotating BTZ black hole has $\mathcal{R}{=}1$, while the charged BTZ hole \cite{Martinez:1999qi} has $\mathcal{R}{<}1$. 

We emphasize  that the inequality (\ref{revisogend}) is written here with~$\mathcal{R}$ defined in terms of the entropy $S$ instead of the horizon area~$A$, as it was originally written in  ref.\,\cite{Cvetic:2010jb}. This is because, in our view,  super--entropicity is a statement about the thermodynamic quantity {\it entropy} (as the title suggests) and not about the outer horizon area\footnote{A similar modification to the reverse isoperimetric inequality was made in ref.~\cite{Feng:2017jub} for black hole solutions of Horndeski theories of gravity.}. Moreover, more general theories of gravity have an entropy that is not proportional to the outer horizon area, but may include contributions from the inner horizon\footnote{In fact, sometimes even in ordinary gravity,  the entropy receives contributions from other objects. See the Taub--NUT and Taub--Bolt examples in refs.~\cite{Chamblin:1998pz,Emparan:1999pm,Mann:1999pc}.}.  


It was  recently observed \cite{Johnson:2019mdp} that several super--entropic black holes are thermodynamically unstable, signified by a negative heat capacity $C_{V}$. It was conjectured there that  super--entropicity may generally imply that $C_V{<}0$, which can be verified analytically for the charged BTZ black hole (as we will show later). Despite this nice interpretation of black hole super-entropicity, it lacks a microscopic interpetation; in fact, the thermodynamic volume also lacks a microscopic interpretation. We find ourselves, then, in a similar position as physicists before us who asked the same question about the microscopic interpretation of black hole entropy.

One interpretation of black hole entropy was to consider black holes in spacetimes with negative $\Lambda$, where the gravitational physics can often be recast in terms of a dual (non--gravitational) field theory in one dimension fewer using the correspondence between anti--de Sitter dynamics and conformal field theory physics (the AdS/CFT correspondence)~\cite{Maldacena:1997re,Gubser:1998bc,Witten:1998qj,Witten:1998zw}, it is natural to ask whether~$V$ has a direct interpretation in the field theory\footnote{See refs.~\cite{Kastor:2009wy,Dolan:2013dga,Johnson:2014yja} for early ideas and remarks, and refs.~\cite{Dolan:2014cja,Couch:2016exn} for some explorations.}. In general, this question is rather hard to explore, since the duality addresses the strongly coupled field theory regime, which is not always easily accessible in traditional field theory terms. Moreover the finite $T$ regime of the AdS/CFT duality is (in general)  rather less well robustly explored than the $T{=}0$ sector. 

Here we point out that progress can be made in the case of three dimensional gravity (with $\Lambda{<}0$), since in that case the duality's dictionary is rather stronger: Asymptotically anti--de Sitter geometries in three dimensions (AdS$_3$) are dual to conformally invariant two dimensional  field theories, which are very tightly constrained in their structure.  Moreover, the finite temperature~$T$ is simply the (inverse) period of a cycle in the two dimensional  Riemann surface the theory is defined on. We will be able to write the thermodynamic volume $V$ in terms of quantities very familiar in the  CFT. With that achieved, it is then straightforward to translate any conditions involving~$V$ into statements in the CFT. 

For example, it is natural in thermodynamics to ask questions about the fixed volume sector. However, in general\footnote{For static black holes with no scalars, $V$ and $S$ are not independent and so in those simple cases fixed~$V$ is simply fixed area. For most cases however, $V$ is a non--geometrical quantity independent of $S$.}, this is  somewhat mysterious from the black  hole thermodynamics perspective---fixed pressure is more natural there since that is simply fixed~$\Lambda$---but with a microscopic dual field theory identification such  as the one presented here, progress can be made in examining the physics of the fixed volume sector. (This may be of use in furthering  recent work~\cite{Johnson:2019vqf,Johnson:2019olt,Johnson:2019mdp} that has uncovered novel and potentially useful physics in the fixed volume sector of black hole thermodynamics.)

We make such progress by  arriving, in an important example, at a microscopic connection between the thermodynamics of the fixed volume sector and, in particular, super-entropicity \cite{Cvetic:2010jb}.
While we do not prove the conjecture about the connection between $C_{V}$ instability and super-entropicity here, we find a microscopic phenomenon that seems to explain (or at least herald) the super-entropicity on the gravity side, and it emerges precisely as a result of our microscopic identification of the thermodynamic volume~$V$ and as a consequence of working in the fixed $V$ sector. It works as follows: The  standard (microscopic) CFT expression for the entropy, $S$, of the black holes  which successfully reproduces \cite{Carlip:1994gc,Strominger:1996sh,Strominger:1997eq,Birmingham:1998jt}  the gravitational Bekenstein-Hawking entropy, is usually the Cardy formula \cite{Cardy:1986ie,Bloete:1986qm} in these dualities, and it turns out to be built out of some of the same quantities as the thermodynamic volume~$V$. What we show is that working at fixed, positive~$V$ places a condition on the CFT sector meaning that the (naive) Cardy formula {\it over--counts} the entropy in the CFT. This is the microscopic herald of the fact that the gravity entropy (as counted by Cardy) is, in a precise sense, ``too much''. 


As a simple first check of our microscopic formalism and our assertion that super--entropicity is connected to the over--counting seen in the CFT, we study a rather large family of examples. These ``generalized exotic'' BTZ black holes~\cite{Carlip:1991zk,Carlip:1994hq,Townsend:2013ela} have a rich extended thermodynamics~\cite{Cong:2019bud} with $C_V{\neq}0$ that can also be written in  two dimensional  CFT terms. While there are sectors that have negative specific heats (both $C_p$ and $C_V$ can be negative for some ranges of parameters, and positive for other ranges) these examples, which are non--unitary in some cases,  are {\it not} super--entropic~\footnote{Here we disagree with the interpretation of  ref. \cite{Cong:2019bud}. They use a definition of super-entropic inherited from the geometrical formula of ref.~\cite{Cvetic:2010jb} that focuses on area $A$, and not entropy, $S$. They therefore conclude that there is a problem with the conjecture connecting super--entropicity to negative $C_V$ since they can find regions with positive $C_V$.  However, we are (as is ref.~\cite{Johnson:2019mdp}) using the entropy-focused interpretation of the term super-entropic as opposed to the (less physical) area-focused usage.}. In the spirit of our methods,  the thermodynamic volume~$V$ can be written in terms of CFT quantities. Doing  so, we see that working at fixed $V$  does {\it not} result in the Cardy formula over-counting the CFT entropy. This therefore fits with our suggestion that super--entropicity is heralded by such an over-count at fixed $V$.



\subsection{CFT and Standard BTZ}
\label{sec:cft-btz}

Two principal quantities in  two dimensional conformal field theory are the energy $E$ and the spin $J$, which  are given in terms of the sum and difference of the eigenvalues, $\Delta,{\bar \Delta}$,  of $L_0$ and ${\bar L}_0$, the zeroth components of the right and left Virasoro generators (which define the conformal algebra):
\begin{equation}
\label{eq:energy-spin}
E = \frac{\Delta+{\bar \Delta}}{\ell}\ , \quad J=\Delta-{\bar \Delta}\ .
\end{equation}
Here $\ell$ is a length scale set by the cosmological constant of the dual gravity theory {\it via} $\Lambda=-1/\ell^2$. The right and left Virasoro algebras have central charges~$c_R$ and $c_L$, which are proportional to $\ell$. Their precise values are example dependent, as we shall see. The values of $E$ and $J$ are computed in the dual gravity theory quite readily, and are the mass~$M$ and angular momentum~$J$ of the black hole spacetime. The entropy on the gravitational side is computed using the Bekenstein--Hawking formula, the quarter of the area of a horizon.  (Note that ``area'' here will mean the circumference of a circle, since there are only two spatial dimensions in the gravity theory. There may be contributions from more than one horizon, as we shall see in later examples.) On the field theory side, this entropy is reproduced in the field theory using~\cite{Carlip:1994gc,Strominger:1996sh,Strominger:1997eq,Birmingham:1998jt} the Cardy formula for the asymptotic degeneracy of states with a given conformal dimension:
\begin{equation}
\label{eq:cardy}
S = \log(\rho(\Delta,{\bar\Delta}))=2\pi\sqrt{\frac{c_R\Delta}{6}}+2\pi\sqrt{\frac{c_L{\bar\Delta}}{6}}\ .
\end{equation}
Crucially, this formula's validity depends upon the key  assumption that the lowest  $L_0,{\bar L}_0$  eigenvalues vanish~\cite{Carlip:1998qw}. We will revisit this issue shortly.

For  the examples discussed in this paper, the spacetime metric will be of the leading Ba\~nados, Teitelbiom and Zanelli (BTZ)~\cite{Banados:1992wn,Banados:1992gq} form:
\begin{eqnarray}
\label{eq:black-hole}
ds^2 &=& -f( r)dt^2
+ f(r)^{-1}dr^2 + r^2 \left(d\varphi-\frac{4j}{r^2}dt\right)^2\ ,\nonumber\\ && f( r) = -8m+\frac{r^2}{\ell^2}+\frac{16j^2}{r^2}+\cdots\ , \label{eq:metric}
\end{eqnarray}
(with one exception we will discuss separately).
The black hole has an outer and inner horizon, at radii denoted $r_{\pm}$, which are the larger and smaller roots of $f(r)=0$. Depending upon the parent gravity theory in question (examples below), the parameters $m$ and $j$ determine the black hole mass $M$ and angular momentum $J$ either directly or in linear combination. The classic  BTZ example has $f(r)$ as written (no extra terms) and $M{=}m$ and $J{=}j$, and together with $S$ they are:
\begin{equation}
\label{eq:gravity-btz}
M=\frac{r_+^2+r_-^2}{8\ell^2}\ ,\quad J=\frac{r_+r_-}{4\ell}\ ,\quad S=\frac{\pi r_+}{2}.
\end{equation}
Comparing the first two quantities to those in equation~(\ref{eq:energy-spin}) gives, after a little algebra:
\begin{equation}
\label{eq:Deltas}
\Delta = \frac{(r_++r_-)^2}{16\ell}\ , \quad{\rm and}\quad{\bar\Delta} = \frac{(r_+-r_-)^2}{16\ell}\ .
\end{equation}
Using these in equation~(\ref{eq:cardy}) with $c_R{=}c_L{=}3\ell/2$ yields the gravity entropy in equation~(\ref{eq:gravity-btz}). 

In extended thermodynamics, the pressure is given by $p{=}1/8\pi\ell^2$, and the mass $M$ is the enthalpy 
\begin{equation}
\label{eq:enthalpy}
H(S,p)=4\pi p \left(\frac{S}{\pi}\right)^2+\frac{\pi^2J^2}{2S^2}\ .
\end{equation}
We will work at fixed $J$ henceforth, treating it as a parameter. The first law remains as in equation~(\ref{eq:first-law}).
Hence, the thermodynamic volume and temperature turn out to be 
\begin{equation}
\label{eq:temp-vol}
V\equiv\left.\frac{\partial H}{\partial p}\right|_S=\pi r_+^2\ ,\quad
 T\equiv\left.\frac{\partial H}{\partial S}\right|_p=\frac{r_+^2-r_-^2}{2\pi\ell^2 r_+}\ ,
\end{equation}
the latter agreeing with either a surface gravity computation or the requirement of regularity of the Euclidean section~\cite{Gibbons:1976ue}.

We can go a step further. The CFT/gravity relations~(\ref{eq:Deltas}) can be inverted to give $r_\pm$ in terms of $\Delta$ and ${\bar \Delta}$, and so we can write $V$ in terms of CFT quantities as\footnote{Here we have cheated a little bit by setting $G=1$. Here we are missing a factor of $G$, which we can subsequently replace $c_{R}=c_{L}=3\ell/2G$. This does not change our overall findings.}:
 \begin{equation}
 \label{eq:volume-microscopic}
 V=\frac{8\pi}{3}\left(\sqrt{c_R\Delta}+\sqrt{c_L{\bar\Delta}}\right)^2\ .
 \end{equation}
 We propose that this relationship should be read in an analogous manner to how the Cardy formula in equation~(\ref{eq:cardy}) is read. States can be constructed in the CFT in the usual manner, acting on the vacuum with the left and right (negatively moded) Virasoro generators as creation operators. Then $L_0$ and ${\bar L}_0$ measure $\Delta$ and ${\bar \Delta}$. For given values of these quantities, equation~(\ref{eq:volume-microscopic}) defines a quantity $V$ that has the interpretation as the thermodynamic value in the gravity theory. Since it is made from (the square of) the same combination of CFT quantities that $S$ is built from, there is not much more to learn from this example. Questions about $V$ are equivalent to questions about~$S$, as they are not independent quantities.


\subsection{Charged BTZ Black Holes}
\label{sec:charged-btz}

Our first example where something new arises is the charged BTZ black hole with no angular momentum, a solution of Einstein--Maxwell in three dimensions~\cite{Martinez:1999qi}. Now, we have $J{=}0$ and  the metric function to use in equation~(\ref{eq:metric}) is instead $f(r){=}-8M+\frac{Q^2}{2}\log\left({r}/{\ell}\right)+{r^2}/{\ell^2}$, where~$Q$ is the $U(1)$ charge of the solution and $M$ is the mass. There is also a gauge field $A_t = Q\log\left({r}/{\ell}\right)$. From the point of view of the two dimensional CFT, $Q$ is merely a deformation parameter, a global charge, which will be kept fixed here. The extended thermodynamics gives~\cite{Frassino:2015oca}:
\begin{eqnarray}
\label{eq:thermodynamic-quantities}
H&=&\frac{4pS^2}{\pi} - \frac{Q^2}{32}\log\left(\frac{32pS^2}{\pi}\right)\ ,\quad S=\frac{\pi}{2}r_+\ ,\nonumber\\
T&=&\frac{8pS}{\pi}-\frac{Q^2}{16S}\ ,\quad \quad V= \frac{4S^2}{\pi} -\frac{Q^2}{32p}\ ,
\end{eqnarray}
and the first law is again equation~(\ref{eq:first-law}). The internal energy of the system is given by $U{\equiv}H{-}pV{=}(Q^2/32)[1-\log(32pS^2/\pi)]$. 

Note that the presence of the charge  $Q$ introduces a $\log(r/\ell)$ term in the metric function $f(r)$. Consequently, the asymptotic symmetry group of the geometry is deformed, hiding the action~\cite{Brown:1986nw} of the Virasoro algebra. Crucially, we regard Virasoro as hidden, but not absent. We propose that the conformal field theory will still have the structure that we saw in the previous example, and below we will find strong evidence in support of this.

 To make Virasoro explicit requires a different approach. The boundary conditions on the metric and gauge field can be modified by enclosing the entire black hole system inside some radius~$r_{0}$ and introducing a renormalized mass according to $M(r_{0})=M+\frac{Q^{2}}{16}\log\left(r_{0}/\ell\right)$, such that the manifest asymptotic Virasoro symmetry  is restored \cite{Cadoni:2007ck}.  This alternative scheme rearranges the thermodynamic quantities (both traditional and extended). In the resulting extended thermodynamics (which requires promoting the scale $r_0$ to a dynamical variable in order to have a consistent first law~\cite{Frassino:2015oca}) the thermodynamic volume $V$ loses its $Q$ dependence, becoming the geometric volume $\pi r_+^2$, and since $S{=}\pi r_+/2$, we have  $C_{V}{=}0$. Hence,  we will not study this renormalized scheme 
 and instead focus our attention on the thermodynamic quantities as presented in equations~(\ref{eq:thermodynamic-quantities}), which yield an interesting case study.  We will revisit the renormalized scheme in later discussion.

Notice that~$V$ and $S$ in equation~(\ref{eq:thermodynamic-quantities}) are now independent. The requirement that the temperature be positive results in the restriction $Q^2{\leq}4\eta$, where $\eta{=}32pS^2/\pi$. Since $V{=}TS/2p$, this also translates into positivity of the volume $V$. The parameter $\eta$ also appears in the internal energy $U$, and requiring that $U{>}0$ gives $\eta{\leq}1$. So, just from the gravity side, we get the bound~$Q^2{\leq}4$.

Turning to the  CFT quantities, $c_L{=}c_R{=}3\ell/2{=}c$ as before, and since $J{=}0$ we have  $\Delta{=}{\bar\Delta}$. The Cardy formula gives the entropy as before: $S=4\pi\sqrt{c\Delta/6}$, but now the thermodynamic volume $V$, written in terms of CFT quantities, is:
\begin{equation}
\label{eq:thermodynamic-volume-charged-btz}
V=\frac{32\pi c}{3}\left(\Delta-\frac{Q^2c}{96}\right)\ .
\end{equation} 
Positivity of $V$ (following from positivity of $T$) translates into a non--trivial statement: The lowest $\Delta$ can be is $\Delta_0{=}Q^2c/96$. Recall that an assumption underlying the Cardy formula~(\ref{eq:cardy}) is that $\Delta_0{=}0$. In fact, when $\Delta_0{\neq}0$, the correct formula to use for the (logarithm of the) asymptotic density of states replaces $c$ by $c_{\rm eff}{\equiv}c{-}24\Delta_0$, resulting in (for positive $\Delta_0$) a reduction of the entropy count \cite{Carlip:1998qw}. For us, $c_{\rm eff}{=}c(1{-}Q^2/4)$, and we recover two interesting pieces of information. The first is that the gravity entropy, which corresponds to the naive Cardy formula, {\it over--counts the number of degrees of freedom of the theory}. The second is that there is a unitarity bound  of $Q^2{\leq}4$, the same bound we obtained by independent gravity requirements that~$T$ and $U$ are positive!

That we have recovered precisely the same condition on $Q$ using two very different considerations (gravity and CFT) is strong support for our proposal for writing a microscopic/CFT formula for $V$. It also strongly suggests that we were correct to  use  the AdS$_3$/CFT$_2$ map for this charged black hole despite the fact that the asymptotic algebra is deformed by the presence of~$Q$.

The  over--counting of the entropy discovered here suggests that something is wrong with the equilibrium thermodynamics suggested by the variables in equation~(\ref{eq:thermodynamic-quantities}). We propose that  it is in fact  a herald of the phenomenon called super--entropicity, discussed next.

\subsection{Super--Entropicity and Instability}
\label{sec:super-entropic}

The charged BTZ solution is the simplest example of a super--entropic black hole \cite{Frassino:2015oca}, as it violates the reverse isoperimetric inequality (\ref{revisogend}),
\beq 4S^{2}>\pi V\;.\eeq


It was  recently observed \cite{Johnson:2019mdp} that several super-entropic black holes are thermodynamically unstable, signified by a negative heat capacity $C_{V}$. It was conjectured there that  super-entropicity may generally imply that $C_V{<}0$, following from the fact that for  a charged BTZ black hole this can be verified analytically: The temperature $T$ and $C_{V}$ take the form:
\begin{eqnarray}
 T&=&\frac{\pi V}{16S}\frac{Q^{2}}{(4S^{2}-\pi V)}\;,\quad  C_{V}=-S\left(\frac{4S^{2}-\pi V}{12S^{2}-\pi V}\right)\;.\label{TCVchargedbtz}
\end{eqnarray}
The temperature is positive when $4S^{2}{>}\pi V$, which is equivalent to the $d{=}3$ super--entropicity condition~$\mathcal{R}{<}1$. Moreover, this is precisely when the charged BTZ solution has  $C_{V}{<}0$, {\it i.e.,} it is thermodynamically unstable.
(Showing that $C_{V}{<}0$ when~$\mathcal{R}{<}1$   was also verified numerically in ref.~\cite{Johnson:2019mdp} for a class of  ultra--spinning Kerr-AdS black holes in various higher dimensions. Analytic counterparts to the above $d{=}3$ demonstration were not obtained however.) 

Positivity of $T$ ensuring a connection between super--entropicity and instability is strongly reminiscent of what we saw in the previous section, when making connections to the CFT.  When the dual CFT is unitary, we may translate $c_{\text{eff}}{>}0$ into $4S_{\text{CFT}}^{2}{>}\pi V$, where  $S_{\text{CFT}}{=}4\pi\sqrt{c_{\text{eff}}\Delta/6}$. Then, since $S{>}S_{\text{CFT}}$, we have $4S^{2}{>}\pi V$. Therefore, super--entropicity reflects that the gravitational entropy over--counts the number of degrees of freedom of the underlying microscopic theory. 

The over--counting is also accompanied by the negativity of $C_{V}$, which itself suggests an instability, a movement in solution space to some new set of thermodynamic quantities for which $C_V$ is no longer negative. It is tempting to speculate that the extended thermodynamics yielded~\cite{Frassino:2015oca} by studying  the renormalized scheme of ref.~\cite{Cadoni:2007ck} (reviewed briefly below equations~(\ref{eq:thermodynamic-quantities})) is the endpoint of the instability. One suggestion of our observations here is that there is another framework (different from the renormalization scheme  recalled below equation~(\ref{eq:thermodynamic-quantities})) in which the asymptotic Virasoro algebra is restored, but in which the central charge is modified to our effective central charge $c_{\rm eff}{=}c(1{-}Q^2/4)$. It would be interesting to find such a framework, and to see whether the resulting thermodynamic quantities produce a super-- or sub--entropic system.


\subsection{Generalized Exotic BTZ Black Holes}
\label{sec:exotic-btz}

As a final example we consider the family of ``generalized exotic BTZ'' black holes \cite{Carlip:1991zk,Carlip:1994hq,Townsend:2013ela}. The relevant gravity theory is a linear combination of the Einstein--Hilbert action and the gravitating Chern--Simons action, $I=\alpha I_{\rm EM}+\gamma I_{\rm GCS}$, where $\gamma=1{-}\alpha$. The metric is again given in equation~(\ref{eq:black-hole}), with no extra terms for $f(r)$, but this time the mass and angular momentum mix the parameters $m$ and $j$: $M=\alpha m{+}\gamma j/\ell$, $J=\alpha j {+} \gamma \ell m$. The case of $\alpha{=}1$ is the standard BTZ black hole, while $\gamma{=}1$ is the exotic BTZ black hole. General $0\leq\alpha\leq1$ interpolates between these two extremes. The thermodynamic variables are given by:
\beq
\begin{split}
\label{eq:thermodynamics-exotic}
&M=\frac{\alpha(r_+^2+r_-^2)}{8\ell^2} + \frac{\gamma r_+r_-}{4\ell^2}\ ,\quad J=\frac{\alpha r_+r_-}{4\ell}+\frac{\gamma(r_+^2+r_-^2)}{8\ell}\ ,\quad \Omega=\frac{r_-}{r_+\ell}\ , \\
&T=\frac{r_+^2-r_-^2}{2\pi\ell^2 r_+}\ ,\quad S=\frac{\pi}{2}(\alpha r_++\gamma r_-)\ , \quad V=\alpha\pi r_+^2+\gamma\pi r_-^2\left(\frac{3r_+}{2r_-}-\frac{r_-}{2r_+}\right)\ ,
\end{split}
\eeq
where $\Omega$ is the angular velocity. 

Recently it was shown that generalized exotic BTZ solutions can have $C_{V}$ both positive and negative \cite{Cong:2019bud}. Specifically, for $\alpha<1/2$, $C_{V}$ is positive for large enough $r_{+}$. In the regions where $C_{V}>0$, however, the heat capacity at constant pressure $C_{p}$ will be negative, indicating that they are generally unstable. Notice that for the  inequality (\ref{revisogend}), we have 
\begin{equation}
\mathcal{R}=\frac{1}{2(\alpha+\gamma x)}\sqrt{4\alpha+6\gamma x-2\gamma x^{3}}\;\ ,
\end{equation}
where $x\equiv r_{-}/r_{+}$ ranges between $0$ and $1$. For the defined range of non--zero $\alpha$, we find $\mathcal{R}>1$, and thus these generalized exotic BTZ black holes form a class of {\it sub--entropic} black holes. Had we instead used the form of $\mathcal{R}$ first written  in \cite{Cvetic:2010jb}, we would have found $\mathcal{R}<1$  and concluded that these solutions are super--entropic, as ref. \cite{Cong:2019bud} does. However, as we have already stated, we are using the entropy--focused interpretation of the term super--entropic as opposed to the (less physical) area--focused usage. In this sense, in the spirit of ref.'s~\cite{Johnson:2019mdp} conjecture and what we've seen in the previous two sections, there is no super--entropicity and hence $C_V$ does not need to become negative, since the solution does not need to somehow shed the extra entropy.

Turning to the dual conformal field theory, some algebra shows that variables $M$, $J$, and $S$  fit the CFT form given in equations~(\ref{eq:energy-spin}) and~(\ref{eq:cardy}), (with    factors  $\alpha{+}\gamma{=}1$ for  right--moving quantities  and $\alpha{-}\gamma{=}2\alpha{-}1$  for left--moving):
\begin{eqnarray}
\Delta &=& \frac{1}{16\ell}(r_+^2+r_-^2),\;\;
{\bar \Delta} = \frac{2\alpha-1}{16\ell}(r_+^2-r_-^2), \;\; c_R = \frac{3\ell}{2},\;\; c_L  =  \frac{3\ell}{2} (2\alpha-1)\ .
\end{eqnarray}
We may recast the thermodynamic volume $V$ (\ref{eq:thermodynamics-exotic}) in terms of these CFT parameters. The resulting expression is:
\beq
\begin{split}
\label{eq:volexoticcft}
& \frac{3V}{4\pi c_{R}}=\left(1+\frac{1}{\epsilon}\right)\left(\sqrt{\Delta}+\sqrt{\epsilon\bar{\Delta}}\right)^{2} +\left(1-\frac{1}{\epsilon}\right)\left(\frac{\sqrt{\Delta}-\sqrt{\epsilon\bar{\Delta}}}{\sqrt{\Delta}+\sqrt{\epsilon\bar{\Delta}}}\right)\left[\Delta+\epsilon\bar{\Delta}+4\sqrt{\epsilon\Delta\bar{\Delta}}\right]\;,
\end{split}
\eeq
where 
$\epsilon{\equiv}{c_{R}}/{c_{L}}$.
Note that  $c_R{=}c_L$ when $\alpha{=}1, \gamma=0$, {\it i.e.}, we have the usual BTZ solution of section~\ref{sec:cft-btz}, and our expression  (\ref{eq:volexoticcft}) reduces to the thermodynamic volume given in equation~(\ref{eq:volume-microscopic}). 

The key observation from (\ref{eq:volexoticcft}) is that, unlike the charged BTZ case,  requiring  positivity of~$V$ does not lead to a shift away from zero for  the lowest value of $\Delta$ or $\bar{\Delta}$. As such, the gravitational entropy (as given by the Cardy formula) does not over--count the number of microscopic degrees of freedom.  This fits with the observation above that there is no super--entropicity in these examples (using the entropy--focused definition of ${\cal R}$ in equation~(\ref{revisogend})).


\section*{Summary and Future Work}
\label{sec:conclusion}
In conclusion, we have shown how to microscopically interpret (using AdS$_3$/CFT$_2$ duality) the thermodynamic volume of extended black hole thermodynamics, by writing formulae for it in terms of CFT quantities. For simple black holes where $V$ and $S$ are not independent, such a formula is no more useful than the Cardy formula for $S$. However, deploying the interpretation in the charged BTZ example where $V$ is independent of $S$, we  uncovered that the naive Cardy formula over--counts the entropy of the theory. We interpret this as a microscopic herald  of the super--entropicity phenomenon associated to some solutions in extended thermodynamics.

Independent conditions derived from gravity and CFT gave precisely the same bound on $Q$, the black hole charge: $Q^2{\leq}4$, suggesting  internal consistency of our methods. These methods included using the CFT dual of the charged black hole solution even though the presence of $Q$ deforms the asymptotic symmetry (Virasoro) algebra. This might suggest that there  is another framework (different from the renormalization scheme  recalled below equation~(\ref{eq:thermodynamic-quantities})) in which the asymptotic Virasoro algebra is restored, but in which the central charge is modified to our effective central charge $c_{\rm eff}{=}c(1{-}Q^2/4)$. It would be interesting to find such a framework, and to see whether the resulting thermodynamic quantities produce a super- or sub-entropic system.

It would  also be interesting to find a similar microscopic understanding of super--entropicity of ultra--spinning black holes~\cite{Hennigar:2014cfa,Hennigar:2015cja}. These solutions exist for $d\,{\geq}\,4$, where we can no longer use $\text{AdS}_{3}/\text{CFT}_{2}$ duality. Instead, however, we could consider Kerr/CFT duality \cite{Guica:2008mu}, along the lines of ref.~\cite{Sinamuli:2015drn}, and see if constraints imposed by the gravitational thermodynamics lead to any requirements on the dual CFT. We leave this  for future work. 

Another line of investigation could be to develop further a  characterization of how super--entropicity may result in the $C_V{<}0$ instability for other black holes, and in other dimensions.  As conjectured in ref.~\cite{Johnson:2019mdp}, a consequence  of super--entropicity is  negativity of $C_{V}$. (Note again that this is not the same as saying that negativity of $C_V$ implies super--entropicity.) For the charged BTZ case this was shown directly in equation~(\ref{TCVchargedbtz}), where the form and sign of $C_{V}$ depends solely on the ability to write the temperature as
$T{=}{\mathcal{F}(S,V,Q)}/{(1{-}\mathcal{R})}\ ,$
where $\mathcal{F}$ is a function we wish to characterize further, and the $\mathcal{R}$ in the denominator is  given in equation~(\ref{revisogend}).
 Not every black hole solution will have a temperature that can be written in this form, as we see in the cases of the  uncharged and exotic BTZ black holes. Moreover, we know of sub--entropic solutions whose temperature does take this form, {\it e.g.}, the $d{=}4$ Kerr--AdS black hole \cite{Dolan:2011xt}. 
Nonetheless, we might attempt to learn something about  a sub--class of super--entropic black hole solutions by demanding the temperature take the form given above. If they have negative $C_{V}$, it implies conditions on $\mathcal{F}$. Our special form of $T$ together with the fact that $T{=}f'(r_{+})/{4\pi}$ for a gravity solution with metric function $f(r)$  might characterize enough about the properties of  $f(r)$ to use it as a diagnostic tool for a wide variety of solutions. 



\newpage

\section{THE EXTENDED FIRST LAW OF ENTANGLEMENT IN ARBITRARY DIMENSIONS}  \label{sec:extfirstlawhighlow}

The first law of entanglement,
\beq \delta S_{EE}=\delta\langle H_{A}\rangle\;,\label{firstlaweegen}\eeq
where $S_{EE}$ is the entanglement entropy across a subsystem $A$, with a state $\rho_{A}=Z^{-1}e^{-H_{A}}$ described in terms of a modular Hamiltonian $H_{A}$, is a natural generalization of the first law of thermodynamics that applies to non-equilibrium states. As first shown in \cite{Blanco:2013joa}, it is a consequence of positivity of relative entropy, and determines the first order variation of entanglement entropy under state perturbations. Its most interesting application is arguably given in \cite{Lashkari:2013koa,Faulkner:2013ica}, where it plays a crucial role in deriving the bulk linearized Einstein's equations about a perturbed AdS background from boundary entanglement correlations of the CFT. 

 Motivated by extended black hole thermodynamics \cite{Kastor:2009wy,Dolan:2010ha,Kubiznak:2016qmn}, where the cosmological constant $\Lambda$ is interpreted as a thermodynamic pressure $p\equiv -\Lambda/8\pi G$, an extension of the first law of entanglement was proposed in  \cite{Kastor:2014dra}, which includes not only variations of the state but also of the CFT itself. It can be written as
\begin{equation}\label{eq:ext}
\delta S_{EE}=\delta \langle K_B \rangle+
  \frac{S_{EE}}{a_d^\ast}
  \delta a_d^\ast\ ,
\end{equation}
where now $S_{EE}$ is the vacuum entanglement entropy associated to a ball in Minkowsk spacei and $K_B$ its modular hamiltonian. The constant $a_d^\ast$ is defined for an arbitrary CFT as
\begin{equation}\label{eq:32}
a_d^\ast=
  \begin{cases} 
  \qquad \qquad \,\,\,\,\,
  A_d
  \qquad \quad \,\,\,\, 
  \ ,
  & {\rm for\,\,d\,\,even} \vspace{6pt}\\
  \,\,(-1)^{\frac{d-1}{2}}\ln[
  Z(S^d)]/2\pi\ ,
  &  {\rm for\,\,d\,\,odd}\ . \\
 \end{cases}
\end{equation}
Here $A_d$ is the coefficient in the trace anomaly proportional to Euler's density, while for odd dimensions $a_d^\ast$ is determined by the partition function of the CFT placed on a unit sphere $S^{d}$ (see \cite{Pufu:2016zxm} for some examples in free theories).  Since $a_d^\ast$ has a monotonous behavior under renormalization group flows \cite{Casini:2017vbe}, we can interpret it as counting the number of degrees of freedom in the CFT. The generalized central charge $a^{\ast}_{d}$ has appeared in a number of holographic $c$-theorems in arbitrary dimensions and higher curvature theories of gravity \cite{Myers:2010tj}.

The first term in (\ref{eq:ext}) is the ordinary contribution to the first law obtained by perturbing the state, while the second gives the behavior of the entanglement entropy when varying the CFT. We must emphasize that this second contribution is \textit{not} equivalent to a renormalization group flow, since the variation continuously interpolates between CFTs. It simply gives the dependence of the entanglement entropy on the CFT data.

The extended first law (\ref{eq:ext}) was initially derived in \cite{Kastor:2014dra} for a holographic CFT dual to Einstein gravity, and later generalized to specific higher curvature gravity theories in \cite{Kastor:2016bph,Caceres:2016xjz,Lan:2017xcl}. These derivations start by considering a particular Killing horizon in pure AdS and deriving an extended bulk first law which considers variations of the cosmological constant, using either Hamiltonian perturbation theory \cite{Kastor:2016bph} or the Iyer-Wald formalism \cite{Caceres:2016xjz}. The horizon entropy associated to this Killing horizon is then identified as the entanglement entropy of the boundary CFT, while the variation of the cosmological constant maps to changing the generalized central charge $a^{\ast}_{d}$. 

Given the importance and wide range of applications of the first law of entanglement, we should take any reasonable generalization seriously, as it has the potential of providing new insights into the structure of space-time and entanglement in QFTs. In this work we explore the extended first law of entanglement (\ref{eq:ext}) by generalizing previous derivations to include arbitrary theories of gravity, clarifying some of its subtle features and studying its low dimensional limit.

The outline is as follows.  We start in section \ref{sec:2} by showing that a remarkably simple argument allows us to derive the bulk analog of (\ref{eq:ext}) for perturbations of any Killing horizon in pure AdS. Contrary to previous derivations, our computation is novel in its simplicity and the fact that it holds for arbitrary bulk gravity theories and Killing horizons in pure AdS, finding no need to resort to technical calculations as in \cite{Kastor:2014dra,Kastor:2016bph,Caceres:2016xjz,Lan:2017xcl}. We discuss how each of the bulk quantities is mapped to the boundary CFT, carefully analyzing some subtleties previously overlooked. Applying our construction to certain bulk Killing horizons, we derive the extended first law (\ref{eq:ext}) for the vacuum state of a CFT reduced to the following regions: a ball and the half-space in Minkowski, a spherical cap in the Lorentzian cylinder $\mathbb{R}\times S^{d-1}$ and de Sitter, and a ball in AdS$_d$. The method used to find the appropriate bulk Killing horizons crucially relies on the freedom to choose conformal frames at the AdS boundary.


We continue in section \ref{sec:2Dlaw}, where we revisit the calculations from  section \ref{sec:2} but carefully analyzing the case in which the bulk theory is two-dimensional. Although this was not considered in previous work, we find no obstructions for the extended first law for Killing horizons in pure AdS$_2$. Motivated by earlier work in extended thermodynamics in two dimensions \cite{Frassino:2015oca}, we point out some connections with Einstein-dilaton theories, where there are certain Einstein-dilaton theories in which the end result takes a different form. We illustrate this for Jackiw-Teitelboim gravity \cite{Teitelboim:1983ux,Jackiw:1984je}, where we show the extended first law for Killing horizons takes a different form.

In section \ref{sec:3Dlaw} we show that in three dimensional gravity an extended first law can be derived for Killing horizons in space-times that are locally but not globally AdS. This allows us to obtain an extended first law for the boundary CFT$_2$ that is analogous to (\ref{eq:ext}) but involving thermal instead of entanglement entropy. From the bulk perspective we find some interesting results for  extended black hole thermodynamics, where we obtain a curious formula for the thermodynamic volume (see Eq. (\ref{eq:36})), the conjugate variable to the pressure $p$.


We conclude in section \ref{sec:4} by expanding some discussions on the calculations in the main text. We clarify some aspects regarding the structure of divergences in the extended first law of entanglement (\ref{eq:ext}) and critically analyze the extent to which it can hold for arbitrary regions and CFTs. We briefly comment on the bulk constraints implied by assuming both the RT holographic entropy formula \cite{Ryu:2006bv} and the extended first law of entanglement holds for arbitrary setups in the boundary CFT. We also discuss additional potential applications of the quantum-corrected extended first law of entanglement in the context of JT gravity. Finally, we discuss some interesting aspects of the thermodynamic volume in three dimensional gravity and its connection to the microscopic interpretation of black hole super-entropicity \cite{Cvetic:2010jb}.

\subsection{Killing Horizons in Pure AdS and Extended (Bulk) First Law}
\label{sec:2}

In this section we present a derivation of the extended first law of entanglement for holographic CFTs described by arbitrary covariant theories of gravity in the bulk. There are essentially three steps to deriving the extended first law of entanglement: (i) we start with a bulk Killing horizon in pure AdS and derive the extended bulk first law, relating variations of the horizon entropy to variations of the conserved charge associated with the Killing symmetry and coupling constants of the theory; (ii) we then take the boundary limit of the bulk space, defining the boundary spacetime on which the CFT lives, and map each of the quantities appearing in the extended bulk first law to a boundary field theory statement, and (iii) finally, we make the connection to boundary CFT entanglement by considering a specific Killing horizon in pure AdS and and show that its boundary limit has an entanglement entropy intepretation. 

Here we present a derivation of the extended bulk first law for holographic CFTs described by arbitrary covariant theories of gravity in the bulk
\begin{equation}\label{genaction}
I[\lambda_i,g_{\mu \nu}]=
  \int d^{d+1}x\,\sqrt{-g}\,
  \mathcal{L}\left(g_{\mu \nu},
  \mathcal{R}_{\mu \nu\rho \sigma},\nabla_\lambda \mathcal{R}_{\mu \nu\rho \sigma},\dots
  \right),
\end{equation}
where $\mathcal{R}_{\mu \nu \rho \sigma}$ is the Riemann tensor. Each theory is characterized by a family of coupling constants $\left\lbrace \lambda_i \right\rbrace$ that are chosen such that the action admits a pure AdS vacuum solution of radius $L$. This length scale is a non-trivial function of the coupling constants of the theory~$L=L(\lambda_i)$, and the pure AdS metric only depends on $\left\lbrace \lambda_i \right\rbrace$ through $L$. Although we could also add some matter to the action, for the most part we consider pure gravity and set matter fields to zero. We will present an illustrative example momentarily.

Consider a Killing vector $\xi^\mu$ of the pure AdS metric $g^{\rm AdS}_{\mu \nu}(L)$ which is time-like over some region
\begin{equation}\label{eq:15}
\xi^2\equiv g_{\mu \nu}^{\rm AdS}
  \xi^\mu \xi^\nu\le 0
  \qquad \Longleftrightarrow \qquad
  {\rm Some\,\,region\,\,of\,\,AdS}\ .
\end{equation}
The surface in which the vector vanishes defines a Killing horizon. One of the central quantities characterizing this horizon is its entropy, that for an arbitrary theory is computed from Wald's functional\footnote{The entanglement entropy of CFTs dual to higher derivative theories of gravity is famously not given by the Wald entropy, but instead the Jacobson-Myers entropy \cite{Hung:2011xb,Dong:2013qoa}. However, these two proposals match when the bulk surface of integration is a bifurcate Killing as in the case we are considering here.} according to \cite{Wald:1993nt,Iyer:1994ys}
\begin{equation}\label{Waldent}
S_{\xi}
  \left[
  g_{\mu \nu}^{\rm AdS}(L),\lambda_i
  \right]=
  -2\pi \int_{\Sigma}
  dV
  \left[
  \frac{\delta \mathcal{L}}
  {\delta \mathcal{R}^{\mu \nu}_{\,\,\,\,\,\,\rho \sigma}}
  n^{\mu \nu}
  n_{\rho \sigma}
  \right],
\end{equation}
where the integral is over the bifurcation Killing surface $\Sigma$ with induced volume element $dV$. The anti-symmetric tensor $n^{\mu \nu}$ is the binormal to the horizon normalized so that $n^{\mu\nu}n_{\mu \nu}=-2$. Our aim is to study the behavior of this entropy functional under general perturbations and to determine its consequences for the boundary CFT.

Let us start by considering the behavior of the entropy under metric perturbations $g_{\mu \nu}^{\rm AdS}(L)\rightarrow  g_{\mu \nu}^{\rm AdS}(L)+\delta g_{\mu \nu}$.\footnote{The perturbation $\delta g_{\mu \nu}$ can be any metric which satisfies the equations of motion obtained from (\ref{genaction}) linearized around pure AdS.} Since we are working with a Killing horizon we can apply the same methods used to study black hole thermodynamics. The first order variation of~(\ref{Waldent}) was computed in \cite{Iyer:1994ys} and shown to be given by 
\begin{equation}\label{Waldvar1}
\delta S_\xi=
  \frac{2\pi}{\kappa}
  \delta Q_\xi\ ,
  \qquad \qquad \quad
  \kappa^2=-\frac{1}{2}\left(\nabla^\mu \xi^\nu\right)\left(\nabla_\mu \xi_\nu\right)\ ,
\end{equation}
where $\kappa$ is the surface gravity and $Q_\xi$ the conserved charge associated to the symmetry generated by $\xi^\mu$. 

We now consider another type of perturbation obtained by changing the gravitational theory itself, \textit{i.e.} $\mathcal{L}\rightarrow \mathcal{L}+\delta \mathcal{L}$, implemented by slightly changing the coupling constants of the theory $\lambda_i\rightarrow \lambda_i+\delta \lambda_i$. Since the pure AdS metric $g_{\mu \nu}^{\rm AdS}(L)$ is a function of $\lambda_i$ through~${L=L(\lambda_i)}$, the perturbation induces a variation of the metric. If we did not take this metric variation into account, the perturbed metric would not be a solution of the perturbed Lagrangian. Hence, the first order variation of Wald's functional is explicitly given by
\begin{equation}\label{eq:Waldvar2}
\delta S_\xi=
  S_\xi\left[g_{\mu \nu}^{\rm AdS}(\lambda_i+\delta \lambda_i),
  \lambda_i+\delta \lambda_i\right]-
  S_\xi\left[g_{\mu \nu}^{\rm AdS}(\lambda_i),
  \lambda_i\right].
\end{equation}

From the definition of Wald's entropy in (\ref{Waldent}) we can compute this in full generality, the key feature being that both terms are evaluated in the pure AdS metric of each theory. Since AdS is maximally symmetric, the integrand in (\ref{Waldent}) can be evaluated explicitly \cite{Myers:2010tj} and written as\footnote{To obtain this general expression, all that is required is that the metric is locally AdS. Then (\ref{eq:24}) comes from computing the equations of motion for an arbitrary theory evaluated in a local AdS background. See section 5.2 of \cite{Myers:2010tj} for details. The observation of requiring only local AdS will prove useful in Sec. \ref{sec:3Dlaw}, where it allows us to extend some of our results beyond pure AdS in three dimensional gravity.}
\begin{equation}\label{eq:24}
\left.
\frac{\delta \mathcal{L}}
{\delta \mathcal{R}^{\mu \nu}_{\quad \rho \sigma}}
\right|_{\rm AdS}=
-\frac{L^2}{4d}
 \left(
 \delta_{\mu}^{\rho}
 \delta_{\nu}^{\sigma}-
 \delta_{\mu}^{\sigma}
 \delta_{\nu}^{\rho}
 \right)
 \left.
 \mathcal{L}
 \right|_{\rm AdS}\ ,
\end{equation}
where $\mathcal{L}\big|_{\rm AdS}$ is the Lagrangian density (\ref{genaction}) evaluated in the pure AdS solution. Using this, we can evaluate Wald's functional and write it as
\begin{equation}\label{eq:2}
S_{\xi}
  \left[
  g_{\mu \nu}^{\rm AdS}(\lambda_i),\lambda_i
  \right]=
  \frac{4\pi a_d^\ast(\lambda_i)}
  {{\rm Vol}(S^{d-1})}
  \widetilde{\mathcal{A}}_{\rm horizon}
  \ ,
\end{equation}
where $\widetilde{\mathcal{A}}_{\rm horizon}$ is the horizon area $\mathcal{A}_{\rm horizon}$ divided by the AdS radius $L^{d-1}$. We have identified~$a_d^\ast$ according to \cite{Myers:2010tj,Casini:2011kv}
\begin{equation}\label{eq:9}
a_d^\ast(\lambda_i)=-
  \frac{1}{2d}
  {\rm Vol}(S^{d-1})L^{d+1}
  \mathcal{L}
  \big|_{\rm AdS}\ ,
\end{equation}
where ${\rm Vol}(S^{d-1})=2\pi^{d/2}/\Gamma(d/2)$. The coefficient $a_d^\ast=a_d^\ast(\lambda_i)$, which is the generalization of the coefficient of the $A$-type trace anomaly of the energy-momentum tensor for even $d$-dimensional CFTs in a curved background, is in general a complicated function of the coupling constants of the theory. Using (\ref{eq:2}) we can easily evaluate the variation in (\ref{eq:Waldvar2}) and find
\begin{equation}\label{Waldvar3}
\delta S_\xi=
  \frac{S_\xi}{a_d^\ast}\delta a_d^\ast\ ,
  \qquad \qquad
  \delta a_d^\ast(\lambda_i)=\sum_i
  \left(
  \frac{\partial a_d^\ast}{\partial \lambda_i}
  \right)
  \delta \lambda_i\ .
\end{equation}
This expression relies on the fact that the pure AdS metric $g_{\mu \nu}^{\rm AdS}(L)$ is only a function of the length scale ${L=L(\lambda_i)}$, which means the dimensionless horizon area ${\widetilde{\mathcal{A}}_{\rm horizon}=\mathcal{A}_{\rm horizon}/L^{d-1}}$ is independent of $\lambda_i$. In section \ref{sec:3Dlaw} we revisit this when considering more general metrics in three dimensional gravity.

Putting together Eqs. (\ref{Waldvar1}) and (\ref{Waldvar3}), we obtain the following bulk extended first law
\begin{equation}\label{bulkextlaw}
\delta S_\xi=
  \frac{2\pi}{\kappa}
  \delta Q'_\xi+
  \frac{S_\xi}{a_d^\ast}\delta a_d^\ast\ .
\end{equation}
We can already see the similarities of this bulk relation with the extended first law of entanglement (\ref{eq:ext}). For a particular Killing vector $\xi$ in AdS, this result was first obtained in \cite{Kastor:2014dra} for Einstein gravity and later in \cite{Kastor:2016bph,Caceres:2016xjz,Lan:2017xcl} for specific higher curvature gravity theories.\footnote{In some of these papers this relation is not written in terms of the coefficient $a_d^\ast$, but in terms of the coupling constants $\left\lbrace \lambda_i \right\rbrace$ of particular theories.} Our derivation generalizes to arbitrary covariant theories of gravity as well as any Killing horizon in pure AdS. The method is quite simple and follows almost immediately upon evaluating Wald's functional in (\ref{eq:2}).

Finally, let us make an observation regarding the normalization of charge $Q'_\xi$, which describes the `prime' notation. From the derivation of (\ref{Waldvar3}) it is clear that when the variation is only given by ${\lambda_i\rightarrow \lambda_i+\delta \lambda_i}$, the first term in (\ref{bulkextlaw}) vanishes, $\delta Q'_{\xi}=0$, \textit{i.e.} 
\begin{equation}\label{eq:16}
Q_\xi\left[
  g_{\mu \nu}^{\rm AdS}(\lambda_i+\delta\lambda_i),\lambda_i+\delta \lambda_i
  \right]-
  Q_\xi\left[
  g_{\mu \nu}^{\rm AdS}(\lambda_i),\lambda_i
  \right]=0\ .
\end{equation}
Given that there is no reason for these terms to cancel each other for arbitrary values of $\lambda_i$, both must vanish separately. This can be achieved by normalizing $Q'_\xi$  as
\begin{equation}\label{eq:18}
Q'_\xi\left[
  g_{\mu \nu},\lambda_i
  \right]\equiv
  Q_\xi\left[
  g_{\mu \nu},\lambda_i
  \right]-
  Q_\xi\left[
  g_{\mu \nu}^{\rm AdS}(\lambda_i),
  \lambda_i
  \right]\ .
\end{equation}
While this normalization plays no role in (\ref{Waldvar1}) when considering metric perturbations, it gives the appropriate behavior required by (\ref{bulkextlaw}). This prescription is equivalent to subtracting the Casimir energy contribution in pure AdS, that is present for certain foliations of the space-time (see \cite{Emparan:1999pm} for some examples). The procedure is common in extended black hole thermodynamics, where the Casimir energy is not included in the first law \cite{Kastor:2009wy}.

Before analyzing the holographic consequences of the relation (\ref{bulkextlaw}), let us examine the above with a concrete example. Consider Einstein Gauss-Bonnet gravity with Lagrangian
\beq \mathcal{L}_{\text{EGB}}=\left(\frac{\mathcal{R}-2\Lambda}{16\pi G}+\alpha[\mathcal{R}^{2}-4\mathcal{R}^{2}_{\mu\nu}+\mathcal{R}_{\mu\nu\rho\sigma}^{2}]\right)\;,\label{EGBactionex}\eeq
with $\alpha$ being the Gauss-Bonnet coupling. The equation of motion for this action is
\beq G_{\mu\nu}+\Lambda g_{\mu\nu}-8\pi G\alpha\mathcal{L}_{GB}g_{\mu\nu}+32\pi G\alpha\mathcal{H}_{\mu\nu}=0\;,\eeq
with $G_{\mu\nu}$ being the Einstein tensor, $\mathcal{L}_{GB}=\mathcal{R}^{2}-4\mathcal{R}^{2}_{\mu\nu}+\mathcal{R}_{\mu\nu\rho\sigma}^{2}$ being the Gauss-Bonnet Lagrangian, and
\beq \mathcal{H}_{\mu\nu}=\mathcal{R}_{\mu\rho\sigma\kappa}\mathcal{R}_{\nu}^{\;\rho\sigma\kappa}-2\mathcal{R}_{\mu\rho}\mathcal{R}^{\rho}_{\;\nu}-2\mathcal{R}_{\mu\rho\nu\sigma}\mathcal{R}^{\rho\sigma}+\mathcal{R}\mathcal{R}_{\mu\nu}\;.\eeq
Einstein-Gauss-Bonnet gravity admits $\text{AdS}_{d+1}$ as a solution \cite{Hung:2011xb}
\beq ds^{2}=\frac{L^{2}}{z^{2}}(dz^{2}-dt^{2}+d\vec{x}^{2})\;,\eeq
but where the AdS length scale $L$ is related to $\Lambda$, $G$ and $\alpha$ via
\beq L^{2}=-\frac{d(d-1)}{4\Lambda}\left(1+\sqrt{1+\frac{(d-3)(d-2)}{d(d-1)}128\pi G\alpha\Lambda}\right)\;,\eeq
or, in terms of $\Lambda$:
\beq \Lambda=\frac{d(d-1)}{2L^{4}}(16\pi G\alpha(d-2)(d-3)-L^{2})\;.\label{cosmocEGB}\eeq
When $\alpha=0$, we recover the usual relation for Einstein gravity $\Lambda=-d(d-1)/2L^{2}$. 

The Wald entropy (\ref{Waldent}) is 
\beq \frac{1}{4G}\int_{\Sigma}d^{d-1}\sqrt{h}\left[1+32\pi G\alpha\mathcal{R}^{(d-1)}\right]\;,\label{waldentGB2}\eeq
where $\mathcal{R}^{(d-1)}$ is the Ricci scalar of the $(d-1)$-dimensional Killing horizon $\Sigma$, with induced metric $h$.

Alternatively, evaluating the action (\ref{EGBactionex}) with
\beq \mathcal{R}_{\mu\nu\rho\sigma}=-\frac{1}{L^{2}}(g_{\mu\rho}g_{\nu\sigma}-g_{\mu\sigma}g_{\nu\rho})\;,\quad\mathcal{R}_{\mu\nu}=-\frac{d}{L^{2}}g_{\mu\nu}\;,\quad \mathcal{R}=-\frac{d(d+1)}{L^{2}}\;,\eeq
we have
\beq \mathcal{L}_{\text{GB}}=\frac{d(d+1)(d-1)(d-2)}{L^{4}}\;.\eeq
Then using the cosmological constant (\ref{cosmocEGB}), the Lagrangian density (\ref{EGBactionex}) evaluated in pure AdS is 
\beq \mathcal{L}_{\text{EGB}}|_{\text{AdS}}=-\frac{2d}{16\pi GL^{2}}+\frac{4\alpha d}{L^{4}}(d-1)(d-2)\;.\eeq
Then, via (\ref{eq:2})
\beq
\begin{split}
S_{\xi}&=-\frac{2\pi}{d}L^{2}\mathcal{A}_{\text{horizon}}\left[-\frac{2d}{16\pi GL^{2}}+\frac{4\alpha d}{L^{4}}(d-1)(d-2)\right]\\
&=\frac{1}{4G}\left[1+32\pi G\alpha\mathcal{R}^{(d-1)}\right]\mathcal{A}_{\text{horizon}}\;,
\end{split}
\eeq
where, and $\mathcal{A}_{\text{horizon}}=\int_{\Sigma}d^{d-1}x\sqrt{h}$. The second line is the expression for the Wald entropy one would normally find (\ref{waldentGB2}), however,  in our case we may pull the term in brackets out of the integral because we have evaluated the entropy on pure $\text{AdS}_{d+1}$. 

Let's now see what (\ref{Waldvar3}) becomes in this context. Here we have couplings $L, G$, and $\alpha$. Therefore
\beq \delta a_{d}^{\ast}(\lambda_{i})=\left(\frac{\partial a_{d}^{\ast}}{\partial L}\right)\delta L+\left(\frac{\partial a_{d}^{\ast}}{\partial G}\right)\delta G+\left(\frac{\partial a_{d}^{\ast}}{\partial \alpha}\right)\delta \alpha\;,\eeq
with
\beq \frac{\partial a_{d}^{\ast}}{\partial L}=\frac{\text{Vol}(S^{d-1}) L^{d-2}}{16\pi GL^{2}}(d-1)\left[L^{2}-32\pi G\alpha(d-2)(d-3)\right]\;,\eeq
\beq \frac{\partial a_{d}^{\ast}}{\partial G}=-\frac{\text{Vol}(S^{d-1})L^{d-1}}{16\pi G^{2}}\;,\eeq
\beq \frac{\partial a_{d}^{\ast}}{\partial\alpha}=-2\text{Vol}(S^{d-1})L^{d-3}(d-1)(d-2)\;.\eeq
Then, 
\beq \frac{S_{\xi}}{a^{\ast}_{d}}\left(\frac{\partial a_{d}^{\ast}}{\partial L}\right)=S_{\xi}\frac{(d-1)}{L}\left(\frac{L^{2}-32\pi G\alpha(d-2)(d-3)}{L^{2}-32\pi G\alpha(d-1)(d-2)}\right)\equiv S_{\xi}c_{L}\;,\eeq
\beq \frac{S_{\xi}}{a^{\ast}_{d}}\left(\frac{\partial a_{d}^{\ast}}{\partial L}\right)=S_{\xi}\frac{1}{G}\left(\frac{L^{2}}{L^{2}-32\pi G\alpha(d-1)(d-2)}\right)\equiv S_{\xi}c_{G}\;,\eeq
\beq \frac{S_{\xi}}{a^{\ast}_{d}}\left(\frac{\partial a_{d}^{\ast}}{\partial \alpha}\right)=-S_{\xi}\left(\frac{32\pi G(d-1)(d-2)}{L^{2}-32\pi G\alpha(d-1)(d-2)}\right)\equiv S_{\xi}c_{\alpha}\;.\eeq
Altogether, the variation (\ref{Waldvar3}) becomes
\beq \delta S_{\xi}=S_{\xi}\left(c_{L}\delta L+c_{G}\delta G+c_{\alpha}\delta\alpha\right)\;.\eeq
Consequently, the extended bulk first law (\ref{bulkextlaw}) becomes
\beq \frac{2\pi}{\kappa}\delta Q_{\xi}=\delta S_{\xi}-S_{\xi}\left(c_{L}\delta L+c_{G}\delta G+c_{\alpha}\delta\alpha\right)\;.\label{GBextbulklaw}\eeq
Our analysis of and final form of the  extended bulk first law for Einstein-Gauss-Bonnet gravity (\ref{GBextbulklaw}) should be compared to Section 3 of \cite{Caceres:2016xjz}, particularly equation (106), with which we agree.

\subsection{Mapping to Boundary CFT}

We are mainly interested in the first law in (\ref{bulkextlaw}) from the perspective of a holographic CFT$_{d}$ living on the asymptotic boundary of the $\text{AdS}_{d+1}$ bulk. . Taking a bulk coordinate $z$ so that the AdS boundary is located at $z\rightarrow 0$, the $d$-dimensional space-time in which the CFT is defined is given by
\begin{equation}\label{eq:33}
\lim_{z\rightarrow 0}ds_{\rm bulk}^2=
  w^2(x^\mu)ds^2_{\rm CFT}+\dots\ .
\end{equation}
Applying a bulk diffeomorphism or changing the definition of $w^2(x^\mu)$ results in a different boundary space-time. We give several examples momentarily. A particular way of taking this limit corresponds to choosing a conformal frame. We will shortly take advantage of this freedom, which from the CFT perspective is equivalent to a conformal transformation.

What about the quantum state of the boundary CFT? Although the bulk space-time is pure AdS, the CFT is technically not in the vacuum state since there is a horizon and therefore an associated temperature, given by the surface gravity in (\ref{Waldvar1}) according to $\beta=2\pi/\kappa$. This means that the boundary state is thermal with respect to the Killing flow evaluated at the boundary, \textit{i.e.}
\begin{equation}\label{cftstate}
\rho=\frac{1}{Z}\exp\left(-\beta K_{\xi}\right)\ ,
\end{equation}
where the operator $K_{ \xi}$ generates the flow of $\xi^\mu$ as we approach the boundary. It can be written explicitly in terms of the boundary coordinates\footnote{A comment on notation: We reserve Greek indices $\alpha,\beta,...$ for the full $d+1$-dimensional spacetime, and Latin indices $a,b,...$ for the $d$-dimensional boundary.} $x^a$ and the pullback of the Killing vector $\xi^a$ as
\begin{equation}\label{modHam}
K_{\xi}=\int_{\Sigma_\xi} 
   \xi^a T_{ab}dS^b\ ,
\end{equation}
where $T_{ab}$ is the stress tensor of the CFT and the integral is over a boundary codimension one space-like surface $\Sigma_\xi$ where the vector $\xi^a$ is time-like. The directed surface element~$dS^a$ is given by $dS^a=dSn^a$, with $n^a$ a unit vector normal to $\Sigma_\xi$. 

The conserved quantity $Q_\xi$ appearing in the gravitational first law (\ref{bulkextlaw}) is given by the expectation value\footnote{This follows from an application of the equations of motion, see, \emph{e.g.} \cite{Bueno16-1}.} of $K_{ \xi}$ in the state (\ref{cftstate}). The normalization condition for $Q_\xi$ in (\ref{eq:18}) translates into the following normalization of the stress tensor $T_{ab}$
\begin{equation}\label{eq:20}
T_{ab}\equiv
  T_{ab}-\langle T_{ab} \rangle_{\rho}\ ,
\end{equation}
with $\rho$ in (\ref{cftstate}). Since a bulk Killing vector gives a \textit{conformal} Killing vector at the boundary, the operator $K_{\xi}$ does not correspond to the Hamiltonian in general. We shall shortly consider some examples which illustrate this.

Putting everything together, the gravitational first law (\ref{bulkextlaw}) maps to the boundary CFT according to
\begin{equation}\label{extlawent}
\delta S=
  \beta\,\delta \langle K_{\xi} \rangle_{\rho}+
  \frac{S}{a_d^\ast}\delta a_d^\ast\ ,
\end{equation}
where we identified the horizon entropy $S_\xi$ with the Von Neumann entropy ${S(\rho)=-{\rm Tr}(\rho\ln(\rho))}$ of $\rho$ in (\ref{cftstate}). From the field theory perspective it might not be entirely clear what each of these terms corresponds to, so let us write them more explicitly.

For perturbations in which we keep the CFT fixed it is clear that $\delta a_d^\ast=0$ while the state is deformed according to $\rho+\delta \rho$. In this case, the relation (\ref{extlawent}) is similar to the first law of thermodynamics. When ${\delta a_d^\ast\neq 0}$ we must be more careful since in this case the CFT is changing, which in particular implies that the Hilbert space shifts $\mathcal{H}\rightarrow \bar{\mathcal{H}}$. The state $\rho$ cannot remain fixed, meaning that $\delta a_d^\ast \neq 0$ induces a variation of $\rho$ given by 
\begin{equation}
\rho \qquad \longrightarrow \qquad
  \bar{\rho}=\frac{1}{Z}\exp\left(
  -\beta \bar{K}_{\xi}
  \right)\ ,
\end{equation}
where $\bar{\rho}$ and $\bar{K}_{\xi}$ are the same operators but acting on the Hilbert space $\bar{\mathcal{H}}$ instead. In this case the extended first law (\ref{extlawent}) can be written explicitly as
\begin{equation}\label{eq:21}
S(\bar{\rho})-S(\rho)=
  \beta\left[
  \langle \bar{K}_{ \xi}
  \rangle_{\bar{ \rho}}-
  \langle 
  K_{ \xi}
  \rangle_\rho
  \right]+
  \frac{S(\rho)}{a_d^\ast}
  \delta a_d^\ast\ .
\end{equation}
Notice that the first terms on the right-hand side involve operators on different Hilbert spaces. Moreover, the normalization of $K_{\xi}$ given in (\ref{eq:20}) (and an analogous expression for $\bar{K}_{\xi}$) implies that both terms between square brackets vanish independently. This is equivalent to the gravitational case, where we obtained (\ref{Waldvar3}).

Putting everything together, the most general perturbation of the Von Neumann entropy of $\rho$ is given by
\begin{equation}\label{eq:12}
S(\bar{\rho}+\delta \bar{\rho})-
  S(\rho)=
  \beta\,
  {\rm Tr}\left(
  \bar{K}_\xi \,\delta \bar{\rho}
  \right)
  +
  \frac{S(\rho)}{a_d^\ast}
  \delta a_d^\ast\ ,
\end{equation}
where we have used $\langle K_{ \xi}\rangle_\rho=\langle \bar{K}_{ \xi}
  \rangle_{\bar{\rho}}=0$. This expression considers the simultaneous variations~$a_d^\ast\rightarrow a_d^\ast+\delta a_d^\ast$ and $\rho \rightarrow\bar{\rho}+\delta \bar{\rho}$, and clarifies the precise meaning of (\ref{extlawent}), which without any explanation is rather obscure.

\subsection{Extended First Law of Entanglement}
\label{sec:3}

So far we have shown that (\ref{extlawent}) follows from AdS/CFT when studying Killing horizons in pure AdS. We now consider particular horizons that will allow us to identify this relation as the extended first law of entanglement. Let us start with the simplest example of a Killing horizon in AdS, obtained by writing pure AdS in a hyperbolic slicing\footnote{Our line element (\ref{adshyp}) arises from us writing AdS in the usual hyperbolically sliced coordinate \cite{Casini:2011kv} with time coordinate $\tilde{\tau}$ and then further making the identification $\tilde{\tau}=L^{2}\tau/R^{2}$.}
\begin{equation}\label{adshyp}
ds^2=-\left(\frac{\rho^2-L^2}{R^2}\right)d\tau^2+
  \left(\frac{L^2}{\rho^2-L^2}\right)d\rho^2+
  \rho^2dH^2_{d-1}\ ,
\end{equation}
where $R$ is an arbitrary positive constant and $dH_{d-1}$ is the line element of a unit hyperbolic plane, 
\beq dH^{2}_{d-1}=du^{2}+\sinh^{2}(u)d\Omega^{2}_{d-2}\;,\eeq
where $d\Omega_{d-2}$ is the line element of a unit sphere $S^{d-2}$. This space-time is often referred as Rindler-AdS since it describes a section of anti-de Sitter. It also describes a massless AdS-Schwarzschild black with hyperbolically sliced horizon, located at  $\rho_{+}=L$. The AdS boundary is located at $\rho\to\infty$. The vector $\xi=\partial_\tau$ trivially satisfies Killing's equation and is time-like over the whole patch $\rho\ge L$, generating a horizon at $\rho=L$. It therefore satisfies all the conditions leading to the first law in (\ref{bulkextlaw}) and (\ref{eq:12}). 

A simple computation shows that the surface gravity is $\kappa=1/R$, while the boundary metric\footnote{We find the boundary metric by pulling a factor of $\rho^{2}/L^{2}$ out of (\ref{adshyp}) such that
$$ds^{2}=\left(\frac{\rho^{2}}{L^{2}}\right)\left[-\frac{L^{2}}{R^{2}}V(\rho)d\tau^{2}+\frac{L^{4}}{\rho^{4}}V^{-1}(\rho)d\rho^{2}+L^{2}dH^{2}_{d-1}\right]$$
with $V(\rho)=1-L^{2}/\rho^{2}$, and then taking the $\rho\to\infty$ limit and identifying $R=L$, giving us (\ref{bdrymetric1}), where we have also dropped the overall conformal factor $\rho^{2}/L^{2}$.}
 is given by
\beq ds^2_{\rm CFT}=-d\tau^2+R^2dH^2_{d-1}\equiv \mathbb{R}\times \mathbb{H}^{d-1}\;.\label{bdrymetric1}\eeq
From this we see that $\xi=\partial_\tau$ is also a Killing vector of $ds^2_{\rm CFT}$, so that $K_\xi$ in (\ref{modHam}) is equal to the Hamiltonian and can be written as
\beq K_\xi=
  \int_{\tau=0}
  T_{\tau \tau}dS^\tau \equiv
  H_\tau\ .\eeq
This means the boundary state is an ordinary thermal state ${\rho_\beta\propto \exp(-\beta H_\tau)}$, where the inverse temperature is fixed by the surface gravity to ${\beta=2\pi R}$. The extended first law (\ref{extlawent}) then becomes
\begin{equation}\label{eq:55}
\delta S(\rho_\beta)=\beta \,
  \delta \langle 
  H_\tau
  \rangle+\frac{S(\rho_\beta)}{a_d^\ast}\delta a_d^\ast\ .
\end{equation}
While the first term is nothing more than the first law of thermodynamics, the second contribution is unique to the case of inverse temperature $\beta=2\pi R$. This is clear from the holographic perspective, since moving away from this temperature is equivalent to leaving pure AdS, where the analysis of the previous section is no longer valid. In section \ref{sec:3Dlaw} we show that for $d=2$ this expression remains valid for arbitrary values of $\beta$. Although (\ref{eq:55}) is not the extended first law of entanglement (since it involves a thermal state in $\mathbb{R}\times \mathbb{H}^{d-1}$), this simple example will be very useful in what follows.

\subsection{Shifting Conformal Frames}

Building on the canonical example we just described, we can obtain the more complicated setups we are actually interested in. To obtain the extended first law of entanglement we take advantage of the freedom present when taking the boundary limit in (\ref{eq:33}). Different ways of taking this limit correspond to distinct conformal frames and result in different setups for the boundary CFT. We still consider the bulk Killing vector $\xi=\partial_\tau$, but written in a different set of coordinates corresponding to distinct conformal frames.


\textbf{Ball in Minkowski}

 Let us first show how we can recover the extended first law of entanglement for the Minkowski vacuum reduced to a ball. We first apply a change of coordinates on the Rindler-AdS metric~(\ref{adshyp}), which is given in Eq. (4.7) of Ref.~\cite{Rosso:2019lsm}:
\beq
\begin{split}
&\rho=\frac{L}{2Rz}\sqrt{(R+\hat{r}_{+})(R+\hat{r}_{-})+z^{2}}\sqrt{(R-\hat{r}_{+})(R-\hat{r}_{-})+z^{2}}\\
&\tanh(\tau/R)=\frac{R(\hat{r}_{+}-\hat{r}_{-})}{R^{2}-(\hat{r}_{+}\hat{r}_{-}+z^{2})}\;,\quad \tanh(u)=\frac{R(\hat{r}_{+}+\hat{r}_{-})}{R^{2}+(\hat{r}_{+}\hat{r}_{-}+z^{2})}\;,
\end{split}
\eeq
where $\hat{r}_{\pm}=r\pm t$,  with $\hat{r}\ge 0$, so that the bulk metric(\ref{adshyp}) becomes
\beq ds^{2}=\frac{L^{2}}{z^{2}}(dz^{2}-dt^{2}+dr^{2}+r^{2}d\Omega^{2}_{d-2})\;,\eeq
the  Poincar\'e patch coordinates. Further writing $(z,r)=\hat{r}(\sin\psi,\cos\psi)$, we have
\begin{equation}\label{eq:26}
ds^2=
  \left(\frac{L}{\hat{r}\sin(\psi)}\right)^2
  \left[
  -dt^2+d\hat{r}^2+
  \hat{r}^2\left(d\psi^2+d\Omega^2_{d-2}\right)
  \right],
\end{equation}
where $\psi\in [0,\pi/2]$. It is also useful to know the inverse bulk coordinate transformation:
\beq z=\frac{RL}{\rho\cosh(u)+\sqrt{\rho^{2}-L^{2}}\cosh(\tau/R)}\;,\quad \hat{r}_{\pm}=R\frac{\rho\sinh(u)\pm\sqrt{\rho^{2}-L^{2}}\sinh(\tau/R)}{\rho\cosh(u)+\sqrt{\rho^{2}-L^{2}}\cosh(\tau/R)}\;.\eeq
At the boundary $\psi \rightarrow 0$ we recover $d$-dimensional Minkowski space-time with $\hat{r}=r$ the spatial radial coordinate. We use the convention in which the boundary coordinate $r$ refers to the bulk coordinate $\hat{r}$ when $\psi\rightarrow 0$. 
 This same notation is used in the following examples.

It is straightforward to write the Killing vector $\xi=\partial_\tau$ in these new coordinates. We have by the chain rule
\beq \partial_{\tau}=\frac{\partial \hat{r}_{+}}{\partial\tau}\partial_{\hat{r}_{+}}+\frac{\partial \hat{r}_{-}}{\partial\tau}\partial_{\hat{r}_{-}}\;,\eeq
where $\partial \tau/\partial \hat{r}_{+}=(\partial\hat{r}_{+}/\partial\tau)^{- 1}$, such that
\beq \partial_{\hat{r}_{\pm}}\text{arctanh}\left[\frac{R(\hat{r}_{+}-\hat{r}_{-})}{R^{2}-(\hat{r}_{+}\hat{r}_{-}+z^{2})}\right]\biggr|_{z=0}=\pm\frac{2R}{R^{2}-\hat{r}_{\pm}^{2}};.\eeq
Therefore,
\beq
\begin{split}
\xi&=\left(\frac{R^2-\hat{r}_+^2}{2R^2}\right)
  \partial_{\hat{r}_+}-
  \left(\frac{R^2-\hat{r}_-^2}{2R^2}\right)
  \partial_{\hat{r}_-}\\
&=\frac{1}{4R^{2}}(\hat{r}_{-}^{2}-\hat{r}_{+}^{2})\partial_{\hat{r}}+\frac{1}{4R^{2}}(2R^{2}-\hat{r}_{+}^{2}-\hat{r}_{-}^{2})\partial_{t}\;.
\end{split}
\label{eq:23}\eeq
 The important difference with respect to the hyperbolic example is that this Killing vector is time-like only in a section of the metric (\ref{eq:26}), given by $|\hat{r}_\pm|\le R$. For the Minkowski boundary this corresponds to the causal domain of a ball of radius $R$. The operator generating the flow of $\xi$ inside the ball can be written from (\ref{modHam}) as (where we work on the $t=0$ slice)
\begin{equation}\label{eq:22}
K_{\xi}=
  \int_{r\le R}
  \left(
  \frac{R^2-r^2}{2R^2}
  \right)T_{tt}\,
  dS^t\ .
\end{equation}

While this is clearly not the Hamiltonian generating $t$ translations in Minkowski, it is proportional to the modular hamiltonian characterizing the Minkowski vacuum reduced to the ball \cite{Casini:2011kv}. The proportionality constant missing to make the identification is given by~${K_{\rm Ball}=2\pi R K_\xi}$, that is precisely the inverse temperature $\beta=2\pi R$ obtained from the surface gravity of the bulk Killing vector (\ref{eq:23}). Altogether, the quantum state $\rho$ in (\ref{cftstate}) is exactly given by the Minkowski vacuum reduced to the ball. The Von Neumann entropy is equivalent to the entanglement entropy, so that (\ref{extlawent}) becomes the extended first law of entanglement (\ref{eq:ext}).


\textbf{Half-Space in Minkowski} 

Another interesting case is obtained by applying the change of coordinates given in Eq. (4.4) of \cite{Rosso:2019lsm} (see also \cite{Emparan:1999gf}) to the Rindler-AdS space-time, so that the bulk metric (\ref{adshyp}) becomes
\begin{equation}
ds^2=\left(L/z\right)^2
  \left(
  dz^2-dt^2+dx^2+d\vec{y}.d\vec{y}\,
  \right)\ ,
\end{equation}
where $(x,\vec{y}\,)\in \mathbb{R}\times \mathbb{R}^{d-2}$. Once again we recognize the Poincar\'e patch of AdS, so that we recover a $d$-dimensional Minkowski boundary when $z\rightarrow 0$. The Killing vector $\xi=\partial_\tau$ in these coordinates is given by
\begin{equation}\label{eq:25}
\xi=
  (x_+/R)\partial_{x_+}-
  (x_-/R)\partial_{x_-}\ ,
\end{equation}
where $x_\pm=x\pm t$. This vector is time-like when $x_\pm \ge 0$, which from the boundary perspective corresponds to the Rindler region, \textit{i.e.} the causal domain of the half space $x\ge 0$. Using (\ref{modHam}) to compute the operator generating the Killing flow at the boundary we find
\beq K_\xi=
  \int_{x>0}(x/R)T_{tt}\,dS^t\ .\eeq
Since the surface gravity of (\ref{eq:25}) is still given by $\kappa=1/R$, the inverse temperature is~${\beta=2\pi R}$ and we recognize ${\rho\propto \exp(-\beta K_\xi)}$ as the Minkowski vacuum reduced to Rindler~\cite{Bisognano:1976za,Unruh76-1}. Similarly to the previous case, (\ref{extlawent}) becomes the extended first law of entanglement (\ref{eq:ext}) but in this case, for the Minkowski vacuum reduced to the half-space.


\textbf{Spherical Cap in Lorentzian Cylinder} 

Let us now show how we can obtain the extended first law of entanglement for holographic CFTs defined on curved backgrounds. Consider the following change of coordinates on the AdS metric (\ref{eq:26})
\begin{equation}\label{eq:27}
\hat{r}_\pm(\hat{\theta}_\pm)=R\frac{\tan(\hat{\theta}_\pm/2)}{\tan(\theta_0/2)}\ ,
\end{equation}
where $\hat{\theta}_\pm=\hat{\theta}\pm \sigma/R$ and $\theta_0\in[0,\pi]$ is a fixed parameter. The metric (\ref{eq:26}) becomes
\begin{equation}\label{eq:28}
ds^2=\left[
  \frac{L/R}{\sin(\psi)\sin(\hat{\theta})}
  \right]^2 
  \left(
  -d\sigma^2
  +R^2d\hat{\theta}^2
  \right.  \left.
  + R^2\sin^2(\hat{\theta})
  \left(
  d\psi^2
  +\cos^2(\psi)d\Omega^2_{d-2}
  \right)
  \right)\ ,
\end{equation}
where $\sigma\in \mathbb{R}$ is the time coordinate and $\hat{\theta}$ is restricted to $\hat{\theta}\in[0,\pi]$. As we take the boundary limit $\psi\rightarrow 0$ and remove the conformal factor between square brackets we find that the CFT is defined in the Lorentzian cylinder $\mathbb{R}\times S^{d-1}$ with metric~${ds^2_{\rm CFT}=-d\sigma^2+R^2d\Omega^2_{d-1}}$. The bulk coordinate $\hat{\theta}$ becomes the polar angle $\hat{\theta}=\theta$ on the spatial sphere~$S^{d-1}$, with~$\theta=0,\pi$ corresponding to the North and South poles respectively.

\begin{figure}
\centering
\includegraphics[scale=0.80]{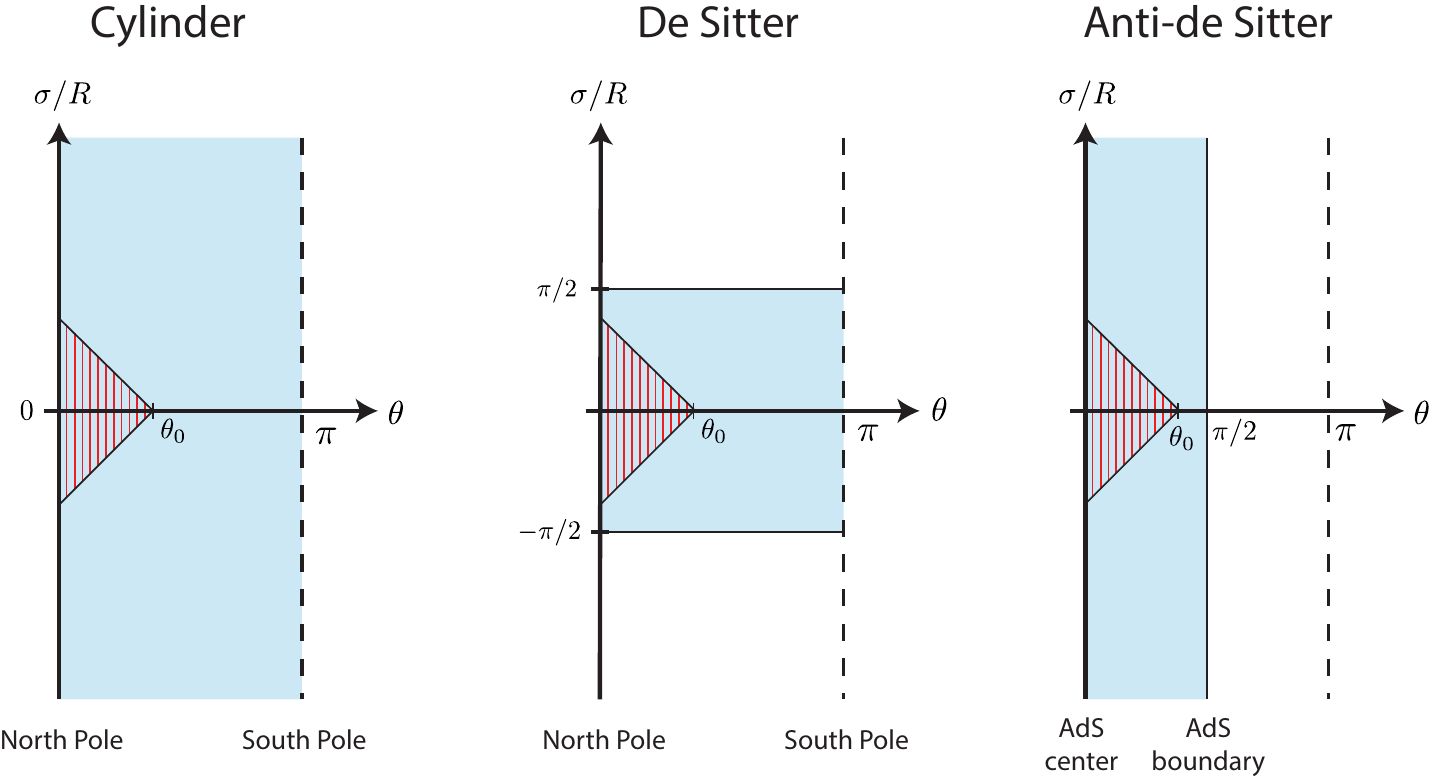}
\caption[Boundary Spacetimes via Different Conformal Frames.]{Boundary space-times represented in the $(\sigma/R,\theta)$ plane. The blue region corresponds to the section of the $(\sigma/R,\theta)$ plane covered by the boundary metrics (\ref{eq:28}) (in the limit $\psi\rightarrow 0$ and without the conformal factor), (\ref{eq:11}) and (\ref{eq:51}). In red we see the region in which the boundary vector $\xi^a$ is time-like and therefore the extended first law of entanglement applies.}\label{fig:1}
\end{figure}

The Killing vector $\xi$ in (\ref{eq:23}) can be written in these coordinates as
\begin{equation}\label{eq:19}
\xi=
  \bigg(
  \frac{\cos(\hat{\theta}_+)-\cos(\theta_0)}
  {R\sin(\theta_0)}
  \bigg)\partial_{\hat{\theta}_+}-
  \bigg(
  \frac{\cos(\hat{\theta}_-)-\cos(\theta_0)}
  {R\sin(\theta_0)}
  \bigg)
  \partial_{\hat{\theta}_-}.
\end{equation}
Computing its magnitude we see that the bulk region in which this vector is time-like is given by $|\hat{\theta}_\pm|<\theta_0$. For the boundary CFT in the Lorentzian cylinder, this corresponds to the causal domain of a spherical cap on the spatial $S^{d-1}$ given by $\theta\in[0,\theta_0]$ at $\sigma=0$. Plotting this region in the $(\sigma/R,\theta)$ plane we obtain the left diagram in Fig. \ref{fig:1}. The whole infinite strip in blue corresponds to the Lorentzian cylinder $\mathbb{
R}\times S^{d-1}$, with the North and South pole located at $\theta=0,\pi$. 

The operator generating the Killing flow at the boundary is computed from (\ref{modHam}) as 
\begin{equation}\label{eq:29}
K_\xi=\int_{\theta\le \theta_0}
  \bigg(
  \frac{\cos(\theta)-\cos(\theta_0)}
  {R\sin(\theta_0)}
  \bigg)
  T_{\sigma \sigma}\, dS^\sigma\ .
\end{equation}
In a similar way to the previous case, we recognize the state ${\rho \propto \exp\left(-\beta K_\xi\right)}$ with $\beta=2\pi R$ as the vacuum state of the cylinder reduced to the spherical cap \cite{Casini:2011kv}. This gives the extended first law of entanglement for a CFT in the Lorentzian cylinder (\ref{eq:ext}).

\textbf{Spherical cap in de Sitter} 

Using the same coordinates as in (\ref{eq:28}) we can obtain a CFT defined on a de Sitter background by taking the limit $\psi \rightarrow 0$ and choosing the conformal factor so that the boundary metric is given by
\begin{equation}\label{eq:11}
ds^2_{\rm CFT}=
  \frac{-d\sigma^2+R^2d\Omega^2_{d-1}}
  {\cos^2(\sigma/R)}\ .
\end{equation}
This is $d$-dimensional global de Sitter space-time, as can be seen by changing the time coordinate to $\cosh(t_s/R)=1/\cos(\sigma/R)$, so that we get
\begin{equation}\label{eq:40}
ds^2_{\rm CFT}=
  -dt_s^2+R^2\cosh^2(t_s/R)d\Omega^2_{d-1}
  \ .
\end{equation}

It is convenient to work in the time coordinate $\sigma$, since the Killing vector $\xi$ has the simple form given in (\ref{eq:19}) and is time-like when $|\theta_\pm|\le \theta_0$. Plotting this region in the~${(\sigma/R,\theta)}$ plane for the boundary metric (\ref{eq:11}), we obtain the center diagram in Fig.~\ref{fig:1}. The main difference with respect to the case of the Lorentzian cylinder is that the full de Sitter space-time (blue region) is constrained to $|\sigma/R|\le \pi/2$ due to the denominator in~(\ref{eq:11}). Since the topology of dS is the same as the cylinder $\mathbb{R}\times S^{d-1}$, the region in which $\xi^a$ is time-like also corresponds to the causal domain of a spherical cap ${\theta\in[0,\theta_0]}$, but with $\theta_0$ restricted to $\theta_0\le \pi/2$.

The operator generating the flow of the Killing vector at the boundary is still given by~(\ref{eq:29}),\footnote{The only difference with respect to the case of the cylinder is given by the induced surface element $dS^\sigma$, which is now computed from (\ref{eq:11}).} which is equivalent to the modular hamiltonian of the dS vacuum after multiplying by $\beta=2\pi R$. Altogether, this results in the extended first law of entanglement (\ref{eq:ext}) for the de Sitter vacuum reduced to a spherical cap.


\textbf{Ball in anti-de Sitter} 

Finally, we can obtain a CFT defined in an AdS$_d$ space-time by taking the limit $\psi\rightarrow 0$ in~(\ref{eq:28}) and choosing the conformal factor so that we get
\begin{equation}\label{eq:51}
ds^2_{\rm CFT}=
  \frac{-d\sigma^2+
  R^2(d\theta^2+
  \sin^2(\theta)
  d\Omega^2_{d-2})}
  {\cos^2(\theta)}
  \ .
\end{equation}
Changing coordinates to $\varrho=R\tan(\theta)\ge 0$ we recognize global AdS$_{d}$, with $\varrho$ the usual radial coordinate. Similar to the dS case, it is convenient to describe the AdS$_d$ boundary in terms of the $(\sigma,\theta)$ coordinates, where the Killing vector $\xi$ and operator $K_\xi$ are still given by (\ref{eq:19}) and (\ref{eq:29}). The main difference is that the region in which $\xi$ is time-like $|\theta_\pm|\le \theta_0$, now corresponds to the causal domain of a ball in AdS$_d$ of radius $\varrho_{\rm max}=R\tan(\theta_0)$. We plot this in the right diagram of Fig. \ref{fig:1}, where $\theta=0,\pi/2$ in (\ref{eq:51}) now correspond to the AdS center and boundary. The entanglement entropy associated to the vacuum state reduced on this ball satisfies the extended first law of entanglement in (\ref{eq:ext}).

\subsection{Killing Horizons in Pure Two-Dimensional AdS}
\label{sec:2Dlaw}

Our calculations so far have been in the context of the AdS$_{d+1}$/CFT$_d$ correspondence for~${d\ge 2}$, where the duality is well understood. In this section we revisit the construction for the case in which $d=1$, where the gravity theory is highly constrained and there is no clear holographic picture. 

Let us start by briefly reviewing some basic notions of two dimensional gravity (see \cite{Strobl:1999wv} for a comprehensive review). In two space-time dimensions the most general scalar curvature invariant is built from the Ricci scalar $\mathcal{R}$ and contractions of its covariant derivatives, \textit{e.g.} $(\nabla \mathcal{R})^2=(\nabla_\mu \mathcal{R})(\nabla^\mu \mathcal{R})$. Both the Riemann and Ricci tensor are fixed by $\mathcal{R}$ and~$g_{\mu \nu}$ according to
\begin{equation}\label{eq:14}
\mathcal{R}_{\mu \nu \rho \sigma}=
  \frac{\mathcal{R}}{2}
  \left(
  g_{\mu \rho}g_{\nu \sigma}-
  g_{\mu \sigma}g_{\nu \rho}
  \right)
  \ ,
  \qquad \qquad
  \mathcal{R}_{\mu \nu}=\frac{\mathcal{R}}{2}g_{\mu \nu}\ .
\end{equation}
This means there is a single gravitational degree of freedom, determined by $\mathcal{R}$. Similarly to the general $d$ case in (\ref{genaction}), the most general two dimensional gravity theory is given by
\begin{equation}\label{eq:7}
I[g_{\mu \nu},\lambda_i]=
  \int d^2x\sqrt{-g}\,
  \mathcal{L}(\mathcal{R},\nabla_\mu \mathcal{R},\dots)\ ,
\end{equation}
where the coefficients $\lambda_i$ are the coupling constants of the theory. The only constraint we impose is that there is a pure AdS solution with some radius $L=L(\lambda_i)$. Notice that the relations in (\ref{eq:14}) imply that the Einstein tensor $G_{\mu \nu}=\mathcal{R}_{\mu \nu}-g_{\mu\nu}\mathcal{R}/2$ vanishes for every two dimensional metric, so that $\mathcal{L}=\mathcal{R}$ gives a trivial theory.

Just as in the higher dimensional case, let us consider a Killing vector $\xi^\mu$ of pure AdS$_2$ which is time-like over some region and generates a horizon (\ref{eq:15}). The associated entropy is computed from Wald's functional (\ref{Waldent}), which in the two dimensional case is given by
\begin{equation}\label{eq:45}
S_\xi[g_{\mu \nu}(L),\lambda_i]=
  -2\pi\left[
  \frac{\delta \mathcal{L}}
  {\delta R^{\mu \nu}_{\,\,\,\,\,\rho \sigma}}
  n^{\mu \nu}n_{\rho \sigma}
  \right]_{\rm Horizon}\ ,
\end{equation}
where there is no integral since the bifurcate horizon is a single point. Evaluating in pure AdS we can use (\ref{eq:24}) to write this as
\begin{equation}\label{eq:17}
S_\xi[g_{\mu \nu}^{\rm AdS}(L),\lambda_i]=
  2\pi a_1^*(\lambda_i)\ ,
  \qquad {\rm where} \qquad
  a_1^*(\lambda_i)=
  -L^2\mathcal{L}\big|_{\rm AdS}\ .
\end{equation}
An important difference with respect to the higher dimensional case, is that in two dimensions this expression is always finite and only depends on the global features of the theory, \textit{i.e.}, it is insensitive to the details of the Killing vector $\xi^\mu$. The entropy in (\ref{eq:17}) only depends on the pure AdS$_2$ radius and the Lagrangian density evaluated on AdS$_2$. Altogether, there is no obstruction in applying the same reasoning as in higher dimensions and write the extended first law for Killing horizons in pure AdS exactly as in (\ref{bulkextlaw})
\begin{equation}\label{eq:49}
\delta S_\xi=
  \frac{2\pi}{\kappa}
  \delta Q_\xi+
  \frac{S_\xi}{a_1^*}\delta a_1^*\ .
\end{equation}

\begin{figure}
\centering
\includegraphics[scale=0.8]{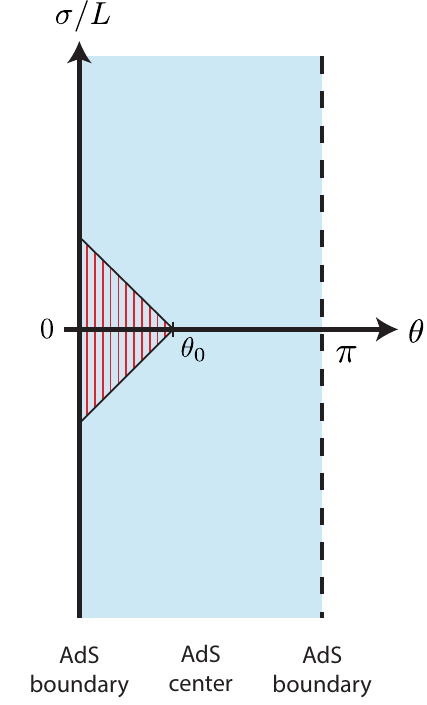}
\caption[An Illustration of AdS$_{2}$.]{The blue region corresponds to AdS$_2$ space-time represented in the $(\sigma/L,\theta)$ plane, with the two boundaries at $\theta=0,\pi$. In red we see the region in which the bulk Killing vector~$\xi^\mu$ (\ref{eq:52}) is time-like and therefore the extended first law in (\ref{eq:49}) applies.}\label{fig:2}
\end{figure}

Let us construct a concrete example by first writing pure AdS$_2$ in global coordinates
\begin{equation}\label{eq:13}
ds^2=
  \frac{-d\sigma^2+L^2d\theta^2}
  {\sin^2(\theta)}\ ,
\end{equation}
where $\sigma\in \mathbb{R}$ and $\theta\in[0,\pi]$. Notice that the notation is different from the previous section, since $\theta$ is now a bulk coordinate and the boundary is just described by $\sigma$. Two-dimensional AdS is distinct from higher dimensions, since there are two disjoint boundaries at $\theta=0,\pi$. A sketch of its Penrose diagram is given in Fig. \ref{fig:2}.

We can easily check that the following is a Killing vector
\begin{equation}\label{eq:52}
\xi^\mu=\left(
  \frac{\cos(\theta_+)-\cos(\theta_0)}
  {L\sin(\theta_0)}
  \right)\partial_{\theta_+}-
  \left(
  \frac{\cos(\theta_-)-\cos(\theta_0)}
  {L\sin(\theta_0)}
  \right)\partial_{\theta_-}\ ,
\end{equation}
with surface gravity $\kappa=1/L$. From its norm we see that it is time-like in the domain of dependence of the bulk surface $(\sigma=0,\psi)$ with $\psi\in[0,\theta_0]$, meaning that the boundary time coordinate is restricted to $|\sigma/L|\le \theta_0$. This corresponds to the red region in Fig. \ref{fig:2}.

As an example, let us compute the horizon entropy explicitly for a particular gravity theory, that we take as
\begin{equation}\label{eq:31}
\mathcal{L}=f(\mathcal{R})=
  \lambda_0+\lambda_2\mathcal{R}^2\ .
\end{equation}
The AdS radius $L$ is determined by solving the equations of motion evaluated at $\mathcal{R}=-2/L^2$, which can be written as
\begin{equation}\label{eq:30}
\nabla_\mu \nabla_\nu 
  f'(\mathcal{R})+
  \frac{1}{2}g_{\mu\nu}
  \left(
  \mathcal{R}f'(\mathcal{R})-f(\mathcal{R})
  \right)=0
  \qquad \Longrightarrow \qquad
  L^4=
  \frac{4\lambda_2}{\lambda_0}\ .
\end{equation}
Using this we can evaluate Wald's entropy in (\ref{eq:17}) as
\beq S_\xi\big[
  g_{\mu \nu}^{\rm AdS}(L),\lambda_i
  \big]=
  2\pi\left(
  -8\lambda_2/L^2
  \right)\ ,\eeq
where between parenthesis we identify the factor $a_1^*$, which is positive if and only if $\lambda_2<0$.

This raises the question regarding the holographic interpretation of the extended first law as written in (\ref{eq:49}), since $a_1^*$ is supposed to capture the number of degrees of freedom of the boundary theory. The usual AdS/CFT correspondence for a two dimensional bulk does not yield a clear picture as in the higher dimensional case. Although there has been very interesting work on the subject (see \cite{Strominger:1998yg,Cadoni:1999ja,Hartman:2008dq,Castro:2008ms,Alishahiha:2008tv,Cvetic:2016eiv}), there continues to be debate about what is meant by the dual ``$\text{CFT}_{1}$", whether it is conformal quantum mechanics or the chiral sector of a two-dimensional CFT. Moreover in the context of Jackiw-Teitelboim (JT) gravity \cite{Teitelboim:1983ux,Jackiw:1984je} it is understood that the boundary is not a single theory but an ensemble average \cite{Saad:2019lba}. For these reasons, we refrain from giving a boundary interpretation of the extended first law and leave this aspect to future investigations.

\subsubsection{Einstein-Dilaton Theories}
\label{D2GRlimit}

So far we have considered two dimensional theories of gravity in which the only field is given by the metric $g_{\mu \nu}$. We now discuss the extended first law for Einstein-dilaton theories, which are widely studied in the context of two dimensional gravity.

One disadvantage of the pure gravity action considered in (\ref{eq:7}) is that since non-trivial theories must have $\mathcal{L}\sim \mathcal{O}(\mathcal{R}^2)$, the equations of motion for the metric are at least fourth order differential equations. This issue can be avoided by the introduction of an auxiliary dilaton field $\phi(x^\mu)$ coupled to ordinary Einstein gravity
\begin{equation}\label{eq:8}
I_\phi[g_{\mu \nu},\lambda_i]=
  \int d^2x\sqrt{-g}\left[
  \phi\mathcal{R}-
  V(\phi)
  \right]\ .
\end{equation}
The equations of motion obtained from this action are second order. In particular, varying with respect to the dilaton field we get the algebraic constraint $\mathcal{R}=V'(\phi)$. If the potential has non-vanishing second derivative, one can invert this relation and substitute back into the action (\ref{eq:8}) to obtain a purely gravitational theory of the type $\mathcal{L}=f(\mathcal{R})$. As an example, if we take $V(\phi)=\phi^2/4\lambda_2-\lambda_0$, the equation of motion for $\phi$ sets $\phi_0=2\lambda_2\mathcal{R}$ and we get
\begin{equation}\label{eq:46}
I_{\phi=\phi_0}[g_{\mu\nu},\lambda_i]=
  \int d^2x\sqrt{-g}
  \left[
  \lambda_0+\lambda_2\mathcal{R}^2
  \right]\ ,
\end{equation}
which is the gravity theory previously considered in (\ref{eq:31}). This allows us to study two dimensional gravity from the simpler action (\ref{eq:8}). We should interpret the dilaton field as a gravitational degree of freedom, which gets non-trivial dynamics from varying (\ref{eq:8}) with respect to the metric
\begin{equation}\label{eq:48}
\nabla_\mu \nabla_\nu 
  \phi
  =\frac{1}{2}
  g_{\mu \nu}
  V(\phi)\ .
\end{equation} 
Since the Einstein-dilaton theories in (\ref{eq:8}) (with $V''(\phi)\neq 0$) are equivalent to the purely gravitational action previously considered in (\ref{eq:7}), the results obtained for the extended first law also hold in this setup. We should mention that while JT gravity is given by (\ref{eq:8}) with~$V(\phi)\propto \phi$, it cannot be written as a purely gravitational theory since $V''(\phi)=0$ and the dilaton equation simply fixes the curvature to a constant $\mathcal{R}={\rm const}$.

There are more general Einstein-dilaton actions than (\ref{eq:8}) that yield interesting two dimensional theories. For instance, there is a particular way of taking the two-dimensional limit of higher dimensional Einstein gravity which results in the following action \cite{Mann:1992ar}
\begin{equation}\label{eq:47}
I_\phi\left[g_{\mu \nu},\Lambda_2\right]=
\int d^{2}x\sqrt{-g}\left[\phi\mathcal{R}+\frac{1}{2}(\nabla\phi)^{2}-2\Lambda_{2}\right]\;,
\end{equation} 
where $\Lambda_2$ is a coupling constant. This theory was studied in \cite{Frassino:2015oca} from the perspective of extended black hole thermodynamics. Although this action is clearly different from (\ref{eq:8}), if we redefine the metric according to\footnote{Such that $\sqrt{-\tilde{g}}=e^{D\Phi/2}\sqrt{-g}$ and
 $$\tilde{\mathcal{R}}=e^{-\Phi}\left(\mathcal{R}-(D-1)g^{\mu\nu}\nabla_{\mu}\nabla_{\nu}\Phi-\frac{1}{4}(D-2)(D-1)g^{\mu\nu}\nabla_{\mu}\Phi\nabla_{\nu}\Phi\right)\;.$$} $\tilde{g}_{\mu \nu}=e^{\phi/2}g_{\mu \nu}$ it can be written as
\begin{equation}
I_\phi\left[\tilde{g}_{\mu \nu},\Lambda_2\right]=
\int d^{2}x\sqrt{-\tilde{g}}
  \big[
  \phi\tilde{\mathcal{R}}
  -V(\phi)\big]\ ,
  \qquad {\rm where} \qquad
  V(\phi)=
  2\Lambda_{2}e^{-\phi/2}\ .
\end{equation}
Once we have the action in this form, we can solve the dilaton field equation and substitute it back into the action to get a purely gravitational theory for the metric $\tilde{g}_{\mu \nu}$
\begin{equation}\label{eq:50}
I_{\phi=\phi_0}
\left[\tilde{g}_{\mu \nu},\Lambda_2\right]=
\int d^{2}x\sqrt{-\tilde{g}}
  f(
  \tilde{\mathcal{R}})\ ,
  \qquad {\rm where} \qquad
  f(x)=2x
  \big(1-\ln(-x/\Lambda_2)\big)
  \ .
\end{equation}

This raises the question of which is the ``physical" gravitational metric, either $g_{\mu\nu}$ or $\tilde{g}_{\mu \nu}$.\footnote{See \cite{Faraoni:1999hp,Postma:2014vaa} for a discussion around a similar issue.} The distinction between the frames is important as the solutions obtained in either case are very different. For instance, if we consider a constant curvature solution for $\tilde{g}_{\mu \nu}$, the equation of motion from (\ref{eq:50}) is given by
\beq \tilde{\mathcal{R}}f'(\tilde{\mathcal{R}})-
  f(\tilde{\mathcal{R}})=0
  \qquad \Longrightarrow \qquad
  \tilde{\mathcal{R}}=0\ .\eeq
From (\ref{eq:14}), this implies that the metric $\tilde{g}_{\mu \nu}$ vanishes, so that the theory does not admit a pure AdS$_2$ solution and we cannot consider the extended first law in (\ref{eq:49}). 

On the other hand, working in the frame with the metric $g_{\mu \nu}$ the action (\ref{eq:47}) allows a pure AdS$_2$ solution \cite{Frassino:2015oca}. This means it is sensible to consider the extended first law for the metric $g_{\mu \nu}$, although the derivation leading to (\ref{eq:49}) does not apply. An extended first law of black hole thermodynamics (which studies the behavior of the black hole entropy under variations of the cosmological constant) was derived in \cite{Frassino:2015oca} for the Einstein-dilaton theory in (\ref{eq:47}). In order to obtain a sensible result, the authors of \cite{Frassino:2015oca} use an unconventional approach that involves rescaling Newton's constant according to $G_{d+1}=\frac{(1-d)}{2}G_2$. Starting from the results in \cite{Kastor:2014dra}, this procedure can also be applied to derive an extended first law for perturbations of Killing horizons in the AdS$_2$ metric $g_{\mu \nu}$. We show this in detail in Appendix \ref{app:lowdentchem}. 

The overall lesson here is that a non-trivial extended bulk first law can be formulated, in principle for pure theories of gravity in $1+1$-dimensions. Moreover, since any Einstein-dilaton theory of gravity with a dilaton potential that has a non-vanishing second derivative can be recast as a pure theory of gravity, the extended bulk first law can be formulated for Einstein-dilaton theories. We note, however, not every Einstein-dilaton theory will satisfy the criterion $V''(\phi)\neq0$, \emph{e.g.}, JT gravity, and so it is unclear how to formulate a bulk first law for such theories. Moreover, even when we have an Einstein-dilaton theory that satisfies the aforementioned criteria, it might be unclear whether pure $\text{AdS}_{2}$ is a solution to such a theory, in which case the bulk first law would be trivial. 

\subsubsection{Jackiw-Teitelboim gravity}

In this subsection we consider the extended first law in the context of Jackiw-Teitelboim gravity \cite{Teitelboim:1983ux,Jackiw:1984je}, that correspond to an Einstein-dilaton theory that cannot be written as a purely gravitational theory of the type $\mathcal{L}=f(\mathcal{R})$. The action defining the theory can be written as
\begin{equation}\label{eq:190}
I_{JT}=I_\phi[g_{\mu \nu};\phi_0,L]=
\int d^2x\sqrt{-g}
\left[
\phi_0\mathcal{R}+
\phi(x)(\mathcal{R}+2/L^2)
\right]\ .
\end{equation}
The dilaton field $\phi(x)$ is dimensionless and there are two coupling constants that define the theory $\lambda_i=(\phi_0,L)$. As usual, the action must be supplemented with appropriate boundary terms to yield a well defined variational problem. The equations of motion can be easily computed and written as
\begin{equation}\label{eq:193}
\begin{aligned}
\mathcal{R}+2/L^2&=0\\
\left[
\nabla_\mu \nabla_\nu-\frac{g_{\mu \nu}}{L^2}
\right]\phi(x) &=0\ .
\end{aligned}
\end{equation}
The first equation fixes the Ricci scalar to a negative constant value and since the theory is two dimensional, it completely determines the Riemann tensor (\ref{eq:14}). This means the \textit{only} metric solution in JT gravity is pure ${\rm AdS}_2$. The analysis of the extended first law in JT gravity is extremely simple given that all we have to do is analyze the thermodynamic behavior of Killing horizons in pure ${\rm AdS}_2$. The theory does not admit any real black hole solution.\footnote{While the classical theory is almost trivial, interesting dynamics arise by introducing a fluctuating boundary. These boundary effects give one loop contributions to the Euclidean partition function \cite{Maldacena:2016upp,Harlow:2018tqv} and therefore lie beyond the semi-classical analysis captured by horizon thermodynamics.}

Writing the metric in global coordinates $(\sigma,\theta)$ as in (\ref{eq:13}) the only Killing horizon is generated by the vector in (\ref{eq:52}), which is time-like in the region $\theta_\pm<\theta_0\in (0,\pi)$, sketched in figure \ref{fig:2}. The equation of motion of the dilaton $\phi(x)$ can be easily solved in global coordinates and written as
\begin{equation}\label{eq:194}
\phi(\sigma,\theta)=\phi_h 
\frac{\cos(\sigma/L)\sin(\theta_0)}{\sin(\theta)}\ ,
\end{equation}
where $\phi_h> 0$ is an integration constant that gives the value of the dilaton at the horizon. The full solution is parametrized by the value of the single constant $\phi_h$.\footnote{While it seems the solution also depends on $\theta_0\in(0,\pi)$, we can use the isometries of ${\rm AdS}_2$ to fix $\theta_0=\pi/2$.}

To compute the horizon entropy we use Wald's functional (\ref{eq:45}) together with the fact that the Riemann tensor is fixed by $\mathcal{R}$ (\ref{eq:14})
\begin{equation}\label{eq:191}
S_\xi=
4\pi
\left.\frac{\delta \mathcal{L}}{\delta R}
\right|_{\rm Horizon}=4\pi\phi_0+
4\pi \phi(x)\big|_{\theta_\pm=\theta_0}=
4\pi(\phi_0+\phi_h)\ .
\end{equation}
This agrees with the result obtained from the semi-classical computation of the Euclidean path integral \cite{Harlow:2018tqv}. The extended first law involves computing the entropy variation with respect to the coupling constants of the theory $\lambda_i=(\phi_0,L)$ and checking whether it can be written as
\begin{equation}\label{eq:195}
\delta_{\lambda_i} S_\xi=\frac{S_\xi}{a_1^\ast}
\delta_{\lambda_i} a_1^\ast\ ,
\end{equation}
where $a_1^\ast$ is some function of the coupling constants $a_1^\ast=a_1^\ast(\phi_0,L)$. In this setup we have no natural definition of $a_1^\ast$ in terms of the on-shell Lagrangian (\ref{eq:17}), so in principle we can allow any function that depends exclusively on the coupling constants $(\phi_0,L)$. However, since $a_1^\ast$ and $\phi_0$ are dimensionless quantities and $L$ has dimensions of length we have it can only depend on $\phi_0$.\footnote{Note that if we naively apply the definition of $a_1^\ast$ in (\ref{eq:17}), we get $a_1^\ast=2\phi_0$.} From the simple expression of the entropy given in (\ref{eq:191}) we can compute the entropy variation explicitly and find it is not compatible with the extended first law as written in (\ref{eq:195}) for any definition of $a_1^\ast(\phi_0)$
\begin{equation}
\delta_{\lambda_i} S_\xi=4\pi \delta \phi_0\neq \frac{S_\xi}{a_1^\ast}\delta_{\lambda_i} a_1^\ast\ .
\end{equation}
This means the form of the extended first law for JT gravity is not the same as in the previous cases we studied so far. The difference is that the solution in JT gravity depends on the additional parameter $\phi_h$, that appears in the horizon entropy and is not related to the AdS radius $L$. In the previous derivations in section \ref{sec:2} we used the fact that the pure AdS solution only depends on the radius $L$. 

We expect a similar situation for other Einstein-dilaton theories that cannot be written as pure gravity theories. For any particular theory one can still compute the variation of the horizon entropy on pure ${\rm AdS}_2$ as in (\ref{eq:191}), but there is no guarantee there exists a function $a_1^\ast=a_1^\ast(\lambda_i)$ such that it can be written as in the extended first law (\ref{eq:195}).

\subsection{Beyond Pure AdS in Three Dimensional Gravity}
\label{sec:3Dlaw}

Given that all our calculations so far have been for Killing horizons in pure AdS, a natural question is whether these results can be extended to horizons in more general space-times. Crucial to our derivation was that pure AdS has \emph{local} AdS symmetry. In general, arbitrary spacetimes are not locally AdS. There is a special case, however, in $2+1$ dimensions, where certain black hole solutions have local $\text{AdS}_{3}$ symmetry. In this section we investigate this in the context of three dimensional gravity, making contact with some concepts in extended black hole thermodynamics \cite{Kubiznak:2016qmn}.

Consider a general three dimensional metric $g_{\mu\nu}$ which solves the equations of motion obtained from (\ref{genaction}) and admits a time-like Killing horizon generated by the vector $\xi^\mu$. The horizon entropy is obtained from Wald's functional (\ref{Waldent}) evaluated on $g_{\mu \nu}$, which for a general metric we cannot evaluate explicitly. However, three dimensional gravity theories admit interesting black hole solutions which are locally but not globally AdS, \textit{i.e.}, which satisfy
\begin{equation}\label{eq:53}
\mathcal{R}_{\mu \nu\rho \sigma}=
  -\frac{1}{L^2}
  \left(
  g_{\mu \rho}g_{\nu \sigma}-
  g_{\mu \sigma}g_{\nu \rho}
  \right)\ .
\end{equation}
For this class of black holes we can evaluate the integrand in Wald's functional using (\ref{eq:24}) and find
\begin{equation}\label{eq:38}
S_{\xi}
  \left[
  g_{\mu \nu},\lambda_i
  \right]=
  2a_2^*(\lambda_i)
  \widetilde{\mathcal{A}}
  \ ,
\end{equation}
where $\widetilde{\mathcal{A}}=\mathcal{A}_{\rm horizon}/L^{d-1}$ and $a_2^*$ in (\ref{eq:9}) is proportional to the Virasoro central charge $c$ of the dual $\text{CFT}_{2}$. This expression is equivalent to the pure AdS relation (\ref{eq:2}) evaluated at~$d=2$. 

Let us now consider the behavior of the entropy under deformations of the theory, \textit{i.e.},~${\lambda_i\rightarrow \lambda_i+\delta \lambda_i}$ in (\ref{genaction}). In this case, apart from the obvious contribution given by the coefficient $a_2^*(\lambda_i)$ in (\ref{eq:38}), we must take into account the variation of the dimensionless horizon area $\widetilde{\mathcal{A}}$. For the pure AdS metric, $\widetilde{\mathcal{A}}$ is independent of $\lambda_i$ since the metric $g_{\mu \nu}^{\rm AdS}(L)$ only depends on the dimensionful parameter $L$, so that dimensional analysis implies $\mathcal{A}_{\rm horizon}\propto L^{d-1}$. This is no longer true for more general metrics which satisfy (\ref{eq:53}) but are not globally pure AdS, as the metric can also depend on some integration constants $\left\lbrace c_j \right\rbrace$ (\textit{e.g.} mass, angular momentum, charge, etc.) so that the horizon area $\mathcal{A}_{\rm horizon}$ is no longer proportional to $L^{d-1}$. Altogether, the variation of (\ref{eq:38}) is now given by
\begin{equation}\label{eq:34}
\delta S_\xi=
  S_\xi\,\delta\left[
  \ln(a_2^*)+\ln(\widetilde{\mathcal{A}})
  \right]\ .
\end{equation}
As we will shortly see in a simple example, computing this extra variation for a particular solution is straightforward. However, while the first term involving $a_2^*$ has a clear meaning in the boundary CFT (given in (\ref{eq:32})), this is not the case for $\widetilde{\mathcal{A}}$. Only by restricting ourselves to black holes in which $\delta \widetilde{\mathcal{A}}=0$, the boundary CFT satisfies the extended first law given by
\begin{equation}\label{eq:39}
\delta \widetilde{\mathcal{A}}=0
  \qquad \Longrightarrow \qquad
  \delta S(\rho_\beta)=\beta\, \delta
  \langle H \rangle+
  \frac{S(\rho_\beta)}{a_2^*}\delta a_2^*\ ,
\end{equation}
where $\rho_\beta$ is a thermal state and we have included the usual energy term $(2\pi/\kappa)\delta Q_\xi$ in (\ref{eq:34}) which maps to $H$, the hamiltonian of the CFT. Additional conserved quantities such as angular momentum or charges, can be added to this relation in the usual way. The first law in~(\ref{eq:39}) is similar to the one obtained for the thermal state at temperature $\beta=2\pi R$ in the background~${\mathbb{R}\times \mathbb{H}^{d-1}}$ (\ref{eq:55}), with the crucial difference that $\beta$ in this case is unconstrained.  

Let us illustrate how everything works by considering a simple example in Einstein gravity
\begin{equation}\label{eq:43}
I[g_{\mu \nu};G,L]=
  \frac{1}{16\pi G}
  \int d^3x\,\sqrt{-g}\left(
  \mathcal{R}+\frac{2}{L^2}
  \right)\ .
\end{equation}
The coupling constants of the theory are $\left\lbrace \lambda_i\right\rbrace=\left\lbrace G,L \right\rbrace$, where $L$ is also the radius of the pure AdS solution. The rotating BTZ black hole solution satisfies (\ref{eq:53}) and is given by \cite{Banados:1992wn}
\begin{equation}\label{eq:44}
ds^2=-f(r)dt^2+\frac{dr^2}{f(r)}+
  r^2\Big(
  d\theta-\frac{GJ}{2r^2}dt
  \Big)^2\ ,
\end{equation}
where $f(r)=-8G M+(r/L)^2+(JG/2r)^2$. Different black holes are labeled by the integration constants ${\left\lbrace c_j \right\rbrace=\left\lbrace M,J \right\rbrace}$, which also give the global charges associated to the Killing vectors~$\partial_t$ and $\partial_\theta$ respectively.

The outer horizon radius $r_+$ is obtained from $f(r_+)=0$ and is a non-trivial function of~$(G,L,M,J)$. We can easily write the dimensionless horizon area $\widetilde{\mathcal{A}}$ in terms of $r_+$
\begin{equation}\label{eq:35}
\widetilde{\mathcal{A}}=
  \frac{2\pi r_+}{L}=
  4\pi \sqrt{MG}
  \left[
  1+\sqrt{1-\left(
  \frac{J}{8ML}
  \right)^2}\,
  \right]^{1/2}.
\end{equation}
This expression depends explicitly on both $G$ and $L$, meaning that the second term in (\ref{eq:34}) gives a non-trivial contribution, which we can easily write explicitly. However, if we consider the static black hole $J=0$ we get $\widetilde{\mathcal{A}}=4\pi\sqrt{2MG}$, which is independent of $L$. Therefore, if we restrict to variations of $L$ (while keeping $G$ fixed), we obtain the extended first law given in (\ref{eq:39}).

\subsubsection{Extended Thermodynamics and Volume}
 
Let us now restrict to a particular type of theory deformation, in which we take the radius of the pure AdS solution $L$ as one of the coupling constants defining the theory and consider~$\delta(\lambda_i,L)=(0,\delta L)$. This corresponds to the variations studied in the extended black hole thermodynamics \cite{Kubiznak:2016qmn}, in which the thermodynamic pressure is identified with $L$ according to~${p\equiv d(d-1)/(16\pi G L^2)}$. Its conjugate variable is referred as the volume $V$ and can be defined from the entropy as
\begin{equation}\label{eq:36}
V\equiv -T \frac{\partial S_\xi}{\partial p}=
  -T S_\xi
  \frac{\partial }{\partial p}
  \left[
  \ln(a_2^*)+\ln(\widetilde{\mathcal{A}})
  \right]\ .
\end{equation}
where the second equality is obtained from (\ref{eq:34}). The $p$ derivative is computed while keeping all the remaining parameters fixed. 

This volume formula holds for locally AdS black holes in any three dimensional theory of gravity. Similar to (\ref{eq:34}), there are two distinct contributions to the volume. While the variation of $a_2^*$ has a natural boundary interpretation in terms of the number of degrees of freedom, the dimensionless area $\widetilde{\mathcal{A}}$ does not. For cases in which $\widetilde{\mathcal{A}}$ is independent of $L$, the thermodynamic volume takes the following simple form
\begin{equation}\label{eq:37}
\frac{\partial \widetilde{\mathcal{A}}}
{\partial L}=0
\quad \Longrightarrow \quad
V=-\left(\frac{T S_\xi}{a_2^*}\right)
  \frac{\partial a_2^*}{\partial p}\ .
\end{equation}
This gives a class of three dimensional black holes whose thermodynamic volume is directly related to changing the central charge of the boundary CFT. Since the meaning of $V$ for the boundary theory is not completely understood (see \cite{Dolan:2013dga,Johnson:2014yja,Dolan:2014cja,Kastor:2014dra,Caceres:2016xjz,Couch:2016exn,Johnson:2019wcq}), this formula might help give further insights. Let us use it in some concrete examples to compute the volume of some black hole solutions.

\subsection*{Thermodynamic volume in Einstein gravity}

Consider the simple setup of a BTZ black hole (\ref{eq:44}) in Einstein gravity (\ref{eq:43}). As previously noted, for the static black hole $J=0$ the dimensionless horizon area $\widetilde{\mathcal{A}}$ in (\ref{eq:35}) is independent of $L$, meaning that we can directly use the volume formula in (\ref{eq:37}). Simple calculations give~$a_2^*=L/8G$ and $T=r_+/2\pi L^2$, so that we can compute the volume as
\begin{equation}
V_{J=0}=-\left(\frac{T S_\xi}{a_2^*}\right)
  \frac{\partial a_2^*}{\partial p}=
   \pi r_+^2\ .
\end{equation}
which agrees with the result obtained from a more standard approach in extended thermodynamics \cite{Frassino:2015oca}. 

For the rotating BTZ solution with $J\neq 0$ the dimensionless horizon area $\widetilde{\mathcal{A}}$ in (\ref{eq:35}) is a non-trivial function of $L$, meaning that we must use the more general volume formula in~(\ref{eq:36}). Although the calculation in this case is slightly more involved, the final result is again very simple and given by
\begin{equation}
V_{J\neq 0}=
  -T S_\xi
  \frac{\partial }{\partial p}
  \left[
  \ln(a_2^*)+\ln(\widetilde{\mathcal{A}})
  \right]=
  \pi r_+^2\ ,
\end{equation}
in agreement with the previously known relation \cite{Frassino:2015oca}. It is interesting to see that the extra variation with respect to $\widetilde{\mathcal{A}}$ is exactly what is needed in order to obtain this simple final answer. An interesting microscopic analysis of this expression was recently given in \cite{Johnson:2019wcq}.\footnote{We should mention that while the charged BTZ black hole in Einstein-Maxwell theory \cite{Martinez:1999qi} is not locally AdS (\ref{eq:53}), if we naively apply the volume formula in (\ref{eq:37}) we obtain ${V=\pi r_+^2-\pi(QL/2)^2}$, which agrees with the previously known result \cite{Frassino:2015oca}. The reason it works is due to the fact that in Einstein gravity Wald's entropy functional always reduces to the Bekenstein-Hawking area expression, \textit{i.e.} $S_\xi=\mathcal{A}/4G$. For higher curvature theories we do not expect the volume formula (\ref{eq:36}) to reproduce the correct result for the charged black hole.}

\subsection*{Thermodynamic volume in higher curvature theories}

Since the volume formula (\ref{eq:36}) is particularly powerful in the context of higher curvature gravity theories, let us apply it in an example by considering the following generalization of new massive gravity \cite{Bergshoeff:2009hq,Bergshoeff:2009aq,Sinha:2010ai}
\beq I[g_{\mu \nu}]=\frac{1}{16\pi G}\int d^{3}x\sqrt{-g}\left(\mathcal{R}+\frac{2}{\ell^{2}}+\ell^{2}\mathcal{R}_{2}+\ell^{4}\mathcal{R}_{3}\right)\;,\label{NMGextaction}\eeq
where 
\beq
\begin{split}
& \mathcal{R}_{2}=4(\lambda_{1}\mathcal{R}_{\mu \nu}\mathcal{R}^{\mu \nu}+\lambda_{2}\mathcal{R}^{2})\;,\\
& \mathcal{R}_{3}=\frac{17}{12}
(\mu_{1}\mathcal{R}^{\nu}_{\;\mu}\mathcal{R}^{\rho}_{\;\nu}\mathcal{R}^{\mu}_{\;\rho}
+\mu_{2}\mathcal{R}_{\mu \nu}\mathcal{R}^{\mu \nu}\mathcal{R}
+\mu_{3} \mathcal{R}^{3})\;.\label{curves}
\end{split}
\eeq
The coupling constants of the theory are given by $\left\lbrace G,\ell,\lambda_1,\lambda_2,\mu_i \right\rbrace$ with $i=1,2,3$, where new massive gravity \cite{Bergshoeff:2009hq,Bergshoeff:2009aq} is obtained by setting $\mu_i=0$ and $\lambda_{2}=-3\lambda_{1}/8$. 

To apply the volume formula in (\ref{eq:36}) we must first compute the $a_2^*$ factor, which depends on the pure AdS solution of the theory. We can find such solution by varying the action~(\ref{NMGextaction}) with respect to the metric, which gives the following equations of motion \cite{Sinha:2010ai}
\begin{equation}\label{eqnsofmotNMG}
\mathcal{R}_{\mu \nu}
-\frac{1}{2}\mathcal{R}g_{\mu \nu}-
\frac{1}{\ell^{2}}g_{\mu \nu}-
H_{\mu \nu}=0\;,
\end{equation}
where
\begin{equation}
\begin{split}
H_{\mu \nu}&=4\ell^{2}
\biggr[\lambda_{1}
\left(-2\mathcal{R}^{\rho}_{\;\mu}\mathcal{R}_{\rho \nu}
+\frac{1}{2}g_{\mu \nu}
\mathcal{R}_{\rho \sigma}
\mathcal{R}^{\rho \sigma}\right)
+
\lambda_{2}\left(-2\mathcal{R}\mathcal{R}_{\mu \nu}+\frac{1}{2}g_{\mu \nu}\mathcal{R}^{2}\right)
\biggr]\\
&+\frac{17}{12}\ell^{4}
\biggr[
\mu_{1}
\left(-3\mathcal{R}_{\mu \rho}\mathcal{R}^{\rho}_{\;\sigma}\mathcal{R}^{\sigma}_{\;\nu}
+\frac{1}{2}g_{\mu \nu}\mathcal{R}^{\rho}_{\;\sigma}\mathcal{R}^{\alpha}_{\;\rho}\mathcal{R}^{\sigma}_{\;\alpha}\right)
+\mu_{3}
\left(-3\mathcal{R}^{2}\mathcal{R}_{\mu \nu}+\frac{1}{2}g_{\mu \nu}\mathcal{R}^{3}\right)
\\
&+
\mu_{2}
\left(-\mathcal{R}^{\rho}_{\;\sigma}\mathcal{R}^{\sigma}_{\;\rho }\mathcal{R}_{\mu \nu}-2\mathcal{R}\mathcal{R}_{\mu \rho}\mathcal{R}^{\rho}_{\;\nu }+\frac{1}{2}g_{\mu \nu}\mathcal{R}
\mathcal{R}_{\rho \sigma}
\mathcal{R}^{\rho \sigma}\right)
\biggr]
+\mathcal{O}(\nabla^{2}\mathcal{R},\nabla^{2}\mathcal{R}^{2},...)\; ,
\end{split}
\end{equation}
and we are omitting derivative terms that do not contribute to the pure AdS solution. 

We can evaluate these complicated terms in a pure AdS metric $g_{\mu \nu}^{\rm AdS}(L)$ of some radius~$L$ using that it is a maximally symmetric space-time (\ref{eq:53}). Taking the trace of (\ref{eqnsofmotNMG}) and writing the AdS radius as $L=\ell/\sqrt{f_\infty}$ we obtain the following algebraic constraint for the factor $f_\infty$
\beq 
L=\ell/\sqrt{f_\infty}
\qquad \Longrightarrow \qquad
1-f_{\infty}-8f_{\infty}^{2}(\lambda_{1}+3\lambda_{2})+17f_{\infty}^{3}(\mu_{1}+3\mu_{2}+9\mu_{3})=0\;.\label{polyfinf}\eeq

To arrive to this expression we used the helpful fact that in pure AdS
\beq \mathcal{R}_{\mu\nu}=-\frac{2}{L^{2}}g_{\mu\nu}\;,\quad \mathcal{R}=-\frac{6}{L^{2}}\;,\eeq
such that 
\beq g^{\mu\nu}H_{\mu\nu}=-\frac{24\ell^{2}}{L^{4}}(\lambda_{1}+3\lambda_{2})+\frac{51\ell^{4}}{L^{6}}(\mu_{1}+3\mu_{2}+9\mu_{3})\;,\eeq
and so the trace of (\ref{eqnsofmotNMG}) is
\beq \frac{3}{L^{2}}-\frac{3}{\ell^{2}}+\frac{24\ell^{2}}{L^{4}}(\lambda_{1}+3\lambda_{2})-\frac{51\ell^{4}}{L^{6}}(\mu_{1}+3\mu_{2}+9\mu_{3})=0\;.\label{traceeqnsmot}\eeq
Setting $\ell=L\sqrt{f_{\infty}}$, expression (\ref{traceeqnsmot}) becomes (\ref{polyfinf}). When we set $\lambda_{2}=-\frac{3}{8}\lambda_{1}$, $\mu_{1}=\frac{64}{17}\mu_{3}$, and $\mu_{2}=-\frac{72}{17}\mu_{3}$, we find (\ref{polyfinf}) is in agreement with \cite{Sinha:2010ai}. The solution $f_\infty$ of this algebraic equation that is smoothly connected to Einstein gravity determines the pure AdS radius $L$. 

We will use the polynomial constraint (\ref{polyfinf}) to help us determine $a_{2}^{\ast}$. This is done by evaluating the Lagrangian density (\ref{NMGextaction}) in AdS, so that we find
\beq
\begin{split}
 \mathcal{L}|_{\text{AdS}}&=\frac{1}{16\pi G}\left[-\frac{6}{L^{2}}+\frac{2}{\ell^{2}}+\frac{48\ell^{2}}{L^{4}}(\lambda_{1}+3\lambda_{2})-\frac{34\ell^{4}}{L^{6}}(\mu_{1}+3\mu_{2}+9\mu_{3})\right]\\
&=-\frac{1}{4\pi G L^{2}}\left[1-16f_{\infty}(\lambda_{1}+3\lambda_{2})+17f_{\infty}^{2}(\mu_{1}+3\mu_{2}+9\mu_{3})\right]
\end{split}
\label{NGMLageval}
\eeq
where we used
\beq
\begin{split}
 \mathcal{R}_{2}&=\frac{48}{L^{4}}(\lambda_{1}+3\lambda_{2})\;,\;\;\mathcal{R}_{3}=-\frac{34}{L^{6}}\left(\mu_{1}+3\mu_{2}+9\mu_{3}\right)\;.
\end{split}
\eeq
Therefore, $2a_{2}^{\ast}=-\pi L^{3}\mathcal{L}|_{\text{AdS}}$ gives
\beq
\begin{split}
 a^{\ast}_{2}&=\frac{L}{8 G}\left[1-16f_{\infty}(\lambda_{1}+3\lambda_{2})+17f_{\infty}^{2}(\mu_{1}+3\mu_{2}+9\mu_{3})\right]\;.\label{a2NGM}
\end{split}
\eeq
Using the same identifications of $\lambda_{2},\mu_{1},\mu_{2}$ as before, we find (\ref{a2NGM}) agreement with \cite{Sinha:2010ai}, and is interpreted as the $d=2$ Weyl anomaly associated with the Euler density for our six dervative theory (\ref{NMGextaction}). When we turn off the cubic contributions $\mu_{i}=0$, (\ref{a2NGM}) is simply the $d=2$ limit of the Weyl anomaly associated with Einstein-Gauss-Bonnet gravity in higher dimensions \cite{Myers:2010tj}.

We can now consider a black hole solution for this theory. Given that the BTZ black hole in (\ref{eq:44}) is locally AdS, it solves the equations of motion in (\ref{eqnsofmotNMG}) as long as we take~$L$ according to (\ref{polyfinf}). The horizon entropy is obtained from (\ref{eq:38}) with $a_2^*$ and $\widetilde{\mathcal{A}}$ as given in (\ref{eq:35}), where 
\beq S_{\xi}=\frac{\mathcal{A}_{\mathcal{H}}}{4 G}\left[1-16f_{\infty}(\lambda_{1}+3\lambda_{2})+17f_{\infty}^{2}(\mu_{1}+3\mu_{2}+9\mu_{3})\right]\;.\label{BTZNGMent}\eeq
This matches the expression found using Wald's formula (\ref{eq:13}), given explicitly in \cite{Sinha:2010ai}. 

For the rotating solution with $J\neq 0$ we can now use the volume formula in~(\ref{eq:36}) and find
\begin{equation}\label{eq:56}
V_{J\neq 0}=
  \pi r_{+}^{2}
  \left[1-16f_{\infty}(\lambda_{1}+3\lambda_{2})+f^{2}_{\infty}(\mu_{1}+3\mu_{2}+9\mu_{3})\right]\;.
\end{equation}

\section*{Summary and Future Work}
\label{sec:4}

The extended first law of entanglement has been previosuly derived for the Minkowski vacuum reduced to a ball by considering particular gravity theories in the bulk \cite{Kastor:2014dra,Kastor:2016bph,Caceres:2016xjz,Lan:2017xcl}. In this work, we have shown a novel and simple procedure that generalizes the proof to arbitrary gravity theories in the bulk and new setups in the boundary CFT. From the bulk perspective we have found no obstructions in working in two dimensional gravity and also obtain some intriguing results concerning extended black hole thermodynamics in three dimensions. Let us discuss some additional aspects regarding the calculations above.


\textbf{Divergent terms in the extended first law of entanglement}

One important feature of the ordinary first law of entanglement $\delta S_{EE}=\delta \langle K_B \rangle$ is that although the entanglement entropy always diverges, the left-hand side is well defined since the difference between entropies associated to different states is finite.\footnote{As shown in \cite{Marolf:2016dob} this is not entirely true, since there are cases in which the entanglement entropy acquires state dependent divergences, so that $\delta S_{EE}$ diverges. However, the relative entropy remains finite.} For the extended first law of entanglement this is no longer the case. Let us consider a variation of the theory without perturbing the state, so that the first term on the right-hand side of (\ref{eq:12}) drops out and we are left with
\begin{equation}\label{eq:41}
S_{EE}(\bar{\rho})-S_{EE}(\rho)=
  \frac{S_{EE}(\rho)}{a_d^\ast}
  \delta a_d^\ast\ .
\end{equation}
Both sides of this equality diverge, the left-hand side due to the fact that the divergences of the entanglement entropies corresponding to different theories do not cancel each other. This raises the question regarding how we should interpret (\ref{eq:41}), which seems to depend on the regularization procedure.

Let us illustrate the issue by considering the simple case of the Minkowski vacuum reduced to a ball of radius $R$ in $d=3$, where the entanglement entropy is \cite{Casini:2011kv}
\begin{equation}\label{eq:42}
S_{EE}(\rho_B)=
  \mu_1\frac{R}{\epsilon}-2 \pi a_3^*\ ,
\end{equation}
with $\mu_1$ a dimensionless and non-universal constant and $a_3^*$ given by (\ref{eq:32}). The short distance cut-off $\epsilon$ can be properly defined using mutual information, see \cite{Casini:2015woa}. If we consider the same setup but for a CFT in which ${\bar{a}_3^*=a_3^*-\delta a_3^*}$, the entanglement entropy is given by
\beq S_{EE}(\bar{\rho}_B)=
  \bar{\mu}_1\frac{R}{\bar{\epsilon}}-2 \pi \bar{a}_3^*\ ,\eeq
where the cut-off $\bar{\epsilon}$ and the constant $\bar{\mu}_1$ are not necessarily related to the ones appearing in~(\ref{eq:42}). 

How should we understand (\ref{eq:41}) in this context? A practical approach is to simply ignore the non-universal contributions to the entanglement entropy and regard (\ref{eq:41}) as a relation between the universal terms, where it is clear that the extended first law is satisfied. A different procedure is instead given by relating the cut-offs of each theory in a particular way such that the extended first law is satisfied to every order. Assuming there is a relation~$\epsilon=\epsilon(\bar{\epsilon})$ which can be expanded around the origin as
\beq \epsilon(\bar{\epsilon})=
  \bar{\epsilon}\left(
  b_0+b_2(\bar{\epsilon}/R)^2+
  b_4(\bar{\epsilon}/R)^4+\dots
  \right)\ ,\eeq
we can fix the coefficients $b_{2n}$ such that (\ref{eq:41}) is satisfied to every order. For the case of a ball in three dimensional Minkowski we find
\beq \epsilon(\bar{\epsilon})=
  \bar{\epsilon}\,
  \frac{\mu_1}{\bar{\mu}_1}
  \left(1-\delta \ln(a_3^*)\right)+\dots\ ,\eeq
where higher order terms are unconstrained. An analogous construction can be considered for the higher dimensional case and other setups in the CFT. This subtle aspect regarding the extended first law of entanglement has not been previously discussed in the literature.


\textbf{Extended first law of entanglement for general setups}

Given that we have shown that the extended first law of entanglement holds in a wide variety of setups, a natural question is whether it is valid for arbitrary CFTs, regions and states. While the ordinary first law follows from positivity of relative entropy \cite{Blanco:2013joa} and therefore holds in full generality, the extended version can only be formulated for CFTs since the coefficient $a_d^\ast$ in even dimensions is only defined for conformal theories (\ref{eq:32}). Although trying to directly prove the extended first law for arbitrary CFTs seems a complicated task, we can check whether the results for the entanglement entropy present in the literature are consistent with (\ref{eq:ext}), which essentially implies $S_{EE}\propto a_d^\ast$ to first order in $a_d^\ast$.

Let us consider two dimensional CFTs, where $a_2^\ast$ is proportional to the Virasoro central charge $c$. For the vacuum entropy associated to any number of disjoint intervals of a holographic CFT in Minkowski space, \cite{Ryu:2006ef,Hartman:2013mia,Faulkner:2013yia} showed that $S_{EE}\propto a_2^\ast$. The same is true for a thermal state reduced to an interval \cite{Calabrese:2004eu} and analogous setups in curved backgrounds \cite{Cardy:2016fqc}. For more general situations, the entanglement entropy is only known for particular CFTs, mostly free theories. In each of these cases the entropy depends on the details of the theory  in a complicated way. However, we are not aware of any result where the entanglement entropy in two dimensions is not proportional to the central charge and, consequently, in contradiction with (\ref{eq:ext}).

For space-time dimensions larger than two, it becomes evident that the extended first law of entanglement as written in (\ref{eq:ext}) cannot hold in full generality. The simplest example is to consider the Minkowski vacuum in $d=4$ reduced to a cylinder. Here the entanglement entropy is independent of $a_d^\ast$ and is instead proportional to the coefficient appearing in the square of the Weyl tensor in the trace anomaly \cite{Solodukhin:2008dh}. For more complicated regions the entropy is a combination of these coefficients. While this shows the extended first law as written in (\ref{eq:ext}) cannot hold in general for $d=4$, it suggests the following generalization might still be true\footnote{We thank an anonymous referee for suggesting this generalization.}
\begin{equation}
\delta S_{EE}=\delta\langle K_B \rangle+
\sum_{i}\frac{S_{EE}}{a_i}\delta a_i\ ,
\end{equation}
where $B$ is a region in four-dimensional Minkowski and $a_i$ are the coefficients of the terms appearing in the trace anomaly (see for example \cite{Myers:2010tj}). This generalization has a better chance of applying to more general regions.

It would be interesting to understand how holography is able to capture the extended first law of entanglement in these more general cases where it is expected to hold. The $d=2$ case stands out as the simplest one in which concrete progress might be possible, maybe using similar techniques as the ones developed in \cite{Faulkner:2013yia}. This deserves further study, in order to determine whether a general derivation of the extended first law of entanglement in this context is possible. 


\textbf{Bulk constraints from extended first law of entanglement}

Assuming the RT holographic formula for entanglement entropy together with the ordinary extended first law of entanglement in the boundary, implies Einstein's bulk equations about a perturbed AdS background. What are the consequences of assuming the extended first law of entanglement instead?\footnote{We thank an anonymus referee for suggesting this question} 

Let us address this question in the simplest setup of ${\rm AdS}_3/{\rm CFT}_2$, where the bulk theory is described by Einstein gravity, so that the coupling constants are $\lambda_i=(G,L)$. Let us assume (the non-trivial statement that) the extended first law of entanglement holds in the boundary CFT for arbitrary states $\rho$ and regions $B$, together with the RT formula
\begin{equation}\label{eq:201}
\delta S_{EE}=\delta \langle K_B \rangle+
\frac{S_{EE}}{c}\delta c\ ,
\qquad \qquad
S_{EE}=
\frac{A(\gamma_{\rm ext})}{4G} \ ,
\end{equation}
where $\gamma_{\rm ext}$ is an extremal bulk curve homologous to the region $B$ at the boundary. Using that in Einstein gravity the central charge $c$ is given by $c=3L/2G$, the ``extended" contribution of the first law of entanglement on the bulk becomes
\begin{equation}\label{eq:200}
\delta_{\lambda_i}\left(
\frac{A(\gamma_{\rm ext})}{4G}
\right)=
\frac{A(\gamma_{\rm ext})}{4G}
\delta_{\lambda_i}\ln(L/G)
\qquad \Longrightarrow \qquad
A(\gamma_{\rm ext})\propto L\ .
\end{equation}
The extended first law of entanglement translates into the statement that the length of the extremal curve on the bulk is proportional to the AdS radius $L$. 

If the boundary state is the vacuum $\ket{0}$ the bulk metric is pure ${\rm AdS}_3$, which only depends on $L$, and $A(\gamma_{\rm ext})\propto L$ immediately follows from dimensional analysis. The constraint becomes more interesting when considering excited states at the boundary, such as a thermal state $\rho(\beta)$ with inverse temperature $\beta$. In this case we can easily compute $A(\gamma_{\rm ext})$ and find the non-trivial statement $A(\gamma_{\rm ext})\propto L$ is indeed true \cite{Ryu:2006bv}. For more general setups this gives a bulk constraint coming from the boundary extended first law of entanglement.

It is also interesting to consider the inverse logic. We can directly compute $A(\gamma_{\rm ext})$ for complicated holographic setups and check whether the end result is proportional to $L$. This could help to understand in which situations the extended first law of entanglement holds for the boundary theory. These  questions would be interesting to investigate in future work.


\textbf{Extended first law in a single dimension}

Despite the fact that two dimensional gravity theories are highly constrained, we have obtained a non-trivial extended first law in the bulk. Our derivation holds for a wide class of pure gravity and Einstein-dilaton theories. Since the holographic correspondence in AdS$_2/{\rm CFT}_1$ is not as well established as in higher dimensions, we have not been provided with a compelling boundary interpretation. It should be interesting to further explore this in a simple case where there is some control on both sides of the duality.

An interesting setup is given by JT gravity, which is an Einstein-dilaton theory known to provide a holographic description of the SYK model \cite{Maldacena:2016hyu,Maldacena:2016upp,Jensen:2016pah}. Interestingly, our bulk derivation of the extended first law does not hold for JT gravity, as it is an Einstein-dilaton theory that cannot be recast as pure gravity.\footnote{Our derivation in section \ref{sec:2Dlaw} does not apply to any Einstein-dilaton theory in (\ref{eq:8}) with $V''(\phi)=0$. JT gravity falls in this category, as it contains a linear potential $V(\phi)\propto \phi$.} Given the recent interest in this system, this is an area that deserves further study as it may prove useful into better understanding JT gravity and SYK, and perhaps, $\text{AdS}_{2}/\text{CFT}_{1}$ more broadly. 

JT gravity also offers us a chance to study quantum effects in the extended first law of entanglement. For general holographic CFTs, the leading $1/N$ correction to the boundary entanglement entropy is dual to a bulk entanglement entropy between two bulk regions separated by the Ryu-Takayanagi entangling surface \cite{Faulkner:2013ana}. In general it is difficult to explicitly calculate the bulk entanglement contributions coming from the $1/N$ corrections. One exception to this is in 1+1 dimensions; specifically, recently the quantum corrected entanglement entropy with the bulk entropy term was computed explicitly in JT gravity \cite{Jafferis:2019wkd}. For this case, it might be possible to write down an extended first law with the bulk entanglement corrections. Moreover, it might even be possible to apply this generalized first law to dynamical spacetimes, such as an evaporating black hole, where the bulk entanglement can be computed explicitly and follows the Page curve, as shown in \cite{Penington:2019npb}. We leave these interesting questions for future work.


\textbf{Three dimensional gravity and thermodynamic volume}

For three dimensional bulk duals we have derived a modification of the extended first law~(\ref{eq:39}) that holds for space-times that are not necessarily (globally) pure AdS, such as the BTZ black hole. In the context of extended black hole thermodynamics, we obtain a curious formula for the thermodynamic volume (\ref{eq:36}), which we verified gives the correct expressions found using standard means. In particular, we obtain a result for the thermodynamic volume of the BTZ black hole in a higher curvature theory of gravity (\ref{eq:56}).

It would be interesting to see whether the formula for the thermodynamic volume in~(\ref{eq:36}) provides anything new to the field of extended thermodynamics. Particularly, it would be beneficial to see if it gives another microscopic viewpoint of $V$, along the lines of \cite{Johnson:2019wcq}. In \cite{Johnson:2019wcq} it was shown that the thermodynamic volume sometimes constrains the number of available CFT states dual to $\text{AdS}_{3}$ gravity, revealing that the Bekenstein-Hawking entropy (given by the Cardy formula) overcounts the number of CFT degrees of freedom. This chain of reasoning provides a microscopic explanation for black hole super-entropicity, a designation for black holes whose entropy exceeds that of Schwarzschild-AdS, and violate the reverse isoperimetric inequality  \cite{Cvetic:2010jb}. In three space-time dimensions, the reverse isoperimetric inequality takes the form
\beq \pi V\geq 4S^{2}G^{2}\;.\label{reviso3d}\eeq
When we input our expression for the volume in (\ref{eq:36}), the reverse isoperimetric inequality imposes a lower bound on the $L$ derivative of $\log(a_{2}^{\ast})$,
\beq \frac{\partial}{\partial L}\left[\log(a^{\ast}_{2})+\log(\tilde{\mathcal{A}})\right]\geq\frac{SG}{\pi^{2}L^{3}T}\geq0\;.\eeq
Black holes which satisfy this inequality, \emph{e.g.}, rotating BTZ, are said to be sub-entropic. Super-entropic black holes, such as the charged BTZ, violate the inequality (\ref{reviso3d}) and impose the following upper bound
\beq \frac{\partial}{\partial L}\left[\log(a^{\ast}_{2})+\log(\tilde{\mathcal{A}})\right]\leq\frac{SG}{\pi^{2}L^{3}T}\;.\eeq
Since $a^{\ast}_{2}$ relates to the number of degrees of freedom of the dual $\text{CFT}_{2}$, these bounds are expected to tell us something about the availability of CFT microstates to be counted by the Cardy formula. It would be interesting to study these bounds in further detail, where $\tilde{\mathcal{A}}$ might acquire a boundary interpretation.


\newpage

\section{FINAL REMARKS}  \label{sec:conclusion}

Black holes lie at the intersection of quantum and classical gravity. As such, black holes provide the best testing ground to better understand the nature of quantum gravity. Starting from the observation that black holes may be treated as genuine thermal systems, we have shown that this provides insights into the nature of gravity. Specifically, by way of spacetime thermodynamics, we illustrated that the second law of thermodynamics applied to local lightsheets in an arbitrary spacetime -- whose entropy is assumed to go as the cross-sectional area -- gives rise to the Ricci convergence condition $R_{ab}k^{a}k^{b}\geq0$, and, via the Einstein equations, the (classical) null energy condition $T_{ab}k^{a}k^{b}\geq0$. Therefore, an \emph{ad hoc} assumption about the behavior of matter in a spacetime has its origins in spacetime thermodynamics. Moreover, we further showed that when the form of the entropy includes logarithmic area corrections (just as 1-loop quantum corrected black hole entropy), the Ricci convergence condition still arises from the second law (the form of the null energy condition, however, is obscured as now the equations of motion are no longer Einstein's equations). 

We then showed that the null energy condition is not the only classical aspect of spacetime which arises from a local holographic thermodynamic principle. By constructing a timelike congruence of  radial boost vectors (the stretched future lightcone) about every point in an arbitrary spacetime, we found a simple statement about thermal equilibrium, namely, the Clausius relation $Q=T\Delta S$, constrains the dynamics of the classical spacetime, equivalent to the gravitational field equations. Depending on what entropy we attribute to the cross-sections of the stretched lightcone, we attain a different type of theory of gravity. We also demonstrated that the techniques can be applied to the past of local causal diamonds, where we find a similar result. Crucial to both derivations was to recognize the entropy change $\Delta S$ include only \emph{reversible} entropy changes (done by subtracting out the natural geometric expansion of a lightone or contraction of a causal diamond). Collectively, we found that when stretched future lightcones or causal diamonds are treated as equilibrium thermodynamic systems, their local, holographic thermodynamics encodes information about the classical dynamics of the spacetime they live on.

Our derivation of the equations of motion via lightcone thermodyamics led us to a local first law of gravity -- a hybrid equation connecting spacetime and matter thermodynamics. Importantly, unlike the first law of black holes which depends on the global structure of horizons, our law is genuinely local, holding about each point in spacetime. Moreover, the local first law includes a pressure-volume `work' term typically absent from the first law of black hole thermodynamics. 

We then changed focus and studied the entanglement of stretched lightcones. Motivated by the entanglement equilibrium proposal -- originally formulated for causal diamonds and says that the vacuum is in a maximally entangled state -- we extended the proposal to stretched lightcones. Applying the proposal and studying constant volume variations of the lightcone entanglement entropy, we uncovered that, with the aid of the first law of entanglement entropy, entanglement equilibrium is equivalent to the gravitational field equations being satisfied about every point in spacetime. In other words, spacetime entanglement generates classical dynamics of a spacetime. A particular feature of the calculation was observing that the condition of constant volume variations of the entropy translates to considering reversible entropy changes in the Clausius relation, mapping entanglement equilibrium to  reversible equilibrium thermodynamics. In summary, this collection of aforementioned work provides a throughline from quantum to classical gravity: how the entanglement structure of spacetime encodes a thermodynamic interpretation of classical aspects of gravity. 


We then concentrated on spacetimes that are asymptotically AdS, where the physics is greatly enriched. AdS-black holes, for example, now come equipped with a thermodynamic  pressure proportional to the cosmological constant, and have a somewhat mysterious `thermodynamic volume'. Using $\text{AdS}_{3}/\text{CFT}_{2}$, we provided a microscopic representation of the volume for BTZ black holes, cast purely in terms of CFT quantities, namely, the central charge, length scale $L$, and eigenvalues of the zero-mode Virasoro generators of the conformal algebra dual to the asymptotic symmetry group of the spacetime. In the case of a charged BTZ black hole we showed the positivity of volume restricts the number of accessible CFT degrees of freedom. Consequently, the gravitational entropy, given by the stastical Cardy formula, is overcounting the number of states. This gives the first microphysical explanation of black hole \emph{super-entropicity}, explaining on a microscopic level why the charged BTZ black hole unexpectedly has more entropy than its static and rotating counterparts.

Finally, we concluded with presenting a collection of novel aspects of the extended first law of entanglement. This AdS/CFT statement -- found originally by taking the bulk first law of entanglement for spherical entangling surfaces on the boundary and including variations of the cosmological constant -- was generalized to arbitrary theories of gravity as well as a slew of boundary regions. We also paid close attention to its lower dimensional limits, where we found a non-trivial statement in $1+1$-dimensions, while  in $2+1$-dimensions we could apply it to black hole systems and derive a new expression for the thermodynamic volume in terms of the generalized central charge of the dual CFT. Our analysis presents another input in the AdS/CFT dictionary, and further insight into the microphysics of extended black hole thermodynamics.


To summarize, the study of black holes is the study of spacetime. Whether it is realizing that classical spacetime can be understood as a type of hydrodynamic limit of some more fundamental quantum theory, or that gravitational entropy and volume have are born from entanglement, black holes continue to offer new glimpses of fundamental physics. While it is unclear at this stage which theory of quantum gravity will come out on top, one thing remains clear: black holes will guide us in our attempts to better understand nature.


\newpage

\appendix

\section{FUNDAMENTALS OF SPACETIME THERMODYNAMICS} \label{app:thermofund}


To keep this work self-contained, we include brief reviews on topics that are crucial to the study of emergent gravity, holography, and entanglement. We divide up the necessary background material broadly into two sections: concepts necessary for (i) spacetime thermodynamics in Appendix \ref{app:thermofund}, and (ii) spacetime entanglement in Appendix \ref{app:entfund}. The reader familiar with these concepts may skip these sections. 


\subsection{Geodesic Congruences}
\noindent


Consider a one-parameter family of geodesics $\gamma_{s}(\lambda)$ where $s\in\mathbb{R}$ and $\lambda$ is some affine parameter. A collection of these curves will define a two-dimensional surface embedded in a higher dimensional manifold $M$. We can describe our surface with the set of coordinates $x^{\mu}(s,\lambda)$. Immediately we find two vector fields: (i) the tangent vector field to the family of geodesics $U^{\mu}=dx^{\mu}/d\lambda$, and (ii) the deviation vector $V^{\mu}=dx^{\mu}/ds$ -- where $V^{\mu}$ points from one geodesic to a neighboring one. The deviation vector suggests a \emph{relative velocity of geodesics}
\beq S^{\mu}=\frac{DV^{\mu}}{d\lambda}\equiv U^{\rho}\nabla_{\rho}V^{\mu}\;,\eeq
and a \emph{relative acceleration of geodesics}
\beq A^{\mu}=U^{\rho}\nabla_{\rho}S^{\mu}\;.\eeq
Using the fact that $U$ and $V$ form a basis set adapted to a coordinate system, we have
\beq
\frac{D^{2}}{d\lambda^{2}}V^{\mu}=R^{\mu}_{\;\nu\rho\sigma}U^{\nu}U^{\rho}V^{\sigma}\;,\label{geodeveqn}
\eeq
the geodesic deviation equation, telling us the relative acceleration between two neighboring geodesics is proportional to the curvature. 

The idea behind deriving the geodesic deviation equation (\ref{geodeveqn}) was to consider initially parallel geodesic curves, and then imagine traveling along the trajectories to determine how they behaved.  We can generalize this idea by considering a multidimensional set of neighboring geodesics, a \emph{congruence}, and see how the congruence evolves with respect to some affine parameter. 

We first begin with a four-dimensional timelike geodesic congruence. Let $U^{\mu}=dx^{\mu}/d\tau$ be a tangent vector field to our congruence, from which we see that the affine parameter $\lambda$ is identified with the proper time $\tau$. The velocity $U$ is normalized to $U^{2}=-1$, and satisfies the geodesic equation $U^{\lambda}\nabla_{\lambda}U^{\mu}=0$. Consider a deviation vector $V^{\mu}$ pointing between neighboring geodesics satisfying
\beq \frac{DV^{\mu}}{d\tau}=U^{\nu}\nabla_{\nu}V^{\mu}\equiv B^{\mu}_{\nu}V^{\nu}\;,\eeq
where $B^{\mu}_{\nu}=\nabla_{\nu}U^{\mu}$. The tensor $B$ quantifies which geodesics in the congruence deviate from being perfectly parallel. 

Let us now construct three vectors orthogonal to our timelike geodesics. That is, we can consider the vectors living in the tangent space $T_{p}M$ that are orthogonal to $U^{\mu}$ for each $p\in M$. Any vector in our tangent space can be projected into a subspace via the \emph{projection tensor} $P_{\mu\nu}$ \cite{Carroll04-1}:
\beq P^{\mu}_{\;\nu}=\delta^{\mu}_{\;\nu}+U^{\mu}U_{\nu}\;.\eeq
Since $U^{\mu}B_{\mu\nu}=U^{\nu}B_{\mu\nu}=0$,
we find that $B_{\mu\nu}$ lives in this normal subspace. Since any $(0,2)$ tensor can be decomposed into an antisymmetric part and a symmetric part, which can be further decomposed into a trace and trace free part, we define
\beq \theta=P^{\mu\nu}B_{\mu\nu}=\nabla_{\mu}U^{\mu}\;,\eeq
the trace of $B_{\mu\nu}$,
\beq \sigma_{\mu\nu}=B_{(\mu\nu)}-\frac{1}{3}\theta P_{\mu\nu}\;,\eeq
a symmetric and traceless tensor, and
\beq \omega_{\mu\nu}=B_{[\mu\nu]}\;,\eeq
an antisymmetric tensor. It is simple to check  the correct decomposition of $B_{\mu\nu}$ is
\beq B_{\mu\nu}=\frac{1}{3}\theta P_{\mu\nu}+\sigma_{\mu\nu}+\omega_{\mu\nu}\;.\eeq
We call $\theta$ the \emph{expansion} of the congruence, describing the change in ``spherical volume'' of our congruence; $\sigma_{\mu\nu}$ is the \emph{shear}, representing the distortion from a sphere to an ellipsoid, and $\omega_{\mu\nu}$ is the \emph{rotation}.

Analogous to the idea behind the geodesic deviation equation, we wish to study the evolution of the congruence by calculating the covariant derivative $D/d\tau$ of the expansion, shear, and rotation. To do this, we first compute the covariant derivative of $B_{\mu\nu}$ and take the correct decomposition to find the other covariant derivatives of interest. First, 
\beq\frac{D}{d\tau}B_{\mu\nu}=U^{\sigma}\nabla_{\sigma}B_{\mu\nu}=U^{\sigma}\nabla_{\sigma}\nabla_{\nu}U_{\mu}=U^{\sigma}\nabla_{\nu}\nabla_{\sigma}U_{\mu}-U^{\sigma}R^{\lambda}_{\mu\nu\sigma}U_{\lambda}\;,\eeq
where we used the fact that the commutator of covariant derivatives is proportional to the Riemann curvature tensor. Then, using the product rule
and making use of the geodesic equation, we arrive to
\beq \frac{D}{d\tau}B_{\mu\nu}=-B^{\sigma}_{\;\nu}B_{\mu\sigma}-R_{\lambda\mu\nu\sigma}U^{\sigma}U^{\lambda}\;.\eeq
Taking the trace leads to \emph{Raychaudhuri's equation}
\beq
\frac{d\theta}{d\tau}=-\frac{1}{3}\theta^{2}-\sigma_{\mu\nu}\sigma^{\mu\nu}+\omega_{\mu\nu}\omega^{\mu\nu}-R_{\mu\nu}U^{\mu}U^{\nu}\;,
\eeq
quantifying the evolution of the expansion of a timelike geodesic congruence.

Let's now move to the evolution of null geodesic congruences.  Deriving an equivalent Raychaudhuri equation for null geodesics is more difficult because the tangent vector to a null curve is orthogonal to itself -- disallowing us to study the evolution of vectors in a three-dimensional subspace normal to $U^{\mu}$. In the case of null geodesics, instead we care about the evolution of vectors living in a 2-D subspace of spatial vectors orthogonal to the null tangent vector field $k^{\mu}=dx^{\mu}/d\lambda$. Then, choosing an auxiliary null vector $\ell^{\mu}$ satisfying
\beq \ell^{\mu}\ell_{\mu}=0\,,\quad \ell^{\mu}k_{\mu}=-1\,,\quad k^{\mu}\nabla_{\mu}\ell^{\nu}=0\eeq
we define a modified projection tensor
\beq Q_{\mu\nu}=g_{\mu\nu}+k_{\mu}\ell_{\nu}+k_{\nu}\ell_{\mu}\;.\eeq
From here, we essentially follow the previous derivation for timelike geodesic congruences, leading to the Raychaudhuri equation for null geodesic congruences:
\beq
\frac{d\theta}{d\tau}=-\frac{1}{2}\theta^{2}-\hat{\sigma}_{\mu\nu}\hat{\sigma}^{\mu\nu}+\hat{\omega}_{\mu\nu}\hat{\omega}^{\mu\nu}-R_{\mu\nu}k^{\mu}k^{\nu}\;,\label{rayeqnapp}
\eeq
where
\beq \theta=Q^{\mu\nu}\hat{B}_{\mu\nu}\,,\quad \hat{\sigma}_{\mu\nu}=\hat{B}_{(\mu\nu)}-\frac{1}{2}\theta Q_{\mu\nu}\,,\quad \hat{\omega}_{\mu\nu}=\hat{B}_{[\mu\nu]}\;,\eeq
with
\beq \hat{B}_{\mu\nu}=\frac{1}{2}\theta Q_{\mu\nu}+\hat{\sigma}_{\mu\nu}+\hat{\omega}_{\mu\nu}\;.\eeq
In this case, the Raychaudhuri equation descibes the evolution of the expansion of null congruences. 

Let's now state the focusing theorem for null geodesic congruences. Consider a  null congruence that is hypersurface orthogonal, \emph{i.e,} $\omega_{\mu\nu}=0$. Then, assuming the null energy condition $T_{\mu\nu}k^{\mu}k^{\nu}\geq0$, such that we have the Ricci convergence condition $R_{\mu\nu}k^{\mu}k^{\nu}\geq0$ by way of Einstein's equations, the Raychaudhuri equation (\ref{rayeqnapp})
\beq\frac{d\theta}{d\lambda}\leq0\;.\label{focusthmnull}\eeq
That is, the geodesics forming the congruence are focused during the evolution of the congruence. Integrating $\frac{d\theta}{d\lambda}=-\frac{1}{2}\theta^{2}$ gives 
\beq \theta^{-1}\geq\theta_{0}^{-1}+\frac{\lambda}{2}\;,\eeq
with $\theta_{0}=\theta(0)$. Therefore, if the congruence is initially converging, $\theta_{0}<0$, then the null geodesics converge $\theta(\lambda)\to-\infty$ in an affine `time' $\lambda\leq\frac{2}{|\theta_{0}|}$, signaling the development of a caustic where the geodesics intersect. 




\subsection{Black Hole Thermodynamics}
\noindent



In 1973, Bardeen, Carter, and Hawking developed the ``four laws of black hole mechanics'' \cite{Bardeen73-1}:

\hspace{2mm}

\noindent {\bf Zeroth Law:} \emph{The horizon for a stationary black hole has constant surface gravity $\kappa$}. 

\hspace{2mm}

 \noindent {\bf First Law:} \emph{The change in energy $E$ for a stationary black hole is related to the change in horizon area $A$, angular momentum $J$, and charge $Q$}
\beq dE=\frac{\kappa}{8\pi G}dA+\Omega dJ+\Phi dQ\;,\label{firstlawBHapp}\eeq 
where $\Omega$ is the angular velocity, and $\Phi$ is the electrostatic potential. 

\hspace{2mm}

 \noindent {\bf Second Law:} \emph{Assuming the weak energy condition, the horizon area is a non-decreasing function of time}
\beq \frac{dA}{dt}\geq0\;.\eeq

\hspace{2mm}

\noindent {\bf Third Law:} \emph{It is not possible to form a black hole with vanishing surface gravity.}.

\hspace{2mm}

Compare these laws to the four laws of ordinary thermodynamics:

\hspace{2mm}

\noindent {\bf Zeroth Law:} \emph{A system in thermal equilibrium is at constant temperature $T$.}

\hspace{2mm}

\noindent {\bf First Law:} \emph{For a thermodynamic system of temperature $T$, entropy $S$, internal energy $E$, and confined to a container of volume $V$ at pressure $P$, the change in internal energy for processes with no matter transfer is given by}
\beq dU=TdS+PdV\;.\label{1stlawordapp}\eeq

\hspace{2mm}

\noindent {\bf Second Law:} \emph{The change in entropy $S$ of an isolated system (over time) will be nonnegative for a spontaneous process:}
\beq \frac{dS}{dt}\geq0\;.\eeq

\hspace{2mm}

\noindent {\bf Third Law:} \emph{The entropy of a closed system in thermodynamic equilibrium will approach a constant value as its temperature approaches absolute zero.}

\hspace{2mm}

The four laws of black hole mechanics indeed remind us of the four laws of thermodynamics, however, the exact connection to black hole thermodynamics wasn't made possible until after Bekenstein postulated the existence of black hole entropy, further confirmed by the discovery of Hawking radiation \cite{Hawking74-1}. Specifically, the surface gravity $\kappa$ of a black hole is to be interpreted as temperature $T$ via $T=\frac{\kappa}{2\pi}$; the horizon area is proportional to the entropy\footnote{Restoring physical units, $S_{\text{BH}}=\frac{A}{4}\left(\frac{c^{3}k_{B}}{G\hbar}\right)=\frac{A k_{B}}{4\ell_{P}^{2}}$. For a Schwarzschild black hole, where $A=4\pi r_{h}^{2}$, with $r_{h}=2MG/c^{2}$, we have that the entropy is $S=4\pi M^{2}\frac{G k_{B}}{\hbar c}\sim(3.7\times 10^{-7}\frac{J\cdot K^{-1}}{kg^{2}})M^{2}$ and temperature $T_{H}=\frac{\hbar c^{3}}{8\pi k_{B}GM}\sim\frac{1.22\times10^{23}}{M}kg\cdot K$. For a solar mass black hole (the smallest stellar black holes are thought to be roughly three solar masses), we have an entropy of $S\sim 1.5\times 10^{54} J\cdot K^{-1}$ or $S\sim10^{77}$, and a temperature $T_{H}\sim 10^{-7} K$. This should be compared to the entropy of the Sun, $S\sim 10^{55}$, and the average temperature of the universe, at around $2.7 K$.}, $S\propto A$, and as the surface gravity tends to zero, so does the entropy\footnote{The third law of black hole thermodynamics isn't always true. Extremal black holes, while having non-zero entropy, have been shown to have vanishing surface gravity \cite{Kallosh92-1}.}.

With the knowledge that black holes carry a thermodynamic entropy, Bekenstein further postulated the generalized second law, \cite{Bekenstein74-1}
\beq \delta\left( S_{\text{matter}}+\frac{A}{4}\right)\geq0\;.\eeq

Below we shall outline a proof for the second law and present a derivation of the first law of black hole thermodynamics.



\subsection*{The Second Law}

Another useful way to write the expansion $\theta$ is in terms of the (null) congruence's cross-sectional area $A$ \cite{Poisson04-1}:
\beq \theta=\frac{1}{A}\frac{dA}{d\lambda}\eeq
where $\lambda$ is some affine parameter along the geodesic. We see then that the expansion describes the fractional rate of change of the cross-sectional area of the null congruence. As noted by Penrose, event horizons are generated by null geodesics with non-terminating endpoints. That is, once null geodesics enter the horizon (through, perhaps, a caustic) they can never again leave the horizon, or cross another null geodesic (for an illustrative proof, see \cite{Misner73-1}). 

In other words, the null generators forming the horizon cannot run into caustics. Assuming the null energy condition, by the focusing theorem (\ref{focusthmnull}) we have that $\theta\geq0$. This must hold everywhere on the event horizon of a stationary black hole, therefore, the area will not decrease in time, assuming the null energy condition holds \cite{Hawking71-1}
\beq \frac{dA}{d\lambda}\geq0\;.\eeq
This is the second law of black hole mechanics. For Einstein gravity, where $S\propto A$, we recognize that Hawking's area theorem is a statement about the second law of black hole thermodynamics:
\beq \frac{dS}{d\lambda}\geq0\;.\eeq




\subsection*{Smarr Relation and the First Law}
\indent

Here we present a derivation of the first law of black hole thermodynamics for a $D$-dimensional Schwarzschild black hole of the form
\beq ds^{2}=-fdt^{2}+f^{-1}dr^{2}+r^{2}d\Omega_{D-2}^{2}\;,\quad f(r)=1-\frac{\tilde{M}}{r^{D-3}}\;,\label{metSchBHD}\eeq
with
\beq \tilde{M}=\frac{16\pi GM}{(D-2)\Omega_{D-2}}\;,\eeq
where $\Omega_{D-2}$ is the unit volume of a $(D-2)$-dimensional sphere. We begin by deriving the Smarr relation. 

Notice that the metric (\ref{metSchBHD}) has a static Killing vector $(\partial/\partial t)^{a}$, and the only non-vanishing components of $\nabla^{a}\xi^{b}$ are
\beq \nabla^{r}\xi^{t}=-\nabla^{t}\xi^{r}=\frac{(D-3)\tilde{M}}{2r^{D-2}}\;.\label{nabxiBH}\eeq

The Smarr relation essentially comes from evaluating Komar integral formulae at infinity and at the black hole horizon:
\beq \frac{(D-2)}{8\pi G}\int_{\partial\Sigma_{\infty}}dS_{ab}\nabla^{a}\xi^{b}-\frac{(D-2)}{8\pi G}\int_{\partial\Sigma_{h}}dS_{ab}\nabla^{a}\xi^{b}=0\;,\label{komar1BH}\eeq
where $\partial\Sigma_{\infty}$ is a closed co-dimension-$2$ surface at $r\to\infty$, and $\partial\Sigma_{h}$ is the co-dimensional cross-section of the event horizon. As shorthand, we will express (\ref{komar1BH}) as
\beq \frac{(D-2)}{8\pi G}\oint_{\partial\Sigma}dS_{ab}\nabla^{a}\xi^{b}=0\;,\label{komar2BH}\eeq
Here $dS_{ab}$ is the volume element normal to the co-dimension 2 surface $\partial\Sigma$, and can be specified in more detail by writing out Gauss' law for $A^{c}=\nabla_{b}B^{bc}$, as
\beq \int_{\Sigma}dvn_{c}A^{c}=\int_{\partial\Sigma_{\infty}}dar_{b}n_{c}B^{bc}-\int_{\partial\Sigma_{h}}dar_{b}n_{c}B^{bc}\;,\eeq
where $n_{a}$ is the unit normal to $\Sigma$ and $r_{b}$ is the unit normal to $\partial\Sigma$ within $\Sigma$ taken to point towards infinity. Therefore, we have the surface volume element $dS_{bc}=dar_{[b}n_{c]}$. Here we take $n^{a}$ to be future pointing. Specifically, for the geometry (\ref{metSchBHD}) under consideration we have
\beq dS_{ab}=dar_{[a}n_{b]}=\frac{1}{2}da(r_{a}n_{b}-r_{b}n_{a})\;\Rightarrow dS_{ab}\nabla^{a}\xi^{b}=dar_{a}n_{b}\nabla^{a}\xi^{b}\equiv da_{a}n_{b}\nabla^{a}\xi^{b}\eeq
where we have defined $da_{a}=dar_{a}=d\Omega_{D-2}r^{D-2}r_{a}$. Here $r_{a}$ and $n_{a}$ are spacelike and timelike unit normals respectively. In particular, 
\beq r^{a}=\mathcal{N}\delta^{a}_{\;r}\;,\quad n^{a}=\mathcal{N}\delta^{a}_{\;t}\;,\eeq
such that $\mathcal{N}$ are normalization vectors; specifically,
\beq r^{a}=\sqrt{f}\delta^{a}_{\;r}\;,\quad n^{a}=\frac{1}{\sqrt{f}}\delta^{a}_{\;t}\;.\eeq
As such, $r_{a}=\frac{1}{\sqrt{f}}\delta_{ar}$ and $n_{a}=-\sqrt{f}\delta_{at}$. 

Then, the Komar integral (\ref{komar2BH}) becomes
\beq 
\begin{split}
0&=\frac{(D-2)}{8\pi G}\oint_{\partial\Sigma}dS_{ab}\nabla^{a}\xi^{b}=\frac{(D-2)}{8\pi G}\oint_{\partial\Sigma}da_{a}n_{b}\nabla^{a}\xi^{b}\\
&=-\frac{(D-2)}{8\pi G}\Omega_{D-2}\frac{(D-3)}{2}\tilde{M}\biggr|^{r\to\infty}_{r\to r_{h}}\;.
\end{split}
\label{closeintBH1}\eeq
Here $r_{h}$ is the horizon radius. 


Let $I_{h}$ and $I_{\infty}$ be the components of the integral in the Komar integral (\ref{komar2BH}) at the horizon and infinity respectively. From (\ref{closeintBH1}), we have that $I_{\infty}$ is given by,
\beq I_{\infty}=-(D-3)M\;.\eeq
Meanwhile, $I_{h}$ is
\beq \frac{(D-2)}{8\pi G}\int_{\partial\Sigma_{h}}da_{a}n_{b}\nabla^{a}\xi^{b}=\frac{(D-2)}{8\pi G}\kappa A\;,\label{IhBHapp}\eeq
where we used $r_{a}n_{b}\nabla^{a}\xi^{b}=-\kappa$ (a constant over the horizon) and $\int_{\partial\Sigma_{h}}da=A$. Combined, $I_{\infty}-I_{h}=0$ gives us the Smarr relation \cite{Smarr:1972kt}
\beq (D-3)M=\frac{(D-2)}{8\pi G}\kappa A\;. \label{smarr1BH}\eeq

More generally, we need only the asymptotic conditions on the metric and properties of the black hole horizon -- without completely specifying the  metric. The required fall conditions are
\beq ds^{2}\approx g_{tt}dt^{2}+g_{rr}dr^{2}+Hr^{2}d\Omega_{D-2}^{2}\;,\label{falloffs}\eeq
with asymptotic metric functions 
\beq g_{tt}=-f_{0}+\frac{c_{t}}{r^{D-3}}\;,\quad g_{rr}=\frac{1}{f_{0}}\;,\quad H=1\;,\quad f_{0}=1\;.\eeq
For the inverse metric, we have
\beq g^{tt}=-f_{0}^{-1}\;,\quad g^{rr}=f_{0}-\frac{c_{r}}{r^{D-3}}\;.\eeq
Here we can take constants $c_{t}=c_{r}=\tilde{M}$. We have, asymptotically at large finite radius,
\beq dar_{b}n_{c}(\nabla^{b}\xi^{c})\approx d\Omega_{d-2}\left(-\frac{(D-3)}{2}\tilde{M}\right)\;.\eeq
Combining the boundary integrals according to the Komar relation, we again arrive to the Smarr formula above (\ref{smarr1BH}). This relation can be generalized to charged and rotating black holes, such that,
\beq (D-3)M=\frac{(D-2)\kappa}{8\pi G}A+(D-2)\Omega J+(D-3)\Phi Q\;,\eeq
where $J$ is the black hole's angular momentum, $\Omega$ its angular velocity, $Q$ its charge and $\Phi$ its electrostatic potential at the horizon radius. 

Let us now move on and derive the first law of black hole thermodynamics. Our method is to use Hamiltonian perturbation techniques \cite{Sudarsky:1992ty,Traschen:2001pb}. As before, we let $\Sigma$ be a family of spacelike surfaces, with unit timelike normal $n_{a}$. Further, let $g_{ab}$ be the spacetime metric and $s_{ab}$ the induced metric on $\Sigma$, such that $g_{ab}=-n_{a}n_{b}+s_{ab}\;,\quad n_{c}n^{c}=-1\;,\quad n^{c}s_{cb}=0$.
The Hamiltonian variables are the spatial metric $s_{ab}$ and its conjugate momentum $\pi^{ab}$. Solutions to the Einstein equations with energy density $\rho=T^{ab}n^{a}n^{b}$ and momentum density $J_{a}=T_{bc}n^{b}s^{c}_{a}$ must satisfy the  Hamiltonian and momentum constraint equations
\beq H=-16\pi G\rho=-2G_{ab}n^{a}n^{b}\;,\quad H_{a}=-16\pi GJ_{a}=-2G_{bc}n^{b}s^{c}_{\;a}\;,\eeq
For a vanishing stress tensor, the constraint equations imply:
\beq H=0\;,\quad H_{a}=0\;.\label{constrqnspure}\eeq

Let $\xi^{a}=Fn^{a}+\beta^{a}$, such that $n^{c}\beta_{c}=0$ is the lapse vector field. The Hamiltonian density for evolution along $\xi^{a}$ in Einstein gravity is given by 
\beq \mathcal{H}=\sqrt{s}[FH+\beta^{a}H_{a}]\;.\eeq
Let $s^{(0)}_{ab}$ and $\pi^{ab}_{(0)}$ be a solution to the vacuum Einstein equations with Killing vector $\xi^{a}$. Now consider perturbations
\beq s_{ab}=s^{(0)}_{ab}+h_{ab}\;,\quad \pi^{ab}=\pi^{ab}_{(0)}+p^{ab}\;,\label{perts1}\eeq
where $h_{ab}=\delta s_{ab}$, and $p_{ab}=\delta \pi_{ab}$. It follows from Hamilton's equations for the zeroth order spacetime that the linearized constraint operators $\delta H$ and $\delta H_{a}$ combine to form a total derivative 
\beq F\delta H+\beta^{a}\delta H_{a}=D_{c}B^{a}\;,\eeq
where $D_{a}$ is the covariant derivative operator on $\Sigma$ compatible with metric $s^{(0)}_{ab}$, and the spatial vector $B^{a}$ is given by\footnote{Here the boundary term arises in the case we have non-vanishing extrinsic curvature, given by $(\sqrt{|s|})^{-1}(\beta^{b}(\pi^{cd}h_{cd}s^{a}_{\;b}-2\pi^{ac}h_{bc}-2p^{a}_{\;b})$. }
\beq B^{a}=F(D^{a}h-D_{b}h^{ab})-hD^{a}F+h^{ab}D_{b}F+\text{bdry term}\;.\eeq
Here $h=h_{ab}s^{ab}$. If the perturbations are taken to be solutions to the linearized Einstein's equations, then we have
\beq D_{c}B^{c}=0\;.\eeq
We study the boundary integral of this divergence:
\beq \oint_{\partial\Sigma}da_{c}B^{c}=0\;.\label{boundintpert1}\eeq

We derive the first law by evaluating the boundary terms above when $g_{ab}^{(0)}$ is a static, asymptotically Schwarzschild black hole with bifurcate Killing horizon. Consider perturbations about the metric $g_{ab}^{(0)}$. Assume that $\xi^{a}$ approaches $(\partial/\partial t)^{a}$ at infinity in the asymptotic coordinates used above. The spacelike hypersurface $\Sigma$ is taken to extend from a boundary $\partial\Sigma_{h}$ at the bifurcation sphere of the horizon to a boundary $\partial\Sigma_{\infty}$ infinity, chosen such that $n_{a}=-F\nabla_{a}t$. With these choices, the terms proportional to the vector $\beta^{a}$ in the boundary term vanish. 

Following the above outline let us write $I_{\infty}-I_{h}=0$. First consider the boundary term at infinity 
At large radius it is sufficient to consider both the background metric and the perturbations to have the Schwarzschild form. Near infinity:
\beq h_{rr}=\delta s_{rr}\approx-\frac{1}{f^{2}}\delta f=-\frac{1}{f^{2}}\frac{\delta\tilde{M}}{r^{D-3}}\;,\quad F\approx \sqrt{f}\;,\quad da_{r}\approx \frac{1}{\sqrt{f}}r^{D-2}d\Omega_{D-2}\;,\eeq
Then, 
\beq
\begin{split}
\int_{\partial\Sigma_{\infty}}da_{c}B^{c}&=\int_{\partial\Sigma_{\infty}}da_{c}s^{ab}s^{cd}F\left[(\partial_{d}h_{ab}-\partial_{b}h_{ad})+(\Gamma^{f}_{\;ab}h_{fd}-\Gamma^{f}_{ad}h_{fb})\right]\\
&+\int_{\partial\Sigma_{\infty}}da_{c}h_{ad}D_{b}F(s^{ab}s^{cd}-s^{bc}s^{ad})\;,
\end{split}
\label{intmedstep1}\eeq
where we used 
\beq F(D^{c}h-D_{b}h^{bc})=s^{ab}s^{cd}F\left[(\partial_{d}h_{ab}-\partial_{b}h_{ab})+(\Gamma^{f}_{\;ab}h_{fd}-\Gamma^{f}_{ad}h_{fb})\right]\;,\eeq
and
\beq h^{bc}D_{b}F-hD^{c}F=h_{ad}D_{b}F(s^{ab}s^{cd}-s^{bc}s^{ad})\;.\eeq
Here $\Gamma^{f}_{\;ab}$ is the Christoffel symbol associated with the metric $s_{ab}$, which has the zeroth order solution plus the perturbation. Since we work to linear order in perturbations, it is the Christoffel symbol with respect to $s_{ab}^{(0)}$, which has the asymptotic form as Schwarzschild at large $r$. It is straightforward to show the second term in (\ref{intmedstep1}) vanishes.

Therefore, we are only interested in:
\beq 
\begin{split}
\int_{\partial\Sigma_{\infty}}da_{c}s^{ab}s^{cd}F\left[(\partial_{d}h_{ab}-\partial_{b}h_{ad})+(\Gamma^{f}_{\;ab}h_{fd}-\Gamma^{f}_{ad}h_{fb})\right]&=\int_{\partial\Sigma_{\infty}}da_{r}s^{\theta_{i}\theta_{i}}s^{rr}F\Gamma^{r}_{\;\theta_{i}\theta_{i}}h_{rr}\\
&=-16\pi G\delta M\;.
\end{split}
\label{intmedstep2}\eeq
Hence, 
\beq I_{\infty}=-16\pi G\delta M\;.\label{IinftBHpert1}\eeq


Now consider the boundary term (\ref{boundintpert1}) at the horizon.  On the bifurcation surface of the horizon, $\xi^{a}$ vanishes, leaving us with
\beq I_{h}=\int_{\partial\Sigma_{h}}da_{c}(-hD^{c}F+h^{cb}D_{b}F)=-2\kappa\delta A\;.\label{IhBHpert1}\eeq
Combining  (\ref{IinftBHpert1})  and (\ref{IhBHpert1}), we find the first law of black hole mechanics.
\beq \delta M=\frac{\kappa}{8\pi G}\delta A\;.\label{firstlawBHpert1}\eeq
Making the identifications $\kappa/2\pi=T$ and $A/4G=S$, we have the first law of black hole thermodynamics.


\subsection{Extended Black Hole Thermodynamics}
\noindent

A noteworthy difference between the first law of black hole thermodynamics (\ref{firstlawBHapp}) and the ordinary first law (\ref{1stlawordapp}), is a missing pressure-volume contribution in the black hole context. This is in part because for general black hole spacetimes it is unclear what we mean by ``pressure" or ``volume". Progress can be made, however, if we embed black holes into spacetimes with a \emph{dynamical} cosmological constant $\Lambda$. Spacetimes with a dynamical cosmological constant have been studied before by Henneaux and Teitelboim \cite{Henneaux:1984ji,Teitelboim:1985dp,Henneaux:1989zc}, and continue to be studied (\emph{e.g.,} \cite{Kaloper:2013zca,Kaloper:2015jra,Svesko:2018cbo}). We then identify the pressure $p$ to be proportional to $\Lambda$; specifically, for black holes embedded in AdS, we have
\beq p=-\frac{\Lambda}{8\pi G}=-\frac{(D-2)(D-1)}{16\pi GL^{2}}\;,\eeq
where $L$ is the AdS length scale and we see $p\geq0$.  Since we now have a dynamical $\Lambda$, we can study how the Smarr formula and first law of black hole thermodynamics change. This was considered in \cite{Kastor:2009wy}, which we follow here. Below we will explore this in detail.

Let's begin by writing down the Smarr relation for an AdS-Schwarzschild black hole. Now the metric is
\beq ds^{2}=-fdt^{2}+f^{-1}dr^{2}+r^{2}d\Omega_{D-2}^{2}\;,\quad f(r)=1-\frac{\tilde{M}}{r^{D-3}}-\tilde{\Lambda}r^{2}\;,\eeq
with
\beq \tilde{M}=\frac{16\pi GM}{(D-2)\Omega_{D-2}}\;,\quad \tilde{\Lambda}=\frac{2\Lambda}{(D-1)(D-2)}\;,\label{valsMLam}\eeq
and $\Omega_{D-2}$ is the volume of a $D-2$ sphere. 
The only non-vanishing components of $\nabla^{a}\xi^{b}$ are now
\beq \nabla^{r}\xi^{t}=-\nabla^{t}\xi^{r}=\frac{(D-3)\tilde{M}}{2r^{D-2}}-\tilde{\Lambda}r\;.\eeq
The linear term in $r$, as we will see, leads to a divergent contribution to the boundary integral in the Komar relation (\ref{komar2BH})
\beq \frac{(D-2)}{8\pi G}\int_{\partial\Sigma}dS_{ab}\left(\nabla^{a}\xi^{b}+\frac{2}{D-2}\Lambda\omega^{ab}\right)=0\;,\label{komar2BHads}\eeq
with Killing potential 
\beq \xi^{b}=\nabla_{a}\omega^{ab}\;.\eeq
For the static Killing vector $(\partial/\partial t)^{a}$, $\omega^{ab}$ is not uniquely determined. We will consider the 1-parameter family of Killing potentials for $\partial/\partial t$:
\beq \omega^{rt}=-\omega^{tr}=\frac{r}{(D-1)}+\alpha r_{h}\left(\frac{r_{h}}{r}\right)^{D-2}\;.\eeq
The linear term in $r$ yields a second divergent contribution to the boundary term at infinity in the Komar relation (\ref{komar2BHads}). The arbitrary constant $\alpha$ reflects the freedom of adding a closed, but not exact, term to the Killing potential. In the case of pure AdS, the second term is not allowed because of its singularity at $r=0$. So, for later use, we take the Killing potential $\omega^{ab}_{AdS}$ for pure AdS to have non-zero components:
\beq \omega^{rt}_{AdS}=-\omega^{tr}_{AdS}=\frac{r}{D-1}\;.\eeq


The same steps which led us to the Komar relation in flat space (\ref{komar2BH}) will allow us to evaluate (\ref{komar2BHads}). Splitting this up, we have

\beq 
\begin{split}
\frac{(D-2)}{8\pi G}\int_{\partial\Sigma}da_{a}n_{b}\nabla^{a}\xi^{b}=-\frac{(D-2)}{8\pi G}\Omega_{D-2}\left(\frac{(D-3)}{2}\tilde{M}-\tilde{\Lambda}r^{D-1}\right)
\end{split}
\eeq
and
\beq 
\begin{split}
\frac{(D-2)}{8\pi G}\int_{\partial\Sigma}da_{a}n_{b}\frac{2\Lambda}{(D-2)}\omega^{ab}&=-\frac{(D-2)}{8\pi G}\Omega_{D-2}\left(\frac{2\Lambda}{(D-1)(D-2)}r^{D-1}+\frac{2\Lambda}{(D-2)}\alpha r_{h}^{D-1}\right)\;.
\end{split}
\eeq

So, (\ref{komar2BHads}) becomes 
\beq 
\begin{split}
0&=-\frac{(D-2)}{8\pi G}\Omega_{D-2}\left(\frac{(D-3)}{2}\tilde{M}+\left[\frac{2\Lambda}{(D-1)(D-2)}-\tilde{\Lambda}\right]r^{D-1}+\frac{2\Lambda\alpha}{(D-2)} r_{h}^{D-1}\right)\;.
\end{split}
\eeq
Plugging in our expressions for $\tilde{M}$ and $\tilde{\Lambda}$ (\ref{valsMLam}), we see that the term which would be divergent as $r\to\infty$ limit vanishes. We are left with
\beq 
\begin{split}
0=\frac{(D-2)}{8\pi G}\int_{\partial\Sigma}da_{a}n_{b}\left(\nabla^{a}\xi^{b}+\frac{2}{D-2}\Lambda\omega^{ab}\right)&=-(D-3)M-\frac{2\Lambda\alpha}{8\pi G}\Omega_{D-2}r_{h}^{D-1}\biggr|^{r\to\infty}_{r\to r_{h}}\;.
\end{split}
\eeq
At $r\to\infty$
\beq I_{\infty}=-(D-3)M-\frac{2\Lambda\alpha}{8\pi G}\Omega_{D-2}r_{h}^{D-1}\;.\eeq
Meanwhile, at the horizon, we still have (\ref{IhBHapp}), but
also
\beq \frac{(D-2)}{8\pi G}\int_{\partial\Sigma_{h}}da_{a}n_{b}\frac{2\Lambda}{(D-2)}\omega^{ab}=-\frac{(D-2)}{8\pi G}\Omega_{D-2}\left(\frac{2\Lambda}{(D-1)(D-2)}r_{h}^{D-1}+\frac{2\Lambda\alpha}{(D-2)}r_{h}^{D-1}\right)\;.\eeq
So, at the horizon, 
\beq I_{h}=-\frac{(D-2)}{8\pi G}\kappa A-\frac{2\Lambda}{8\pi G}\Omega_{D-2}\left(\frac{r_{h}^{D-1}}{(D-1)}+\alpha r_{h}^{D-1}\right)\;.\eeq
Altogether, 
\beq 
\begin{split}
0&=I_{\infty}-I_{h}=-(D-3)M+\frac{(D-2)}{8\pi G}\kappa A+\frac{2\Lambda}{8\pi G}\frac{\Omega_{D-2}r_{h}^{D-1}}{(D-1)}\;,
\end{split}
\eeq
such that we attain the Smarr formula for Schwarzschild-AdS black holes \cite{Kastor:2009wy}
\beq (D-3)M=\frac{(D-2)}{8\pi G}\kappa A+\frac{2\Lambda}{8\pi G}V\;,\label{smarrext1}\eeq
where
 \beq V\equiv \frac{\Omega_{D-2}r_{h}^{D-1}}{(D-1)}\;.\label{Thermovol1}\eeq. 
Identifying the pressure $p=-\Lambda/8\pi G$, we attain
\beq (D-3)M=(D-2)TS-2PV\;.\label{smarrext2}\eeq
We will provide a physical interpretation of $V$ momentarily, but as one might guess, it is known as the \emph{thermodynamic volume}.

As in the Schwarzschild case, we need only the asymptotic conditions on the metric. The fall conditions (\ref{falloffs}) are now
\beq ds^{2}\approx g_{tt}dt^{2}+g_{rr}dr^{2}+Hr^{2}d\Omega_{D-2}^{2}\;,\eeq
with asymptotic metric functions 
\beq g_{tt}=-f_{0}+\frac{c_{t}}{r^{D-3}}\;,\quad g_{rr}=\frac{1}{f_{0}}\left(1-\frac{c_{r}}{\tilde{\Lambda}r^{D-1}}\right)\;,\quad H=1+\tilde{\Lambda}\frac{c_{\theta}}{r^{D-1}}\;,\quad f_{0}=1-\tilde{\Lambda}r^{2}\;.\eeq
For the inverse metric, we have
\beq g^{tt}=f_{0}^{-1}\left(-1+\frac{c_{t}}{\tilde{\Lambda} r^{D-1}}\right)\;,\quad g^{rr}=f_{0}-\frac{c_{r}}{r^{D-3}}\;.\eeq
Here we can take constants $c_{t}=c_{r}=\tilde{M}$ and $c_{\theta}=0$, and
\beq dar_{b}n_{c}(\nabla^{b}\xi^{c})\approx d\Omega_{d-2}\left(\tilde{\Lambda}r^{D-1}-\frac{(D-3)}{2}\tilde{M}\right)\;.\eeq

We now want to process the Killing potential term at infinity such that we can leave the form of the Killing potential general, but still provide the cancellation of the divergence at $r\to\infty$. We do this by both adding and subtracting the divergent term $\omega^{ab}_{AdS}$ to and from the Killing potential. So, 
\beq dar_{b}n_{c}\left(\frac{2\Lambda}{D-2}\omega^{bc}\right)\approx -d\Omega_{D-2}(\tilde{\Lambda}r^{D-1})+dar_{b}n_{c}\left(\frac{2\Lambda}{D-2}[\omega^{bc}-\omega^{bc}_{AdS}]\right)\;.\eeq
Therefore, for the boundary integral at the horizon, one has
\beq  I_{h}=-(D-2)\frac{\kappa A}{8\pi G}+\int_{\partial\Sigma_{h}}dS_{ab}\omega^{ab}\;.\eeq
Combining the boundary integrals according to the Komar relation, then we find the Smarr formula as before, but this time with a more general expression for the thermodynamic volume $V$:
\beq V=-\left[\int_{\partial\Sigma_{\infty}}dS_{ab}(\omega^{ab}-\omega^{ab}_{AdS})-\int_{\partial\Sigma_{h}}dS_{ab}\omega^{ab}\right]\;.\eeq

Let's now derive the first law of black hole mechanics with the inclusion of variations for $\Lambda$. Our strategy is as before, where we use Hamiltonian perturbation theory. 
For a cosmological constant stress energy, the constraint equations (\ref{constrqnspure}) now become
\beq H=-2\Lambda\;,\quad H_{a}=0\;.\eeq
The Hamiltonian density for evolution along $\xi^{a}$ in Einstein gravity with cosmological constant $\Lambda$ is given by 
\beq \mathcal{H}=\sqrt{s}[F(H+2\Lambda)+\beta^{a}H_{a}]\;.\eeq
And now let $s^{(0)}_{ab}$ and $\pi^{ab}_{(0)}$ be a solution to the Einstein equations with cosmological constant $\Lambda_{(0)}$ and with a Killing vector $\xi^{a}$. For perturbations (\ref{perts1}), it follows from Hamilton's equations for the zeroth order spacetime that the linearized constraint operators $\delta H$ and $\delta H_{a}$ combine to form a total derivative satisfying
\beq D_{c}B^{c}=2F\delta\Lambda\;.\eeq
We may rewrite the cosmological term as a total derivative making use of the Killing potential $F=-n_{a}\xi^{a}=-D_{c}(n_{a}\omega^{ca})$, such that,
\beq \int_{\partial\Sigma}da_{c}(B^{c}+2\omega^{cd}n_{d}\delta\Lambda)=0\;.\label{komarintadsbh}\eeq


Now we evaluate (\ref{komarintadsbh}) at infinity and at the horizon. We begin with the contribution at infinity, $I_{\infty}$. At large radius it is sufficient to consider both the background metric and the perturbations to have the Schwarzschild-AdS form. Near infinity we have
\beq h_{rr}=\delta s_{rr}\approx-\frac{1}{f^{2}}\delta f=-\frac{1}{f^{2}}\left[\frac{\delta\tilde{M}}{r^{D-3}}+\delta\tilde{\Lambda}r^{2}\right]\;,\quad F\approx \sqrt{f}\;,\quad da_{r}\approx \frac{1}{\sqrt{f}}r^{D-2}d\Omega_{D-2}\;.\eeq
Then, as in the Schwarzschild case (\ref{intmedstep1}), 
\beq
\begin{split}
\int_{\partial\Sigma_{\infty}}da_{c}B^{c}&=\int_{\partial\Sigma_{\infty}}da_{r}s^{\theta_{i}\theta_{i}}s^{rr}F\Gamma^{r}_{\;\theta_{i}\theta_{i}}h_{rr}\\
&=-16\pi G\delta M-\lim_{r\to\infty}\left(\frac{2r^{D-1}\Omega_{D-2}}{(D-1)}\right)\delta\Lambda
\end{split}
\eeq

Evaluating the second term at the boundary integral (\ref{komarintadsbh}) at infinity using the asymptotic form of $\omega^{ab}$ leads to 
\beq 2\int_{\partial\Sigma_{\infty}}da_{c}\omega^{cd}n_{d}\delta\Lambda=\lim_{r\to\infty}\left(\frac{2r^{D-1}\Omega_{D-2}}{D-1}\right)\delta\Lambda+2\left(\int_{\partial\Sigma_{\infty}}da_{c}(\omega^{cd}-\omega^{cd}_{AdS})n_{d}\right)\delta\Lambda\;.\eeq
We are then left with
\beq I_{\infty}=16\pi G\delta M-2\left(\int_{\partial\Sigma_{\infty}}da_{c}(\omega^{cd}-\omega^{cd}_{AdS})n_{d}\right)\delta\Lambda\;.\label{Inftadspert}\eeq

Now consider (\ref{komarintadsbh}) at the horizon. Evaluation of the first term is still (\ref{IhBHpert1}), such that altogether 
\beq I_{h}=-2\kappa\delta A+2\left(\int_{\partial\Sigma_{h}}da_{c}\omega^{cd}n_{d}\right)\delta\Lambda\;.\label{Ihpertads1}\eeq
Plugging in (\ref{Inftadspert}) and  (\ref{Ihpertads1}) into (\ref{komarintadsbh}), we arrive to the first law of black hole mechanics with varying cosmological constant \cite{Kastor:2009wy}
\beq \delta M=\frac{\kappa}{8\pi G}\delta A+\frac{V}{8\pi G}\delta\Lambda\;.\label{firstlawextpert}\eeq
Or, in terms of thermodynamic variables, 
\beq \delta M=T\delta S+V\delta p\;,\label{firstlawextthermoapp}\eeq
we have the first law of extended black hole thermodynamics. 

The quantity $V$ is called the thermodynamic volume because, as observed above, it is simply the volume of a co-dimension-2 sphere in $D$ spacetime dimensions. Also notice by direct computation and the form of the first law that $V$ is given as the pressure derivative of $M$:
\beq V\equiv\left(\frac{\partial M}{\partial p}\right)_{S}\;.\label{thermovolgenapp}\eeq

The inclusion of $p-V$ together with the first law (\ref{firstlawextthermoapp}) motivates us to reinterpret $M$ as the gravitational version of chemical enthalpy, \emph{i.e.,} the total energy of a system including both its internal energy $E$, and the energy $pV$ required to displace the vacuum energy of the environment: $M=E+pV$. Another way of putting it, $M$ is the enegy required to create a black hole and place it in an AdS environment. Because we are now dealing with enthalpies, often the subject of extended black hole thermodynamics is referred to as \emph{black hole chemistry}. 

The Smarr relation (\ref{smarrext2}) actually arises from an application of Euler's formula for the homogenous function $M=M(A,\Lambda)$, plus the scaling relation $M(A,\Lambda)\to M(\gamma^{D-2}A,\gamma^{-2}\Lambda)=\gamma^{D-3}M(A,\Lambda)$. Upon taking the derivative with respect to $\gamma$ we arrive to the Smarr relation (\ref{smarrext2}) \cite{Kastor:2009wy}. To emphasize, in the event the mass $M$ depends on $\Lambda$, in order to have a well-defined Smarr relation we \emph{must} include $p-V$. 

Practically speaking, this is the recipe for extended thermodynamics: Impose the Smarr relation, assuming the scaling dimensions of $M$, $A$, and any other parameter, such as charge $Q$ and rotation $J$ be that for ordinary black hole thermodynamics (without $\Lambda$). This will fix the scaling dimension for $\Lambda$. By demanding that the first law be of the form (\ref{firstlawextthermoapp}) --  with additional fixed parameters like $Q$, $J$, etc. -- we then fix $V$ to be (\ref{thermovolgenapp}) formally. Note that this has consequences for non-static black holes in $D\geq4$ dimensions. Namely, the volume $V$ is generally not the naive geometric volume of the black hole horizon. Rather, it is a thermodynamic variable in its own right, and makes the interpretation of $V$ a bit mysterious (see \cite{Kubiznak:2016qmn} for a longer discussion on the interpretation of $V$, as well as many other aspects of extended thermodynamics).



\subsection{The Einstein Equation of State: A Review}
\noindent

Here we present a review of Jacobson's derivation of the Einstein equation of state \cite{Jacobson:1995ab}, as this thesis is heavily motivated by the original work. The set-up is as follows: Pick an arbitrary point $p$ in an arbitrary $D$-dimensional spacetime $\mathcal{M}$ with arbitrary metric $g_{ab}$. We will restrict ourselves to a small enough region such that we can define a spacelike foliation with respect to a time coordinate labeled $t$. Let $p$ be located on a spacelike codimension-1 hypersurface $\Sigma_{1}$ at some time $t_{1}$. We then consider a codimenson-2 (nearly) flat spacelike surface $\mathcal{P}_{1}$ containing our point $p$. By nearly flat we just mean that the null congruences emanating from and normal to $\mathcal{P}_{1}$ have initial vanishing expansion $\theta$ and shear $\sigma_{ab}$ at $p$ to first order in a distance from $p$ (by distance, we mean up to leading order in a Riemann normal coordinate expansion, \emph{i.e.}, $g_{ab}\approx \eta_{ab}+\frac{1}{3}R_{abcd}(p)x^{b}x^{d}+...$). Let $A_{1}$ be the area of $\mathcal{P}_{1}$, such that the expansion is $\theta=\frac{1}{A_{1}}\frac{dA_{1}}{d\lambda}$. 

Now we fix a closed orientable smooth spacelike codimension-2 surface $\mathcal{B}_{1}$ containing $\mathcal{P}_{1}$ and choose a future-directed inward null direction normal to $\mathcal{B}_{1}$, defining a null congruence emanating from $\mathcal{B}_{1}$.  The affine parameter along the congruence is denoted $\lambda$, and the null congruence has a tangent vector $k^{a}=(\frac{d}{d\lambda})^{a}$. The expansion is then $\theta=\nabla_{a}k^{a}$. At $p$ we set $\lambda=0$, and increase toward the future. The points of the congruence generate a \emph{lightsheet} $\mathcal{H}$ emanating from $\mathcal{P}_{1}$.  A spacelike region of $\Sigma_{1}$ that lies inside $\mathcal{B}_{1}$ is labeled $R_{1}$. 

\begin{figure}[t]
\centering
 \includegraphics[width=9.3cm]{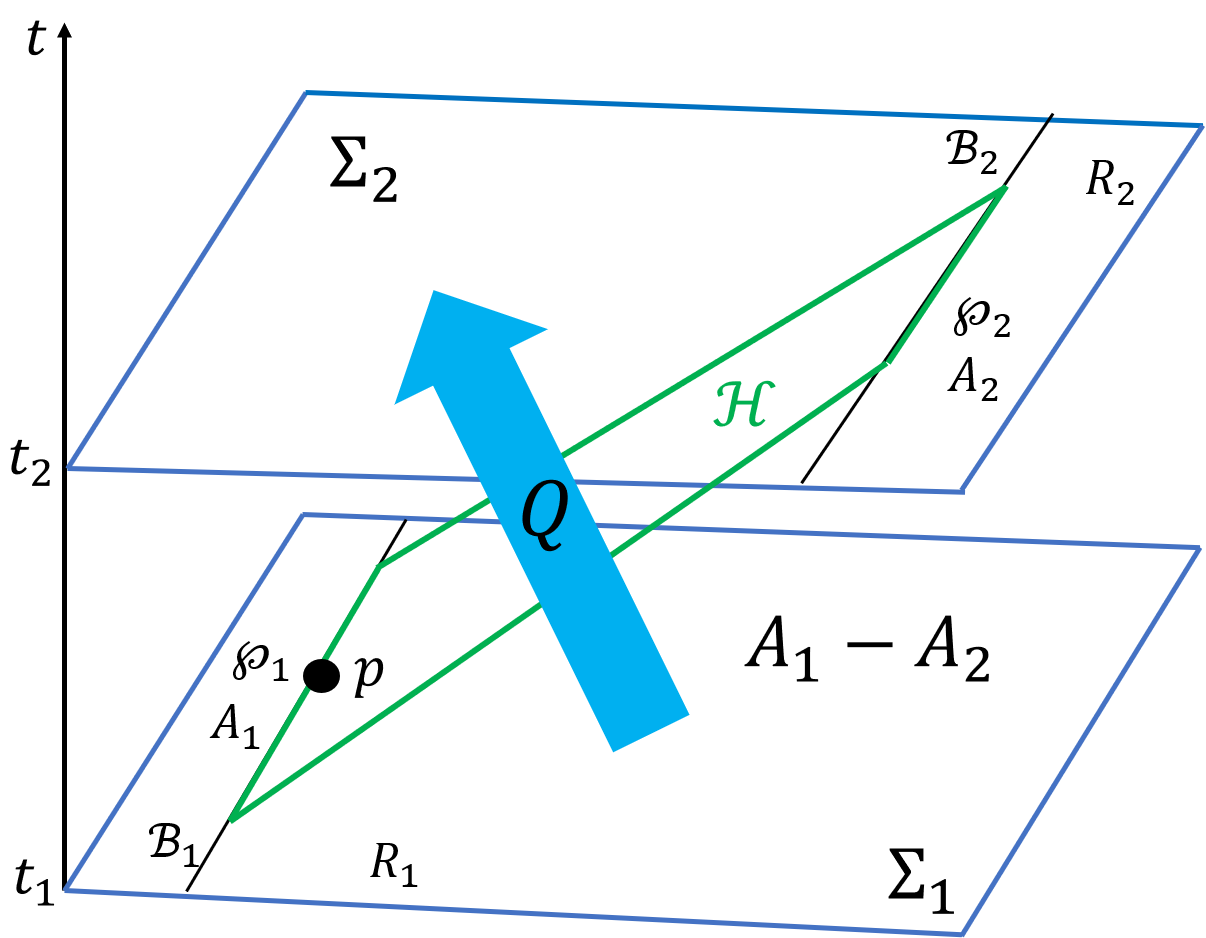}
 \caption[Construction of Local Rindler Horizon.]{Construction of local Rindler horizon $\mathcal{H}$ about an arbitrary point $p$ \cite{Jacobson:1995ab}, following the conventions of \cite{Carroll16-1}. A heat flux exits $\mathcal{P}_{1}$ through the horizon $\mathcal{H}$ generated by a local boost Killing vector, resulting in an area deficit $\Delta\mathcal{A}=A_{1}-A_{2}$. Via the assumptions of local holographic thermodynamics, one has $Q=T\Delta S\Rightarrow G_{ab}(p)+\Lambda g_{ab}(p)=8\pi G T_{ab}(p)$.
}
 \label{localrindsetup}\end{figure}

We follow the flow of the congruence along $\lambda$ to some later time $t_{2}$ (but not much longer), where the null congruence intersects with a spacelike hypersurface $\Sigma_{2}$, defining a codimension-2 surface $\mathcal{B}_{2}$, such that $\mathcal{P}_{1}$ has evolved to $\mathcal{P}_{2}$ contained in $\mathcal{B}_{2}$. Let $A_{2}$ be the area of $\mathcal{P}_{2}$ and $R_{2}$ a spacelike region inside of $\mathcal{B}_{2}$. By the time we evolve to $\Sigma$, the expansion of the lightsheet is $\theta=\frac{1}{A_{2}}\frac{dA_{2}}{d\lambda}$. See Figure \ref{localrindsetup} for a pictorial representation of this set-up.

Since our spacetime locally appears flat, due to the Riemann normal coordinate expansion, we retain local isometries of flat space, including the Lorentz boosts. Of course, Lorentz boosts can be seen as Rindler time translations for a locally accelerating observer. The local Rindler observer will have a local Rindler horizon, which we identify as the lightsheet $\mathcal{H}$. Since the Killing vector is only Killing within $\mathcal{O}(x^{2})$ of the Riemann normal coordinate expansion, the boost Killing vector 
\beq \xi^{b}=a\lambda k^{b}\;,\eeq
is said to be approximately Killing. Here $a$ is the proper acceleration of the associated Rindler worldline. Our local Rindler observers will measure a constant, uniform Unruh-Davies temperature proportional to their acceleration $a$:
\beq T=\frac{\hbar a}{2\pi}\;.\eeq
As in the case of global Rindler spacetime, our local Rindler horizon is a constant temperature system, and is thus in thermal equilibrium.

We now imagine some matter accompanied with energy-momentum tensor $T_{ab}(p)$ leaving region $R_{1}$ through the the lightsheet. The resulting energy-flux through the local horizon, measured with respect to the local Rindler observer is 
\beq Q\equiv\int_{\mathcal{H}} d\Sigma^{a}\xi^{b}T_{ab}(p)=a\int_{\mathcal{H}}d\lambda d\mathcal{A}\lambda k^{a}k^{b}T_{ab}(p)\;,\label{intengfluxapp}\eeq
where surface area element for the local Rindler horizon is $d\Sigma^{a}=k^{a}d\lambda d\mathcal{A}$, with $d\mathcal{A}$ as the codimension-2 spacelike cross-sectional area element. In ordinary thermodynamic systems, \emph{heat} $Q$ is interpreted as the energy-flux that flows into macroscopically unobservable degrees of freedom. Since the Rindler observers are out of causal contact with the region behind the lightsheet, the integrated energy-flux (\ref{intengfluxapp}) is thus interpreted as heat. 

We are therefore considering a thermodynamic process where an amount of matter is exiting region $R_{1}$ at $t_{1}$, going into some non-accessible region from the viewpoint of the local Rindler observers, such that the energy associated with the matter is heat $Q$. Consequently, there should be some change in entropy of the system, if we continue to interpret our local geometric set-up as an ordinary thermal system. The question now is what is the entropy given by. Motivated by black hole physics, the only geometric quantity which can really change under this physical process is the area of the lightsheet, which will in fact \emph{decrease} as the matter passes through. We denote this change in area by 
\beq \Delta \mathcal{A}\equiv A_{1}-A_{2}=-\int_\mathcal{H}\theta d\lambda d\mathcal{A}\;.\label{areadefapp}\eeq
Here $A_{i}=\int_{\mathcal{P}_{i}}d\mathcal{A}$. To attain (\ref{areadefapp}) we used Stokes' theorem. 

Due to the heat flux, we are in effect studying the evolution of a null geodesic congruence. Such evolution is described by Raychaudhuri's equation:
\beq \frac{d\theta}{d\lambda}=-\frac{1}{(D-2)}\theta^{2}-\sigma_{ab}^{2}+\omega_{ab}^{2}-R_{ab}(p)k^{a}k^{b}\;.\eeq
In our set-up, we are considering geodesics which are hypersurfce orthogonal, and therefore by Frobenius' theorem, the twist $\omega_{ab}=0$. Moreover, we have chosen the local Rindler horizon to be instantaneously stationary at $\mathcal{P}_{1}$, such that to leading order $\theta$ and $\sigma$ vanish, such that, approximately $\frac{d\theta}{d\lambda}=-R_{ab}(p)k^{a}k^{b}$, so the expansion is just
\beq \theta=-\lambda R_{ab}(p)k^{a}k^{b}\;.\eeq
Thus, the area deficit becomes
\beq \Delta\mathcal{A}=\int_{\mathcal{H}}R_{ab}(p)k^{a}k^{b}\lambda d\lambda d\mathcal{A}\;.\label{areadef2app}\eeq

 Everything we have said thus far is purely geometric, aside from our occasional motivations from black hole thermodynamics. Now we input the two critical assumptions of spacetime thermodynamics: (i) \emph{Local holography}: For the constructed lightsheet $\mathcal{H}$, we assume the entropy change $\Delta S$ of our thermal system is proportional to the change in the area $\Delta\mathcal{A}$, up to a universal constant $\eta$:
\beq \Delta S=\eta\Delta\mathcal{A}\;.\label{entchangapptg}\eeq
This assumption is also well-motivated by black hole thermodynamics, namely the Bekenstein-Hawking area relation, for which we would write $\eta=\frac{1}{4 G\hbar}$. 

(ii) \emph{Clausius relation}: We assume that there is an entropy change associated with the flow of heat $Q$ through the lightsheet, which in local thermodynamic equilibrium is given by 
\beq Q=T\Delta S\;.\label{claurel}\eeq
We should be careful in using the Clausius relation. More precisely, the Clausius relation is really an inequality $Q\leq T\Delta S$, including both \emph{reversible} and \emph{irreversible} entropy changes to entropy. We attain equality when there are no irreversible changes to the entropy. Thus, if we want equality, the thermodynamic process we consider should be a reversible one. So is the process we have described above a reversible one? It turns out it is. This is because if we did not deposit any matter into the system, there would be no area deficit, and thus no entropy change. Moreover, when we consider the heat exchange, we imagine it is done slowly enough such that the exchange is totally reversible (such as slowly heating up a box). Therefore, the only entropy change is due to a reversible thermodynamic process\footnote{There is another way of seeing we are dealing with a reversible thermodynamic process, though it changes the method of the derivation. One instead uses a Noetheresque approach, where Killing's equation and Killing's identity are required. For an approximate Killing vector, neither of these geometric relations are satisfied globally, but will hold true to some order in the Riemann normal coordinate expansion. It was shown in \cite{Parikh:2017aas} that irreversible processes correspond to the failure of Killings identity at order $\mathcal{O}(x^{-1})$, which only occurs for approximate `Killing' vectors that are not Killing in flat space, \emph{e.g.}, radial boosts.  In the current set-up, however, the (Cartesian) boost Killing vector is a Killing in pure flat space, and does not have any contributions to Killing's identity at order $\mathcal{O}(x^{-1})$. Therefore, no irreversible contributions appear, illustrating the process is purely reversible in the thermodynamic sense.}.

Putting everything together with (\ref{areadef2app}), (\ref{intengfluxapp}) and our assumptions (\ref{entchangapptg}) and (\ref{claurel}), respectively, we have
\beq \frac{\eta \hbar a}{2\pi}\int_{\mathcal{H}}d\lambda d\mathcal{A}\lambda R_{ab}(p)k^{a}k^{b}=a\int_{\mathcal{H}}d\lambda d\mathcal{A}T_{ab}(p)k^{a}k^{b}\;.\eeq
We now invoke the freedom we had to choose $k^{a}$, allowing us to set the integrands together:
\beq \frac{\hbar \eta}{2\pi}R_{ab}(p)k^{a}k^{b}=T_{ab}(p)k^{a}k^{b}\;.\eeq
This holds for all null vectors $k^{a}$, allowing us to drop the vectors at a cost of introducing some unknown scalar function $f$:
\beq R_{ab}(p)+fg_{ab}(p)=\frac{2\pi}{\hbar \eta}T_{ab}(p)\;.\eeq
Now we imposing that the energy-momentum tensor be conserved, $\nabla^{a}T_{ab}=0$. Using the Bianchi identity, $2\nabla^{a}R_{ab}=\nabla_{b}R$, we identify $f$ to be $f=-\frac{1}{2}R+\Lambda$, for some constant $\Lambda$, leading us to 
\beq R_{ab}(p)-\frac{1}{2}R(p)g_{ab}(p)+\Lambda g_{ab}(p)=8\pi GT_{ab}(p)\;.\eeq
In the last step we made the identification $\eta=1/4\pi \hbar G$, consistent with the Bekenstein-Hawking formula. We emphasize that we have arrived to the non-linear Einstein's holding about point $p$. However, $p$ is completely arbitrary, and so we have that about any point our construction holds, \emph{i.e.,} as long as we are not at any caustic points or singularities, we have that Einstein's equations will held throughout the entire spacetime. 

In summary, we have shown that the local Einstein's equations are a  geometric consequence of applying thermodynamic principles to local horizons in any spacetime. This shows that, just as with the hydrodynamic limit of water, classical spacetime dynamics arises from some more fundamental microscopic theory of spacetime.


\subsection{Horizon Entropy, Noether Charge, and Beyond}
\noindent

\subsection*{Horizon Entropy as Conserved Charge}
\noindent

The Bekenstein-Hawking area formula
\beq S_{\text{BH}}=\frac{A}{4G}\;,\label{Bekhawkapp}\eeq
gives the horizon entropy for spacetimes whose dynamics are controlled by Einstein's general relativity. However, string theory, and, more generally, quantum field theory in curved space, suggests that there are other more general theories of gravity with actions of the form
\beq I=\int d^{D}x\sqrt{-g}L(g^{ab},R^{abcd},\nabla_{k}R^{abcd},...)\;.\label{arbaction}\eeq
Wald \cite{Wald:1993nt} showed that the horizon entropy for systems whose dynamics are controlled by general theories of gravity will have an entropy different from the Bekenstein-Hawking entropy (\ref{Bekhawkapp}). The resulting entropy is known as the Wald entropy functional, given by the Noether charge associated with the diffeomorphism invariance of the theory\footnote{We point out the Wald entropy is \emph{classical} in nature; we have not included any quantum corrections. There are a plethora of ways to compute quantum corrections, (see \emph{e.g.} \cite{Banerjee:2010qc,Denef:2009kn,David:2009xg,Keeler:2018lza,Keeler:2019wsx,Martin:2019flv,Martin:2020api} and references therein), and the Wald entropy functional can be used to compute semi-classical contributions to entropy (as the gravity theory is still classical in that context).}.

Here we will take some time to show this result, and present a few examples of the Wald formalism for higher derivative theories of gravity. Rather than using Wald's original notation, we will instead make use of the notation used in, \emph{e.g.}, \cite{Padmanabhan07-1,Vollick07-1}. We start by showing that the Bianchi identity is a consequence of general covariance. Take the action (\ref{arbaction}) and compute the local variation of the Lagrangian with repect to the metric $g_{ab}$ under the diffeomorphism $x^{a}\to x^{a}+\xi^{a}(x)$, such that 
\beq \delta_{\xi}g^{ab}=g^{'ab}(x)-g^{ab}(x)=\nabla^{a}\xi^{b}+\nabla^{b}\xi^{a}=\mathcal{L}_{\xi} g^{ab}\;.\eeq
Then, 
\beq
\begin{split}
 \int d^{D}x\delta_{\xi}(\sqrt{-g}L)&=\int d^{D}x\sqrt{-g}[E_{ab}(\nabla^{a}\xi^{b}+\nabla^{b}\xi^{a})+\nabla_{a}(\delta_{\xi}v^{a})]\\
&=\int d^{D}x\sqrt{-g}[2\nabla_{a}(E^{ab}\xi_{b})-2\nabla_{a}E^{ab}\xi_{b}+\nabla_{a}(\delta_{\xi} v^{a})]\;.
\end{split}
\label{deltaxiL}\eeq
For comparison, if we were to simply compute the variation of the action with respect to $g_{ab}$, then $E_{ab}$ is the tensor contracted with $\delta g^{ab}$, such that we recognize $2E_{ab}=T_{ab}$ as the equations of motion for the general theory, while we also attain $\nabla_{a}\delta v^{a}$ which leads to some surface term. What we have done above is not computing the field equations -- we merely noted the variation with respect to $\xi$ gives  the higher derivative generalization of the Einstein tensor, $E_{ab}$. 

Let's now rewrite the left hand side of (\ref{deltaxiL}) using
\beq \delta_{\xi}(\sqrt{-g}L)=-\sqrt{-g}\nabla_{a}(L\xi^{a})\;,\eeq
we find (\ref{deltaxiL}) becomes
\beq
\begin{split}
 \int d^{D}x\sqrt{-g}2(\nabla_{a}E^{ab})\xi_{b}&=\int d^{D}x\sqrt{-g}\nabla_{a}(2E^{ab}\xi_{b}+L\xi^{a}+\delta_{\xi}v^{a})\\
&=\int d^{D-1}\sigma_{a}\sqrt{-g}(2E^{ab}\xi_{b}+L\xi^{a}+\delta_{\xi}v^{a})\;,
\end{split}
\eeq
where we used Gauss' law to turn the second line into an integral over the boundary. We are imposing diffeomorphism invariance, such that the variation of the metric together with its derivative vanish on the boundary, \emph{i.e.}, the right hand side will vanish. Since $\xi^{a}$ is arbitrary and the volume of spacetime over which the integration is being performed is arbitrary, the integrand of the left hand side must also vanish:
\beq \nabla_{a}E^{ab}=0\;.\label{genbianchi}\eeq
This is just the Bianchi identity for more general theories of gravity, and is a direct consequence of general covariance. Importantly, note that the Bianchi identity is an off-shell geometric identity -- we never had to use the equations of motion, only the form of the variation of the action. Recall that for the case of Einstein gravity, where $E_{ab}=G_{ab}$, the Bianchi identity $\nabla_{a}G^{ab}=0$ is really just constraining a tensor with the same algebraic properties of the Riemann curvature tensor; in this case, $R^{a}_{\;b[cd;k]}=0$.

Let's now demonstrate that general covariance and an application of the Bianchi identity (\ref{genbianchi}) leads to a conserved currend $J^{a}$. We can write down $J^{a}$ explicitly by writing the local variation of $\sqrt{-g}L$ under $x^{a}\to x^{a}+\xi^{a}(x)$ in two different ways. Namely, 
\beq \delta_{\xi}(\sqrt{-g}L)=-\sqrt{-g}\nabla_{a}(L\xi^{a})\;,\label{varLagden1}\eeq
and
\beq \delta_{\xi}(\sqrt{-g}L)=\sqrt{-g}\nabla_{a}(2E^{ab}\xi_{b}+\delta_{\xi}v^{a})\;,\label{varLagden2}\eeq
where in the second expression we used the generalized Bianchi identity $\nabla_{a}E^{ab}=0$. Equating (\ref{varLagden1}) and (\ref{varLagden2}), we find
\beq \nabla_{a}(2E^{ab}\xi_{b}+L\xi^{a}+\delta_{\xi}v^{a})=0\;.\eeq
Therefore, we introduce the conserved current
\beq J^{a}=2E^{ab}\xi_{b}+L\xi^{a}+\delta_{\xi}v^{a}\;,\quad \nabla_{a}J^{a}=0\;.\label{conservcurrent}\eeq
We emphasize that the continuity equation $\nabla_{a}J^{a}$ and the Bianchi identity $\nabla_{a}E^{ab}=0$ are both off-shell relations. In the event the equations of motion are satisfied, the off-shell conserved current (\ref{conservcurrent}) is equivalent to the on-shell Noether current.

With a conserved current comes a conserved charge $Q$, coming from integrating $\nabla_{a}J^{a}$ over a proper volume integral $d^{D}x\sqrt{-g}$:
\beq \int_{V}d^{D}x\sqrt{-g}\nabla_{a}J^{a}=\int d^{D}x\partial_{a}(\sqrt{-g}J^{a})=\int_{\partial V}d\sigma_{a}J^{a}\;,\eeq
where we used $\nabla_{a}A^{a}=\frac{1}{\sqrt{-g}}\partial_{a}(\sqrt{-g}A^{a})$, and applying Gauss' theorem, where $d\sigma_{a}$ is a $(D-1)$-dimensional volume element of the boundary $\partial V$. We define our conserved charge to be
\beq Q_{\xi}\equiv\int_{\partial V}d\sigma_{a}J^{a}=\int_{\partial V}d\sigma_{a}\nabla_{b}J^{ab}\;\eeq

Here we introduced an antisymmetric conserved potential $J^{ab}$ via $J^{a}=\nabla_{b}J^{ab}$. The potential $J^{ab}$ is antisymmetric because it must satisfy $\nabla_{a}\nabla_{b}J^{ab}=0$ via the continuity equation. Applying  Stokes' theorem for any antisymmetric second rank tensor, we arrive to
\beq Q_{\xi}=\frac{1}{2}\int_{\Sigma}dS_{ab}J^{ab}\;.\label{conservedcharge}\eeq
Here $dS_{ab}=(n_{a}u_{b}-n_{b}u_{a})dA$ being the binormal surface area element on the codimenion-2 surface $\Sigma$, with $n_{a}$ ($u_{a}$) being a unit spacelike (timelike) normal vector to the surface. Sometimes $dS_{ab}=dA\epsilon_{ab}$, such that $\epsilon_{ab}\epsilon^{ab}=-2$. We emphasize that $Q$ is evaluated at some constant time $t$-slice of the manifold. In the case of black holes, the surface $\Sigma$ is taken to be the black hole horizon. The factor of $\frac{1}{2}$ is conventional and can be absorbed into the definition of the binormal surface area element\footnote{We should also point out that the final equality here is true up to a sign, depending on which direction we have our timelike normal vector facing; here we take $u$ to be pointing outward, giving us a positive sign.}.

Let's now write things more explicitly by considering the action
\beq I=\int d^{D}x\sqrt{-g}L(g^{ab},R^{a}_{\;bcd})\;.\label{action2app}\eeq
A theory with action (\ref{action2app}) includes, for example, $f(R)$ models, Lovelock gravity, and arbitrary curvature squared theories of gravity.

The variation of $I$ is
\beq \delta I=\int d^{D}x\sqrt{-g}\left[\left(\frac{\partial L}{\partial g^{ab}}-\frac{1}{2}g_{ab}L\right)\delta g^{ab}+P_{a}^{\;bcd}\delta R^{a}_{\;bcd}\right]\;,\eeq
where we have defined the \emph{Wald tensor}\footnote{If we consider theories that include higher derivatives of the Riemann tensor, $P^{abcd}$ becomes
$$P^{abcd}\equiv\frac{\partial L}{\partial R_{abcd}}-\nabla_{a_{1}}\frac{\partial L}{\partial\nabla_{a_{1}}R_{abcd}}+...+(-1)^{m}\nabla_{(a_{1}...}\nabla_{a_{m})}\frac{\partial L}{\partial\nabla_{(a_{1}...}\nabla_{a_{m})}R_{abcd}}$$.}
\beq P_{a}^{\;bcd}\equiv\frac{\partial L}{\partial R^{a}_{\;bcd}}\;.\label{waldtensorapp}\eeq
 We will work out the Wald tensor for a number of models momentarily. Importantly, note $P^{abcd}$ has the same algebraic symmetries as the Riemann tensor. 

Continuing with the variation $\delta I$ following the usual procedure, we find
\beq 
\begin{split}
\delta I&=\int d^{D}x\sqrt{-g}\left[E_{ab}\delta g^{ab}+\nabla_{a}\delta v^{a}\right]
\end{split}
\eeq
where
\beq E_{ab}=P_{b}^{\;kij}R_{akij}-\frac{1}{2}g_{ab}L-2\nabla^{m}\nabla^{n}P_{amnb}\eeq
and
\beq \delta v^{a}=[2P^{\ell bad}\nabla_{b}-2(\nabla_{b}P^{\ell abd})]\delta g_{d\ell}\;.\eeq
This procedure can be straightforwardly generalized to include the more general theories given by the action (\ref{arbaction}), however, we won't go through the details here. 

We may also write down an explicit form for the conserved current $J^{a}$ and potential $J^{ab}$ in this theory. Note that 
\beq 2E^{ak}\xi_{k}=2P^{adb\ell}R^{k}_{\;db\ell}\xi_{k}-L\xi^{a}-4\nabla_{d}\nabla_{b}P^{adbk}\xi_{k}\;,\label{eqnpartofJ}\eeq
and the boundary contribution can be cast as
\beq \delta_{\xi} v^{a}=2P_{\ell}^{\;bad}\delta_{\xi}\Gamma^{\ell}_{\;bd}-2\nabla_{b}P^{\ell bad}\delta_{\xi}g_{d\ell}\;, \label{boundarycontri1}\eeq
with $\delta_{\xi}g_{d\ell}=-\nabla_{(d}\xi_{\ell)}$. The second term above can be rewritten as
\beq -2\nabla_{b}P^{\ell bad}\delta_{\xi}g_{d\ell}=2\nabla_{b}(P^{\ell bad}+P^{dba\ell})\nabla_{d}\xi_{\ell}\;,\eeq
where we have used algebraic symmetries of $P^{abcd}$ and performed some index gymnastics. 

The first term in the boundary contribution involves a variation of a Christoffel symbol, which can be written as
\beq \delta_{\xi}\Gamma^{\ell}_{\;bd}=\frac{1}{2}R^{\ell}_{\;(bd)k}\xi^{k}-\frac{1}{2}\nabla_{(b}\nabla_{d)}\xi^{\ell}\;.\eeq
With some additional massaging, the first term becomes
\beq 2P_{\ell}^{\;bad}\delta_{\xi}\Gamma^{\ell}_{\;bd}=2P^{adb\ell}\nabla_{d}\nabla_{b}\xi_{\ell}-2P^{adb\ell}R_{kdb\ell}\xi^{k}\;.\eeq
Collectively then, the boundary contribution (\ref{boundarycontri1}) is 
\beq \delta_{\xi} v^{a}=-2\nabla_{b}(P^{adb\ell}+P^{a\ell bd})\nabla_{d}\xi_{\ell}+2P^{adb\ell}\nabla_{d}\nabla_{b}\xi_{\ell}-2P^{adb\ell}R_{kdb\ell}\xi^{k}\;.\label{boundarycontri2}\eeq

Substituting (\ref{eqnpartofJ}) and (\ref{boundarycontri2}) into the conserved current (\ref{conservcurrent}), we find, upon using the symmetries of $P^{abcd}$ and $R_{abcd}$:
\beq J^{a}=-2\nabla_{b}(P^{adb\ell}+P^{a\ell bd})\nabla_{d}\xi_{\ell}+2P^{adb\ell}\nabla_{d}\nabla_{b}\xi_{\ell}-4\nabla_{d}\nabla_{b}P^{adb\ell}\xi_{\ell}\;.\label{conservedcurr2}\eeq

At this point, we can `guess' the form of the potential $J^{ab}$. We do this choosing the ansatz,
\beq J^{ab}=A^{abd\ell}\nabla_{d}\xi_{\ell}+B^{ab\ell}\xi_{\ell}+C^{ab}\;,\eeq
with $\nabla_{b}C^{ab}=0$. Then, 
\beq \nabla_{b}J^{ab}=\nabla_{b}A^{abd\ell}\nabla_{d}\xi_{\ell}+A^{abd\ell}\nabla_{b}\nabla_{d}\xi_{\ell}+\nabla_{b}B^{ab\ell}\xi_{\ell}+B^{ab\ell}\nabla_{b}\xi_{\ell}\;.\eeq
Comparing to (\ref{conservedcurr2}), we find
\beq A^{abd\ell}=2P^{abd\ell}\;,\quad \nabla_{b}B^{ab\ell}\xi_{\ell}=-4\nabla_{b}\nabla_{d}P^{abd\ell}\xi_{\ell}\;.\eeq
The second of these implies 
\beq B^{ab\ell}=-4\nabla_{d}P^{abd\ell}+V^{ab\ell}\;,\quad \nabla_{b}V^{ab\ell}=0\;.\eeq
We also have that the following identity must hold:
\beq \nabla_{b}A^{adb\ell}+B^{ad\ell}=-2\nabla_{b}(P^{adb\ell}+P^{a\ell bd})\;.\eeq
Plugging in our expressions for $A^{adb\ell}$ and $B^{ad\ell}$, we find this forces $V^{ab\ell}=0$. Altogether then, we find the potential $J^{ab}$ associated with the conserved current (\ref{conservedcurr2}) is:
\beq J^{ab}=-2P^{abcd}\nabla_{c}\xi_{d}+4\xi_{d}\nabla_{c}P^{abcd}+C^{ab}\;.\label{conservpotentialapp}\eeq
Due to the presence of $C^{ab}$, we have that $J^{ab}$ is not unique, though we always recover the same conserved current. With the potential $J^{ab}$, we may write down the conserved charge (\ref{conservedcharge}):
\beq Q=\frac{1}{2}\int_{\Sigma} dS_{ab}(-2P^{abcd}\nabla_{c}\xi_{d}+4\xi_{d}\nabla_{c}P^{abcd})\;.\label{conserQ2}\eeq

Let's study this form of the conserved charge in the case $\Sigma$ is a Killing horizon. Then $\xi^{a}$ is a Killing vector, satisfying Killing's equation $\nabla_{(a}\xi_{b)}=0$ and Killing's identity $\nabla_{a}\nabla_{b}\xi_{c}=R^{k}_{\;abc}\xi_{k}$, such that the conserved current (\ref{conservedcurr2}) reduces to 
\beq J^{a}_{\text{Kill}}=2P^{adb\ell}R^{k}_{\;db\ell}\xi_{k}-4\nabla_{d}\nabla_{b}P^{adbk}\xi_{k}=(2E^{ak}+Lg^{ak})\xi_{k}\;.\eeq
That is, the boundary term $\delta_{\xi}v^{a}$ vanishes when $\xi^{a}$ is a Killing vector. Moreover, if $\Sigma$ is a bifurcate Killing horizon\footnote{Recall that the bifurcation surface $B$ of a Killing horizon is a $(D-2)$-dimensional spacelike cross section on which the Killing field generating the horizon vanishes. The bifurcation surface lies at the intersection of the two null hypersurfaces that comprise the full Killing horizon. For example, $B$ is the 2-sphere at the origin of Kruskal $U-V$ coordinates in the eternal Schwarzschild black hole \cite{Poisson04-1}.}, as in the case of a black hole horizon, $\xi^{a}=0$, then (\ref{conserQ2}) becomes
\beq Q_{\xi}=-\int_{\Sigma}dS_{ab}P^{abcd}\nabla_{c}\xi_{d}\;.\label{conservchargebifK}\eeq

On a (bifurcation surface of a) black hole horizon, where the timelike Killing vector $\xi^{a}$ goes null, we may express $\xi^{a}$ in terms of the timelike normal $u^{a}$ via $\xi^{a}=\kappa u^{a}$ where $\kappa$ is the surface gravity, which is constant over $\Sigma$. Then, using $\nabla_{c}u_{d}=\epsilon_{cd}$ on $\Sigma$, we may write (\ref{conservchargebifK}) in a more conventional manner,
\beq Q_{\xi}=-\kappa\int_{\Sigma}dAP^{abcd}\epsilon_{ab}\epsilon_{cd}\;.\label{noetherchargefinalapp}\eeq
Here $dA$ is just the induced volume element of the codimension-2 spatial surface, $dA=d^{D-2}x\sqrt{h}$. 

As we will see momentarily, the horizon entropy is a simple scaling of the conserved (Noether) charge \cite{Wald:1993nt}:
\beq S_{\text{W}}=\frac{2\pi}{\kappa}Q_{\xi}=-2\pi\int_{\Sigma}dAP^{abcd}\epsilon_{ab}\epsilon_{cd}\;.\label{waldentapp}\eeq
The factor of $2\pi$ is conventional, depending on the choice of units. We have chosen $L(g_{ab},R^{abcd})$ to include the gravitational couplings, \emph{e.g.}, for Einstein gravity $L=\frac{1}{16\pi G} R$.


\subsection*{Examples of Wald Entropy}

Let us now work out the Wald tensor $P^{abcd}$ and Wald entropy for a few illustrative examples, beginning with Einstein gravity. We have then
\beq
\begin{split}
P^{abcd}_{\text{GR}}&=\frac{1}{16\pi G}\frac{\partial R}{\partial R_{abcd}}=\frac{1}{16\pi G}\frac{\partial}{\partial R_{abcd}}(g^{\mu\nu}g^{\alpha\beta}R_{\alpha\mu\beta\nu})\\
&=\frac{1}{16\pi G}\frac{1}{2}\frac{\partial}{\partial R_{abcd}}[g^{\mu\nu}g^{\alpha\beta}(R_{\alpha\mu\beta\nu}-R_{\mu\alpha\beta\nu})]\\
&=\frac{1}{32\pi G}(g^{ac}g^{bd}-g^{ad}g^{bc})\;,
\end{split}
\label{PabcdGRapp}\eeq
where we used $\frac{\partial R_{\alpha\beta\gamma\delta}}{\partial R_{abcd}}=\delta^{a}_{\alpha}\delta^{b}_{\beta}\delta^{c}_{\gamma}\delta^{d}_{\delta}$.  We see then $\nabla_{a}P^{abcd}_{\text{GR}}=0$, 
such that the Wald entropy functional (\ref{waldentapp}) for Einstein gravity is 
\beq S_{\text{W}}=-2\pi\int_{\Sigma}dA\frac{1}{32\pi G}(g^{ac}g^{bd}-g^{ad}g^{bc})\epsilon_{ab}\epsilon_{cd}=\frac{A_{\Sigma}}{4G}\;,\eeq
where we used $\epsilon_{cd}^{2}=-2$ and $\int_{\Sigma}dA=A_{\Sigma}$. We see that the Wald entropy recovers the Bekenstein-Hawking entropy formula in the Einstein limit. 

Given the Wald tensor for Einstein gravity (\ref{PabcdGRapp}), it is straightforward to workout $P^{abcd}$ for $f(R)$ theories of gravity,
\beq P^{abcd}_{f(R)}=\frac{f'(R)}{32\pi G}(g^{ac}g^{bd}-g^{ad}g^{bc})\;,\eeq
where $f'(R)=df/dR$. The gravitational entropy -- when $\Sigma$ is a bifurcate Killing horizon -- is
\beq S_{\text{W}}=\frac{1}{4G}\int_{\Sigma}dAf'(R)\;.\eeq
Note that when $\Sigma$ is not a bifurcate Killing horizon, the above expression will be modified by needing to include the $\nabla_{c}P^{abcd}$, which in the case of $f(R)$ gravity is non-zero generically. 

Let's now move on to a reasonably generic  quadratic theory of gravity, with Lagrangian
\beq L_{\text{quad}}=\frac{1}{16\pi G}(R-2\Lambda)+\alpha_{1}R^{2}+\alpha_{2}R^{2}_{ab}+\alpha_{3}R^{2}_{abcd}\;.\eeq
For Einstein-Gauss-Bonnet gravity, we select $\alpha_{1}=\alpha_{3}=\alpha'$ and $\alpha_{2}=-4\alpha'$. We have
\beq \frac{\partial R^{2}}{\partial R_{abcd}}=R(g^{ac}g^{bd}-g^{ad}g^{bc})\;,\quad \frac{\partial R^{2}_{\alpha\beta\gamma\delta}}{\partial R_{abcd}}=2R^{abcd}\;.\eeq
If we want the algebraic symmetries of $P^{abcd}$ to be manifest, we must be careful with the $R^{2}_{ab}$ term. We have
\beq
\begin{split}
\frac{\partial R_{\mu\nu}}{\partial R_{abcd}}&=\frac{1}{2}\frac{\partial}{\partial R_{abcd}}(R_{\mu\nu}+R_{\nu\mu})=\frac{1}{2}g^{\rho\sigma}\frac{\partial}{\partial R_{abcd}}[R_{\sigma\mu\rho\nu}+R_{\rho\nu\sigma\mu}]\\
&=\frac{1}{4}g^{\rho\sigma}\frac{\partial}{\partial R_{abcd}}[R_{\sigma\mu\rho\nu}+R_{\mu\sigma\nu\rho}-R_{\nu\rho\sigma\mu}-R_{\rho\nu\mu\sigma}]\\
&=\frac{1}{4}(g^{ac}\delta^{b}_{\mu}\delta^{d}_{\nu}+g^{bd}\delta^{a}_{\mu}\delta^{c}_{\nu}-g^{bc}\delta^{a}_{\nu}\delta^{d}_{\mu}-g^{ad}\delta^{b}_{\nu}\delta^{a}_{\mu})\;,
\end{split}
\eeq
where to get to the second line we used the symmetries of the Riemann tensor,
\beq R_{\rho\nu\sigma\mu}=-\frac{1}{2}(R_{\nu\rho\sigma\mu}+R_{\rho\nu\mu\sigma})\;,\quad R_{\sigma\mu\rho\nu}=\frac{1}{2}(R_{\sigma\mu\rho\nu}+R_{\mu\sigma\nu\rho})\;.\eeq
Therefore, 
\beq
\begin{split}
 \frac{\partial R^{2}_{\mu\nu}}{\partial R_{abcd}}&=2R^{\mu\nu}\frac{\partial R_{\mu\nu}}{\partial R_{abcd}}=\frac{1}{2}(g^{ac}R^{bd}-g^{bc}R^{ad}+g^{bd}R^{ac}-g^{ad}R^{bc})\\
&=(R^{a[c}g^{d]b}+R^{b[d}g^{c]a})\;.
\end{split}
\eeq
Therefore, 
\beq P^{abcd}_{\text{quad}}=\left(\frac{1}{32\pi G}+\alpha_{1}R\right)2g^{a[c}g^{d]b}+\alpha_{2}(R^{a[c}g^{d]b}+R^{b[d}g^{c]a})+2\alpha_{3}R^{abcd}\;,\eeq
from which we can compute the Wald entropy. 

As a final example, let us work out Wald's entropy for Lovelock theories of gravity \cite{Lovelock71-1}. Recall that Lovelock gravity is characterized by having higher derivative contributions, but added in such a way that their gravitational equations of motion include only second derivatives of the metric. This is granted by imposing that $\nabla_{a}P^{abcd}=0$ for Lovelock theories, just as we observed for Einstein gravity.  In this way, Lovelock theories are the most natural extension of general relativity\footnote{It is often said general relativity is the unique pure theory of gravity in four dimensions with second order equations of motion; all higher derivative Lovelock terms are purely topological in four dimensions. Recently, however, so called `novel' pure theories of gravity have been written down \cite{Glavan:2019inb,Lu:2020iav,Hennigar:2020lsl,Easson:2020mpq}, where a dimensional rescaling of, \emph{e.g.}, the Gauss-Bonnet coupling leads to a theory which influences the local dynamics.}. More than that, in 2-dimensions, the Einstein-Hilbert action is the Euler density for a $2$-manifold, so too is Lovelock gravity for $2p$-dimensional manifolds, \emph{e.g.}, Gauss-Bonnet is the Euler density in 4-dimensions. 

In fact, we can use the structure of Einstein's general relativity, and the corresponding Wald tensor $P^{abcd}$ to build the Lovelock action. First note that the Einstein-Hilbert Lagrangian (dropping factors of $1/16\pi G$ for now) can be written in terms of the Wald tensor (\ref{PabcdGRapp})
\beq L_{\text{EH}}=P_{a}^{\;bcd}R^{a}_{\;bcd}=\delta^{cd}_{\;ab}R^{ab}_{\;\;\; cd}\;,\eeq
where we have the (2-dimensional) generalized Kronecker delta symbol
\beq \delta^{cd}_{ab}=\delta^{c}_{[a}\delta^{d}_{b]}=\frac{1}{2}(\delta^{c}_{a}\delta^{d}_{b}-\delta^{d}_{a}\delta^{c}_{b})=\frac{1}{2}\text{det}\begin{pmatrix}\delta^{c}_{a}&\delta^{c}_{b}\\ \delta^{d}_{a}&\delta^{d}_{b}\end{pmatrix}\;.\eeq

We generalize the Einstein-Hilbet term by simply considering higher-dimensional generalized Kronecker delta symbols, 
\beq \delta^{a_{1}a_{2}...a_{m}}_{b_{1}b_{2}...b_{m}} = \frac{1}{m!}\begin{pmatrix} 
    \delta^{a_{1}}_{b_{1}} & \dots & \delta^{a_{1}}_{b_{m}} \\
    \vdots & \ddots & \\
   \delta^{a_{m}}_{b_{1}} &        & \delta^{a_{m}}_{b_{m}}
    \end{pmatrix}
\;,\eeq
to be contracted with additional Riemann tensors. We would find that the next allowed choice, which maintains $\nabla_{a}P^{abcd}=0$ is the Gauss-Bonnet term, 
\beq L_{\text{GB}}=\frac{1}{2^{2}}\delta^{cdeg}_{abfh}R^{ab}_{\;\;\;cd}R^{fh}_{\;\;\;eg}=R^{2}-4R^{2}_{ab}+R^{2}_{abcd}\;.\eeq

We may further generalize this to 
\beq \mathcal{L}_{2p}(R)=\frac{1}{2^{p}}\delta^{a_{1}a_{2}...a_{2p-1}a_{2p}}_{b_{1}b_{2}...b_{2p-1}b_{2p}}R^{b_{1}b_{2}}_{\quad\; a_{1}a_{2}}...R^{b_{2p-1}b_{2p}}_{\quad\quad\quad a_{2p-1}a_{2p}}\;,\eeq
such that $p=1$ gives $\mathcal{L}_{2}=L_{\text{EH}}$, and $\mathcal{L}_{4}=L_{\text{GB}}$. When $p=D/2$, $\mathcal{L}_{2p}$ is purely topological, and for $p>D/2$ the $\mathcal{L}_{2p}$ vanishes. 

The Lovelock action is thus given by
\beq I=\frac{1}{16\pi G}\int d^{D}x\sqrt{-g}\sum_{p=2}^{[\frac{D}{2}]}c_{p}\mathcal{L}_{2p}(R)\;.\eeq
Here $[D/2]$ denotes the integer part of $D/2$ and $c_{p}$ are dimensionless coupling constants for the higher curvature terms.

Now notice that 
\beq 
\begin{split}
(P_{cd}^{\;\;\;ef})_{\text{GB}}&=\frac{\partial\mathcal{L}_{4}(R)}{\partial R^{cd}_{\;\;\;ef}}=\frac{1}{4}\delta^{efa_{3}a_{4}}_{cdb_{3}b_{4}}R^{b_{3}b_{4}}_{\quad\;\; a_{3}a_{4}}+\frac{1}{4}\delta^{a_{1}a_{2}ef}_{b_{1}b_{2}ed}R^{b_{1}b_{2}}_{\quad\;\; a_{1}a_{2}}\\
&=\frac{2}{4}\delta^{efa_{3}a_{4}}_{cdb_{3}b_{4}}R^{b_{3}b_{4}}_{\quad\;\; a_{3}a_{4}}\\
&=2(P_{cd}^{\;\;\; ef})_{\text{EH}}R\;.
\end{split}
\eeq
Recursively, one finds
\beq (P_{cd}^{\;\;\;ef})_{\mathcal{L}_{2p}}=\frac{\partial \mathcal{L}_{2p}(R)}{\partial R^{cd}_{\;\;\;ef}}=p(P_{cd}^{\;\;\;ef})_{EH}\mathcal{L}_{2p-2}(R)\;.\eeq
By construction we have $\nabla_{a}P^{abcd}=\nabla_{b}P^{abcd}=...=0$. Consequently, Lovelock theories only have second order equations of motion.

Let us now compute the Wald entropy for Einstein-Lovelock gravity using (\ref{waldentapp}) 
\beq 
\begin{split}
S_{\text{W}}&=-2\pi\int_{\Sigma}d^{D-2}x\sqrt{h}\frac{\partial\mathcal{L}}{\partial R^{cd}_{\;\;\;ef}}\epsilon^{cd}\epsilon_{ef}=\frac{1}{4G}\int_{\Sigma}d^{D-2}x\sqrt{h}\left[1+\sum_{p=2}^{[\frac{D}{2}]}c_{p}p\mathcal{L}_{2p-2}(R^{||})\right]\;.
\end{split}
\label{waldentloveapp}\eeq
where we $R^{||}$ denotes the components of the curvature tensor projected onto the horizon:
\beq [R^{||}]^{ab}_{\;\;\;cd}=h^{a}_{\;a'}h^{b}_{\;b'}h_{c}^{\;c'}h_{d}^{\;d'}R^{a'b'}_{\;\;\;c'd'}\;.\label{curvtenonhor}\eeq




\subsection*{Jacobson-Myers Entropy}
\noindent

Seemingly crucial to the computation of the Wald entropy is that we are working on the bifurcation surface of a Killing horizon; indeed, this was critical in Wald's original proof using the Noether charge \cite{Wald:1993nt}, where he also only considered stationary black holes. However, not every stationary black hole has a bifurcation surface. In particular, an asymptotically stationary black hole formed from gravitational collapse does not have a bifurcation surface. Moreover, the zeroth law of black hole mechanics in general relativity says that the surface gravity $\kappa$ is constant over the entire horizon, where the proof requires one to invoke the equations of motion and the dominant energy condition. For higher curvature theories this proof is not readily extended \emph{except} when one assumes the existence of a bifurcation surface. This begs the question as to whether Wald's formalism holds for more general black hole systems, namely, those without a bifurcation surface. 

Jacobson, Kang and Myers (JKM) showed how to extend Wald's `black hole entropy as Noether charge' to arbitrary horizon cross sections for asymptotically stationary black holes \cite{Jacobson:1993vj}. They did this by first showing the difference between the Noether charge evaluated at two cross-sections of a stationary  horizon is by an integral of the Noether current which, when pulled back to a stationary horizon, vanishes. Therefore, the Wald entropy formula holds for any cross-section of the horizon -- not just at the bifurcation surface. 

All of the above arguments considered a stationary black hole with a regular Killing horizon. There are, however, black holes which are nonstationary -- what is the black hole entropy given by then? As discussed at length in \cite{Jacobson:1993vj}, there are three potential choices: (i) the entropy which depends on the full potential $J_{1}^{ab}$, which may depend on arbitrarily high order derivatives of the vector field $\xi^{a}$; (ii) the entropy depends on a potential $J^{ab}_{2}$ depending only on $\xi^{a}$ and $\nabla^{[a}\xi^{b]}$, and (iii) the entropy depends on $J^{ab}_{3}$, characterized entirely by the binormal $\epsilon_{ab}$, dropping all reference to $\xi^{a}$. All three of these possibilities are equivalent when the surface over which one integrates is a bifurcate Killing horizon. For nonstationary black holes, there is no preferred choice of a vector field with which the expressions (i) and (ii) can be obtained unambiguously. Due to this, and because all three choices appear to be consistent with the first law of black hole mechanics, Wald \cite{Wald:1993nt} and JKM \cite{Jacobson:1993vj} cautiously conclude that the third option, where one writes the entropy solely in terms of $P^{abcd}$ and the binormal, is the most natural candidate. 

There is another candidate expression for black hole entropy that is different than Wald's proposal \cite{Wald:1993nt}, developed by Jacobson and Myers (JM) \cite{Jacobson:1993xs}, which holds for nonstationary black holes. In fact, this formula predates Wald's Noether charge method, and was uncovered using the Hamiltonian perturbation techniques developed in \cite{Sudarsky:1992ty} to write down an expression for the gravitational entropy for Einstein- Lovelock theories, given by
\beq S_{\text{JM}}=\frac{1}{4G}\int_{\Sigma}d^{D-2}x\sqrt{h}\left[1+\sum_{p=2}^{[\frac{D}{2}]}pc_{p}\mathcal{L}_{2p-2}(\mathcal{R})\right]\;,\label{JMentlove}\eeq
where $\mathcal{R}_{\alpha\beta\gamma\delta}$ are the components of the intrinsic curvature tensor of $\Sigma$.  In particular, note that by the Gauss-Codazzi equations we may relate intrinsic curvature $\mathcal{R}$ to the projection of the full spacetime curvature $R^{||}$ (\ref{curvtenonhor}) via \cite{Poisson04-1}
\beq [R^{||}]_{abcd}=\mathcal{R}_{abcd}-\sum_{i=1}^{2}\eta_{ij}(K^{i}_{ac}K^{j}_{bd}-K^{i}_{ad}K^{j}_{bc})\;,\eeq
where $\eta^{ij}=n^{i}_{a}n^{ja}$ is the Minkowski metric in the tranverse tangent space spanned by unit vectors $n^{i}_{a}$ orthogonal to the surface and each other and $K^{i}_{ab}$ is extrinsic curvature. 

Compare the Jacobson-Myers entropy to (\ref{JMentlove}) to the Wald entropy for Einstein-Lovelock theories (\ref{waldentloveapp}). Specifically, consider $D=5$, for which we have Einstein-Gauss-Bonnet gravity, and

\beq S_{\text{W}}=\frac{1}{4G}\int d^{3}x\sqrt{h}\left(1+2c_{2}\mathcal{L}_{2}(R^{||})\right)\;,\quad  S_{JM}=\frac{1}{4G}\int d^{3}x\sqrt{h}\left(1+2c_{2}\mathcal{L}_{2}(\mathcal{R})\right)\;.\eeq
Since $\mathcal{L}_{2}=R$ and using
\beq R^{||}=\mathcal{R}-\eta_{ij}(K^{i}K^{j}-K^{i\alpha\beta}K^{j}_{\alpha\beta}),\eeq
we have 
\beq S_{\text{JM}}=S_{\text{W}}+\frac{1}{4G}\int d^{3}x\sqrt{h}2c_{2}\eta_{ij}(K^{i}K^{j}-K^{iab}K^{j}_{ab})\;.\eeq
In the event $\Sigma$ is a Killing horizon for a stationary (and even non-stationary) black hole, one finds $K^{i}_{ab}=0$, such that the Wald entropy (\ref{waldentloveapp}) and the JM entropy (\ref{JMentlove}) are in agreement. 

A natural question to ask is which of the two proposals for horizon entropy is more fundamental. From the perspective of black hole physics, there is no clear argument to prefer one over the other. One is derived elegantly via the principle of general covariance and applies to any diffeomorphism invariant theory of gravity, and the other from a more complicated Hamiltonian method which is difficult  to generalize to higher derivative theories. From the perspective of holographic entanglement entropy, however, the JM entropy is considered to be more fundamental. In this set-up, the codimension-2 surface $\Sigma$ being integrated over is not a Killing horizon of a black hole, but instead some minimal bulk surface, which typically has non-vanishing extrinsic curvature. It was found in \cite{Hung:2011xb} that Wald's formula does not reproduce the correct CFT entanglement entropy, however, in the case of Lovelock gravity dual to CFTs in four and six dimensions, the JM entropy correctly computes the entanglement entropy. This was later derived using a generalization of the Euclidean method of squashed cones in \cite{Dong:2013qoa}, where, for CFTs dual to a general higher derivative theory of gravity, the holographic entanglement entropy is given by 
\beq 
\begin{split}
S_{EE}&=2\pi\int_{\Sigma}d^{D-2}x\sqrt{g}\biggr\{-\frac{\partial L}{\partial R_{abcd}}\epsilon_{ab}\epsilon_{cd}+\sum_{\alpha}\left(\frac{\partial^{2}L}{\partial R_{a_{1}b_{1}c_{1}d_{1}}\partial R_{a_{2}b_{2}c_{2}d_{2}}}\right)_{\alpha}\frac{2K_{\ell_{1}b_{1}d_{1}}K_{\ell_{2}b_{2}d_{2}}}{q_{\alpha}+1}\\
&\times[(n_{a_{1}a_{2}}n_{c_{1}c_{2}}-\epsilon_{a_{1}a_{2}}\epsilon_{c_{1}c_{2}})n^{\ell_{1}\ell_{2}}+(n_{a_{1}a_{2}}\epsilon_{c_{1}c_{2}}+\epsilon_{a_{1}a_{2}}n_{c_{1}c_{2}})\epsilon^{\ell_{1}\ell_{2}}\biggr\}\;.
\end{split}
\label{dongsEE}\eeq
Here $K_{abc}$ is the extrinsic curvature for the codimension-2 minimal bulk surface, and $q_{\alpha}$ is some weighting factor that is unimportant for the present discussion. We see that the entanglement entropy is given by the Wald entropy plus extrinsic curvature corrections, similar to the JM entropy. Indeed, when it is assumed the bulk theory is governed by Lovelock gravity, (\ref{dongsEE}) reduces to (\ref{JMentlove}). The formula for the entanglement entropy also tells us that when $\Sigma$ is a Killing horizon, which occurs when the minimal surface wraps entirely around a black hole in the bulk, the JM entropy reduces to the Wald entropy. This suggests, according to AdS/CFT, black hole entropy is a measure of entanglement entropy, a topic which we will explore in more detail in the next appendix.


\newpage


\section{FUNDAMENTALS OF SPACETIME ENTANGLEMENT} \label{app:entfund}


\subsection{Gibbs and the Thermofield Double}
\indent

Consider a system $A$ in a Gibbs state $\rho^{\text{Gibbs}}_{A}$, \emph{i.e.}, $\rho_{A}$ is expressed as a thermal density matrix\footnote{Since a density operator $\rho_{A}$ is Hermitian and semi-positive definite, we can always express $\rho_{A}$ in Gibbs form, $\rho_{A}=\frac{1}{Z}e^{-H_{A}}$, for some \emph{modular Hamiltonian} $H_{A}$.},
\beq \rho^{\text{Gibbs}}_{A}=\frac{1}{Z}e^{-\beta H_{A}}=\frac{1}{Z}\sum_{n=0}^{\infty}e^{-\beta E_{n}}|n\rangle_{A}\langle n|\;.\label{Gibbsstate}\eeq
Here $|n\rangle_{A}$ is the energy eigenstate of the Hamiltonian $H_{A}$, $H_{A}|n\rangle_{A}=E_{n}|n\rangle_{A}$, $\beta$ is the inverse temperature, and $Z=\text{tr}(e^{-\beta H_{A}})$ is the thermal partition function. The von Neumann entropy gives the usual thermodynamic entropy in the canonical ensemble
\beq S_{A}=-\text{tr}\rho_{A}^{\text{Gibbs}}\log\rho_{A}^{\text{Gibbs}}=-\frac{1}{Z}\sum_{n=0}^{\infty}e^{-\beta E_{n}}\log\left(Z^{-1}e^{-\beta E_{n}}\right)=(1-\beta\partial_{\beta})\log Z\;.\eeq

Note that we can represent the ground state $|0\rangle_{A}\langle0|$ as the $\beta\to\infty$ limit of $\rho_{A}$:
\beq \lim_{\beta\to\infty}\rho^{\text{Gibbs}}_{A}=\lim_{\beta\to\infty}\frac{e^{-\beta E_{0}}|0\rangle_{A}\langle 0|+e^{-\beta E_{1}}|1\rangle_{A}\langle1|+...}{e^{-\beta E_{0}}+e^{-\beta E_{1}}+...}\approx |0\rangle_{A}\langle0|\;,\eeq
where we used $E_{0}<E_{1}<E_{2}<...$. Put another way, the ground state can be cast as the infinite temperature limit of a Gibbs state, 
\beq |0\rangle_{A}\langle0|=\lim_{\beta\to\infty}\frac{1}{Z}e^{-\beta H_{A}}\;.\label{groundstatelim}\eeq
Since the thermal partition function can be cast as a path integral in Euclidean time over loops in space, we can understand the ground state $|0\rangle\langle 0|$ as the infinite Euclidean time limit of the path integral (we will study an example of this momentarily). 

From the form of the Gibbs state (\ref{Gibbsstate}) we see that system $A$ is mixed -- but mixed with what other system? We can introduce an ancilliary system $B$ for $A$ to mix with and write down the joint system $AB$ in a pure state. This technique is formally known as \emph{purification}\footnote{Generically, for a state $\rho_{A}=\sum_{a}p_{a}|a\rangle_{A}\langle a|$ of a single system $A$, we purify $A$ by constructing a large system $AB$, with Hilbert space $\mathcal{H}_{B}$ have a dimension at least the rank of $\rho_{A}$, and with an orthonormal basis $\{|a\rangle_{B}\}$, such that $|\psi\rangle_{AB}=\sum_{a}\sqrt{p_{a}}|a\rangle_{A}\otimes|a\rangle_{B}$ is a pure state for the joint system $AB$.}. We simply introduce a system $B$ with Hilbert space $\mathcal{H}_{B}$ to be a copy of $\mathcal{H}_{A}$, such that the purification of $\rho_{A}^{\text{Gibbs}}$ is
\beq |\text{TFD}\rangle=\frac{1}{\sqrt{Z}}\sum_{n=0}^{\infty}e^{-\beta E_{n}/2}|n\rangle_{A}\otimes|n\rangle_{B}\;.\label{thermofielddouble}\eeq
Indeed, taking the partial trace over states $\{|n\rangle_{B}\}$ returns (\ref{Gibbsstate}). The state (\ref{thermofielddouble}) is known as the \emph{thermofield double}. Though we won't go through the details here, the thermofield double can be computed via a Euclidean path integral represented by a semicircle with length $\beta/2$. 

The thermofield double state appears often in black hole thermodynamics. Specifically, the Hartle-Hawking state of a double-sided (eternal) AdS-Schwarzschild black hole is a thermofield double, where each AdS boundary is dual to a CFT \cite{Maldacena:2001kr}. Even though the two CFTs are not coupled (the Hamiltonians of $\text{CFT}_{1}$ and $\text{CFT}_{2}$ simply add) the Einstein-Rosen bridge connecting the two sides of the black hole is created via the two CFTs being entangled with one another. In other words, \emph{entanglement generates geometry}, which forms the basis of  `ER=EPR' \cite{Maldacena:2013xja}. Below we see how the thermofield double appears when studying entanglement entropy in Rindler spacetime.


\subsection{Rindler Entanglement}
\noindent

Let us now consider one of the simplest non-trivial examples of entanglement in a relativistic field theory. Here we will work with a field theory in $1+1$-dimensional Minkowski space in the Minkowski vacuum, $\rho=|0\rangle\langle 0|$, reduced to the half-line. That is, we want to compute $\rho_{A}$ for the region $A$ to be the set $A=\{t=0,x\geq0\}$. The metric is the standard one 
\beq ds^{2}=-dt^{2}+dx^{2}\;,\label{globalmink}\eeq
and the causal domain $D(A)=\{|t|\leq x\}$ is known as the (right) \emph{Rindler wedge}, or Rindler spacetime. The right wedge is described by coordinates $(\chi,r)$ with $r\geq0$, and $\chi\in(-\infty,\infty)$, and are related to Minkowski coordinates via
\beq x=r\cosh\chi\;,\quad t=r\sinh\chi\;,\eeq
such that 
\beq ds^{2}=-r^{2}d\chi^{2}+dr^{2}\;.\label{Rindler2app}\eeq

\begin{figure}[t]
\centering
 \includegraphics[width=9.3cm]{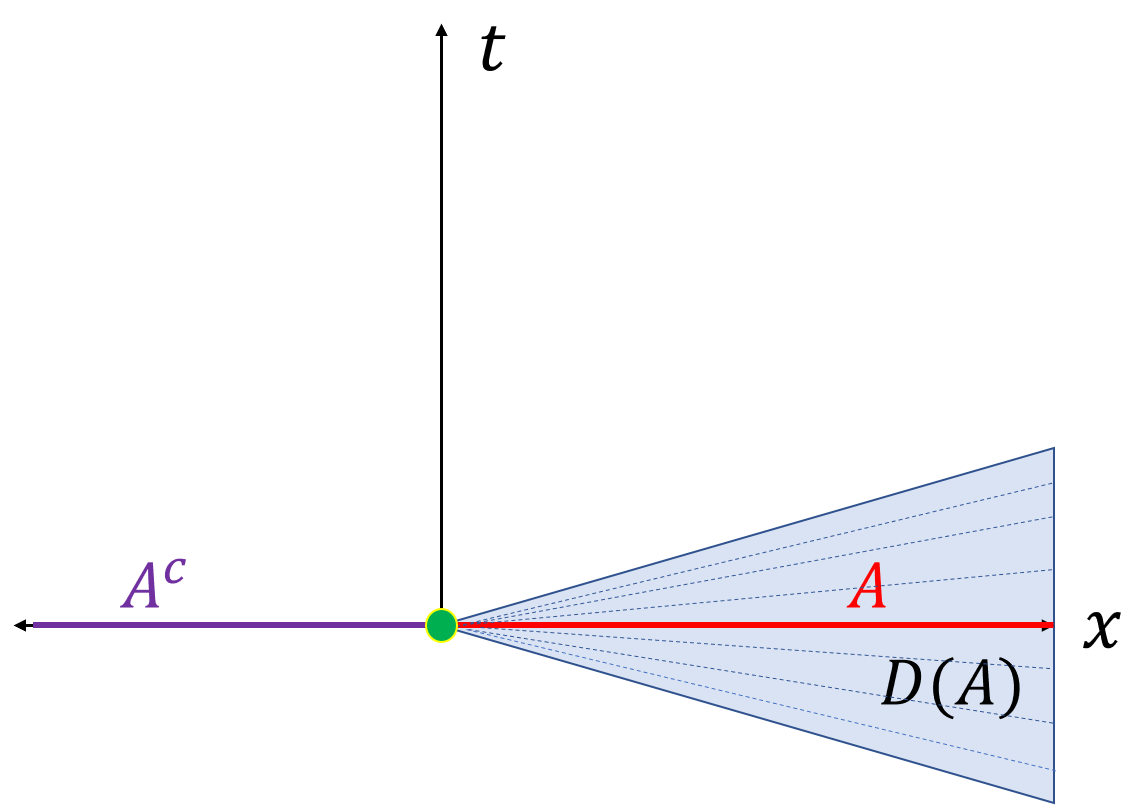}
 \caption[A Depiction of the Right Rindler Wedge.]{A depiction of the right Rindler wedge, the causal domain $D(A)$ of the half-line $A$ ($D(A)$ is the shaded region). The complement $A^{c}$ lives in the left Rindler wedge. In Minkowski coordinates, the entire line $t=0$ is a Cauchy slice, \emph{i.e.}, a set for which any two points cannot be connected by a causal curve and whose domain is the entire spacetime manifold. The lines of constant $\chi$ (drawn on the right hand side) are Cauchy slices for Rindler spacetime. The entangling surface is the point located at the origin $(x,t)=(0,0)$. We associate $\rho_{A}$, the Hilbert space $\mathcal{H}_{A}$, and the entropy $S(A)$ with the causal domain $D(A)$, a notable difference between entanglement in field theories and ordinary quantum mechanics.
}
 \label{rindlerwedgepic}\end{figure}

The field theory lives on the whole spacetime, however, we will only be interested in those field degrees of freedom which reside in $A$ \footnote{We should be more careful here when we talk about subsystems in field theories. First, when we talk about the Hilbert space for a field restricted to some region $A$, we really imagine subdividing the entire system in ``lattice sites" for which each point has its own local Hilbert space. Then, $\mathcal{H}_{A}$ is formally given by the tensor product of all of the local Hilbert spaces of each lattice point living inside of $A$. For relativistic field theories we further care about our observables obeying the causal structure of the background spacetime, and so our subsystem should really be the causal domain of $A$, $D(A)$, not $A$ by itself. Therefore, any regions of different Cauchy slices, but whose causal domains are the same, can be described by the same single state, have the same Hilbert space, and the same entropy.}, and thus we want to trace out the field degrees of freedom residing in the complement of $A$, denoted by $A^{c}$. The entangling surface is taken to be the point $(x,t)=(0,0)$. A pictorial representation of our set-up is given in Figure \ref{rindlerwedgepic}.

\begin{figure}[t]
\centering
 \includegraphics[width=14cm]{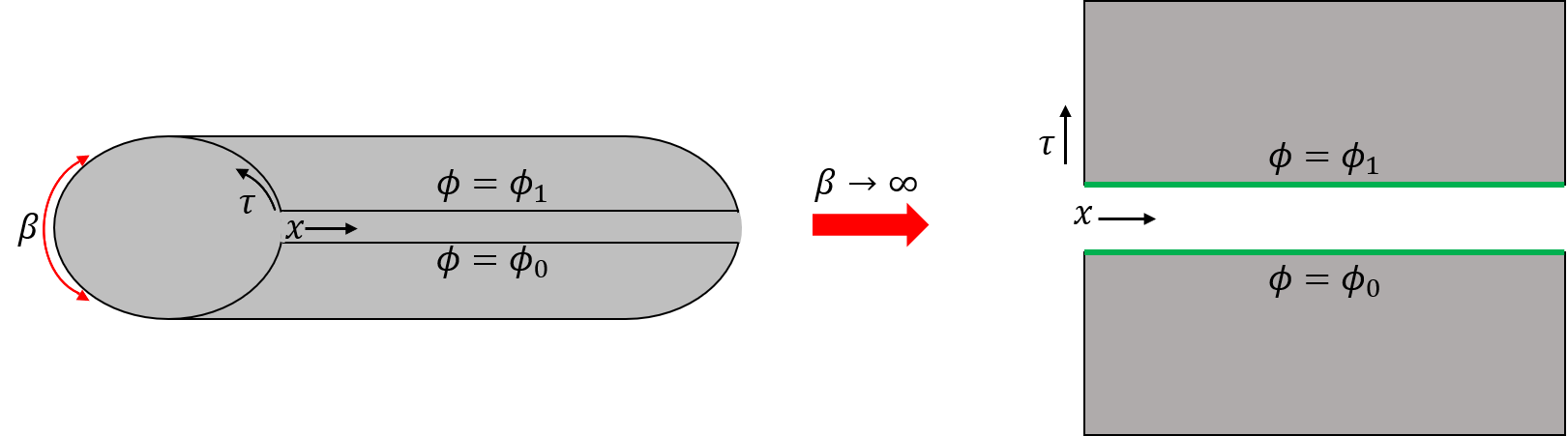}
 \caption[Density Matrix as Euclidean Path Integral.]{On the left we have represented the Euclidean path integral (\ref{Eucpathintapp}) as a cylinder with radius given by the the period $\beta$ of Euclidean time $\tau$ cut along $\tau=0$. The vacuum $\rho=|0\rangle\langle0|$ is given as the $\beta\to\infty$ limit of this transition amplitude, up to a factor of $1/Z$ to normalize $\rho$, depicted on the right. The state $\langle\phi_{0}|0\rangle$ is given by the integral only over the lower half-plane.
}
 \label{transvacEuc}\end{figure}

We want to compute $\rho_{A}$ explicitly. Our first task is to express the vacuum $\rho=|0\rangle\langle 0|$. We will do this using Euclidean path integrals, following the approach presented in \cite{Headrick:2019eth}. Let $\phi$ denote a full set of some fields, with $|\phi_{0}\rangle$ representing a field configuration on a fixed time slice. We start by writing the vacuum density matrix $|0\rangle\langle0|$ as the $\beta\to\infty$ limit of a Euclidean path integral
\beq\langle \phi_{0}|e^{-\beta H}|\phi_{1}\rangle=\mathcal{N}\int_{\phi_{1}}^{\phi_{0}}\mathcal{D}\phi e^{-I_{E}[\phi(\tau)]}\;,\quad Z=\text{tr}(e^{-\beta H})=\mathcal{N}\int_{\phi_{1}=\phi_{0}}\mathcal{D}\phi e^{-I_{E}}\;,\label{Eucpathintapp}\eeq
where $\mathcal{N}$ is some normalization. We have Wick rotated global Minkowski space (\ref{globalmink}) via $t=-i\tau$, such that $ds^{2}_{E}=d\tau^{2}+dx^{2}$, and $H$ is the Hamiltonian responsible for $\tau$ translations. In the language of transition amplitudes, (\ref{Eucpathintapp}) is the transition amplitude for a field with initial condition $|\phi_{1}\rangle$ to transition to a field with final condition $|\phi_{0}\rangle$. The transition amplitude can be represented pictorially as a cylinder with a cut along the $\tau=0$ axis, as shown in Figure \ref{transvacEuc}. The partition function $Z$ is understood to be the closed cylinder. 

Then, the matrix elements of the vacuum $|0\rangle\langle0|$ in this basis of fields is the $\beta\to\infty$ limit of (\ref{Eucpathintapp}):
\beq \langle\phi_{0}|0\rangle\langle0|\phi_{1}\rangle=\lim_{\beta\to\infty}\frac{1}{Z}\langle \phi_{0}|e^{-\beta H}|\phi_{1}\rangle\;,\eeq
which can pictured as a $(x,\tau)$ plane cut along $\tau=0$ (see Figure \ref{transvacEuc}).  We see that the we may factorize our above expression such that $\langle\phi_{0}|0\rangle$ is represented as the path integral only over the lower half-plane
\beq \langle\phi_{0}|0\rangle=\frac{1}{\sqrt{Z}}\int_{\phi(\tau=-\infty)=0}^{\phi(\tau=0)=\phi_{0}}\mathcal{D}\phi e^{-I_{E}}\;.\eeq

\begin{figure}[t]
\centering
 \includegraphics[width=9cm]{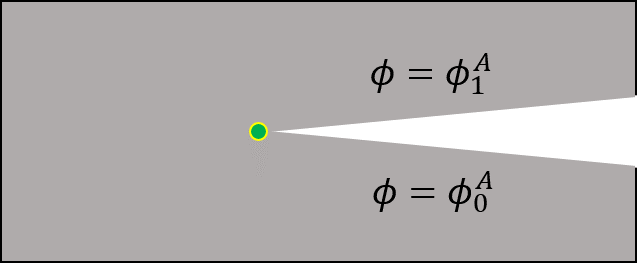}
 \caption[The Matrix Elements of the Vacuum Reduced to the Half-Line.]{The matrix elements of the vacuum $|0\rangle\langle0|$ reduced to the half-line $A$, $\langle\phi^{A}_{0}|\rho_{A}|\phi^{A}_{1}\rangle$. 
}
 \label{reducedrhorind}\end{figure}

Let us now restrict ourselves to (the causal domain of) $A$ and compute the reduced density matrix $\rho_{A}$. This means we must trace $\rho$ over the field degrees of freedom living in the complement $A^{c}$, namely, those states in the Hilbert space $\mathcal{H}_{A^{c}}$. We do this by writing the basis field $|\phi_{0}\rangle$ as a tensor product over those fields living in the right Rindler wedge $|\phi^{A}_{0}\rangle_{A}$ and left Rindler wedge $|\phi^{A^{c}}_{0}\rangle_{A^{c}}$, \emph{i.e.}, $\phi^{A}_{0}=\{\phi_{0}(x)\;|\;x\geq0\}$ and so forth; 
\beq |\phi_{0}\rangle=|\phi^{A}_{0}\rangle_{A}\otimes |\phi^{A^{c}}_{0}\rangle_{A^{c}}\;.\eeq
The state $\rho_{A}$ arises from summing over the $\phi^{A^{c}}_{0}$ field degrees of freedom, which has matrix elements
\beq 
\begin{split}
{}_{A}\langle\phi_{0}^{A}|(\text{tr}_{A^{c}}(0\rangle\langle0|)|\phi^{A}_{1}\rangle_{A}&=\int\mathcal{D}\phi_{0}^{A^{c}}({}_{A^{c}}\langle\phi^{A^{c}}_{0}|\otimes{}_{A}\langle\phi^{A}_{0}|)|0\rangle\langle0|(|\phi^{A^{c}}_{0}\rangle_{A^{c}}\otimes|\phi_{0}^{A}\rangle_{A})\\
&=\frac{1}{Z}\int_{\phi^{A}_{1},\phi^{A^{c}}_{0}=\phi^{A^{c}}_{1}}^{\phi^{A}_{0}}\mathcal{D}\phi e^{-I_{E}}\\
&={}_{A}\langle\phi_{0}^{A}|\rho_{A}|\phi^{A}_{1}\rangle_{A}\;.
\end{split}
\label{redrhorindapp}\eeq
The act of summing over the field degrees of freedom in $A^{c}$ pictorially glues the top and bottom sheets over the (left) half-line, up to the location of the entangling surface $(0,0)$, leaving a path integral on the plane cut along the half-line $A=\{\tau=0,x\geq0\}$, with the specified boundary conditions  (see Figure \ref{reducedrhorind}).

Since we are studying fields in a Euclideanized Minkowski background, let's consider Euclideanized Rindler spacetime, by Wick rotating the Rindler time $\chi$ to Euclidean `time' $\theta=i\chi$, such that (\ref{Rindler2app}) becomes
\beq ds^{2}_{E}=r^{2}d\theta^{2}+dr^{2}\;.\label{rindEucapp}\eeq
We recognize the line element as flat space in polar coordinates $(r,\theta)$, with $r\in[0,\infty)$ and $\theta\in[0,2\pi]$. Unlike the usual polar coordinates, where we identify $(r,0)\sim(r,2\pi)$, the cut along $\tau=0$ for $x\geq0$ makes them distinct. When we interpret $\theta$ as a Euclidean time coordinate, we recognize the cut plane represents a time interval of length $2\pi$ (rather than $\beta$) over the half-line $r\geq0$ via  the metric (\ref{rindEucapp}). 

Let's now compare the matrix elements of our reduced state $\rho_{A}$ (\ref{redrhorindapp}) to the Gibbs state (\ref{Gibbsstate}). The matrix elements of the  Gibbs state can be cast as a Euclidean path integral
\beq {}_{A}\langle x_{0}|\rho_{A}^{\text{Gibbs}}|x_{1}\rangle_{A}=\frac{1}{Z}{}_{A}\langle x_{0}|e^{-\beta H_{A}}|x_{1}\rangle_{A}\;.\eeq
This can be represented pictorially as a cut circle of length $\beta$ with endpoints $x_{0}$ and $x_{1}$. Comparing to the Minkowski vacuum reduced to the right Rindler wedge (\ref{redrhorindapp}), we see that we may interpret (\ref{redrhorindapp}) as a Gibbs state, with inverse temperature $\beta=2\pi$. In other words, $\rho_{A}$, \emph{in the Lorentzian picture}, is
\beq \rho_{A}=\frac{1}{Z}e^{-2\pi K}\equiv\frac{1}{Z}e^{-H_{A}}\;,\eeq
where we have defined the modular Hamiltonian\footnote{For a totally generic quantum state $\rho_{A}$, the modular Hamiltonian $H_{A}$ is not known explicitly and is typically a non-local operator. Nonetheless, it is important as the associated unitary $U(s)=e^{-iH_{A}s}$ generates a symmetry via $\text{tr}(\rho_{A}U(s)\mathcal{O}U(-s))=\text{tr}(\rho_{A}\mathcal{O})$ for any operator $\mathcal{O}$ localized inside $A$. The symmetry group of $U(s)$ transforms the operators inside the causal domain $D(A)$ into itself, and is known as the modular group \cite{Casini:2011kv}. In the case we consider here, $H_{A}$ is a local operator, and its modular flow is a local geometric flow.} $H_{A}$ associated with the modular flow (in this case given by the a $\chi$ translation) in the Rindler wedge to be $H_{A}=2\pi K$. In the Lorentzian picture, $K$ is the generator of the (Lorentzian) Rindler time translations $\partial_{\chi}$. According to the Rindler metric (\ref{Rindler2app}), $\xi^{a}=(\partial_{\chi})^{a}$ is a Killing vector, and $K$ is the associated conserved charge\footnote{Generally, given a Killing vector $\xi^{a}$ and a Cauchy slice $\mathcal{S}$, the associated conserved charge is $K=\int_{\mathcal{S}}d^{d-1}x\sqrt{h}\xi^{a}n^{b}T_{ab}$, with $h$ being the induced metric on the Cauchy slice.}, namely, 
\beq K=\int_{\chi=0,r}dr\xi^{a}n^{b}T_{ab}=\int^{\infty}_{0}dr\frac{1}{r}T_{\chi\chi}(\chi=0,r)\;,\eeq
where we are integrating over the Cauchy slice $(t=0,x)$, $n^{a}$ is a unit normal, and $T_{ab}$ is the energy-momentum tensor with respect to the quantum fields living in the background.

We can rewrite $K$ in terms of the original Minkowski coordinates using the fact that Rindler time translations are just Lorentz boosts in the $x-t$ plane $\partial_{\chi}=x\partial_{t}+t\partial_{x}$. The Cauchy slice we are integrating over is now the line $(x,t=0)$, leaving us with
\beq K=\int^{\infty}_{0}dx xT_{tt}(t=0,x)\;.\eeq
This result of the modular Hamiltonian for the right Rindler wedge can be proven more rigorously using algebraic quantum field theory, and is a consequence of the Bisognano-Wichmann theorem. 

To summarize briefly, we have shown that the vacuum state of a quantum field reduced to the right Rindler wedge $\rho_{A}$ is given by a Gibbs state with temperature $\beta=2\pi$ and modular Hamiltonian $H_{A}$ given as the conserved charge with respect to the Lorentz boost symmetry. We know, however, that the Gibbs state can be found by reducing the thermofield double $|\text{TFD}\rangle$. Consequently, we see that the Minkowski vacuum $|0\rangle$ is the thermofield double of the Rindler state $\rho_{A}$, with the left Rindler wedge $D(A^{c})$ being the purifying system. That is, heuristically,
\beq |0\rangle=\frac{1}{\sqrt{Z}}\sum_{i}e^{-\pi\omega_{i}}|i\rangle_{A}\otimes|i\rangle_{A^{c}}\;,\eeq
such that 
\beq \rho_{A}=\text{tr}_{A^{c}}|0\rangle\langle0|=\frac{1}{Z}\sum_{i}e^{-2\pi\omega_{i}}|i\rangle_{A}\langle i|\;.\eeq
Because $\rho_{A}$ is a thermal state, the von Neumann (entanglement) entropy $S_{A}=-\text{tr}(\rho_{A}\log\rho_{A})$ is a \emph{thermal} entropy. 

Note that we have in fact uncovered the \emph{Unruh effect} \cite{Unruh76-1}: An observer confined to the Rindler wedge will observe the state $\rho_{A}$, or, equivalently, the Minkowski vacuum, as a thermal state with respect to the boost generator at a temperature $T=(2\pi)^{-1}$. Such an observer is one who moves along a constant $r$ worldline (a hyperbola in the left wedge). According to inertial coordinates, such a trajectory is described by $x(t)=\sqrt{r^{2}+t^{2}}$, with constant proper acceleration $a=1/r$. The associated physical temperature of this observer is just the redshifted (Rindler) temperature:
\beq T_{\text{phys}}(r)=\frac{1}{\sqrt{-g_{\chi\chi}}}T=\frac{1}{2\pi r}=\frac{a}{2\pi}\;.\eeq
The physical inverse temperature $\beta_{\text{phys}}$ is just the proper length of the Euclidean time circle of constant $r$, which has circumference $2\pi r$. Notice that as observers get close to the entangling surface, where $r=0$, the physical temperature diverges. Moreover, the thermal entropy as measured by the Rindler observers is the entanglement entropy due to field degrees of freedom correlated between between the left and right Rindler wedges.


\section*{Spherical Entangling Surfaces}
\noindent

We saw above that the vacuum of a generic QFT reduced to the half-space $A$ allows us to express the state $\rho_{A}$ as a Gibbs state, where the modular Hamiltonian happens to be a local expression whose modular flow in the Rindler wedge corresponds to Rindler time translations. Another example where we can explicitly write down a local modular Hamiltonian is considering a CFT in vacuum $|0\rangle\langle0|$ in $d$-dimensional Minkowski space reduced to a ball $B$ of radius $R$. We consider a Cauchy slice of Minkowski space to be  the $t=0$ $(d-1)$-dimensional hypersurface, where the region $A$ is the ball $B(R)$ centered at $t=0,x=0$, described by state $\rho_{R}$. The complement $A^{c}$ is everything outside of the ball, where the spherical boundary is the entangling surface. As noted earlier, really we associate the state $\rho_{R}$ with the causal domain of ball, $D(B)$, which in this case is the causal diamond -- the intersection of the future of a past vertex and the past of a future vertex, and has a conformal isometry and spherical symmetry.

Our task then is to write down $\rho_{R}=\frac{1}{Z}e^{-K_{R}}$ for modular Hamiltonian $K_{R}$. There are actually a number of ways to write down $K_{R}$ explicitly. One sophisticated approach, utilized in \cite{Casini:2011kv}, is to note that the Rindler wedge (the causal domain of the half-space) can be mapped to the causal diamond via a special conformal transformation $K_{t}$ and time translation $P_{t}$ in Minkowski coordinates, such that the Killing vector associated with the conformal isometry of the causal diamond is given as
\beq \xi_{B}=\frac{i\pi}{R}(R^{2}P_{t}+K_{t})\;,\label{confkillBapp}\eeq
\beq iP_{t}=\partial_{t}\;,\quad iK_{t}=-[t^{2}+|\vec{x}|^{2}]\partial_{t}-2tx^{k}\partial_{k}\;.\eeq
The vector $\xi_{B}$ is in fact a conformal Killing vector in the original Minkowski space. 

The generator of the flow $\xi_{B}$ is then
\beq K_{R}=\int_{\mathcal{S}=B(R,t=0,x=0)}d^{d-1}xn^{a}\xi^{b}_{B}T_{ab}=2\pi\int_{B(R,t=0,x=0)}d^{d-1}x\frac{R^{2}-|\vec{x}|^{2}}{2R}T_{tt}(0,\vec{x})\;.\label{modhamball1}\eeq
Here $T_{\mu\nu}$ is the stress-tensor of the CFT. 

We can also consider a more direct approach in calculating $\xi_{B}$, following appendix B of \cite{Jacobson16-1}. We begin by writing the Minkowski line element in spherical polar coordinates $ds^{2}=-dt^{2}+dr^{2}+r^{2}d\Omega^{2}$ and introduce null coordinates $u=t-r$ and $v=t+r$, such that 
\beq ds^{2}=-dudv+r^{2}d\Omega^{2}\;.\label{Minknullapp}\eeq
Now we wish to determine the flow $\xi_{B}$ which preserves the conformal isometry and spherical symmetry of the causal diamond. We begin by noting that any vector field of the form 
\beq \xi^{a}=A(u)\partial^{a}_{u}+B(v)\partial^{a}_{v}\;\eeq
is a conformal isometry of the null coordinates of the Minkowski line element (\ref{Minknullapp}).  That is, 
\beq \mathcal{L}_{\xi}g_{uv}=[A'(u)+B'(v)]g_{uv}\;,\eeq
for Lie derivatie $\mathcal{L}$. The vector $\xi$ will be a conformal isometry of the full Minkowski metric provided we also have
\beq \mathcal{L}_{\xi}r^{2}=[A'(u)+B'(v)]r^{2}\;,\eeq
as then $\mathcal{L}_{\xi}g_{ab}=[A'(u)+B'(v)]g_{ab}$. 

Using $r=(v-u)/2$, we have $\mathcal{L}_{\xi}r^{2}=\xi^{a}\partial_{a}(r^{2})=(B-A)r$. So, $\xi$ is a conformal Killing vector provided
\beq [A'(u)+B'(v)]\frac{(v-u)}{2}=B(v)-A(u)\;.\eeq
Notice that at $u=v$, we have $B(v)=A(v)$, and consequently, at $v=0$ the above becomes
\beq [A'(u)+A'(0)]\frac{u}{2}=A(u)-A(0)\;.\eeq
We can solve this differential equation in general to find
\beq A(u)=B(u)=a+bu+cu^{2}\;.\eeq

The group generated by $\xi$ is $SL(2,\mathbb{R})$. For us to map the diamond onto itself (to preserve the conformal structure of the diamond), $\xi$ must leave invariant the boundaries at $u=-R$ and $v=R$. This tells us that $A(\pm R)=0$, so
\beq A(u)=a\left(1-\frac{u^{2}}{R^{2}}\right)\;,\eeq
for $a$ a constant. We can fix the constant by demanding $\xi$ be normalized such that it has a surface gravity of $\kappa=1$. Making this substitution for $A(u)$, we find
\beq \xi_{B}^{a}=\frac{1}{2R}\left[(R^{2}-u^{2})\partial^{a}_{u}+(R^{2}-v^{2})\partial^{a}_{v}\right]\;.\eeq
Or, back in Minkowski coordinates, 
\beq \xi_{B}^{a}=\frac{1}{2R}\left[(R^{2}-|\vec{x}|^{2}-t^{2})\partial^{a}_{t}-2tx^{k}\partial_{k}\right]\;,\eeq
matching (\ref{confkillBapp}) up to a factor of $\pi$ (as they choose a different normalization for $\kappa$. We still attain the same modular Hamiltonian (\ref{modhamball1}). We will analyze the entanglement entropy for a CFT vacuum state in Minkowski space reduced to a ball in more detail below.


\subsection{The CHM Map}
\noindent

The Ryu-Takayanagi (RT) formula, 
\beq S^{EE}_{A}=\frac{\mathcal{A}(\gamma_{A})}{4G^{(d+2)}}\;,\eeq
relates the entanglement entropy of holographic CFTs -- holographic entanglement entropy (HEE) -- to the area of a $d$-dimensional (static) minimal surface $\gamma_{A}$ in $AdS_{d+2}$ whose boundary is homologous to the boundary of a region $A$ in the CFT. Casini, Huerta, and Myers (CHM) \cite{Casini:2011kv} provided an early attempt to derive the RT formula. Their derivation involved reducing the ground state of a CFT in Minkowski space to a ball, conformally mapping this ground state to a thermal state of a massless hyperbolic black hole, and then computed the entanglement entropy of the CFT ground state from the thermal entropy of the hyperbolic black hole using the Bekenstein-Hawking relation. In other words, the thermal entropy of a (hyperbolic, massless) black hole is equivalent to the vacuum entanglement entropy of a CFT. 

Let us describe the CHM map in some detail.  The map is comprised of essentially two steps, the first of which requires no reference to gravity. Consider a CFT in $d$-dimensional Minkowski space in spherical coordinates. We now reduce the state to a ball $B$ of radius $R$.  A CFT ground state $\rho_{R}$ in Minkowski space ($\mathbb{R}\times\mathbb{R}^{d-1}$) reduced to a ball of radius $R$ can be written in terms of a `modular Hamiltonian' $K_{R}$ as \cite{Casini:2011kv}
\beq \rho_{R}=e^{-K_{R}}\;,\quad K_{R}=2\pi\int_{|\vec{x}|\leq R}d^{d-1}x\left(\frac{R^{2}-|\vec{x}|^{2}}{2R}\right)T_{00}(\vec{x})\;,\label{modhamil}\eeq
where $T_{00}$ is the energy density of the CFT stress tensor and $K_{R}$ is a local operator generating a flow in the causal domain of the ball\footnote{Modular flows can be defined for any region via Tomita-Takesaki theory, however, the modular flow for the ball is special in that it is local, described by a timelike coordinate $x^{0}$.}. 

We now recast the Minkowski line element in a different way by performing the following change of coordinates
\beq (t,r)=\frac{R}{\cosh(u)+\cosh(\tau/R)}\left(\sinh(\tau/R),\sinh(u)\right)\;,\eeq
with $\tau\in\mathbb{R}$ and $u\in\mathbb{R}_{+}$. In these new coordinates $(\tau,u)$ only cover the causal domain of the ball such that the flat metric becomes
\beq ds^{2}=\frac{1}{(\cosh(u)+\cosh(\tau/R))^{2}}\left(-d\tau^{2}+R^{2}(du^{2}+\sinh^{2}(u)d\Omega_{d-2}^{2})\right)\;,\eeq
which we recognize  as $\mathbb{R}\times\mathbb{H}^{d-1}$ times a conformal factor. We can perform a conformal transformation to remove the overall factor, thereby mapping the causal region of a ball in flat spacetime to the entire hyperbolic space, where the complement of the ball in Minkowski space gets pushed off to infinity via the conformal transformation. 

While the full ground state is invariant under the above conformal transformation, the reduced state $\rho_{R}$ is not. CHM further showed that, given the unitary operator $U$ acting on the Hilbert space of the CFT which implements the conformal transformation, the reduced state is mapped to
\beq \rho_{R}=e^{-K_{R}}=U^{\dagger}\left(\frac{e^{-\beta H_{\tau}}}{Z}\right)U\;,\label{CHMmap}\eeq
where $H_{\tau}$ is the Hamiltonian in the hyperpolic space generating time translations in $\tau$, and $Z$ represents the partition function of the \emph{thermal} state\footnote{Our localized states are thermal with respect to the modular flow of the causal domain of the ball $D(B)$, and thermal in the sense of the Kubo-Martin-Schwinger (KMS) condition -- a type of boundary condition for correlators in thermal equilibrium. In fact, we should stress that by definition the KMS condition, the state reduced to a ball is also thermal. Therefore, we simply mapped the thermal state of a ball, with respect to a `time' parameter governing the modular flow, to a thermal state $\mathbb{R}\times\mathbb{H}^{d-1}$, with respect to some different time parameter.}  $e^{-\beta H_{\tau}}/Z$, with inverse temperature given by the period of CFT correlators, $\beta^{-1}=1/(2\pi R)$. Now, given that the von Neumann entropy $S_{EE}=-\text{tr}(\rho\log\rho)$ is invariant under a unitary transformation, then the entanglement entropy across the sphere, $S_{EE}=-\text{tr}(\rho_{R}\log\rho_{R})$, is mapped to the thermal entropy on the hyperbolic background. We emphasize that gravity has not yet entered the picture; this is a calculation purely done with a CFT in Minkowski space.


Now we invoke AdS/CFT, \emph{i.e.}, we assume our CFT has a holographic dual. According to the standard lore of AdS/CFT, the ground state of the CFT in a $d$-dimensional flat space (not reduced to a ball) is dual to pure $AdS_{d+1}$ in Poincar\'e coordinates. Moreover, the ground state of the CFT reduced to a ball on the background $\mathbb{R}\times \mathbb{H}^{d-1}$ is dual to the massless, hyperbolic black hole\footnote{The massless hyperbolic black hole is also known AdS in Rindler coordinates} embedded in $AdS_{d+1}$ with boundary $\mathbb{R}\times\mathbb{H}^{d-1}$:
\beq ds^{2}=-\left(\frac{\rho^{2}}{L^{2}}-1\right)d\tilde{\tau}^{2}+\left(\frac{\rho^{2}}{L^{2}}-1\right)^{-1}d\rho^{2}+\rho^{2}(du^{2}+\sinh^{2}(u)d\Omega_{d-2}^{2})\;,\label{masslesshypbh}\eeq
with $L$ the AdS scale, $\rho\in[L,\infty]$, and $u\in\mathbb{R}_{+}$. This spacetime describes a massless ($M=0$) black hole in $AdS_{d+1}$ with a hyperbolic horizon located at $\rho=L$, and thermal temperature and entropy given by 
\beq T^{(M=0)}_{BH}=\frac{1}{2\pi L}\;,\quad S^{(M=0)}_{BH}=\frac{\omega_{d-1}L^{d-1}}{4G}=\frac{L^{d-1}}{4G}\Omega_{d-2}\int^{\infty}_{0}du\sinh^{d-2}(u)\;,\label{TShypBH}\eeq
where $\omega_{d-1}$ is the surface area of the hyperbolic plane with unit radius, and $\Omega^{d-2}$ is the surface area of a unit sphere. The massless black hole (\ref{masslesshypbh}) indeed describes the thermal state obtained (\ref{CHMmap}), as can be shown explicitly by performing the coordinate transformation 
\beq (t,r)=\frac{e^{-\gamma L}}{\cosh(u)+\cosh(\tilde{\tau}/L)}\left(\sinh(\tilde{\tau}/L),\sinh(u)\right)\;,\eeq
on Minkowski space in spherical coordinates, as well as making the identifications $R\to e^{-\gamma }L$ and $\tilde{\tau}=e^{\gamma}\tau$. 

With the thermal entropy of the black hole (\ref{TShypBH}) in hand, we are now in a position to determine the entanglement entropy of the vacuum state reduced to a ball. We first observe that the horizon of the black hole is the infinite extended hyperbolic plane $\mathbb{H}^{d-1}$, leading to an infinite entropy, as seen by the $u$ integral. This divergence is in accordance with the divergent nature of the entanglement entropy, instructing us to introduce a cutoff $u_{max}$ via
\beq x_{max}\equiv\sinh(u_{max})=\sqrt{\left(\frac{R}{\epsilon}\right)^{2}-1}\;,\eeq
leading to the regulated entanglement entropy:
\beq S_{EE}^{reg}=-\text{tr}(\rho_{R}\log\rho_{R})=S^{reg(M=0)}_{BH}=\left(\frac{2\Gamma(d/2)\Omega_{d-2}}{\pi^{d/2-1}}\right)a^{\ast}_{d}\int^{x_{max}}_{0}dx\frac{x^{d-2}}{\sqrt{1+x^{2}}}\;,\label{EEredvac}\eeq
where we have introduced the $L$-dependent generalized central charge $a^{\ast}_{d}$ \cite{Casini:2011kv}
\beq a^{\ast}_{d}=\frac{\pi^{d/2-1}}{8\Gamma(d/2)}\frac{L^{d-1}}{G}\;.\label{gencentralcharge}\eeq
\emph{This verifies that the entanglement entropy of a CFT ground state reduced to a ball in Minkowski space is equivalent to the thermal entropy of a massless hyperbolic black hole embedded in AdS spacetime of one dimension higher.}

We can also use the CHM map to analytically compute holographic R\'enyi entropies \cite{Casini:2010kt,Hung:2011nu}, and their physical generalizations \cite{Johnson:2018bma}. In particular, R\'enyi entropy will undergo phase transitions dual to the black hole transitioning a non-hairy black hole to a hairy one \cite{Dias:2010ma,Belin:2013uta,Belin:2014mva}.

The CHM map was one of the first attempts at deriving the more general statement known as the Ryu-Takayanagi formula, which says that the entanglement entropy of a holographic CFT reduced to a boundary region $A$ is equal to the area of a minimal surface homologous to $A$, such that the boundary of $A$ is identified with the boundary of $m$ \cite{Ryu06-1,Ryu06-2}
\beq S^{\text{CFT}}_{\text{EE}}(A)=\frac{1}{4G_{N}}\text{min}\,\text{area}[m(A)]_{m\sim A}\;.\label{RTformula2}\eeq
In this way, the entanglement entropy of a CFT follows an area law, like the Bekenstein-Hawking formula, except there need not be a Killing horizon present in the bulk. In fact, the Ryu-Takayanagi relation reduces to the Bekenstein-Hawking entropy relation, \emph{e.g.}, for the thermal entropy of a two-sided static asymptotically AdS black hole identified with the entanglement entropy of the thermofield double, where the minimal surface is the bifurcate Killing horizon at the center of the Einstein-Rosen bridge connecting the two conformal boundaries \cite{Maldacena:2001kr}. The Ryu-Takayanagi formula has been shown to obey a number of non-trivial properties of entanglement entropy, including strong subadditivity \cite{Hayden:2011ag}, and is consistent with holographic calculations of Reny\'i entropies using the replica trick in Euclidean quantum gravity \cite{Headrick:2010zt,Faulkner:2013yia}.  The Ryu-Takayanagi formula was proven by computing the holographic entanglement Reny\'i entropy via Euclidean quantum gravity by Lewkowycz and Maldacena \cite{Lewkowycz:2013nqa}. 

Ryu-Takayanagi (\ref{RTformula2}) can be generalized in a number of ways. The original statement required that the bulk spacetime have a time reflection symmetry such that the boundary spatial region $A$ is invariant. This can be generalized to bulk spacetimes that do not have any time reflection symmetry, and for general boundary regions. The area of a minimal surface in the Ryu-Takayanagi relation is then replaced with the area of a minimal bulk extremal spacelike surface, leading to the Hubeny-Rangamani-Takayanagi formula \cite{Hubeny:2007xt}. We can also assume our bulk theory is described not by classical Einstein gravity, as assumed in \cite{Ryu06-1,Ryu06-2}, but instead a more general theory of gravity including higher derivative corrections, where the area law is replaced by a more general entropy functional \cite{Hung:2011xb,Dong:2013qoa}. In the case of Lovelock corrections, the entropy functional is the Jacobson-Myers entropy, \emph{not} the Wald entropy. Finally, we can move away from the classical limit by including $G_{N}$ corrections. At order $G^{0}_{N}$ the Ryu-Takayanagi formula includes bulk entanglement entropy contributions by treating the bulk fields -- including the metric -- as quantum fields on a fixed background \cite{Faulkner13-1}.

Finally, a comment. The CHM map seemingly relied on our ability to map localized thermal states on Minkowski space to hyperbolic black holes. Doing so involved an intermediate step of performing a coordinate and a conformal transformation, in which we mapped the thermal state of the ball $\rho_{B}$ to a thermal state in $\mathbb{R}\times\mathbb{H}^{d-1}$, and then, via AdS/CFT, mapped the thermal state of $\mathbb{R}\times\mathbb{H}^{d-1}$ to a hyperbolic black hole. A similar argument holds for localized thermal states of the half space \cite{Rosso:2019lsm}. Naturally, one might wonder whether this intermediate step is necessary at all, such that the properties of hyperbolic black holes can be mapped directly to the thermal state of the ball. It turns out that one can take the metric of the hyperbolic black hole (\ref{masslesshypbh}), and take a non-standard asymptotic boundary limit $\rho\to\infty$ such that one directly maps the boundary of the hyperbolic black hole metric to the line element describing the causal domain of the ball in Minkowski space. For a particularly illustrative treatment of an analogous non-standard boundary limit applied to the half-space, see \cite{Rosso:2019lsm}.

\subsection{The First Law of Entanglement and its Extension}
\indent

Consider a general quantum system with a subsystem $A$, where $A$ is described by the reduced density matrix $\rho_{A}$. The entanglement between $A$ and its complement $\bar{A}$ is quantified by the von Neumann entropy $S_{A}=-\text{tr}\rho_{A}\log\rho_{A}$. Since $\rho_{A}$ is Hermitian and positive semi-definite, we may always express it in its ``Gibbs form":
\beq \rho_{A}=\frac{e^{-H_{A}}}{\text{tr}e^{-H_{A}}}\;,\eeq
where $H_{A}$ is called the modular Hamiltonian, formally defined through this expression. 

Now consider an infinitesimal state variation, $\rho_{A}\to \rho_{A}+\delta\rho_{A}$. Then, the first order variation of the von Neumann entropy is 
\beq
\begin{split}
\delta S_{A}&=-\text{tr}(\delta\rho_{A}\log\rho_{A})-\text{tr}(\rho_{A}\rho_{A}^{-1}\delta\rho_{A})\\
&=\text{tr}(\delta\rho_{A}H_{A})-\text{tr}(\delta\rho_{A})\;.
\end{split}
\eeq
Since $\text{tr}\rho_{A}=1$, we must have that $\text{tr}\delta\rho_{A}=0$, leaving us with the \emph{first law of entanglement entropy}
\beq \delta S_{A}=\delta\langle H_{A}\rangle\;.\label{firstlawEEapp}\eeq
In the event we started with an actual thermal state, such that $H_{A}=-\beta H$, the first law of entanglement represents an exact \emph{quantum} version of the first law of thermodynamics valid for arbitrary perturbations and arbitrary (including non-equilibrium) states. 

The first law (\ref{firstlawEEapp}) holds for generic quantum systems, including holographic CFTs. The gravitational interpretation of this law has a particularly interesting consequence: it is equivalent to the gravitational constraint for the linearized equations of motion to hold. More precisely, for small perturbations around the CFT vacuum state, the dual gravitational constraint for all ball shaped regions in the CFT are exactly equivalent to imposing the dual geometry satisfy  gravitational equations of motion linearized about pure AdS \cite{Faulkner13-2}.

A CFT can change even if the state is not varied. It can change if we allow for its number of degrees freedom to vary. In  AdS/CFT, the cosmological constant $\Lambda$ is understood to control the number of degrees of freedom, as a varying $\Lambda$ is a varying length $L$, such that the central charge $a^{\ast}$ varies. We have already seen that we can extend the laws of black hole thermodynamics by including a varying cosmological constant. Likewise, we may extend the first law of entanglement entropy so as to include variations of the central charge. This was first accomplished for spherically entangling surfaces in pure AdS in \cite{Kastor:2014dra}. Let us review their derivation in some detail.


Consider a $d$-dimensional CFT in vacuum, and reduce it to a spherical ball $B$ of radius $R$, the same set-up appearing in the CHM map \cite{Casini:2011kv}. The boundary of the ball $\partial\Sigma$ matches the boundary of the minimal bulk entangling surface $\Sigma$. By the Ryu-Takayanagi formula, the entanglement entropy of the vacuum restricted to $B$ is computed exactly via the area of the minimal bulk surface homologous to $B$. 

We can compute the area of $\Sigma$, denoted $A_{\Sigma}$ exactly using the Poincar\'e metric in $D=d+1$-dimensional pure AdS:
\beq ds^{2}_{D}=\frac{L^{2}}{z^{2}}(dz^{2}-dt^{2}+d\vec{x}\cdot d\vec{x})\;,\label{adspoinc}\eeq
where spatial infinity is located at $z=0$. We take $B$ to be centered at the origin and the constant time slice to be $t=0$. The corresponding bulk minimal surface $\Sigma$ on the $t=0$ hypersurface is then given by
\beq z^{2}+r^{2}=R^{2}\;,\quad r^{2}=\vec{x}\cdot\vec{x}\;.\eeq
We see that the surface extends in the bulk to $z=R$, and its area $A_{\Sigma}$ is
\beq A_{\Sigma}=L^{D-2}\Omega_{D-3}\int_{y_{c}}^{1}dy\frac{(1-y^{2})^{\frac{D-4}{2}}}{y^{D-2}}\;,\label{Asigent}\eeq
where $y=z/R$ and a cutoff at $z_{c}$ has been imposed to regularize the area -- due to vacuum fluctuations just across the boundary. The integral may be evaluated in any particular dimension, generically given in terms of hypergeometric functions.
In the case of a ball we can write down the modular Hamiltonian $H_{B}$ explicitly, (\ref{modhamil}). 

Our minimal bulk surface $\Sigma$ has an important feature which we will exploit: it is the bifurcation surface of a bulk Killing horizon, generated by the bulk Killing vector
\beq \xi=-\frac{2\pi}{R}(tz\partial_{z}+tx^{k}\partial_{k})+\frac{\pi}{R}(R^{2}-z^{2}-r^{2}-t^{2})\partial_{t}\;,\label{KVentsurf}\eeq
with norm
\beq \xi^{2}=-\frac{L^{2}\pi^{2}}{z^{2}R^{2}}[(R-t)^{2}-(r^{2}+z^{2})][(R+t)^{2}-(r^{2}+z^{2})]\;.\eeq
The  boundary of $\Sigma$ at $z=0$ includes the causal diamond, where (\ref{KVentsurf}) reduces to the conformal Killing vector  (\ref{confkillBapp}) whose flow preserves the conformal isometry and spherical symmetry of the diamond.
Note the Killing vector $\xi$ vanishes on the minimal surface $\Sigma$ at the $t=0$ hypersurface (this is what makes it a bifurcation surface for this Killing horizon).

The Hamiltonian formalism we used to derive the first law of black hole thermodynamics, including its extension, needed a Killing vector and a bifurcate Killing horizon. Therefore, we can just as easily replace the black hole horizon with the minimal bulk surface, and replace the Killing vector of Schwarzschild-AdS $(\partial/\partial t)^{a}$ with (\ref{KVentsurf}). We take the region of integration to be the volume bounded by the bulk minimal surface $\Sigma$ in the interior, out to the portion of spatial infinity covered by the spherical ball $B$:
\beq \int_{\Sigma}da_{c}(B^{c}-2\delta\Lambda\omega^{cb}n_{b})-\int_{B}da_{c}(B^{c}-2\delta\Lambda\omega^{cb}n_{b})=0\;,\label{komarintentsurf}\eeq
the area element $da_{c}$ is taken to point into the integration region  on the minimal surface $\Sigma$ in the interior and out of the integration region on the ball $B$ at spatial infinity\footnote{This choice leads to a difference in relative sign appearing in (\ref{komarintentsurf}).}. The Killing potential $\omega^{ab}$ is found by combining the trace of Killing's identity $\nabla_{a}\nabla^{a}\xi^{b}=-R^{b}_{c}\xi^{c}$ with Einstein's equation $G_{ab}=-\Lambda g_{ab}$ for the AdS background to get
\beq \omega^{ab}=-\frac{(D-2)}{2\Lambda}\nabla^{a}\xi^{b}\;.\eeq
Specifically, 
\beq \omega=\frac{1}{2}\omega^{ab}\partial_{a}\wedge\partial_{b}=\frac{\pi z}{(D-1)R}\{(R^{2}+z^{2}-t^{2}-r^{2})\partial_{t}\wedge\partial_{z}+2tx^{k}\partial_{z}\wedge\partial_{k}+2zx^{k}\partial_{t}\wedge\partial_{k}\}\;.\eeq

We now want to evaluate the different boundary integrals appearing in (\ref{komarintentsurf}). Let's focus on the contribution at infinity. We will only consider what happens as we allow for variations of $L$. Under such variations, the AdS metric (\ref{adspoinc}) changes as
\beq \delta g_{ab}=\frac{2\delta L}{L} g_{ab}\;,\quad \delta h_{ab}=\frac{2\delta L}{L}h_{ab}\;.\eeq
The variation of $\Lambda$ is just
\beq \delta\Lambda=\frac{(D-1)(D-2)}{L^{3}}\delta L\;.\eeq

The normal component of the Killing vector $\xi^{a}$ in the expression for the boundary vector $B^{a}$ is given by $F=(\pi\ell/Rz)(R^{2}-z^{2}-r^{2})$. The area element at the boundary at spatial infinity points in the $z$-direction and so we only need the $z$-component of $B^{a}$
\beq B^{z}=\frac{2(D-2)\pi\delta L}{RL^{2}}(R^{2}+z^{2}-r^{2})\;.\label{Bz}\eeq
Moreover, the $z$-component of the Killing potential term is 
\beq 2\omega^{zt}n_{t}\delta\Lambda=\frac{2(D-2)\pi\delta L}{RL^{2}}(R^{2}+z^{2}-r^{2})\;,\label{Kilpotz}\eeq
Substituting (\ref{Bz}) and (\ref{Kilpotz}) into the boundary at infinity term in (\ref{komarintentsurf}), we see that infinite contributions from $r\to\infty$ vanish. Here we have used the fact that our background spacetime is unperturbed AdS, so that $\omega^{ab}_{AdS}=\omega^{ab}$. Thus,  the boundary integral at infinity  receives no new contributions from varying the cosmological constant,  and we are left with
\beq \int_{\infty}da_{a}(B^{a}-2\omega^{ab}_{AdS}n_{b}\delta\Lambda)=-16\pi G\delta E_{\xi}\;.\eeq
where $E_{\xi}$ is the ADM charge associated with $\xi^{a}$ (\ref{KVentsurf}).

The integral of the boundary vector $B^{a}$ over the minimal surface $\Sigma$ is again given by 
\beq \int_{\Sigma}da_{a}B^{a}=-2\kappa\delta A_{\Sigma}\;,\eeq
with surface gravity $\kappa=2\pi$ for the Killing vector $\xi^{a}$. 

Combining these results, we then have the extended bulk first law \cite{Kastor:2014dra}
\beq \delta E_{\xi}=\frac{\delta A_{\Sigma}}{4G}-\frac{V\delta\Lambda}{8\pi G}\;,\eeq
where the thermodynamic volume in this case is
\beq V=-\int_{\Sigma}da_{a}\omega^{ab}n_{b}\;.\eeq

We can actually evaluate the volume $V$ explicitly. Using the unit normal to the constant time slice $n=-(L/z)dt$ and $da_{b}=m_{b}da$, where  $m=-\frac{L}{zR}(zdz+\vec{x}\cdot d\vec{x})$ is the outgoing normal to $\Sigma$ within the constant time slice and $da$ is the induced area element, we have
\beq V=\frac{2\pi L^{2}}{D-1}A_{\Sigma}\;,\eeq
where $A_{\Sigma}$ is the area element of the minimal surface $\Sigma$, (\ref{Asigent}).  

We may rewrite the extended first law in the bulk with the thermodynamic volume entirely in terms of the entanglement entropy $S_{\Sigma}$ and the AdS curvature radius $L$:
\beq \delta E_{\xi}=\delta S_{\Sigma}-(D-2) S_{\Sigma}\frac{\delta L}{L}\;.\eeq
So far, our statement is still a bulk relation. We make contact to entanglement of the boundary region by replacing variations in $L$ with variations of the central charge $a_{d}^{\ast}=\frac{\pi^{d/2-1}L^{d-1}}{8\Gamma(d/2)G}$, and identifying $\delta E_{\xi}$ with the variation of the modular Hamiltonian (\ref{modhamil}) $\delta\langle H_{B}\rangle$ \cite{Faulkner13-2},
\beq \delta S_{\Sigma}=\delta \langle H_{B}\rangle+\frac{S_{\Sigma}}{a^{\ast}_{d}}\delta a^{\ast}_{d}\;.\label{firstlawextEEapp}\eeq
This is the extended first law of (holographic) entanglement entropy, specific to spherical entangling surfaces and where the dual bulk theory is governed by Einstein gravity. 

Since $a_{d}^{\ast}$ measures the number of degrees of freedom of the CFT, the extended first law gives the dependence of the entanglement entropy on the number of degrees of freedom. Of course, when we recall elementary thermodynamics, the chemical potential $\mu$ is conjugate to the number of particles. Comparing (\ref{firstlawextEEapp}) to the usual first law of thermodynamics (without $p-V$), $\delta E=T\delta S-\mu \delta N$, it is natural to interpret the new term in (\ref{firstlawextEEapp}) as a chemical potential contribution with $\mu=-S_{\Sigma}/a^{\ast}_{d}$ \cite{Kastor:2014dra}. A similar relation holds in higher even-$d$-dimensional spacetimes, where, however, we must also need to take into account additional coupling constants of the theory \cite{Karch:2015rpa,Caceres:2016xjz}. Just as we introduced $p-V$ into black hole thermodynamics, leading to black hole ``chemistry", our introduction of  $\mu$ allows us to interpret (\ref{firstlawextEEapp}) as the first law of \emph{holographic entanglement chemistry}.




\newpage

\section{FAILURE OF KILLING'S IDENTITY} 


Here review important calculational details of the derivation of gravitational field equations from spacetime thermodynamics in Chapter \ref{sec:gravfromthermo}. 


\subsection{Stretched Lightcones} \label{app:failKILC}

In our derivation of the gravitational equations, we made critical use of the Killing identity even though $\xi_a$ is only an approximate Killing vector. The purpose of this appendix is to justify that step, as well as to eliminate the $\int d \Sigma_a q^a$ term in (\ref{TdStot}). We denote the failure of $\xi_a$ to satisfy Killing's identity via the tensor
\be
f_{bcd} \equiv \nabla_{b}\nabla_{c}\xi_{d}-R^{e}_{\;bcd}\xi_{e} =\frac{1}{2}\left(\nabla_{d}S_{bc}-\nabla_{c}S_{db}-\nabla_{b}S_{cd}\right)
\ee
where $S_{ab} = \nabla_{(a}\xi_{b)}$ \cite{Kothawala:2010bf}. From this we see that $f_{bdc}=-f_{bcd}$. 

In evaluating $\Delta S_{\rm tot}$, we encounter integrals of the form $\int d \Sigma_a P^{abcd} (R_{dcbe} \xi^e + f_{bcd})$, as in (\ref{TdStot}). (For Einstein gravity, $P^{abcd} = \frac{1}{2} (g^{ac} g^{bd} - g^{ad} g^{bc})$.) We would like to discard $n_{a}P^{abcd}f_{bcd}$ but retain $n_{a}P^{abcd}R^{e}_{\;bcd}\xi_{e}$. This latter quantity is, to lowest order, $\mathcal{O}(x^{2})$, since $\xi_a$ and $n_a$ are both of order $x$. Hence all terms in $f_{bcd}$ of $\mathcal{O}(x)$ and lower are problematic.

In general, $f_{bcd}$ has two types of contributions because our $\xi_a$ fails to be a Killing vector in two ways. First, $\xi_a$ generates radial boosts. These are not true isometries even of Minkowski space. This contributes a term to $f_{bcd}$ of ${\cal O}(x^{-1})$ in Riemann normal coordinates. Second, we will see that in a general curved spacetime, $\xi_a$ will have to be redefined to include quadratic and higher terms. These contribute terms to $f_{bcd}$ at ${\cal O}(1)$ and ${\cal O}(x)$. Therefore, in general, $f_{bcd}$ does not vanish at the required order.

Fortunately, we do not actually need $f_{bcd}$ to vanish, as in \cite{Guedens:2012sz,Guedens:2011dy}; rather we require only a much weaker condition, namely that the integral of the contraction $n_a P^{abcd} f_{bcd}$ vanish to ${\cal O}(x^{2})$. We shall use several tricks to deal with nonzero terms in $f_{bcd}$. First, some terms give zero when contracted with $P^{abcd}$, because of symmetry. Second, the vast majority of terms integrate to zero over the spherical spatial sections of $\Sigma$, since the integral of any odd power of a Cartesian spatial coordinate over a sphere is zero. The remaining terms are of two types: there is the $f_{bcd}$ term of ${\cal O}(x^{-1})$ that exists even in Minkowski space,
and there are a small handful of leftover $f_{bcd}$ terms of ${\cal O}(1)$ and ${\cal O}(x)$ in curved space. The integral of the first term does not vanish. However, as we show, it is precisely canceled by subtracting the component of $T\Delta S$ that comes from the natural expansion of $\Sigma$. The other terms can be eliminated by redefinining the higher-order terms in $\xi^a$, as we will show.

Our integrand $\sqrt{g} n_a P^{abcd} f_{bcd}$ will have various order pieces ranging from $\mathcal{O}(1)$ to $\mathcal{O}(x^{2})$, with higher orders negligible. We need to show that the integral at each order either vanishes or can be canceled. Let us first classify each of the terms. We do this by expanding 
 \be
n_{a}\approx n^{(1)}_{a}+n^{(2)}_{a}+n^{(3)}_{a},\quad P^{abcd}\approx P^{abcd}_{(0)}+P^{abcd}_{(1)}+P^{abcd}_{(2)},\quad f_{bcd}\approx f^{{\cal O}(-1)}_{bcd}+f^{(0)}_{bcd}+f^{(1)}_{bcd}
 \ee
where the subscript or superscript indicates the order, in $x$, of the given quantity. We also note that for the integration measure we have $\sqrt{g}\approx\sqrt{\eta}+\sqrt{h}$ which is of $\mathcal{O}(1)+\mathcal{O}(x^{2})$. 

Then the lowest order contribution to the offending term is 
\be
  \frac{1}{4G \hbar}\int_{\Sigma}dA d \tau n^{(1)}_{a}P^{abcd}_{(0)}f^{{\cal O}(-1)}_{bcd}\label{npf0}
\ee
which is of $\mathcal{O}(1)$. The next order terms, of $\mathcal{O}(x)$, are given by
 \be
  \frac{1}{4G \hbar}\int_{\Sigma}dAd \tau \left(n^{(1)}_{a}P^{abcd}_{(1)}f^{{\cal O}(-1)}_{bcd}+n^{(2)}_{a}P^{abcd}_{(0)}f^{{\cal O}(-1)}_{bcd}+n^{(1)}_{a}P^{abcd}_{(0)}f^{(0)}_{bcd}\right)\label{npf1}
   \ee
Last, the highest order term we need consider is
 \be
 \begin{split}
\frac{1}{4G \hbar} \int_{\Sigma}dA d \tau&\biggr\{\sqrt{h}n^{(1)}_{a}P^{abcd}_{(0)}f^{{\cal O}(-1)}_{bcd}+n^{(1)}_{a}P^{abcd}_{(2)}f^{{\cal O}(-1)}_{bcd}+n^{(1)}_{a}P^{abcd}_{(1)}f^{(0)}_{bcd}+n^{(1)}_{a}P^{abcd}_{(0)}f^{(1)}_{bcd}\\
 &+n^{(2)}_{a}P^{abcd}_{(1)}f^{{\cal O}(-1)}_{bcd}+n^{(2)}_{a}P^{abcd}_{(0)}f^{(0)}_{bcd}+n^{(3)}_{a}P^{abcd}_{(0)}f^{{\cal O}(-1)}_{bcd}\biggr\}
 \end{split}
\label{npf2} 
\ee
which is clearly of $\mathcal{O}(x^{2})$. We therefore need to show (\ref{npf0}), (\ref{npf1}), and (\ref{npf2}) vanish for an arbitrary $P^{abcd}$. Let us begin with (\ref{npf0}).


\subsection*{Removing the Natural Expansion of the Hyperboloid}
\indent

Writing out $f_{bcd}$ explicitly, we have
\be
f_{bcd}=\partial_{b}\partial_{c}\xi_{d}+\left(2\Gamma^{f}_{\;b(c}\Gamma^{e}_{\;d)f}-\partial_{b}\Gamma^{e}_{\;cd}\right)\xi_{e}-\left(\Gamma^{e}_{\;bc}\partial_{e}\xi_{d}+2\Gamma^{e}_{\;d(c}\partial_{b)}\xi_{e}\right)-R^{e}_{\;bcd}\xi_{e} \label{fbcd-explicit}  
\ee
Note that $\xi_{a}$, $n_a$, and the Christoffel symbols are all of $\mathcal{O}(x)$. 
Therefore the term $n_{a}2\Gamma^{f}_{\;b(c}\Gamma^{e}_{\;d)f}\xi_{e}$ is of much higher order than the rest of the terms and we can neglect it. Moreover, given that $P^{abcd}$ is antisymmetric in its final two indices and $\Gamma^{e}_{\;cd,b}$ is symmetric in $c$ and $d$, it will not contribute to $n_a P^{abcd} f_{bcd}$. Therefore, we need only consider the reduced expression:
\be
f_{bcd} \approx \partial_{b}\partial_{c}\xi_{d}-2\Gamma^{e}_{\;bc}\partial_{[e}\xi_{d]}-R^{e}_{\;bcd}\xi_{e}
\ee
 
To lowest order, we have
\be
f_{bcd}^{\mathcal{O}(-1)}=\partial_{b}\partial_{c}\xi^{\mathcal{O}(1)}_{d}
\ee
From (\ref{Killingfailure}), we find that Killing's identity, at $\mathcal{O}(x^{-1})$, fails as,
 \be
 \begin{split}
 &f_{tij}^{\mathcal{O}(-1)}=f_{itj}^{\mathcal{O}(-1)}=-f_{ijt}^{\mathcal{O}(-1)}=\frac{1}{r}\left(\delta_{ij}-\frac{x_{i}x_{j}}{r^{2}}\right)\\
& f_{ijk}^{\mathcal{O}(-1)}=-\frac{t}{r^{3}}\left(x_{i}\delta_{jk}+x_{j}\delta_{ik}+x_{k}\delta_{ij}\right)+\frac{3t}{r^{5}}x_{i}x_{j}x_{k}
 \label{fO-1bcd}\end{split}
 \ee
Using the algebraic symmetries of $P^{abcd}$ and $f^{\mathcal{O}(-1)}_{bcd}$, we have
\be
P^{abcd}f^{\mathcal{O}(-1)}_{bcd}=P^{aijk}f_{ijk}^{\mathcal{O}(-1)}+P^{atij}f^{\mathcal{O}(-1)}_{tij}+P^{aitj}f_{itj}^{\mathcal{O}(-1)}+P^{aijt}f_{ijt}^{\mathcal{O}(-1)} =2P^{aitj}f_{itj}^{\mathcal{O}(-1)}
\ee
The undesired term then becomes
\be
\begin{split}
 \frac{1}{4G\hbar}\int_{\Sigma}dAd\tau n_{a}P^{abcd}f_{bcd}^{\mathcal{O}(-1)}&=\frac{1}{4G\hbar}\int_{\Sigma}dAd\tau\left(2n_{t}P^{titj}f^{\mathcal{O}(-1)}_{itj}+2n_{i}P^{tkij}f^{\mathcal{O}(-1)}_{jtk}\right)\\
 &=-\frac{1}{4G\hbar}\int_{\Sigma}dAd\tau\frac{2t}{\alpha r}P^{titj}\left(\delta_{ij}-\frac{x_{i}x_{j}}{r^{2}}\right)
 \end{split}
 \ee
 where in the last step we used spherical symmetry killing off all integrals with parity. Moreover, by parity, this term will vanish for all terms $i\neq j$, keeping only terms with $i=j$. With this fact in mind, and using that $d\tau = dt \alpha/r$, and $\sum x_{i}^{2}=r^{2}$, we have
 \be
\begin{split}
 &\frac{1}{4G\hbar}\int_{\Sigma}dAd\tau n_{a}P^{abcd}f_{bcd}^{\mathcal{O}(-1)}=-\frac{1}{4G\hbar}(D-2)\frac{2\sum_{i}P^{titi}}{\alpha (D-1)}\left(\int d\Omega_{D-2}\right)\int_{0}^{t_{0}}dt\frac{\alpha}{r} r^{D-3}t\\
 &=-\frac{1}{2(D-1)G\hbar}(D-2)\sum_{i}P^{titi}\Omega_{D-2}\int_{0}^{t_{0}}dt\left(\alpha^{2}+t^{2}\right)^{(D-4)/2}t\\
 &=-\frac{1}{2(D-1)G\hbar}\sum_{i}P^{titi}\Omega_{D-2}\left[\left(\alpha^{2}+t^{2}_{0}\right)^{(D-2)/2}-\alpha^{(D-2)}\right] \label{leadingnpf}
 \end{split}
 \ee

Recall that we are applying Clausius' theorem, $T\Delta S_{\rm rev}= Q$, to derive the equations of motion for an arbitrary theory of gravity. But $\Delta S_{\rm tot}$ includes all change in the entropy, not just the change in entropy due to the heat flow through $\Sigma$. In particular, even in the absence of heat flow, the entropy increases because of the natural increase in an area of a congruence of outwardly accelerating observers.

Let us calculate the increase in entropy from the natural background expansion of the hyperboloid. Begin with the Wald entropy,
 \be
S =\frac{1}{8G\hbar}\int_{S}dS_{ab}J^{ab}=-\frac{1}{4G\hbar}\int_{S}dS_{ab}\left(P^{abcd}\nabla_{c}\xi_{d}-2\xi_{d}\nabla_{c}P^{abcd}\right)\;.\label{SW}
  \ee
To leading order we can neglect the $\nabla_{c}P^{abcd}$ term. Substituting in our leading-order expressions for the outward pointing normal $n_{a}$, and $u_{a}=\xi_{a}/\alpha$, we find
 \be 
 \begin{split}
S &=-\frac{1}{4G\hbar}\int_{S}dA\left(n_{t}u_{i}-n_{i}u_{t}\right)\left[P^{titj}2\partial_{t}\xi_{j}+P^{tijk}\partial_{j}\xi_{k}\right]\\
&=-\frac{1}{4G\hbar}\int_{S}dA\frac{x_{i}}{r}\left[2P^{titj}\partial_{t}\xi_{j}+P^{tijk}\partial_{j}\xi_{k}\right]\\
&=-\frac{1}{4G\hbar}\int_{S}dA\left(2P^{titj}\frac{x_{i}x_{j}}{r^{2}}\right)\\
&=-\frac{1}{2(D-1)G\hbar}\sum_{i}P^{titi}\Omega_{D-2}r^{D-2}(t_{0})\;,
 \end{split}
 \ee
where we used parity to move to the final line. We are interested in the change in entropy, $\Delta S_{\rm hyp}$, due to the expansion of the hyperboloid. Using $r_{\rm hyp}(t) = (\alpha^2 + t^2)^{1/2}$, we find
\be
\begin{split}
\Delta S_{\rm hyp} &\equiv S_{\rm hyp}(t_{0})-S_{\rm hyp}(0)=-\frac{1}{2(D-1)G\hbar}\sum_{i}P^{titi}\Omega_{D-2}\left[r_{\rm hyp}^{D-2}(t_{0})-r_{\rm hyp}^{D-2}(0)\right]\\
&=-\frac{1}{2(D-1)G\hbar}\sum_{i}P^{titi}\Omega_{D-2}\left[(\alpha^{2}+t_{0}^{2})^{(D-2)/2}-\alpha^{(D-2)}\right]\;,
\end{split}
\ee
which precisely matches the leading-order part of the term, Eq. (\ref{leadingnpf}), we are trying to eliminate:
 \be 
  \Delta S_{\rm hyp}= \frac{1}{4G\hbar}\int_{\Sigma}dA d \tau n_{a}P^{abcd}f_{bcd}^{\mathcal{O}(-1)}\;.
  \ee
That is, the unwanted term is exactly equal to the entropy due to the natural expansion of the hyperboloid. This term should be subtracted from $\Delta S_{\rm tot}$ before equating it to $Q$. Moreover, note that here we did not specify the exact form of $P^{abcd}$, and therefore this subtraction holds for arbitrary theories of gravity.


\subsection*{Eliminating Higher Order Contributions}

Now we must deal with the higher order contributions, namely $\mathcal{O}(x)$ and $\mathcal{O}(x^{2})$. As alluded to above, in order to eliminate the higher order contributions to $n_{a}P^{abcd}f_{bcd}$, we consider a more generic $\xi_{a}$ and $n_{a}$, namely, 
\be
\begin{split}
\xi_{a}&=\xi^{(1)}_{a}+\xi^{(2)}_{a}+\xi^{(3)}_{a}+...\\
&=-r\delta_{ta}+\frac{tx^{i}}{r}\delta_{ia}+\frac{1}{2!}C_{\mu\nu a}x^{\mu}x^{\nu}+\tilde{C}_{\nu a}rx^{\nu}+\frac{1}{3!}D_{\mu\nu \rho a}x^{\mu}x^{\nu}x^{\rho}+\frac{1}{2!}\tilde{D}_{\mu\nu a}rx^{\mu}x^{\nu}+...
\end{split}
\ee
\be
\begin{split}
\alpha n_{a}&=\alpha(n^{(1)}_{a}+n^{(2)}_{a}+n^{(3)}_{a}+...)\\
&=-t\delta_{at}+x^{i}\delta_{ai}+\frac{1}{2!}C'_{\mu\nu a}x^{\mu}x^{\nu}+\frac{1}{3!}D'_{\mu\nu\rho a}x^{\mu}x^{\nu}x^{\rho}+...
\end{split}
\ee
Here we adopt the notation that $\mu,\nu,\rho...,$ represent the full spacetime index while $i,j,k,\ell,h$ represent spatial components, and where $\xi_{a}^{(\cdot)}$ denotes the order of the component; e.g., $\xi^{(1)}_{a}=-r\delta_{ta}+\frac{tx^{i}}{r}\delta_{ia}$ is of order $\mathcal{O}(x)$. 

Let us substitute our modified $\xi_{a}$ into our expression for $f_{bcd}$, for which we reproduce the simplified version here for convenience:
 \be
  f_{bcd}=\partial_{b}\partial_{c}\xi_{d}-\Gamma^{e}_{\;bc}\partial_{e}\xi_{d}-R^{e}_{\;bcd}\xi_{e}\;.\label{fbcdred}
  \ee
We have already worked out the $f^{\mathcal{O}(-1)}_{bcd}$ terms (\ref{fO-1bcd}).
 
 Next, the only possible term in $f_{bcd}$ of order $\mathcal{O}(1)$ is
 \be
  f^{\mathcal{O}(0)}_{bcd}\equiv\partial_{b}\partial_{c}\xi^{(2)}_{d}=C_{bcd}\;.
  \ee
Now let us work out the term in $f_{bcd}$ of order $\mathcal{O}(x)$. This will include a combination of terms including $\partial_{b}\partial_{c}\xi_{d}^{\mathcal{O}(3)}$, and the remaining terms in (\ref{fbcdred}) of order $\mathcal{O}(x)$, namely,
 \be
 \partial_{b}\partial_{c}\xi_{d}^{(3)}=D_{\nu bcd}x^{\nu}+r\tilde{D}_{bcd}+\tilde{D}_{\nu cd}(\partial_{b}r)x^{\nu}+\tilde{D}_{\nu bd}(\partial_{c}r)x^{\nu}+\frac{1}{2!}\tilde{D}_{\mu\nu d}x^{\mu}x^{\nu}(\partial_{b}\partial_{c}r)
 \ee
 \be
 -2\Gamma^{e}_{\;bc}(h)\partial_{[e}\xi^{\mathcal{O}(1)}_{d]}+\mathcal{O}(x^{2})
 \ee
 \be 
  R^{e}_{\;bcd}(p)\xi^{(1)}_{e}+\mathcal{O}(x^{2})\;,
  \ee
 where 
 \be 
  \Gamma^{e}_{\;bc}(h)\equiv\frac{1}{2}\eta^{ef}\left(\partial_{b}h_{cf}+\partial_{c}h_{bf}-\partial_{f}h_{bc}\right)
=-\frac{x^{\mu}}{3}\eta^{ef}(R_{c\mu fb}+R_{b\mu fc})\;,
  \ee
and we used $h_{ab}=-\frac{1}{3} R_{a\mu b\nu}x^{\mu}x^{\nu}$. Moreover,  since
  \be 
  \partial_{i}\xi_{t}^{\mathcal{O}(1)}=-\frac{x_{i}}{r}=-\partial_{t}\xi^{\mathcal{O}(1)}_{i}\;,
 \ee
  the only nonvanishing contribution to $\partial_{[e}\xi_{d]}$ is $\partial_{[i}\xi_{t]}=-\frac{x_{i}}{r}$.
  Altogether, one finds:
  \be
  \begin{split}
  f_{bcd}^{\mathcal{O}(1)}&=\partial_{b}\partial_{c}\xi^{\mathcal{O}(3)}_{d}-2\Gamma^{e}_{\;bc}(h)\partial_{[e}\xi^{\mathcal{O}(1)}_{d]}-R^{e}_{\;bcd}\xi_{e}^{\mathcal{O}(1)}\;.
  \end{split}
\ee
Note that this is the highest order of $f_{bcd}$ we need to keep since any higher order would give at least an $\mathcal{O}(x^{3})$ contribution to the integrand of the offending term, which we neglect. 
 
Recall that we need to eliminate (\ref{npf0}), (\ref{npf1}), and (\ref{npf2}) for an arbitrary $P^{abcd}$. We have already dealt with  (\ref{npf0}).
Before we go through the minutiae of these calculations, let us first explain the aim of the next two subsections providing us with a tether to hold onto as we work through the details.
 
The general prescription in eliminating the higher order contributions to $n_{a}P^{abcd}f_{bcd}$ is as follows. The integrand will include all sorts of monomial contributions, e.g., $t^{3}x_{i}x_{j}/r^{3}$. Since we care about the integral $\int_{\Sigma}n_{a}P^{abcd}f_{bcd}$ vanishing -- not the integrand -- we see that several of the monomials do not end up contributing to the final result; for example, $t^{3}x_{i}x_{j}/r^{3}$ will vanish for all $i\neq j$ as we are integrating over a sphere. Therefore we need only concern ourselves with, e.g., $t^{3}(x_{i})^{2}/r^{3}$. 
 
While these greatly reduce the number of monomial contributions, we still cannot fully eliminate the entire $\int_{\Sigma}n_{a}P^{abcd}f_{bcd}$. This is why we modify $\xi_{a}$ and $n_{a}$. More specifically, there are only a select few combinations of monomials which will appear in the integrand that do not vanish upon integration over the sphere. By modifying $\xi_{a}$ and $n_{a}$ we do not change the number of monomial contributions. Instead we find our modifications to $\xi_{a}$ and $n_{a}$ give us sets of coefficients that allow us the freedom to eliminate all other monomials, provided we have enough coefficients to do so. In short, we have a counting argument: If the number of nonvanishing monomials is less than the number of coefficients contributing to the same monomial, we can potentially force each monomial contribution to zero, i.e., $\int_{\Sigma}n_{a}P^{abcd}f_{bcd}\to0$ with a judicious choice of coefficients. 
 
In what follows we use this general prescription to separately eliminate monomials of order $\mathcal{O}(x)$ and $\mathcal{O}(x^{2})$. With the benefit of hindsight, we realize that only certain modifications to $\xi_{a}$ and $n_{a}$ will aid us, particularly,
\be
\begin{split}
\xi_{a}&=\xi^{(1)}_{a}+\xi^{(2)}_{a}+\xi^{(3)}_{a}+...\\
&=-r\delta_{ta}+\frac{tx^{i}}{r}\delta_{ia}+\tilde{C}_{\nu a}rx^{\nu}+\frac{1}{3!}D_{\mu\nu\rho a}x^{\mu}x^{\nu}x^{\rho}\;,
\end{split}
\ee
\be
\begin{split}
\alpha n_{a}&=\alpha(n^{(1)}_{a}+n^{(3)}_{a}+...)\\
&=-t\delta_{at}+x^{i}\delta_{ai}+\frac{1}{3!}D'_{\mu\nu\rho a}x^{\mu}x^{\nu}x^{\rho}\;.
\end{split}
\ee
As we will now explicitly show, this will be enough to cancel all undesired contributions coming from $\int_{\Sigma}n_{a}P^{abcd}f_{bcd}$ through $\mathcal{O}(x^{2})$. (Note that although we have set $n_{a}^{(2)}$ to zero, if we insist that $n_{a}$ be orthogonal to $\xi_{a}$ at order $\mathcal{O}(x^{3})$, we should include an $n^{(2)}_{a}$ contribution of the form $\tilde{C}'_{\nu a}tx^{\nu}$. It can be tediously verified that adding such terms to $n_{a}$ does not affect the counting argument, allowing us to leave them off in what follows.)

        
\subsection*{$\mathcal{O}(x)$ Contributions}
 
With the $n_{a}^{\mathcal{O}(2)}$ term being set to zero, the $\mathcal{O}(x)$ term to be eliminated becomes
\be
\frac{1}{4}\int_{\Sigma}dA d \tau\left(n^{\mathcal{O}(1)}_{a}P^{abcd}_{\mathcal{O}(1)}f^{\mathcal{O}(-1)}_{bcd}+n^{\mathcal{O}(1)}_{a}P^{abcd}_{\mathcal{O}(0)}f^{\mathcal{O}(0)}_{bcd}\right)\;.
\ee
Let us first list the various types of monomial contributions which might appear in the integrand:
\be
\mathcal{O}(x):\quad t,\;r,\;\frac{(x_{i})^{2}}{r},\;\frac{t^{2}(x_{i})^{2}}{r^{3}},\;\frac{(x_{i})^{2}(x_{j})^{2}}{r^{3}},\;\frac{(x_{i})^{4}}{r^{3}}\;.
\label{monoO1}\ee
As we will verify explicitly in a moment, only a subset of these monomials appear. Following the outlined prescription above, we need to check that we have enough coefficients to remove each of the monomial contributions. The only coefficients which will appear are those coming from the $f_{bcd}^{\mathcal{O}(0)}$ contribution, specifically $\tilde{C}_{na}$, for which we have $D^{2}$ coefficients. The number of problematic monomials which might appear is $1+1+1+(D-2)+(D-2)+\frac{1}{2}(D-1)(D-2)=D(D+1)/2<D^{2}$, for $D\geq3$. Therefore it already seems plausible that we will in fact have far more than enough coefficients to eliminate all of the monomial contributions appearing in the integrand.  Let us now verify this in detail. 

As was worked out in the previous section, we have
\be
 P^{abcd}f_{bcd}^{\mathcal{O}(-1)}=2P^{aitj}f^{\mathcal{O}(-1)}_{itj}=\frac{2}{r}P^{aitj}\left(\delta_{ij}-\frac{x_{i}x_{j}}{r^{2}}\right)\;.
\ee
Hence
\be
\begin{split}
n^{\mathcal{O}(1)}_{a}P^{abcd}_{\mathcal{O}(1)}f^{\mathcal{O}(-1)}_{bcd}&=\frac{2}{r}\left(\delta_{ij}-\frac{x_{i}x_{j}}{r^{2}}\right)\left[-\frac{t}{\alpha}P^{titj}_{\mathcal{O}(1)}+\frac{x_{k}}{\alpha}P^{kitj}_{\mathcal{O}(1)}\right]\\
&=\frac{2}{\alpha r}x_{k}\delta_{ij}P^{kitj}_{\mathcal{O}(1)}-\frac{2t}{\alpha r}\left(\delta_{ij}-\frac{x_{i}x_{j}}{r^{2}}\right)P^{titj}_{\mathcal{O}(1)}\;.
\end{split}
\ee
Defining
\be
P^{titj}_{\mathcal{O}(1)}\equiv\mathcal{P}_{\mathcal{O}(1),\mu}^{titj}x^{\mu}\quad P^{kitj}_{\mathcal{O}(1)}=\mathcal{P}^{kitj}_{\mathcal{O}(1),\mu}x^{\mu}\;,
\ee
we find that the only contributing terms to the integrand, i.e., those which do not vanish via parity arguments, are
\be
\begin{split}
n^{\mathcal{O}(1)}_{a}P^{abcd}_{\mathcal{O}(1)}f^{\mathcal{O}(-1)}_{bcd}&=-\frac{2}{\alpha r}\left(\delta_{ij}-\frac{x_{i}x_{j}}{r^{2}}\right)t^{2}\mathcal{P}^{titj}_{\mathcal{O}(1),t}+\frac{2}{\alpha r}\delta_{ij}x_{k}x^{\ell}\mathcal{P}^{kitj}_{\mathcal{O}(1),\ell}\;,
\end{split}
\ee
where we have used $x_{k}x_{i}P^{ikcd}=0$ using the symmetries of $P^{abcd}$. 

Generally, then, we see that only certain monomials appear which need to be removed. Specifically, 
\be
\begin{split}
n^{\mathcal{O}(1)}_{a}P^{abcd}_{\mathcal{O}(1)}f^{\mathcal{O}(-1)}_{bcd}&=\frac{A}{\alpha}\frac{t^{2}}{r}+\frac{A^{ii}}{\alpha}\frac{t^{2}(x_{i})^{2}}{r^{3}}+\frac{B^{ii}}{\alpha}\frac{(x_{i})^{2}}{r}\;,
\end{split}
\ee
where we have defined
\be
A\equiv-2\delta_{ij}\mathcal{P}^{titj}_{\mathcal{O}(1),t}\,,\quad A^{ii}\equiv2\mathcal{P}^{titi}_{\mathcal{O}(1),t}\,,\quad B^{k}_{\;\ell}\equiv 2\delta_{ij}\mathcal{P}^{kitj}_{\mathcal{O}(1),\ell}\;.
\ee
We now show that modifying $\xi_{a}$ via 
\be
\xi_{a}^{\mathcal{O}(2)}= r\tilde{C}_{\mu a}x^{\mu}
 \ee
will eliminate all the above undesired contributions. We have
\be 
\begin{split}
\partial_{b}\partial_{c}\xi^{\mathcal{O}(2)}_{d}&=\partial_{b}\left[\tilde{C}_{\mu d}(\partial_{c}r)x^{\mu}+\tilde{C}_{cd}r\right]\\
&=\tilde{C}_{\mu d}(\partial_{b}\partial_{c}r)x^{\mu}+\tilde{C}_{bd}(\partial_{c}r)+\tilde{C}_{cd}(\partial_{b}r)\;.
\end{split}
\ee
Then, using
\be
\partial_{i}r=\frac{x_{i}}{r}\,,\quad \partial_{i}\partial_{j}=\frac{1}{r}\left(\delta_{ij}-\frac{x_{i}x_{j}}{r^{2}}\right)\;,
\ee
we find
\be
\partial_{i}\partial_{j}\xi^{\mathcal{O}(2)}_{d}=\tilde{C}_{\mu d}\frac{x^{\mu}}{r}\left(\delta_{ij}-\frac{x_{i}x_{j}}{r^{2}}\right)+\tilde{C}_{id}\frac{x_{j}}{r}+\tilde{C}_{jd}\frac{x_{i}}{r}\;,
\ee
\be
\partial_{i}\partial_{t}\xi^{\mathcal{O}(2)}_{d}=\tilde{C}_{td}\frac{x_{i}}{r}\,,\quad \partial_{t}^{2}\xi^{\mathcal{O}(2)}_{d}=0\;.
\ee
Using these relations we find that
\be
\begin{split}
n^{\mathcal{O}(1)}_{a}P^{abcd}_{\mathcal{O}(0)}f_{bcd}^{\mathcal{O}(0)}&=\frac{1}{\alpha}\biggr\{-tP^{titj}_{\mathcal{O}(0)}(\partial_{t}\partial_{t}\xi^{\mathcal{O}(2)}_{j})-tP^{tijk}_{\mathcal{O}(0)}(\partial_{i}\partial_{j}\xi^{\mathcal{O}(2)}_{k})-tP^{tijt}_{\mathcal{O}(0)}(\partial_{i}\partial_{j}\xi^{\mathcal{O}(2)}_{t})\\
&+x_{i}P^{ijtk}_{\mathcal{O}(0)}(\partial_{j}\partial_{i}\xi_{k}^{\mathcal{O}(2)})+x_{i}P^{ijk\ell}_{\mathcal{O}(0)}(\partial_{j}\partial_{k}\xi^{\mathcal{O}(2)}_{\ell})+x_{i}P^{ijkt}_{\mathcal{O}(0)}(\partial_{i}\partial_{j}\xi_{t}^{\mathcal{O}(2)})\biggr\}\\
&=\frac{1}{\alpha r}\biggr\{-t^{2}\left(\delta_{ij}-\frac{x_{i}x_{j}}{r^{2}}\right)\left[\tilde{C}_{tk}P^{tijk}_{\mathcal{O}(0)}+\tilde{C}_{tt}P^{tijt}_{\mathcal{O}(0)}\right]\\
&+\left[\tilde{C}_{h\ell}P^{ijk\ell}_{\mathcal{O}(0)}+\tilde{C}_{ht}P^{ijkt}_{\mathcal{O}(0)}\right]\delta_{jk}x_{i}x^{h}+\left[\tilde{C}_{j\ell}P^{ijk\ell}_{\mathcal{O}(0)}+\tilde{C}_{jt}P^{ijkt}_{\mathcal{O}(0)}\right]x_{k}x_{i}\biggr\}\;.
\end{split}
\ee

Combining this with the term we wish to eliminate gives
\be
\left[\frac{A}{\alpha}-\frac{\delta_{ij}}{\alpha}(P^{tijt}_{\mathcal{O}(0)}\tilde{C}_{tt}+\tilde{C}_{tk}P^{tijk}_{\mathcal{O}(0)})\right]\frac{t^{2}}{r}
\ee
and
\be
\left[\frac{A^{ii}}{\alpha}+\frac{1}{\alpha}(\tilde{C}_{tt}P^{tiit}_{\mathcal{O}(0)}+\tilde{C}_{tk}P^{tiik}_{\mathcal{O}(0)})\right]\frac{t^{2}}{r^{3}}(x_{i})^{2}\;,
\ee
and last, 
\be
\left[\frac{B^{ii}}{\alpha}+\frac{1}{\alpha}(\tilde{C}^{i}_{\;\ell}P^{ijk\ell}_{\mathcal{O}(0)}+\tilde{C}^{i}_{\;t}P^{ijkt}_{\mathcal{O}(0)})\delta_{jk}+\frac{1}{\alpha}(\tilde{C}_{j\ell}P^{iji\ell}_{\mathcal{O}(0)}+\tilde{C}_{jt}P^{ijit}_{\mathcal{O}(0)})\right]\frac{(x_{i})^{2}}{r}\;.
\ee
The first two of these gives us $1+(D-2)=(D-1)$ monomials to cancel. But to remove these monomials, we have $1+(D-1)=D$ coefficients to work with, giving us  enough coefficients to cancel all of the undesired terms. Studying the problem at this level has provided us with insight that will prove useful when we study the elimination of $\mathcal{O}(x^{2})$ terms: (i) Not all of the possible monomials appear, and (ii) not all of the possible coefficients we have to work with will appear. Despite this we will still have enough coefficients to achieve our goal of removing $\int_{\Sigma}n_{a}P^{abcd}f_{bcd}$.


\subsection*{$(2+1)$-Dimensional $f(R)$-gravity: A Restrictive Case}

Based on the above calculation, however, it is clear that if one of the quantities multiplying a set of the coefficients vanishes, e.g., $P^{tijk}$, then we might be in trouble as we can no longer use these coefficients. This is precisely the case for $f(R)$ theories of gravity (except Einstein gravity, for which there is no $P^{abcd}_{\mathcal{O}(1)}$ contribution to be canceled and we can set all $\tilde{C}$ coefficients to zero). Thus, the most restrictive case is $(2+1)$-dimensional $f(R)$ gravity. Let us study this particular example explicitly and verify that we still have enough coefficients to eliminate all monomials. 

In $f(R)$ gravity one has
\be
P^{abcd}_{f(R)}=\frac{f'(R)}{2}(g^{ac}g^{bd}-g^{ad}g^{bc})\;.
\ee
So, 
\be
\begin{split}
&P^{abcd}_{f(R),\mathcal{O}(0)}=\frac{f'(R)(p)}{2}(\eta^{ac}\eta^{bd}-\eta^{ad}\eta^{bc})\,,\\
&P^{abcd}_{f(R),\mathcal{O(1)}}=\frac{f'(R)(x)}{2}(\eta^{ac}\eta^{bd}-\eta^{ad}\eta^{bc})\equiv \mathcal{P}^{abcd}_{\mathcal{O}(1),\mu}x^{\mu}\;,
\end{split}
\ee
where $p$ is the spacetime point where these expressions are being evaluated. This tells us that $B^{ii}=0$, leaving
\be
\left[\frac{A}{\alpha}-\frac{\delta_{ij}}{\alpha}P^{tijt}_{\mathcal{O}(0)}\tilde{C}_{tt}\right]\frac{t^{2}}{r}
\ee
and
\be
\left[\frac{A^{ii}}{\alpha}+\frac{1}{\alpha}\tilde{C}_{tt}P^{tiit}_{\mathcal{O}(0)}\right]\frac{t^{2}}{r^{3}}(x_{i})^{2}\;,
\ee
where
\be
A=-2\delta_{ij}\mathcal{P}^{titj}_{\mathcal{O}(1),t}\,,\quad A^{ii}=\mathcal{P}^{titi}_{\mathcal{O}(1),t}\;.
\ee
Expanding our above expressions in a $(2+1)$-dimensional spacetime yields
\be
\frac{1}{\alpha}\left[-2(\mathcal{P}^{txtx}_{\mathcal{O}(1),t}+\mathcal{P}^{tyty}_{\mathcal{O}(1),t})+\tilde{C}_{tt}(P^{txtx}_{\mathcal{O}(0)}+P^{tyty}_{\mathcal{O}(0)})\right]\frac{t^{2}}{r}
\ee
and
\be
\frac{1}{\alpha}\left[2(\mathcal{P}^{txtx}_{\mathcal{O}(1),t}x^{2}+\mathcal{P}^{tyty}_{\mathcal{O}(1),t}y^{2})-\tilde{C}_{tt}(P^{txtx}_{\mathcal{O}(0)}x^{2}+P^{tyty}_{\mathcal{O}(0)}y^{2})\right]\frac{t^{2}}{r^{3}}
\ee

Each of these must vanish separately. Using that 
\be
 P^{txtx}_{\mathcal{O}(0)}=P^{tyty}_{\mathcal{O}(0)}\,,\quad \mathcal{P}^{txtx}_{\mathcal{O}(1),t}=\mathcal{P}^{tyty}_{\mathcal{O}(1),t}\;,
 \ee
 we are led to
 \be
 \frac{1}{\alpha}\left(-4\mathcal{P}^{titi}_{\mathcal{O}(1),t}+2\tilde{C}_{tt}P^{titi}_{\mathcal{O}(0)}\right)\frac{t^{2}}{r}\;,	 \label{measurezeroex}
 \ee
 \be
 \frac{1}{\alpha}\left(2\mathcal{P}^{titi}_{\mathcal{O}(1),t}-\tilde{C}_{tt}P^{titi}_{\mathcal{O}(0)}\right)\frac{t^{2}(x^{2}+y^{2})}{r^{3}}\;.
 \ee
Since $x^{2}+y^{2}=r^{2}$, we find that the above two conditions are in fact the same; miraculously the monomials add in such a way that we need only a single coefficient. (In fact, this feature of two seemingly different conditions becoming one can readily be obtained in this case if one uses the fact that  $P^{titj}_{\mathcal{O}(0)}\left(\delta_{ij}-\frac{x_{i}x_{j}}{r^{2}}\right)=-\frac{f'(R)(p)}{2}(D-2)$ from the start.) Finally, it is possible in principle that, say, $P^{titi}_{\mathcal{O}(0)}$ vanishes while $\mathcal{P}^{titi}_{\mathcal{O}(1),t}$ does not, preventing (\ref{measurezeroex}) from being set to zero. However, inspecting (\ref{measurezeroex}), it is easy to see that this can happen at most on a set of measure zero.

 
\subsection*{$\mathcal{O}(x^{2})$ Contributions}

Let us now move on to the $\mathcal{O}(x^{2})$ contribution to $n_{a}P^{abcd}f_{bcd}$ where the story and prescription are the same, though far more tedious to work out. Setting $n_{a}^{\mathcal{O}(2)}$ to zero means that we must eliminate
 \be
 \begin{split}
\frac{1}{4} \int_{\Sigma}dA d \tau&\biggr\{\sqrt{h}n^{\mathcal{O}(1)}_{a}P^{abcd}_{\mathcal{O}(0)}f^{\mathcal{O}(-1)}_{bcd}+n^{\mathcal{O}(1)}_{a}P^{abcd}_{\mathcal{O}(2)}f^{\mathcal{O}(-1)}_{bcd}+n^{\mathcal{O}(1)}_{a}P^{abcd}_{\mathcal{O}(1)}f^{\mathcal{O}(0)}_{bcd}+n^{\mathcal{O}(1)}_{a}P^{abcd}_{\mathcal{O}(0)}f^{(1)}_{bcd}\\
 &+n^{\mathcal{O}(3)}_{a}P^{abcd}_{\mathcal{O}(0)}f^{\mathcal{O}(-1)}_{bcd}\biggr\}\;.
 \end{split}
\ee
At the $\mathcal{O}(x^{2})$ level, the only monomials which might appear are
\be
 t^{2},\;(x_{i})^{2},\;\frac{t(x_{i})^{2}}{r},\;\frac{t^{5}}{r^{3}},\;\frac{t^{3}(x_{i})^{2}}{r^{3}},\;\frac{t(x_{i})^{4}}{r^{3}},\;\frac{t(x_{i})^{2}(x_{j})^{2}}{r^{3}} \; ,
\label{monoO2}\ee
giving us a total of $1+(D-1)+(D-1)+1+(D-1)+1/2(D-1)(D-2)=D(D+3)/2$. Naively we have far more coefficients to work with; e.g., in $\tilde{D}_{\mu\nu a}$ alone we have $D^{3}$ coefficients to use. However, as observed at the $\mathcal{O}(x)$ level, only a subset of the monomials and coefficients will appear. 

After much tedious algebra, one finds that the $n_{a}P^{abcd}f_{bcd}$ terms at the $\mathcal{O}(x^{2})$ level are
 \be
\begin{split}
&n_{a}P^{abcd}f_{bcd}=\frac{1}{\alpha}\biggr\{X+\frac{1}{2}P^{titj}_{\mathcal{O}(0)}\delta_{ij}\tilde{D}_{ttt}-\frac{1}{2}P^{tijk}_{\mathcal{O}(0)}\tilde{D}_{ttk}+\frac{1}{3}(D'_{tttt}P^{titj}_{\mathcal{O}(0)}\delta_{ij}+D'_{tttk}P^{kitj}_{\mathcal{O}(0)}\delta_{ij})\biggr\}\frac{t^{3}}{r}\\
&+\frac{1}{\alpha}\biggr\{Y^{ii}+\frac{1}{2}P^{tiik}_{\mathcal{O}(0)}\tilde{D}_{ttk}-\frac{1}{2}P^{titi}_{\mathcal{O}(0)}\tilde{D}_{ttt}-\frac{1}{3}\left(D'_{tttt}P^{titi}_{\mathcal{O}(0)}+D'_{tttk}P^{kiti}_{\mathcal{O}(0)}\right)\biggr\}\frac{(x_{i})^{2}t^{3}}{r^{3}}\\
&+\frac{1}{\alpha}\biggr\{Z^{iikk}-\frac{1}{2}\tilde{D}^{kk}_{\;\;\;t}P^{titi}_{\mathcal{O}(0)}-2\tilde{D}^{ki}_{\;\;\;t}P^{titk}_{\mathcal{O}(0)}\\
&-2\left(D'^{kk}_{\;\;\;tt}P^{titi}_{\mathcal{O}(0)}+2D'^{ik}_{\;\;\;tt}P^{titk}_{\mathcal{O}(0)}+D'^{kk}_{\;\;\;t\ell}P^{\ell iti}_{\mathcal{O}(0)}+2D'^{ik}_{\;\;\;t\ell}P^{\ell itk}_{\mathcal{O}(0)}\right)\biggr\}\frac{(x_{k})^{2}(x_{i})^{2}t}{r^{3}}\\
&+\frac{1}{\alpha}\left(\mathcal{X}-P^{tijk}_{\mathcal{O}(0)}\tilde{D}_{ijk}-P^{titj}_{\mathcal{O}(0)}(\tilde{D}_{itj}-\tilde{D}_{ijt})\right)rt+\frac{1}{\alpha}\biggr\{W^{kk}+P^{kjk\ell}_{\mathcal{O}(0)}\tilde{D}_{tj\ell}\\
&+P^{kji\ell}_{\mathcal{O}(0)}\delta_{ij}\tilde{D}^{k}_{\;t\ell}-P^{tktk}_{\mathcal{O}(0)}\tilde{D}_{ttt}-(P^{tkij}_{\mathcal{O}(0)}+P^{tikj}_{\mathcal{O}(0)})\tilde{D}^{k}_{\;ij}-P^{tktj}_{\mathcal{O}(0)}(\tilde{D}^{k}_{\;tj}-\tilde{D}^{k}_{\;jt})\\
&+\frac{1}{2}P^{titj}_{\mathcal{O}(0)}\delta_{ij}\tilde{D}^{kk}_{\;\;\;t}+2\left(D'^{kk}_{\;\;\;tt}P^{titj}_{\mathcal{O}(0)}\delta_{ij}+D'^{kk}_{\;\;\;t\ell}P^{\ell itj}_{\mathcal{O}(0)}\delta_{ij}\right)\biggr\}\frac{(x_{k})^{2}t}{r}\;,
\end{split}
\ee
where $X, Y^{ii}, Z^{iikk}, \mathcal{X}$, and $W^{kk}$ are some messy collection of constants independent of the $\tilde{D}$ and $D'$ coefficients. 

From counting one finds that there are more than enough coefficients to remove all of the undesired monomial expressions for arbitrary theories of gravity, and, even in the most restrictive case of $(2+1)$-dimensional $f(R)$ gravity, we will still find that we have just enough coefficients to remove all of the undesired monomials. 

To see how even the most restrictive case is satisfied, it suffices to study only a single contribution from $n_{a}^{\mathcal{O}(1)}P^{abcd}_{\mathcal{O}(0)}f^{\mathcal{O}(1)}_{bcd}$,
\be
\begin{split}
n^{\mathcal{O}(1)}_{a}P^{abcd}_{\mathcal{O}(0)}f^{\mathcal{O}(1)}_{bcd}&=-\frac{t}{\alpha}\left[P^{tijk}_{\mathcal{O}(0)}f^{\mathcal{O}(1)}_{ijk}+P^{titj}_{\mathcal{O}(0)}(f^{\mathcal{O}(1)}_{itj}-f^{\mathcal{O}(1)}_{ijt})\right]\\
&+\frac{x_{i}}{\alpha}\left[P^{ijk\ell}_{\mathcal{O}(0)}f^{\mathcal{O}(1)}_{jk\ell}+P^{itkt}_{\mathcal{O}(0)}(f^{\mathcal{O}(1)}_{tkt}-f^{\mathcal{O}(1)}_{ttk})+P^{ijtk}_{\mathcal{O}(0)}(f^{\mathcal{O}(1)}_{jtk}-f^{\mathcal{O}(1)}_{jkt})\right]\;.
\end{split}
\ee
In particular, we need only study the first line. After much algebra we find
  \be
 \begin{split}
& -\frac{t}{\alpha}P^{titj}_{\mathcal{O}(0)}(f^{\mathcal{O}(1)}_{itj}-f_{ijt}^{\mathcal{O}(1)})=\frac{1}{\alpha}\left[\mathcal{F}-P^{titj}_{\mathcal{O}(0)}(\tilde{D}_{itj}-\tilde{D}_{ijt})\right]rt\\
&+\frac{1}{2\alpha}\tilde{D}_{ttt}P^{titj}_{\mathcal{O}(0)}\delta_{ij}\frac{t^{3}}{r}-\frac{1}{2\alpha}P^{titi}_{\mathcal{O}(0)}\tilde{D}_{ttt}\frac{(x_{i})^{2}t^{3}}{r^{3}}-\frac{1}{2\alpha}(\tilde{D}^{kk}_{\;\;t}P^{titi}_{\mathcal{O}(0)}+4\tilde{D}^{ki}_{\;\;\;t}P^{titk}_{\mathcal{O}(0)})\frac{(x_{k})^{2}(x_{i})^{2}t}{r^{3}}\\
&-\frac{1}{\alpha}\left[\mathcal{M}^{kk}+P^{tktj}_{\mathcal{O}(0)}(\tilde{D}^{k}_{\;tj}-\tilde{D}^{k}_{\;jt})-\frac{1}{2}P^{titj}_{\mathcal{O}(0)}\delta_{ij}\tilde{D}^{kk}_{\;\;\;t}\right]\frac{(x_{k})^{2}t}{r}\;,
 \end{split}
 \label{n1p0f1-2exp}\ee
 where we have defined
 \be
 \mathcal{M}^{kk}\equiv \frac{4}{3}P^{titj}_{\mathcal{O}(0)}R_{i\;\;\;j}^{\;kk}(p)\,,\quad \mathcal{F}\equiv P^{titj}_{\mathcal{O}(0)}(R_{titj}(p)-R_{tijt}(p))\;.
 \ee
 Consider a $(2+1)$-dimensional spacetime. We immediately see that
 \be
 \frac{1}{2\alpha}\tilde{D}_{ttt}P^{titj}_{\mathcal{O}(0)}\delta_{ij}\frac{t^{3}}{r}-\frac{1}{2\alpha}P^{titi}_{\mathcal{O}(0)}\tilde{D}_{ttt}\frac{(x_{i})^{2}t^{3}}{r^{3}}
 \ee
 cancel each other. This is fine as it only depends on a single coefficient $\tilde{D}_{ttt}$. We have 
 \be
 \begin{split}
 \frac{1}{\alpha}\left[\mathcal{F}-P^{titj}_{\mathcal{O}(0)}(\tilde{D}_{itj}-\tilde{D}_{ijt})\right]rt&=\frac{1}{\alpha}\left[\mathcal{F}-P^{titi}_{\mathcal{O}(0)}\left(\tilde{D}_{xtx}-\tilde{D}_{xxt}+\tilde{D}_{yty}-\tilde{D}_{yyt}\right)\right]rt\;,
 \end{split}
 \ee
 \be
 \begin{split}
 -\frac{1}{2\alpha}(\tilde{D}^{kk}_{\;\;t}P^{titi}_{\mathcal{O}(0)}+4\tilde{D}^{ki}_{\;\;\;t}P^{titk}_{\mathcal{O}(0)})\frac{(x_{k})^{2}(x_{i})^{2}t}{r^{3}}&=-\frac{1}{2\alpha}\biggr\{5\tilde{D}_{xxt}x^{4}+5\tilde{D}_{yyt}y^{4}\\
&+(\tilde{D}_{xxt}+\tilde{D}_{yyt})x^{2}y^{2}\biggr\}\frac{t}{r^{3}}\;,
 \end{split}
 \ee
 and
 \be
 \begin{split}
& -\frac{1}{\alpha}\left[\mathcal{M}^{kk}+P^{tktj}_{\mathcal{O}(0)}(\tilde{D}^{k}_{\;tj}-\tilde{D}^{k}_{\;jt})-\frac{1}{2}P^{titj}_{\mathcal{O}(0)}\delta_{ij}\tilde{D}^{kk}_{\;\;\;t}\right]\frac{(x_{k})^{2}t}{r}=-\frac{1}{\alpha}\left(\frac{4}{3}P^{titi}_{\mathcal{O}(0)}R_{yxxy}(p)\right)rt\\
 &-\frac{1}{\alpha}P^{titi}_{\mathcal{O}(0)}\left[(\tilde{D}_{xtx}-\tilde{D}_{xxt})\frac{x^{2}t}{r}+(\tilde{D}_{yty}-\tilde{D}_{yyt})\frac{y^{2}t}{r}\right]\;.
 \end{split}
 \ee
 Let us now set $\tilde{D}_{kkt}=0$. This choice yields the two expressions
 \be
 \begin{split}
 \frac{1}{\alpha}\left[\mathcal{F}-P^{titj}_{\mathcal{O}(0)}(\tilde{D}_{itj}-\tilde{D}_{ijt})\right]rt&=\frac{1}{\alpha}\left[\mathcal{F}-P^{titi}_{\mathcal{O}(0)}\left(\tilde{D}_{xtx}+\tilde{D}_{yty}\right)\right]rt
 \end{split}
 \ee
 and
 \be
 \begin{split}
& -\frac{1}{\alpha}\left[\mathcal{M}^{kk}+P^{tktj}_{\mathcal{O}(0)}(\tilde{D}^{k}_{\;tj}-\tilde{D}^{k}_{\;jt})-\frac{1}{2}P^{titj}_{\mathcal{O}(0)}\delta_{ij}\tilde{D}^{kk}_{\;\;\;t}\right]\frac{(x_{k})^{2}t}{r}\\
&=-\frac{1}{\alpha}\left(\frac{4}{3}P^{titi}_{\mathcal{O}(0)}R_{yxxy}(p)\right)rt-\frac{1}{\alpha}P^{titi}_{\mathcal{O}(0)}\left[\tilde{D}_{xtx}\frac{x^{2}t}{r}+\tilde{D}_{yty}\frac{y^{2}t}{r}\right]\;.
 \end{split}
 \ee
 Let us further choose that $\tilde{D}_{xtx}=\tilde{D}_{yty}\equiv\tilde{D}$. The second expression then becomes
 \be
 -\frac{1}{\alpha}\left(\frac{4}{3}P^{titi}_{\mathcal{O}(0)}R_{yxxy}(p)\right)rt-\frac{1}{\alpha}P^{titi}_{\mathcal{O}(0)}\tilde{D}rt\;.
 \ee
 Defining $4/3P^{titi}_{\mathcal{O}(0)}R_{yxxy}(p)\equiv\mathcal{M}$, we find that the following combination must be made to vanish:
 \be
 -\frac{1}{\alpha}\left[\mathcal{M}-\mathcal{F}+3P^{titi}_{\mathcal{O}(0)}\tilde{D}\right]rt
 \ee
We have the freedom to choose $\tilde{D}$ such that this monomial vanishes. 
 
 The reason this specific case is enough to show that there are enough coefficients to remove all of the $\mathcal{O}(x^{2})$ monomial contributions to $\int_{\Sigma}n_{a}P^{abcd}f_{bcd}$ is that every type of possible monomial is present. Any additional contributions which come into play can easily be handled by (i) altering the choice of $\tilde{D}_{\mu\nu a}$, and (ii) having the presence of $\tilde{D}'_{\mu\nu\rho a}$ coefficients. The only monomial which might give us pause is that proportional to $t(x_{i})^{2}/r$, as the $\tilde{D}_{ttt}$ happened to exactly cancel. It turns out, however, that there are enough $D'$ coefficients to deal with these monomials.
 
 In summary, by modifying $\xi_{a}$ and $n_{a}$, we have more than enough coefficients to remove all of the monomial contributions to $n_{a}P^{abcd}f_{bcd}$ that do not vanish due to integration over the sphere, through the $\mathcal{O}(x^{2})$ level. Therefore, while there might be $\mathcal{O}(x^{3})$ contributions to the integrand, these terms are sufficiently smaller than those we wish to keep in the equations of motion, allowing us to effectively neglect the undesired contribution $\int_{\Sigma}n_{a}P^{abcd}f_{bcd}$. 

 
 \subsection*{Eliminating $q^{a}$}
\indent
Last, let us discuss how to eliminate another unwanted term,
\be
-\frac{1}{4G\hbar}\int_{\Sigma}dAd\tau n_{a}q^{a} \; ,
\ee
where $q^{a}=\nabla_{b}(P^{adbc}+P^{acbd})\nabla_{c}\xi_{d}$. This term is only present for non-Lovelock theories of gravity, such as non-Einstein $f(R)$ gravity. Only the symmetric parts of $\nabla_{c}\xi_{d}$ survive the contraction. From (\ref{Killingfailure}), we see that the symmetric parts have both ${\mathcal O}(x^2)$ and ${\mathcal O}(1)$ parts. Since $n^a$ is of order $x$, the ${\mathcal O}(x^2)$ part of $q^a$ gives a term in $n_a q^a$ of order $x^3$, and we can therefore neglect it. But the ${\mathcal O}(1)$ $i-j$ contributions cannot be neglected outright:
\be
-\frac{1}{4G\hbar}\int_{\Sigma}d\Sigma_{a}\nabla_{b}(P^{aibj})(\nabla_{i}\xi_{j}+\nabla_{j}\xi_{i})\;.
\label{failkillingeqn}
\ee
To match our approximations we must therefore eliminate this contribution for non-Lovelock theories of gravity. This is indeed possible, as we now show. Because of the form, Eq. (\ref{Killingfailure}), of $\nabla_{(i}\xi_{j)}$, terms with $i\neq j$ integrate to zero in
(\ref{failkillingeqn}). When $i=j$, the integrand is of $\mathcal{O}(x)$ for the combination $n^{(1)}_{t}(\nabla_b P^{tibi}_{\mathcal{O}(0)})\nabla_{i}\xi_{i}$. This yields two types of monomials:
\be
 \frac{t^{2}}{r},\quad \frac{t^{2}(x_{i})^{2}}{r^{3}}\;.
\ee
However, precisely these monomials already appear in (\ref{monoO1}). They can therefore be absorbed in the $\mathcal{O}(x)$ contributions to $n_{a}P^{abcd}f_{bcd}$ that have already been shown to be eliminated; the counting argument discussed at length above is not altered. 
The integrand of (\ref{failkillingeqn}) will be of ${\mathcal O}(x^2)$ in two ways: (i) $n_{a}^{(2)}(\nabla_{b}P^{aibj})^{(0)}\nabla_{(i}\xi_{j)}$, or (ii) $n_{a}^{(1)}(\nabla_{b}P^{aibj})^{(1)}\nabla_{(i}\xi_{j)}$. Together, the only monomials that appear are
\be
\frac{t^{3}}{r},\quad \frac{t^{3}(x_{i})^{2}}{r^{3}}, \quad \frac{t(x_{i})^{2}}{r},\quad \frac{t(x_{i})^{2}(x_{j})^{2}}{r^{3}}
\ee
matching the monomials already appearing in (\ref{monoO2}). In summary, the terms appearing in (\ref{failkillingeqn}) can be readily eliminated by the coefficients we use to dispose of similar terms in $n_{a}P^{abcd}f_{bcd}$, without altering the counting.


\subsection*{Equating Integrands} \label{app:eqlintegrands}
\noindent
We have seen that Clausius' theorem, $Q = \Delta S_{\rm rev}/T$, leads to an equality between integrals of the form
\be
\int_{\Sigma}dAd \tau A_{ab}\xi^{a}n^{b}=\int_{\Sigma}dAd \tau T_{ab}\xi^{a}n^{b} \; . \label{equalintegrals}
\ee
For Einstein gravity, $A_{ab} = \frac{1}{8 \pi G} R_{ab}$, while for general theories of gravity, $A_{ab}$ can be read off from the left-hand side of (\ref{Clausiusgeneral}).
In this appendix, we show that the equality of integrals (\ref{equalintegrals}) implies the equality of their integrands:
\be
A_{ab} \xi^{a}n^{b}=T_{ab}\xi^{a}n^{b}\;.
\ee
Ordinarily, the equality of integrands follows from the equality of integrals if the boundaries of the domain of integration can be suitably varied without affecting the equality of the integrals. 

Defining the symmetric matrix $M_{ab}\equiv A_{ab}-T_{ab}$, and with the proper time element on the hyperboloid given by $d \tau = dt \alpha/r$, we can write (\ref{equalintegrals}) as
\be
0=\int^{\epsilon}_{0}dt \frac{\alpha}{r(t)} \int_{\omega(t)}dAM_{ab}\xi^{a}n^{b}\;.
\ee
We would like to conclude from this that $M_{ab} \xi^a n^b = 0$. Because $\epsilon$ is arbitrary, for this integral to vanish for all values of $\epsilon$, the standard argument from calculus implies that the integrand must itself be zero:
\be
0= \int_{\omega(t)} dA M_{ab}\xi^{a}n^{b} \; ,
\ee
for all spheres $\omega(t)$. However, we cannot apply the same argument to this integral because a sphere has no boundary to vary.

Expanding the integrand gives
\be
 0=\int dA\left[M_{00}rt+M_{0i}tx^{i}\left(1+\frac{t}{r}\right)+M_{ii}\frac{t(x^{i})^{2}}{r}+M_{ij,i\neq j}\frac{tx^{i}x^{j}}{r}\right]\;.
 \ee
Integration over the sphere causes the terms in the integrand proportional to odd powers of $x^i$ to automatically vanish, telling us nothing about $M_{ij,i\neq j}$ and $M_{0i}$. We see, however, that the other components must obey the condition
\be
M_{00}+\frac{1}{(D-1)}\sum_{i}M_{ii}=0\;.\label{cond1}
\ee

To proceed, note that (\ref{equalintegrals}) also holds for a different hyperboloid, $\Sigma'$, obtained by an active Lorentz transformation of $\Sigma$. This active transformation does not affect the matrix $M$, whose elements are evaluated at $p$, but transforms the vectors $\xi$ and $n$ to $\xi'$ and $n'$. We then follow this with a passive Lorentz transformation on the coordinates such that the components of the new $\xi'$ and $n'$ are the same as the original components of the old $\xi$ and $n$. Under a passive Lorentz transformation, $M$ transforms as a matrix, and we have
\be
\begin{split}
& 0=\int_{\Sigma'}dAdt\frac{\alpha}{r} M'_{ab}\xi^{a}n^{b} \Rightarrow\\
& 0 = \int dA\left[M'_{00}rt+M'_{0i}tx^{i}\left(1+\frac{t}{r}\right)+M'_{ii}\frac{t(x^{i})^{2}}{r}+M'_{ij,i\neq j}\frac{tx^{i}x^{j}}{r}\right]
\end{split}
 \ee
from which we find
\be
M'_{00}+\frac{1}{(D-1)}\sum_{i}M'_{ii}=0\;. \label{cond2}
\ee
We now show that (\ref{cond1}) and (\ref{cond2}) are enough to claim $M_{ab}\propto\eta_{ab}$. Perform a Lorentz transformation in the $0-1$ plane. Then applying (\ref{cond1}) and (\ref{cond2}) leads to
\be
M_{00}=-M_{11}-\frac{2\beta\gamma^{2}}{(1-\gamma^{2})}M_{01}\;.
\ee
For this to hold for all $\beta$, we conclude that $M_{01}=0$. Moreover, $M_{00}=-M_{11}$. A similar argument holds for Lorentz boosts in other planes, and therefore, $M_{00}=-M_{11}=-M_{22}=...$, and $M_{0i}=0$. It is also straightforward to show that $M_{ij}=0$ for $i\neq j$ by first performing a rotation on $M_{ab}$, and then a Lorentz boost. In summary, we find that $M_{ab}$ is a diagonal matrix with $M_{00}=-M_{ii}$. Hence $M_{ab}\propto\eta_{ab}$.
But since $\eta_{ab} \xi^a n^b = 0$, we find 
\be
 M_{ab}\xi^{a}n^{b}=0\;,
 \ee
as desired.


\subsection{Causal Diamonds} \label{sec:appendB}
\indent

In our derivation of the gravitational equations of motion via the thermodynamics of causal diamonds, we made use of the conformal Killing equation
\beq \nabla_{a}\zeta_{b}+\nabla_{b}\zeta_{a}=2\Omega g_{ab}\;,\eeq
and the conformal Killing identity
\beq \nabla_{b}\nabla_{c}\zeta_{d}=R^{e}_{\;bcd}\zeta_{e}+(\nabla_{c}\Omega)g_{bd}+(\nabla_{b}\Omega)g_{cd}-(\nabla_{d}\Omega)g_{bc}\;.\eeq
An arbitrary spacetime, however, does not admit a global conformal Killing vector, therefore $\zeta^{a}$ can be understood as an \emph{approximate} conformal Killing vector. More precisely, $\zeta_{a}$ will fail to be a conformal Killing vector to some order in a Riemann normal coordinate expansion of the arbitrary spacetime (\ref{RNC}). The order at which these quantities fail depends on the order of the vector itself. The conformal Killing vector $\zeta^{a}$ we used 
\beq
\begin{split}
 \zeta^{a}&=\left(\frac{\ell^{2}-r^{2}-t^{2}}{\ell^{2}}\right)\partial^{a}_{t}-\frac{2rt}{\ell^{2}}\partial^{a}_{r}\\
& = \left(\frac{\ell^{2}-r^{2}-t^{2}}{\ell^{2}}\right)\partial^{a}_{t}-\frac{2x^{i}t}{\ell^{2}}\partial^{a}_{i}\;,
\end{split}
\eeq
with $\Omega=-2t/\ell^{2}$, was specific to $D$-dimensional Minkowski space, and is of order $\zeta^{a}=\mathcal{O}(0)+\mathcal{O}(x^{2})$, where the $\mathcal{O}(0)$ contribution is a constant. From this one finds that in an arbitrary spacetime $\zeta_{a}$ will fail the conformal Killing equation to order $\mathcal{O}(x)+\mathcal{O}(x^{3})$ and the Killing identity to order $\mathcal{O}(0)+\mathcal{O}(x^{2})$. Note that the term we keep in deriving the equations of motion, namely the integrand of\footnote{Here we ignore the vector $N_{a}$ since it will be contracted with all terms in the integrand, including the higher order contributions we neglected.}
\beq
 \int_{\Sigma}d\Sigma_{a}\left(P^{abcd}R_{ebcd}\zeta^{e}-2\zeta_{d}\nabla_{b}\nabla_{c}P^{abcd}\right)\;, \label{intdes}\eeq
is, $\mathcal{O}(0)+\mathcal{O}(x^{2})$. However, since $d\Sigma_{a}=N_{a}dAd\tau$, with $N_{a}\propto x_{i}/r$, the $\mathcal{O}(0)$ contributions vanish due to the fact we are integrating over a spherical subregion for which $\int_{\partial B}x_{i}dA=0$. Therefore, we need only concern ourselves with the $\mathcal{O}(x^{2})$ contributions coming from the failure of the conformal Killing identity. 

We realize, in fact, that the only contribution of the conformal Killing identity we made use of was the term proportional to the Riemann tensor, $R_{ebcd}\zeta^{e}$ -- we neglected all other contributions. This means that we effectively treated $\zeta^{a}$ as an approximate Killing vector rather than an approximate conformal Killing vector. We therefore find ourselves in a similar situation as the authors of \cite{Parikh:2017aas}: We must remove the higher order contributions coming from the failure of Killing's identity. Specifically, in the integrand (\ref{intdes}), the term $P^{abcd}\nabla_{b}\nabla_{c}\zeta_{d}$ should be replaced with
 \beq P^{abcd}\nabla_{b}\nabla_{c}\zeta_{d}=P^{abcd}R_{ebcd}\zeta^{e}+P^{abcd}f_{bcd}\;,\eeq
with
\beq f_{bcd}=\nabla_{b}\nabla_{c}\zeta_{d}-R_{ebcd}\zeta^{e}-(\nabla_{c}\Omega)g_{bd}+(\nabla_{d}\Omega)g_{bc}\;,\eeq
from which we see that $f_{bdc}=-f_{bcd}$. Here $f_{bcd}$ quantifies the failure of Killing's identity. Our task is therefore to find a way to eliminate
\beq \int_{\Sigma}d\Sigma_{a}P^{abcd}f_{bcd}\;,\label{undesired}\eeq
at least to the order at which we keep the desired contribution $\int_{\Sigma} d\Sigma_{a}P^{abcd}R_{ebcd}\zeta^{e}$. Specifically the integrand we wish to keep
\beq N_{a}P^{abcd}R_{bcde}\zeta^{e}\;,\eeq
goes like $\mathcal{O}(0)+\mathcal{O}(x^{2})$. The $\mathcal{O}(0)$ contribution, as mentioned above, vanishes due to the fact we are integrating over a spherical subregion. Therefore, the order of the integrand we are interested in keeping is $\mathcal{O}(x^{2})$, and we must remove the $\mathcal{O}(x^{2})$ contributions of the undesired term. 

To study this problem we introduce the notation
\beq f_{bcd}=f^{(0)}_{bcd}+f_{bcd}^{(1)}+f_{bcd}^{(2)}+...\;,\eeq
where $f^{(0)}_{bcd}$ denotes the $\mathcal{O}(0)$ contribution to $f_{bcd}$, $f^{(1)}_{bcd}$ the $\mathcal{O}(x)$ contribution, and so forth. We will use this notation to decompose each object appearing in the integrand (\ref{undesired}), i.e., $N_{a}=N_{a}^{(0)}$, and $P^{abcd}=P^{abcd}_{(0)}+P^{abcd}_{(1)}+...$. 

In order to remove contribution (\ref{undesired}) to the desired order, we will follow the method developed in \cite{Parikh:2017aas}, by modifying $\zeta_{a}$ and $N_{a}$, by adding undetermined higher order contributions to $\zeta_{a}$. The algorithm for removing the terms can be described as follows: The integrand of (\ref{undesired}) is a collection of monomials. Because we are integrating over a spherical subregion, many of these monomial contributions will vanish, e.g., when the integrand goes like $tx_{i}/r$. Some terms will remain, however, and the only way to remove these contributions is to add in higher order modifications to $\zeta^{a}$, e.g., 
\beq \zeta_{a}=\left(\frac{\ell^{2}-r^{2}-t^{2}}{\ell^{2}}\right)\partial^{a}_{t}-\frac{2x^{i}t}{\ell^{2}}\partial^{a}_{i}+\frac{1}{3!}D_{a\mu\nu\rho}x^{\mu}x^{\nu}x^{\rho}+...\;,\eeq
where here the greek indices $\mu,\nu$ run over the whole spacetime index. We can likewise modify $N_{a}$. These modifications to $\zeta_{a}$ will include additional contributions to $f_{bcd}$ of the same monomial structure as before. We then choose the undetermined coefficients $D_{a\mu\nu\rho}$, etc. so as to cancel these terms. In essence we add counterterms to $\zeta_{a}$ to remove (\ref{undesired}) to the desired order. One problem which may arise is whether there are enough undetermined coefficients to cancel all of the monomials which may appear. 

Putting all of this together, the lowest order contribution in the integrand of the offending term (\ref{undesired}) is
\beq \int_{\Sigma}dAd\tau n_{a}^{(0)}P^{abcd}_{(0)}f^{(0)}_{bcd}\;.\eeq
As already discussed, this term vanishes via parity arguments. The next order term in the integrand is $\mathcal{O}(x)$,
\beq \int_{\Sigma}dAd\tau\biggr\{N^{(1)}_{a}P^{abcd}_{(0)}f^{(0)}_{bcd}+N^{(0)}_{a}P^{abcd}_{(1)}f^{(0)}_{bcd}+N^{(0)}_{a}P^{abcd}_{(0)}f^{(1)}_{bcd}\biggr\}\;, \label{NPfO1}\eeq
and the $\mathcal{O}(x^{2})$ term we must remove is
\beq 
\begin{split}
&\int_{\Sigma}dAd\tau\biggr\{N^{(0)}_{a}P^{abcd}_{(2)}f_{bcd}^{(0)}+N^{(0)}_{a}P^{abcd}_{(0)}f_{bcd}^{(2)}+N^{(0)}_{a}P^{abcd}_{(1)}f_{bcd}^{(1)}+N_{a}^{(1)}P^{abcd}_{(0)}f^{(1)}_{bcd}\\
&+N^{(1)}_{a}P^{abcd}_{(1)}f_{bcd}^{(0)}+N^{(2)}_{a}P^{abcd}_{(0)}f^{(0)}_{bcd}+\sqrt{h}N_{a}^{(0)}P^{abcd}_{(0)}f_{bcd}^{(0)}\biggr\}\;.
\end{split}
\label{NPfO2}\eeq
As we will see, we can in fact drop the terms proportional to $N^{(1)}_{a}$. 

To summarize the algorithm, in order to say we have achieved in deriving the nonlinear equations of motion for higher derivative gravity, we must show how to eliminate the above two contributions (\ref{NPfO1}) and (\ref{NPfO2}). We do this by modifying the $\zeta$ to include higher order contributions, and count the number of undetermined coefficients to see if we have enough terms to eliminate (\ref{NPfO1}) and (\ref{NPfO2}). At first glance it seems as though this is indeed possible simply by a naive counting of the number of monomials which appear in the integrand, compared to a naive counting of the number of undetermined coefficients that are available.


\subsection*{Removing $\mathcal{O}(x)$ Contributions}
\indent

First we write $f_{bcd}$ in a more useful form
\beq 
\begin{split}
f_{bcd}&=\nabla_{b}\nabla_{c}\zeta_{d}-R_{ebcd}\zeta^{e}-(\nabla_{c}\Omega) g_{bd}+(\nabla_{d}\Omega)g_{bc}\\
&=\partial_{b}\partial_{c}\zeta_{d}+\left(2\Gamma^{f}_{\;b(c}\Gamma^{e}_{\;d)f}-\partial_{b}\Gamma^{e}_{\;cd}\right)\zeta_{e}-\left(\Gamma^{e}_{\;bc}\partial_{e}\zeta_{d}+2\Gamma^{e}_{\;d(c}\partial_{b)}\zeta_{e}\right)-R^{e}_{\;bcd}\zeta_{e}\\
&-(\nabla_{c}\Omega) g_{bd}+(\nabla_{d}\Omega)g_{bc}\;.
\end{split}
\eeq
We can drop the whole second term because it is symmetric in indices $cd$ and is being contracted with $P^{abcd}$. What remains is:
\beq
\begin{split}
 f_{bcd}&=\partial_{b}\partial_{c}\zeta_{d}-\left(\Gamma^{e}_{\;bc}\partial_{e}\zeta_{d}+2\Gamma^{e}_{\;d(c}\partial_{b)}\zeta_{e}\right)-R^{e}_{\;bcd}\zeta_{e}-(\nabla_{c}\Omega) g_{bd}+(\nabla_{d}\Omega)g_{bc}\;.
\end{split}
\eeq
We think about modifying $\zeta_{a}$ in the following way:
\beq
\begin{split}
 \zeta_{a}&=\zeta_{a}^{(0)}+\zeta_{a}^{(2)}+\zeta^{(3)}_{a}+\zeta^{(4)}_{a}+...\\
&=-\frac{1}{\ell^{2}}(\ell^{2}-r^{2}-t^{2})\partial_{a}^{t}-\frac{2tx_{i}}{\ell^{2}}\partial_{a}^{i}+\zeta^{(3)}_{a}+\zeta^{(4)}_{a}+...\;,
\end{split}
\eeq
where the $\zeta_{a}^{(0)}$ contribution is constant. A similar expansion holds for $N^{a}$. 

Let's now classify $f_{bcd}^{(0)}$. Clearly we get a contribution from $\partial_{b}\partial_{c}\zeta_{d}$, and from the $\nabla\Omega$ terms. Specifically, 
\beq
\begin{split}
 f_{bcd}^{(0)}&=\partial_{b}\partial_{c}\zeta_{d}^{(2)}-(\nabla_{c}\Omega) \eta_{bd}+(\nabla_{d}\Omega)\eta_{bc}\\
&=\partial_{b}\partial_{c}\zeta_{d}^{(2)}-\frac{2}{\ell^{2}}(\delta^{t}_{\;d}\eta_{bc}-\delta^{t}_{\;c}\eta_{bd})\;.
\end{split}
\eeq

Let's look at the $\mathcal{O}(x)$ contribution of  which would be present in (\ref{NPfO1}) even without modifying $\zeta_{a}$ or $N_{a}$. This is:
\beq 
\begin{split}
N^{(0)}_{a}P^{abcd}_{(1)}f^{(0)}_{bcd}&=N^{(0)}_{i}P^{ibcd}_{(1)}f^{(0)}_{bcd}=N^{(0)}_{i}P^{itcd}_{(1)}f^{(0)}_{tcd}+N^{(0)}_{i}P^{ijcd}_{(1)}f^{(0)}_{jcd}\\
&=N^{(0)}_{i}P^{itjd}_{(1)}f^{(0)}_{tjd}+N^{(0)}_{i}P^{ittd}_{(1)}f^{(0)}_{ttd}+N^{(0)}_{i}P^{ijtd}_{(1)}f^{(0)}_{jtd}+N^{(0)}_{i}P^{ijkd}_{(1)}f^{(0)}_{jkd}\\
&=N^{(0)}_{i}P^{itjt}_{(1)}f^{(0)}_{tjt}+N^{(0)}_{i}P^{itjk}_{(1)}f^{(0)}_{tjk}+N^{(0)}_{i}P^{ittj}_{(1)}f^{(0)}_{ttj}+N^{(0)}_{i}P^{ijtk}_{(1)}f^{(0)}_{jtk}\\
&+N^{(0)}_{i}P^{ijkt}_{(1)}f^{(0)}_{jkt}+N^{(0)}_{i}P^{ijk\ell}_{(1)}f^{(0)}_{jk\ell}\;.
\end{split}
\eeq
Thus our task is to compute 
\beq f^{(0)}_{tjt},\quad f^{(0)}_{tjk},\quad f^{(0)}_{ttj},\quad f^{(0)}_{jtk},\quad f^{(0)}_{jkt},\quad f^{(0)}_{jk\ell}\;.\eeq
It is straightforward to work out that the only non-zero term is
\beq 
\begin{split}
f^{(0)}_{tjk}&=\partial_{t}\partial_{j}\zeta_{k}-\frac{2}{\ell^{2}}(\delta^{t}_{\;d}\eta_{bc}-\delta^{t}_{\;c}\eta_{bd})|_{b=t,c=j,d=k}\\
&=-\frac{2}{\ell^{2}}\delta_{jk}+0=-\frac{2}{\ell^{2}}\delta_{jk}\;,
\end{split}
\eeq
Therefore, the only non-zero contribution will be:
\beq N_{i}^{(0)}P^{itjk}_{(1)}f^{(0)}_{tjk}\;.\eeq
But this term vanishes because $f^{(0)}_{tjk}$ is symmetric in $jk$ indices, while $P_{(1)}^{itjk}$ is antisymmetric. Thus, the entire contribution:
\beq N_{a}^{(0)}P^{abcd}_{(1)}f^{(0)}_{bcd}=0\;.\eeq
In fact, whenever we have something of the form $N^{(a)}_{(0)}P^{abcd}f^{(0)}_{bcd}$, we see that it vanishes, as we never specified the form of $P^{abcd}$ above. We will therefore be able to drop some terms appearing in the $\mathcal{O}(x^{2})$ contribution (\ref{NPfO2}) as well.

There is another term in (\ref{NPfO1}) which appears due to $\zeta_{a}$ being an approximate (conformal) Killing vector, namely, the one proportional to $f^{(1)}_{bcd}$. Without modifying $\zeta_{a}$, the only contribution to this comes from
\beq (\nabla_{d}\Omega)g_{bc}-(\nabla_{c}\Omega)g_{bd}-R^{e}_{\;bcd}\zeta^{(0)}_{e}\;.\eeq
To leading order, we have $\nabla\Omega g\sim (\nabla\Omega)(p)_{\mu}x^{\mu}\eta$, where $\eta$ is the Minkowski metric. Calling $(\nabla_{d}\Omega)_{\mu}(p)\equiv \Omega_{d\mu}(p)$, and noting that $\zeta^{(0)e}=\delta^{te}$, we find that, without modifying $\zeta_{a}$, we have: 
\beq f^{(1)}_{bcd}=(\Omega_{d\mu}x^{\mu}\eta_{bc}-\Omega_{c\mu}\eta_{bd}x^{\mu})-(R_{tbcd})_{\mu}x^{\mu}\;,\eeq
where it is understood that $(R_{tbcd})_{\mu}$ is evaluted at the point $p$. Now we work to see which of
\beq 
\begin{split}
N^{(0)}_{a}P^{abcd}_{(1)}f^{(0)}_{bcd}&=N^{(0)}_{i}P^{itjt}_{(0)}f^{(1)}_{tjt}+N^{(0)}_{i}P^{itjk}_{(0)}f^{(1)}_{tjk}+N^{(0)}_{i}P^{ittj}_{(0)}f^{(1)}_{ttj}+N^{(0)}_{i}P^{ijtk}_{(0)}f^{(1)}_{jtk}\\
&+N^{(0)}_{i}P^{ijkt}_{(0)}f^{(1)}_{jkt}+N^{(0)}_{i}P^{ijk\ell}_{(0)}f^{(1)}_{jk\ell}\;,
\end{split}
\eeq
must be cancelled. Let's work out each of the $f^{(1)}_{bcd}$. The only non-zero contributions we have include:
\beq f^{(1)}_{tjt}=\Omega_{j\mu}x^{\mu}=-f^{(1)}_{ttj}\;,\eeq
\beq f^{(1)}_{jkt}=\Omega_{t\mu}x^{\mu}\eta_{jk}-(R_{tjkt})_{\mu}x^{\mu}=-f^{(1)}_{jtk}\;,\eeq
\beq f^{(1)}_{jk\ell}=(\Omega_{\ell\mu}\eta_{jk}-\Omega_{k\mu}\eta_{j\ell})x^{\mu}-(R_{tjk\ell})_{\mu}x^{\mu}\;,\eeq
Then, using the symmetries of $P^{abcd}$ and $f^{(1)}_{bcd}$, we have:
\beq 
\begin{split}
N_{a}^{(0)}P^{abcd}_{(0)}f^{(1)}_{bcd}&=N^{(0)}_{i}P^{itjt}_{(0)}(2\Omega_{j\mu}x^{\mu})+N^{(0)}_{i}P^{ijkt}_{(0)}(2\Omega_{t\mu}x^{\mu}\eta_{jk}-2(R_{tjkt})_{\mu}x^{\mu})\\
&+N^{(0)}_{i}P^{ijk\ell}_{(0)}\left[(\Omega_{\ell\mu}\eta_{jk}-\Omega_{k\mu}\eta_{j\ell})x^{\mu}-(R_{tjk\ell})_{\mu}x^{\mu}\right]\;.
\end{split}
\eeq
Using spherical symmetry, and that $N^{(0)}_{i}=x_{i}/r$, we see that the only non-vanishing contributions to this will be when $\mu =m$ -- a spatial index, \emph{i.e.,}
\beq 
\begin{split}
&\int_{\Sigma}dAd\tau\biggr\{2P^{itjt}_{(0)}(\Omega_{jm})+2P^{ijkt}_{(0)}(\Omega_{tm}\eta_{jk}-(R_{tjkt})_{m})+P^{ijk\ell}_{(0)}(\Omega_{\ell m}\eta_{jk}-\Omega_{k m}\eta_{j\ell}\\
&-(R_{tjk\ell})_{m})\biggr\}N^{(0)}_{i}x^{m}\\
&\equiv\int_{\Sigma}dAd\tau\mathcal{M}^{i}_{\;m}N^{(0)}_{i}x^{m}\;,
\end{split}
\label{N0P0f1}\eeq
where
\beq 
\begin{split}
\mathcal{M}^{i}_{\;m}&\equiv \biggr\{2P^{itjt}_{(0)}(\Omega_{jm})+2P^{ijkt}_{(0)}(\Omega_{tm}\eta_{jk}\\
&-(R_{tjkt})_{m})+P^{ijk\ell}_{(0)}(\Omega_{\ell m}\eta_{jk}-\Omega_{k m}\eta_{j\ell}-(R_{tjk\ell})_{m})\biggr\}\;.
\end{split}
\eeq
More precisely, the only non-vanishing contribution occurs when $i=m$, i.e., 
\beq \int_{\Sigma}dAd\tau\sum_{i}\mathcal{M}_{ii}\frac{(x^{i})^{2}}{r}\;.\eeq
We see then that the only type of polynomial we see appearing includes $(x_{i})^{2}/r$ -- or $(D-1)$ such terms for a $D$-dimensional spacetime.

This shows us that we must modify $\zeta_{a}$ such that we can eliminate such contributions. Consider, then, the modification
\beq \zeta^{(3)}_{d}=\frac{1}{3!}C_{\mu\nu\rho d}x^{\mu}x^{\nu}x^{\rho}\;,\eeq
where $C_{\mu\nu\rho d}$ is a collection of $D^{4}$ completely undetermined coefficients. It is easy to see that this will provide a contribution to $f^{(1)}_{bcd}$ only through
\beq \partial_{b}\partial_{c}\zeta^{(3)}_{d}=C_{\mu bcd}x^{\mu}\;.\eeq

Putting this into the integrand (\ref{N0P0f1}) we have
\beq \int_{\Sigma}dAd\tau(\mathcal{M}^{i}_{\;m}+P^{ibcd}_{(0)}C_{mbcd})N^{(0)}_{i}x^{m}\;.\eeq
Or, using spherical symmetry,
\beq \int_{\Sigma}dAd\tau\sum_{i}(\mathcal{M}_{ii}+P_{i,(0)}^{\;\;bcd}C_{ibcd})\frac{(x^{i})^{2}}{r}\;.\eeq
We see then that there are more than enough $C$ coefficients to eliminate the undesired terms. 

The only other contribution in (\ref{NPfO1}) is one which arises form the $N_{a}^{(1)}$ modification. Clearly, this term is unnecessary, and therefore we simply do not modify $N$ at this level. This then takes care of the (\ref{NPfO1}) term -- by modifying $\zeta_{a}$ at $\mathcal{O}(x^{3})$ as shown above, we can remove the undesired (\ref{NPfO1}). Let's  move on to the $\mathcal{O}(x^{2})$ contribution, (\ref{NPfO2}). 

\hspace{2mm}


\subsection*{Removing $\mathcal{O}(x^{2})$ Contributions}
\indent

We first point out some simplifications we can make to (\ref{NPfO2}). Using that $N^{(0)}_{a}P^{abcd}f^{(0)}_{bcd}$ all cancel, we can neglect all such terms. Likewise, we can drop any term proportional to $N^{(1)}_{a}$. Thus, we have
\beq 
\begin{split}
&\int_{\Sigma}dAd\tau\biggr\{n^{(0)}_{a}P^{abcd}_{(0)}f_{bcd}^{(2)}+n^{(0)}_{a}P^{abcd}_{(1)}f_{bcd}^{(1)}+n^{(2)}_{a}P^{abcd}_{(0)}f^{(0)}_{bcd}\biggr\}\;.
\end{split}
\label{NPfO2re}\eeq
A priori we have no reason to drop the $N_{a}^{(2)}$ modification, however, as we will see, we may drop it simply because we have enough coefficients to eliminate all undesired terms, leaving us with two terms. Note that $N^{(0)}_{a}P^{abcd}_{(1)}f^{(1)}_{bcd}$ will include contributions both from the failure of $\zeta$ being a Killing vector, and from us modifying $\zeta_{a}$. This means we bring in a large number of $C$ coefficients, potentially all $D^{4}$ of them. However, $(D-1)$ of these coefficients we potentially used, while many others cannot be used due to the fact we are integrating over a co-dimension-2 sphere. Thus, while there are a handful of remaining $C$ coefficients which can be used to eliminate the $\mathcal{O}(x^{2})$ integrand, we cannot rely on or assume we have each coefficient; we must look to modifying $\zeta_{a}$ by adding a term of the form
\beq \zeta^{(4)}_{a}=\frac{1}{4!}D_{\mu\nu\rho\sigma a}x^{\mu}x^{\nu}x^{\sigma}x^{\rho}\;,\eeq
which we see has $D^{5}$ undetermined coefficients. Therefore, by a naive counting argument we find that we will have more than enough $D$ and remaining $C$ coefficients to eliminate all undesired contributions at the $\mathcal{O}(x^{2})$ level.

Begin with
\beq 
\begin{split}
&N^{(0)}_{a}P^{abcd}_{(0)}f^{(2)}_{bcd}=N^{(0)}_{i}P^{itjt}_{(0)}(f^{(2)}_{tjt}-f^{(2)}_{ttj})\\
&+N^{(0)}_{i}P^{itjk}_{(0)}f^{(2)}_{tjk}+N^{(0)}_{i}P^{ijtk}_{(0)}(f^{(2)}_{jtk}-f^{(2)}_{jkt})+N^{(0)}_{i}P^{ijk\ell}_{(0)}f^{(2)}_{jk\ell}\;,
\end{split}
\eeq
where
\beq
\begin{split}
 f^{(2)}_{bcd}&=\partial_{b}\partial_{c}\zeta^{(4)}_{d}-(\Gamma^{e}_{\;bc}\partial_{e}\zeta^{(2)}_{d}+2\Gamma^{e}_{\;d(c}\partial_{b)}\zeta^{(2)}_{e})-R^{e}_{\;bcd}(p)\zeta^{(2)}_{e}-(\nabla_{c}\Omega)h_{bd}+(\nabla_{d}\Omega)h_{bc}\\
&-\frac{1}{2}(\nabla_{c}\Omega)_{\mu\nu}x^{\mu}x^{\nu}\eta_{bd}+\frac{1}{2}(\nabla_{d}\Omega)_{\mu\nu}x^{\mu}x^{\nu}\eta_{bc}\;,
\end{split}
\eeq
with
\beq h_{bd}=-\frac{1}{3}R_{b\mu d\nu}(p)x^{\mu}x^{\nu}\;.\eeq
Following a similar strategy to remove $\mathcal{O}(x)$ contributions and using \cite{Parikh:2017aas} as a guide, several lines of algebra later show that
\beq 
\begin{split}
&N_{a}^{(0)}P^{abcd}_{(0)}f^{(2)}_{bcd}=N^{(0)}_{i}P^{itjt}_{(0)}\biggr[\left(\frac{1}{2}(D_{\mu\nu tjt}-D_{\mu\nu ttj})+\frac{4}{3\ell^{2}}R_{t\mu j\nu}(p)+\Omega_{j\mu\nu}\right)x^{\mu}x^{\nu}\\
&+\frac{4}{\ell^{2}}R_{ktjt}(p)tx^{k}\biggr]\\
&+N^{(0)}_{i}P^{itjk}_{(0)}\left[\frac{1}{2}D_{\mu\nu tjk}x^{\mu}x^{\nu}+\frac{2}{\ell^{2}}R_{\ell tjk}(p)tx^{\ell}\right]\\
&+N^{(0)}_{i}P^{ijtk}_{(0)}\left[\left(\frac{1}{2}(D_{\mu\nu jtk}-D_{\mu\nu jkt})-\frac{4}{3\ell^{2}}R_{j\mu k\nu}(p)-\Omega_{t\mu\nu}\delta_{jk}\right)x^{\mu}x^{\nu}+\frac{4}{\ell^{2}}R_{\ell jtk}(p)tx^{\ell}\right]\\
&+N^{(0)}_{i}P^{ijk\ell}_{(0)}\left[\left(\frac{1}{2}D_{\mu\nu jk\ell}+\frac{1}{2}(\Omega_{\ell\mu\nu}\delta_{jk}-\Omega_{k\mu\nu}\delta_{j\ell})\right)x^{\mu}x^{\nu}+\frac{2}{\ell^{2}}R_{mjk\ell}(p)tx^{m}\right]\;.
\end{split}
\label{n0P0f2}\;,\eeq
and
\beq 
\begin{split}
&N_{a}^{(0)}P^{abcd}_{(1)}f^{(1)}_{bcd}\\
&=\biggr\{(P^{ibcd}_{(1)})_{\nu}C_{\mu bcd}+(P^{itjt}_{(1)})_{\nu}(2\Omega_{j\mu})+(P^{ijkt}_{(1)})_{\nu}(2\Omega_{t\mu}\delta_{jk}-2(R_{tjkt})_{\mu})\\
&+(P^{ijk\ell}_{(1)})_{\nu}\left[(\Omega_{\ell\mu}\eta_{jk}-\Omega_{k\mu}\eta_{j\ell})-(R_{tjk\ell})_{\mu}\right]\biggr\}N^{(0)}_{i}x^{\mu}x^{\nu}\;.
\end{split}
\label{n0P1f1}\;,\eeq
where we have written $P^{abcd}_{(1)}(x)=(P^{abcd}_{(1)})_{\nu}x^{\nu}$. Since $N_{i}^{(0)}\propto x_{i}$, this fixes what $\mu,\nu$ have to be. Either $\mu =0,\nu=j=i$ or $\mu=j=i,\nu=0$. All other contributions vanish due to integration. 

We would now add together (\ref{n0P0f2}) and (\ref{n0P1f1}) in the integrand (\ref{NPfO2re}). We see that we have enough $D$ coefficients to cancel these terms, without introducing $N^{(2)}_{a}$. This can be explicitly checked in the case of $f(R)$ gravity in $2+1$ dimensions -- the most restrictive example. Since we have more than enough coefficients to account for the above monomial contributions, we need not modify $N_{a}$ at all, and may therefore have eliminated (\ref{NPfO2re}). This completes the derivation of the equations of motion.


\newpage

\section{CAUSAL DIAMONDS AND ENTANGLEMENT EQUILIBRIUM} \label{app:cdmechanicsandEE}


\subsection{First Law of Causal Diamond Mechanics}
\indent

Here we present a slightly different derivation of the first law of causal diamond mechanics (FLCD) for higher derivative theories of gravity than given in \cite{Bueno16-1}. Let us take the minus sign of (\ref{DeltaS2}), when $\Sigma$ is the co-dimension-1 spacelike ball $B$. In this picture, the $\Delta$ is not referring to a comparison of $S_{\text{Wald}}$ at two different time slices, i.e., not a physical process -- all we have done is make use of Stokes' theorem. To make this point clear we drop the $\Delta$. 

 Following the similar steps used for stretched lightcone thermodynamics, we have
\beq
\begin{split}  S_{\text{Wald}}&=-\frac{1}{4G\kappa}\int_{B}dB_{a}\{P^{abcd}R_{ebcd}\zeta^{e}-2\zeta_{d}\nabla_{b}\nabla_{c}P^{abcd}+2P^{abcd}(\nabla_{c}\Omega)g_{bd}\\
&-2\Omega g_{cd}\nabla_{b}P^{adbc}\}\;,
\end{split}
\eeq
where we have chosen to write the volume element of $B$ as $dB_{a}=U_{a}dV$. On $B(t=0)$, $\Omega=0$, leading to:
\beq 
\begin{split}
 S_{\text{Wald}}&=-\frac{1}{4G\kappa}\int_{B}dB_{a}\{P^{abcd}R_{ebcd}\zeta^{e}-2\zeta_{d}\nabla_{b}\nabla_{c}P^{abcd}+2P^{abcd}(\nabla_{c}\Omega)g_{bd}\}\;.
\end{split}
\label{deltaS1}\eeq
The final term is
\beq
\begin{split}
\frac{2K}{4G(D-2)}\int_{B}dVP^{abcd}U_{a}U_{d}h_{bc}\equiv\frac{K}{2G}\bar{W}\;,
\end{split}
\label{Wbar}\eeq
where we used $(\nabla_{c}\Omega)|_{B}=\kappa KU_{c}/(D-2)$, and introduced the induced metric $h_{bc}$ on $B$. This contribution $\bar{W}$ is proportional to a part of the \emph{generalized volume} introduced in \cite{Bueno16-1}: 
\beq W=\frac{1}{(D-2)P_{0}}\int_{B}dV(P^{abcd}U_{a}U_{d}h_{bc}-P_{0})\;.\label{Wvol}\eeq
 Here $P_{0}$ is a theory dependent constant defined by the $P^{abcd}$ tensor in a maximally symmetric solution to the field equations via $P^{abcd}_{MSS}=P_{0}(g^{ac}g^{bd}-g^{ad}g^{bc})$. It can be verified that in the case of Einstein gravity (\ref{Wvol}) is the spatial volume $V$ of the diamond. Our expression $\bar{W}$ does not include the $P_{0}$ term\footnote{We can arrive to the generalized volume (\ref{Wvol}) by subtracting $P^{abcd}_{MSS}$ from $P^{abcd}$ in the expression for the Wald entropy; specifically, replace $P^{abcd}$ with $P^{abcd}-\frac{1}{(D-1)}P^{abcd}_{MSS}$ in $S_{\text{Wald}}$. Repeating the steps that lead to (\ref{deltaS1}) will include an additional term which is precisely the extra term found in $W$, missing from $\bar{W}$.}.

 We observe that, like $W$, $\bar{W}$ is also proportional to the physical volume in the case of Einstein gravity. Specifically, in Einstein gravity, $P^{abcd}=1/2(g^{ac}g^{bd}-g^{ad}g^{bc})$, we find
\beq \bar{W}_{GR}=\frac{(D-1)}{(D-2)}V\;.\eeq
This expression is reminiscent of the Smarr formula for a maximally symmetric ball with a vanishing cosmological constant: $(D-2) A=(D-1)KV$ \cite{Jacobson:2018ahi}. This suggests that $\bar{W}$ is really related to the entropy; indeed, in the body of this report we will find such an interpretation when we study the thermodynamics of causal diamonds. 

Moving on, to linear order in the Riemann normal coordinate expansion, a perturbation about flat space leads to \cite{Bueno16-1}
\beq
\begin{split}
 &\delta \left(S_{Wald}-\frac{K}{2G}\bar{W}\right)=-\frac{U_{a}U_{d}}{4G\kappa}\int_{B}dV\left(P^{abcd}_{GR}\delta R^{d}_{\;bce}-2\partial_{b}\partial_{c}\delta P_{\text{higher}}^{abcd}\right)\left(1-\frac{r^{2}}{\ell^{2}}\right)\;,
\end{split}
\eeq
where we have separated $P^{abcd}=P^{abcd}_{GR}+P^{abcd}_{\text{higher}}$. Introducing the conformal Killing energy $H^{m}_{\zeta}$,
\beq H^{m}_{\zeta}=\int_{B}dVT_{ab}U^{a}\zeta^{b}\;, \label{matHam}\eeq
we find
\beq \delta H^{m}_{\zeta}=\int_{B}dV\delta T_{ab}U^{a}U^{b}\left(1-\frac{r^{2}}{\ell^{2}}\right)\;.\eeq
Notice then that for all timelike unit vectors one finds that 
\beq \frac{\kappa}{2\pi}\delta\left(S_{\text{Wald}}-\frac{K}{2G}\bar{W}\right)=-\delta H^{m}_{\zeta}\;,\label{FLCDnoether}\eeq
is equivalent to the tensor equation \cite{Jacobson16-1}:
\beq  \delta R^{ad}-2\partial_{b}\partial_{c}(\delta P^{abcd}_{\text{higher}})+(\delta X)\eta^{ad}=8\pi G\delta T^{ad}\;,\eeq
where we have introduced the spacetime scalar $X$, an assumption to be explained momentarily. Demanding local conservation of energy leads to 
\beq \delta\left(R^{ad}-\frac{1}{2}\eta^{ad}R+\Lambda \eta^{ad}\right)-2\partial_{b}\partial_{c}(\delta P^{abcd}_{\text{higher}})=8\pi G\delta T^{ad}\;,\label{lineareqnsmot}\eeq
which we recognize as the linearized gravitational equations of motion around flat space. 

More explicitly, suppose that we are only considering higher curvature theories of gravity. Then, following the arguments of  \cite{Bueno16-1}:
\beq
\begin{split}
&\frac{\kappa}{2\pi}\delta\left(S_{\text{Wald}}-\frac{K}{2G}\bar{W}\right)_{\text{higher}}=-\frac{1}{8\pi G}\eta_{bc}U_{a}U_{d}\left(-2\partial_{b}\partial_{c}\delta P^{abcd}_{\text{higher}}(0)\right)\left(\frac{2\Omega_{D-2}\ell^{D-1}}{(D^{2}-1)}\right)\\
&+\mathcal{O}(\ell^{D+1})\;.
\end{split}
\eeq
Meanwhile, 
\beq \delta H^{m}_{\zeta}=\delta T^{ad}U_{a}U_{d}\left(\frac{2\Omega_{D-2}\ell^{D-1}}{(D^{2}-1)}\right)+\mathcal{O}(\ell^{D+1})\;.\eeq
Therefore,
\beq
\begin{split}
&\frac{\kappa}{2\pi}\delta\left(S_{\text{Wald}}-\frac{K}{2G}\bar{W}\right)_{\text{higher}}=-\delta H^{m}_{\zeta}\\
&\Rightarrow -2\partial_{b}\partial_{c}\delta P^{abcd}_{\text{higher}}(0)=8\pi G\delta T^{ad}
\end{split}
\eeq
which exactly matches what is found in appendix C of \cite{Bueno16-1}. The Einstein contribution can be dealt with following the method described in \cite{Jacobson16-1}, and as briefly described above.

The condition (\ref{FLCDnoether}) can be understood as the Iyer-Wald identity for a theory of gravity for the geometric set-up of a causal diamond:
\beq \frac{\kappa}{2\pi}\delta \left(S_{\text{Wald}}-\frac{K}{2 G} \bar{W}\right)+ \delta H^{\zeta}_{m}=\int_{B}\delta C_{\zeta}\;,\label{1stlawcdoff}\eeq
where $\delta C_{\zeta}$ is the linearized constraint that the gravitational field equations hold. 

Following \cite{Bueno16-1} one finds that the first law of causal diamond mechanics can be understood as the Iyer-Wald identity \cite{Iyer:1994ys} in the case of a conformal Killing horizon as opposed to the dynamical horizon of a black hole. In this picture the generalized volume can be interpreted as the variation of the gravitational Hamiltonian. The first two terms on the LHS of (\ref{1stlawcdoff}), moreover, can be combined into a single object, namely, the variation of the Wald entropy keeping $\bar{W}$ held constant, i.e., 
\beq \frac{\kappa}{2\pi}\delta \left(S_{\text{Wald}}-\frac{K}{2 G}\bar{W}\right)=\frac{\kappa}{2\pi}\delta S_{\text{Wald}}|_{\bar{W}}\;,\eeq
leading to 
\beq \frac{\kappa}{2\pi}\delta S_{\text{Wald}}|_{\bar{W}}+ \delta H^{\zeta}_{m}=\int_{B}\delta C_{\zeta}\;.\label{1stlawcdoff2}\eeq

As identified in \cite{Bueno16-1}, the Wald formalism contains (JKM) ambiguities in how the Noether current and Noether charge are defined. In particular we may add an exact form $dY$ that is linear in the field variations and their derivatives to the Noether current, and $Y$ to the Noether charge. This would modify both the entropy $S_{\text{Wald}}$ and $\bar{W}$. However, as verified in \cite{Bueno16-1}, the combined modification cancel, and one may write
\beq \frac{\kappa}{2\pi}\delta S_{\text{Wald}}|_{\bar{W}}=\frac{\kappa}{2\pi}\delta(S_{\text{Wald}}+S_{JKM})|_{\bar{W}'}\;,\label{1stlawcdjkm}\eeq
where $\bar{W}'=\bar{W}+\bar{W}_{JKM}$. This shows that the resolution of the JKM ambiguity yields the same on-shell first law, provided the Wald entropy and generalized volume are modified by an exact form $dY$. 


\subsection{Entanglement Equilibrium}
\indent

Let us now show how the first law of causal diamond mechanics -- an off-shell geometric identity -- is related to a condition on entanglement. In an effective field theory the entanglement entropy can be computed using the replica trick \cite{Calabrese:2009qy}, where one defines the entropy as
\beq S_{EE}=(n\partial_{n}-1)I_{\text{eff}}(n)|_{n=1}\;,\eeq
where the effective action $I_{\text{eff}}(n)$ is evaluated on an orbifold with a conical singularity at the entangling surface with excess angle $2\pi(n-1)$. If a covariant regulator is used to define the theory, the resulting expression for the entanglement entropy is a local integral of diffeomorphism invariant contributions. When the entangling surface is the bifurcation surface of a stationary horizon, the entanglement entropy is simply the Wald entropy. In the case of nonstationary entangling surfaces, the computation can be accomplished used squashed cone techniques \cite{Fursaev:2013fta}, leading to extrinsic curvature modifications of the Wald entropy \cite{Dong:2013qoa} -- the so-called Jacobson-Myers entropy \cite{Jacobson:1993vj}. As discussed in \cite{Bueno16-1}, the extrinsic curvature modifications of the Wald entropy may be identified with the JKM ambiguities mentioned above. Thus, the entanglement entropy is given by the Wald entropy modified by specific JKM terms, i.e., the Jacobson-Myers entropy. 

This realization allows us to relate the entanglement entropy to our off-shell geometric identity (\ref{1stlawcdjkm}). The below discussion closely follows \cite{Jacobson16-1, Bueno16-1}. As briefly described in the introduction, we are performing a simultaneous geometric and quantum state variation of the entanglement entropy in a causal diamond. Therefore, the variation of the entanglement entropy $\delta S_{EE}$ includes a UV, state-independent contribution and an IR state-dependent contribution
\beq \delta S_{EE}=\delta S_{UV}+\delta S_{IR}\;.\eeq
The IR contribution describes states of a QFT in a background spacetime, while the UV contribution represents short distance physics, including quantum gravitational degrees of freedom. We should point out here that we are positing that the Hilbert space of states on $B$ can be factorized into IR and UV contributions, $\mathcal{H}_{B}=\mathcal{H}_{UV}\otimes\mathcal{H}_{IR}$, i.e., entanglement separability -- there is minimal entanglement among degrees of freedom at widely separated energy scales. 

 Upon a UV completion, the entanglement entropy in a spatial region is finite in any state, with leading term proportional to the area of the boundary of the region, and higher order contributions described by the Wald entropy. Therefore, when the geometry is varied, the entanglement entropy in the diamond (which is equivalent to entanglement in $B$) from the UV degrees of freedom near the boundary $\partial B$ will change by 
\beq \delta S_{UV}=\delta S_{\text{Wald}}^{(\epsilon)}\;.\eeq

The scale of UV completion  $\epsilon$ -- which we take to be below the Planck scale -- is such that $\mathcal{H}_{IR}$ and $\mathcal{H}_{UV}$ contain degrees of freedom with energies above and below $\epsilon$. We take the size $\ell$ of the causal diamond to be such that $L_{\text{Planck}}<\ell<1/\epsilon$. The separation between UV and IR degrees of freedom allow us to define the IR vacuum state of the ball $B$
\beq \rho_{IR}=\text{tr}_{UV}\rho\;,\eeq
where $\rho$ is the total quantum state of the diamond. Formally we may write $\rho_{IR}$ as a thermal state
\beq \rho_{IR}=\frac{1}{Z}e^{-H_{\text{mod}}}\;,\eeq
where $H_{\text{mod}}$ is the modular Hamiltonian and $Z$ is the partition function. In Minkowski space, the causal diamond may be  conformally transformed to the (planar) Rindler wedge. The Bisognano-Wichmann theorem then allows us to interpret $\rho_{IR}$ as a true thermal state with respect to the Hamiltonian generating time-translation; in the case of a conformal field theory the modular Hamiltonian will take a specific form in terms of the matter Hamiltonian $H^{m}_{\zeta}$ (\ref{matHam}) \cite{Casini:2011kv}
\beq H_{\text{mod}}=\frac{2\pi}{\kappa}H^{m}_{\zeta}\;,  \label{Hmod}\eeq
i.e., the Hamiltonian generating flow along the CKV $\zeta$. 

The entanglement entropy due to IR degrees of freedom $S_{\text{IR}}=-\text{tr}\rho_{IR}\log\rho_{IR}$ will satisfy the first law of entanglement entropy \cite{Blanco:2013joa,Wong:2013gua}
\beq \delta S_{IR}=\delta\langle H_{\text{mod}}\rangle\;.\eeq
We shall make the further conjecture, and assume that the variation of the modular Hamiltonian will carry an additional term $\delta X$ that is a spacetime scalar such that 
\beq \delta \langle H_{\text{mod}}\rangle=\frac{2\pi}{\kappa}\delta\int_{B}dB_{a}(T^{ab}\zeta_{b}+Xg^{ab}\zeta_{b})\;.\eeq
Such a conjecture was  made in \cite{Jacobson16-1}. There one assumes, to leading order that $\delta\langle H_{\text{mod}}\rangle\propto(\delta \langle T_{00}\rangle+\delta X)$, which has been shown to be a correct assumption  \cite{Casini:2016rwj,Carroll:2016lku}, though $\delta X$ may depend on $\ell$. 

Adding this to our total variation of $\delta S_{EE}$, we have a modified first law of EE
\beq \delta S_{EE}=\delta(S_{\text{Wald}}+S_{JKM})+\delta\langle H_{\text{mod}}\rangle\;. \label{1stlawEEmod}\eeq

We may now postulate the equilibrium condition: A small diamond is in equilibrium if the quantum fields are in a vacuum state and the curvature is that of a MSS, e.g., Minkowski space. Moreover, motivated by the first law of causal diamond mechanics, we require that $B$ has the same $\bar{W}'$ as in vacuum. With this, we substitute (\ref{1stlawEEmod}) into (\ref{1stlawcdjkm}), using (\ref{Hmod}), leading to
\beq \frac{\kappa}{2\pi}\delta S_{EE}|_{\bar{W}'}=\int_{B}\delta C_{\zeta}\;,\eeq
which is valid for minimally coupled, conformally invariant matter fields. 

When the variation of $\delta S_{EE}$ vanishes, we recover (\ref{lineareqnsmot}). We therefore arrive to an equivalence between the following statements: (i) the entanglement entropy $S_{EE}$ is maximal in vacuum for all (small) balls in all frames, and (ii) the linearized higher derivative equations hold everywhere. That is, the entanglement equilibrium condition is equivalent to the linearized higher derivative equations of motion to be satisfied, and vice versa. The verification of this equivalence can be found in the appendix of \cite{Bueno16-1}, which we will not repeat here but was described earlier.



\newpage

\section{IYER-WALD FORMALISM FOR STRETCHED LIGHTCONES} \label{app:IyerWaldformLC}


Here, after reviewing the basic set-up of the Iyer-Wald formalism \cite{Iyer:1994ys}, we consider the Iyer-Wald identity for the geometry of future stretched lightcones. We will closely follow the arguments presented in \cite{Bueno16-1} due to the geometric similarity between the stretched lightcone and causal diamond. 

\subsection{Iyer-Wald Formalism}
\indent

Let $L[\phi]$ be the local spacetime $D$-form Lagrangian of a general diffeomorphism invariant theory, where $\phi$ represents a collection of dynamical fields, e.g., the metric and matter fields. Varying the Lagrangian yields
\beq \delta L=E\cdot\delta\phi+d\theta[\delta\phi]\;,\eeq
where $E$ denotes the equations of moton for all of the dynamical fields, and $\theta$ is the symplectic potential $(D-1)$-form. The antisymmetric variation of $\theta$ leads to the symplectic current, a $(D-1)$-form,
\beq \omega[\delta_{1}\phi,\delta_{2}\phi]=\delta_{1}\theta[\delta_{2}\phi]-\delta_{2}\theta[\delta_{1}\phi]\;,\label{sympcurrent}\eeq
whose integral over a Cauchy surface $B$ gives the symplectic form for the phase description of the theory. Given an arbitrary vector field $\xi^{a}$, evaluating the symplectic form on the Lie derivative $\mathcal{L}_{\xi}\phi$ yields the variation of the Hamiltonian $H_{\xi}$ which generates the flow $\xi^{a}$:
\beq\delta H_{\xi}=\int_{B}\omega[\delta\phi,\mathcal{L}_{\xi}\phi]\;. \label{Hamilxi}\eeq
Now take $B$ to be a ball-shaped region, and let $\xi^{a}$ be a future-pointed, timelike vector that vanishes on the boundary $\partial B$. When the background geometry satisfies the field equations $E=0$, , and $\xi$ vanishes on $\partial B$, we arrive to  Wald's variational identity 
\beq \int_{B}\omega[\delta\phi,\mathcal{L}_{\xi}\phi]=\int_{B}\delta J_{\xi}\;,\label{Waldvarid}\eeq
where we have introduced the Noether current $J_{\xi}$
\beq J_{\xi}=\theta[\mathcal{L}_{\xi}\phi]-i_{\xi}L\;,\eeq
with $i_{\xi}$ representing the contraction of the vector $\xi^{a}$ on the first index of the differential form. Recall that the Noether current $J_{\xi}$ can always be written as \cite{Iyer:1995kg}
\beq J_{\xi}=dQ_{\xi}+C_{\xi}\;,\label{Noethercurrent}\eeq
where $Q_{\xi}$ is the Noether charge $(D-2)$-form and $C_{\xi}$ are the constraint field equations associated with diffeomorphism gauge symmetry. When we assume that the matter equations are imposed, one finds 
\beq C_{\xi}=-2\xi^{a}E_{a}^{\;b}\epsilon_{b}\;,\eeq
where $E^{ab}$ is the variation of the Lagrangian density with respect to the metric, and $\epsilon_{a}$ is the volume form on $B$. Combining (\ref{Hamilxi}), (\ref{Waldvarid}), and (\ref{Noethercurrent}) leads to the Iyer-Wald identity:
\beq -\int_{\partial B}\delta Q_{\xi}+\delta H_{\xi}=\int_{B}\delta C_{\xi}\;.\label{IWid}\eeq
When the linearized constraints hold, $\delta C_{\xi}=0$, the variation of the Hamiltonian is a boundary integral of $\delta Q_{\xi}$. We will show that this off-shell identity leads to the first law of stretched lightcones. Observe that, unlike the case with black hole thermodynamics, $\delta H_{\xi}$ here is non-vanishing; this is because $\xi^{a}$ is not a true Killing vector. 

Let us proceed and evaluate the Iyer-Wald identity (\ref{IWid}) for an arbitrary theory of gravity for the geometric set-up for the stretched lightcone described above. Here we will make the simplifying assumption that the matter fields are minimally coupled, such that the Lagrangian splits into metric and matter contributions
\beq L=L^{g}+L^{m}\;,\eeq
with $L^{g}$ being an arbitrary diffeomorphism-invariant function of the metric, Riemann tensor, and the covariant derivatives of the Riemann tensor\footnote{In our discussion above we did not consider theories of gravity which also depend on derivatives of the Riemann tensor, however, it is easy to modify our arguments to include such theories -- in the case one perturbs around maximally symmetric spacetimes.}. This separation allows us to also decompose the symplectic potential and the Hamiltonian as $\theta=\theta^{g}+\theta^{m}$, and $\delta H_{\xi}=\delta H^{g}_{\xi}+\delta H^{m}_{\xi}$. Therefore, the Iyer-Wald identity (\ref{IWid}) becomes
\beq -\int_{\partial B}\delta Q_{\xi}+\delta H^{g}_{\xi}+\delta H^{m}_{\xi}=\int_{B}\delta C_{\xi}\;.\eeq

We can relate the integrated Noether charge to the Wald entropy via \cite{Wald:1993nt}:
\beq -\int_{\partial B}Q_{\xi}= 4GS_{\text{Wald}}\;.\eeq
where $G$ is Newton's gravitational constant, and the Wald entropy functional $S_{\text{Wald}}$ is
\beq S_{\text{Wald}}=-\frac{1}{4G}\int_{\partial B}dS_{ab}(P^{abcd}\nabla_{c}\xi_{d}-2\xi_{d}\nabla_{c}P^{abcd})\;,\eeq
with $dS_{ab}=\frac{1}{2}(n_{a}u_{b}-n_{b}u_{a})dA$\footnote{A brief comment on notation: For comparison to \cite{Bueno16-1}, we note that there the authors choose the convention where $1/4G \to 2\pi$, and use that the Wald entropy is written as
\beq S_{\text{Wald}}=-2\pi\int_{\partial B}\mu P^{abcd}n_{ab}n_{cd}\;,\eeq
where $\mu$ is the volume form on $\partial B$, which $\epsilon_{ab}=-n_{ab}\wedge\mu$.}. Following, \cite{Iyer:1994ys}, this relationship also holds for first order perturbations
\beq \int_{\partial B}\delta Q_{\xi}=-4G\delta S_{\text{Wald}}\;.\eeq

Our next task is to evaluate the variation of the gravitational Hamiltonian $\delta H_{\xi}^{g}$. As we detail below, this leads us to the derivation of the generalized area of stretched lightcones, analogous to the generalized volume of causal diamonds constructed in \cite{Bueno16-1}.


\subsection{Generalized Area of Stretched Lightcones}
\indent

Here we closely follow the arguments presented in \cite{Bueno16-1} to work out the variation of the gravitational Hamiltonian for an arbitrary theory of gravity in the geometric set-up of the stretched lightcone. In the calculation that follows we will consider the case of looking at perturbations about a maximally symmetric background (MSS), specifically Minkowski space. Along the way we will mention how some of these assumptions might be relaxed. 

For a Lagrangian that depends on the Riemann tensor and its covariant derivatives, the symplectic potential $\theta^{g}$ is given by 
\beq \theta^{g}=2P^{bcd}\nabla_{d}\delta g_{bc}+S^{ab}\delta g_{ab}+\sum_{i=1}^{m-1}T_{i}^{abcda_{1}...a_{i}}\delta\nabla_{(a_{1}}...\nabla_{a_{i})}R_{abcd}\;,\eeq
where we use the notation of \cite{Bueno16-1} such that $P^{bcd}=\epsilon_{a}P^{abcd}$, and $S^{ab}$ and $T_{i}^{abcd...}$ are locally constructed from the metric, its curvature, and covariant derivatives of the curvature. Due to the antisymmetry of $P^{bcd}$ in $c$ and $d$, the symplectic current (\ref{sympcurrent}) takes the form
\beq 
\begin{split}
\omega^{g}&=2\delta_{1}E^{bcd}\nabla_{d}\delta_{2}g_{bc}-2E^{bcd}\delta_{1}\Gamma^{e}_{\;db}\delta_{2}g_{ec}+\delta_{1} S^{ab}\delta_{2}g_{ab}\\
&+\sum_{i=1}^{m-1}\delta_{1}T_{i}^{abcda_{1}...a_{i}}\delta_{2}\nabla_{(a_{1}}...\nabla_{a_{i})}R_{abcd}-(1\leftrightarrow2)\;. 
\end{split}
\label{sympcurrent2}\eeq

Let's now employ the geometric set-up discussed above. We use the fact that we are perturbing around a maximally symmetric background. This allows us to write
\beq R_{abcd}=\frac{R}{D(D-1)}(g_{ac}g_{bd}-g_{ad}g_{bc})\;,\eeq
with a constant Ricci scalar $R$, such that 
\beq \nabla_{e}R_{abcd}=0\;,\quad\mathcal{L}_{\xi}R_{abcd}|_{t=0}=0\;.\eeq
Moreover, since the tensors $P^{abcd}$, $S^{ab}$ and $T_{i}^{abcd...}$ are all constructed from the metric and curvature, they will also have vanishing Lie derivatives along $\xi^{a}$, when evaluated on $B$. 

If we replace $\delta_{2}g_{ab}$ in (\ref{sympcurrent2}) with $\mathcal{L}_{\xi}g_{ab}$, and make use of (\ref{nabLg})
\beq \nabla_{d}(\mathcal{L}_{\xi}g_{ab})|_{t=0}=\frac{2}{N_{\xi}}u_{d}\tilde{g}_{ab}\;,\eeq
with
\beq \tilde{g}_{ab}=\delta_{a}^{i}\delta^{j}_{b}\left(\delta_{ij}-\frac{x_{i}x_{j}}{r^{2}}\right)\;,\eeq
then,
\beq 
\begin{split}
&\omega^{g}[\delta g,\mathcal{L}_{\xi}g]|_{B}=\frac{2}{N_{\xi}}\biggr\{2\tilde{g}_{bc}u_{d}\delta P^{bcd}+P^{bcd}\{u_{b}\tilde{\delta}^{e}_{\;d}\delta g_{ce}\\
&+u_{d}\tilde{\delta}^{e}_{\;b}\delta g_{ce}-u^{e}\tilde{g}_{db}\delta g_{ce}\}\biggr\}\;.
\end{split}
\eeq

Following similar computations performed in \cite{Bueno16-1} we find to leading order in the RNC
\beq \omega[\delta g,\mathcal{L}_{g}]|_{B}=-\delta[\frac{4}{N}\eta P^{abcd}U_{a}u_{d}\tilde{g}_{bc}]\;.\eeq
Showing this takes quite a few lines of algebra, however, when all is said and done, we can take (33) of \cite{Bueno16-1} and simply replace $g_{bc}$ with $\tilde{g}_{bc}$. 

Thus, we are varying the object
\beq \int_{B}dB_{a}\frac{\alpha}{r^{2}}P^{abcd}u_{d}\tilde{g}_{bc}\;.\eeq
However, after converting back to the conventions used in the body of this paper, we find that 
\beq\delta H_{\xi}^{g}=-\frac{1}{2\pi\alpha}\delta \tilde{S}\;,\eeq
 i.e., the entropy due to the natural expansion of the hyperboloid $\bar{S}$ (\ref{genarea}). 

In summary, we have arrived to the off-shell variational identity 
\beq \frac{1}{2\pi\alpha}\delta (S_{\text{Wald}}-\bar{S})+\delta H^{m}_{\xi}=\int_{B}\delta C_{\xi}\;.\eeq
Imposing the linearized constrant $\delta C_{\xi}=0$, this simply becomes the first law of stretched future lightcones for higher derivative gravity.


\newpage

\section{$D\to2$ LIMIT OF THE EXTENDED BULK FIRST LAW}  \label{app:lowdentchem}


The extended bulk first law of entanglement entropy across a ball in Minkowski space was found to be (\ref{firstlawextEEapp})
\beq \delta E_{\xi}=\frac{\delta A_{\Sigma}}{4G}-V\frac{\delta\Lambda}{8\pi G}\;,\label{firstlawextEEapp2}\eeq
with 
\beq V=-\int_{\Sigma}da_{a}\omega^{ab}n_{b}=V=\frac{2\pi L^{2}}{D-1}A_{\Sigma}\;,\label{VintermsofA}\eeq
and
\beq A_{\Sigma}=L^{D-2}\Omega_{D-3}\int_{y_{c}}^{1}dy\frac{(1-y^{2})^{\frac{D-4}{2}}}{y^{D-2}}\;.\label{Aent}\eeq
We may rewrite the extended first law as
\beq \delta E_{\xi}=\delta S_{\Sigma}-(D-2)S_{\Sigma}\frac{\delta L}{L}\;.\eeq

Here we study the $D\to2$ limit of the extended bulk first law (\ref{firstlawextEEapp2}). Naively, from (\ref{firstlawextEEapp2}) it appears as though there cannot be an extended first law in $1+1$ dimensions, as the term proportional to $\delta L$ vanishes, leaving us with $\delta E_{\xi}=\delta S_{E}$. However, just as was the case for the extended first law of black holes shown in \cite{Frassino:2015oca}, the extended bulk first law of entanglement has a non-trivial limit in $1+1$-dimensions.

 Thus, applying the philosophy of \cite{Frassino:2015oca}, we perform a (perhaps \emph{ad hoc}) rescaling of Newton's constant $G_{D}\to(1-\frac{D}{2})G_{2}$, with $G_{2}$ being the two-dimensional Newton's constant, we find that\footnote{We should also note that the sign in front of $V\delta\Lambda$ changes just as in the black hole context},
\beq -V\frac{\delta\Lambda}{8\pi G_{D}}\to+V\frac{\delta\Lambda_{2}}{4\pi G_{2}}\;,\eeq
where $\Lambda_{2}=+\frac{1}{L^{2}}$. So, the $1+1$ dimensional limit of (\ref{firstlawextEEapp2}) is, thus far, 
\beq \delta E_{\xi}=\frac{\delta A_{\Sigma}}{4G_{D}}\biggr|_{D\to2}+V\frac{\delta\Lambda_{2}}{8\pi G_{2}}\;.\label{varD2v1}\eeq
We have not yet evaluated the term proportional to $\delta A_{\Sigma}$, however, we immediately see in $1+1$-dimensions there is a term proportional to the variation of $L$. 

Let us now evaluate the term $\delta A_{\Sigma}$ in the $D\to2$ limit. Defining $\epsilon\equiv D-2$, we have that (\ref{Aent}) is
\beq
\begin{split}
 A_{\Sigma}^{(\epsilon)}&=\frac{2(L\sqrt{\pi})^{\epsilon}}{\Gamma\left(\frac{\epsilon}{2}\right)}\int^{1}_{y_{c}}(1-y^{2})^{\frac{\epsilon-2}{2}}{y^{\epsilon}}\\
&=\frac{2(L\sqrt{\pi})^{\epsilon}}{\Gamma\left(\frac{\epsilon}{2}\right)}\left[\frac{y^{1-\epsilon}}{(1-\epsilon)}\, {}_{2}F_{1}\left(\frac{1-\epsilon}{2}\,,\,1-\frac{\epsilon}{2}\,,\,\frac{3-\epsilon}{2}\,,\,y^{2}\right)\right]\biggr|^{1}_{y_{c}}\\
&=\frac{2(L\sqrt{\pi})^{\epsilon}}{\Gamma\left(\frac{\epsilon}{2}\right)}\biggr\{\frac{\Gamma\left(\frac{3-\epsilon}{2}\right)\Gamma\left(\frac{\epsilon}{2}\right)}{\sqrt{\pi}(1-\epsilon)}-\frac{y_{c}^{1-\epsilon}}{(1-\epsilon)}\, {}_{2}F_{1}\left(\frac{1-\epsilon}{2}\,,\,1-\frac{\epsilon}{2}\,,\,\frac{3-\epsilon}{2}\,,\,y_{c}^{2}\right)\biggr\}\;.
\end{split}
\eeq
 
It is straightforward to verify that $A^{(0)}_{\Sigma}=1$ for any cutoff $y_{c}$. Performing a power series expansion in $\epsilon$ to linear order, we have:
\beq 
\begin{split}
A^{(\epsilon)}_{\Sigma}&\approx1+\epsilon\left(1-\frac{\text{arctanh}(y_{c})}{\sqrt{\pi}}+\log(L\sqrt{\pi})-\frac{1}{2}\psi^{(0)}(\frac{3}{2})\right)\\
&=1+\epsilon\left[1-\frac{\text{arctanh}(y_{c})}{\sqrt{\pi}}+\log\left(2L\sqrt{\pi}e^{\frac{\gamma}{2}-1}\right)\right]\;,
\end{split}
\eeq
where we used\footnote{This comes from writing the digamma function for half-integers: $\psi^{(0)}(n+1/2)=-\gamma-2\log 2+\sum_{k=1}^{n}\frac{2}{2k-1}$.} $\psi^{(0)}(3/2)=-\gamma-2\log 2+2$. Notice that we may safely take the limit $y_{c}\to0$, and so, to leading order, we have the $D\to2$ limit of (\ref{Aent})
\beq A_{\Sigma}\to A^{p}_{\Sigma}+(D-2)\left[1+\log\left(2L\sqrt{\pi}e^{\frac{\gamma}{2}-1}\right)\right]\;.\label{D2limitAsig}\eeq
Here we have defined $A^{p}_{\Sigma}\equiv1$ as the area of a point, following the notation of \cite{Frassino:2015oca}. Defining the area of the minimal bulk `surface' $\tilde{A}^{(2)}_{\Sigma}\equiv2\left[1+\log\left(2L\sqrt{\pi}e^{\frac{\gamma}{2}-1}\right)\right]\equiv -A^{(2)}_{\Sigma}$, we find (\ref{varD2v1}) becomes
\beq \delta E_{\xi}=\frac{\delta A^{(2)}_{\Sigma}}{4G_{2}}+\frac{V\delta\Lambda_{2}}{8\pi G_{2}}\;.\label{D2lawv3}\eeq
From (\ref{VintermsofA}), we have $V=2\pi L^{2}A^{p}_{\Sigma}$, and defining the entanglement entropy of a `point', $S_{E}^{p}\equiv \frac{A^{p}_{\Sigma}}{4G_{2}}=\frac{1}{4G_{2}}$, we reexpress (\ref{D2lawv3}) as
\beq \delta E_{\xi}=\delta S^{(2)}_{E}-2S_{E}^{p}\frac{\delta L}{L}\;.\label{D2bulkEEv1}\eeq
Substituting $S_{E}^{p}=1/4G_{2}$ and our definition for $S_{E}^{(2)}$, we find that the variation of the ADM charge is entirely proportional to the variation of the AdS length $L$: $\delta E_{\xi}=-\frac{1}{G_{2}L}\delta L$. We see that at fixed ADM energy $\delta E_{\xi}=0$, we are necessarily at fixed AdS length $L$.

\newpage

\section{EXTENDED FIRST LAW OF ENTANGLEMENT AND JT GRAVITY}  \label{app:extfirstlawjt}

Here we review an alternative derivation for the first law of entanglement for Jackiw-Teitelboim (JT) gravity. Our derivation will follow the Iyer-Wald formalism developed in \cite{Caceres:2016xjz}. As discussed in \ref{sec:extfirstlawhighlow}, we will find that the extended first law of entanglement for JT gravity is not expressible in the usual way, \emph{i.e.}, the variation of the entropy with respect to the couplings of the theory satisfies $\delta_{\lambda_i} S_\xi=4\pi \delta \phi_0\neq \frac{S_\xi}{a_1^\ast}\delta_{\lambda_i} a_1^\ast$. We provide the alternative derivation using the Iyer-Wald formalism as it reveals some interesting cancellations with respect to the UV divergences arising from calculating the variations of geometric quantities near the asymptotic boundary. 


\hspace{2mm}

\subsection{JT Gravity and Wald Entropy}

Consider the action for JT gravity, following the conventions of \cite{Harlow:2018tqv}, where we drop the Gibbons-Hawking-York boundary terms
\beq I_{JT}=\frac{\phi_{0}}{16\pi G_{N}}\int d^{2}x\sqrt{-g}R+\frac{1}{16\pi G_{N}}\int d^{2}x\sqrt{-g}\phi(R+\frac{2}{L^{2}})\;.\label{JTact}\eeq
This action can be shown to arise from a higher dimensional theory describing the $s$-wave sector of the near horizon limit of a near extremal (magnetically charged) black hole. Here $\phi_{0}$ is a coupling constant multiplying the two-dimensional Euler-characteristic, $\phi$ is a scalar function, \emph{i.e.}, the dilaton, and $L$ is a coupling from the higher-dimensional parent theory from which this action is reduced from and will represent the AdS$_{2}$ radius. Our total Lagrangian density is 
\beq \mathcal{L}_{JT}=\frac{1}{16\pi G_{N}}\left[(\phi_{0}+\phi)R+\frac{2}{L^{2}}\phi\right]\;.\label{JTlag}\eeq
The equations of motion for this theory are
\beq R+\frac{2}{L^{2}}=0\;,\quad (\nabla_{a}\nabla_{b}-\frac{1}{L^{2}}g_{ab})\phi=0\;.\eeq
From the gravitational field equations, we see that the Ricci scalar $R$ is entirely fixed by the cosmological constant, $R=-2/L^{2}$, such that the only spacetime solution for this theory is $\text{AdS}_{2}$, which we express in Poincar\'e patch coordinates:
\beq ds^{2}=\frac{L^{2}}{z^{2}}(-dt^{2}+dz^{2})\;.\eeq
The asymptotic boundary limit occurs when $z\to0$. Since they will be useful later on, the non-vanishing Christoffel symbols are
\beq \Gamma^{t}_{\;tz}=\Gamma^{z}_{\;tt}=\Gamma^{z}_{\;zz}=-\frac{1}{z}\;.\eeq

The solution for the dilaton $\phi(z,t)$ is 
\beq \phi(z,t)=\phi_{\mathcal{H}}\left(\frac{1}{z}+c z-\frac{ct^{2}}{z}\right)\;,\eeq
where $\phi_{\mathcal{H}}$ and $c$ are integration constants. The $\phi_{\mathcal{H}}$ constant, we we will see momentarily, is naturally interpreted as the value of the dilaton at the horizon $\mathcal{H}$. The constant $c$ comes from analyzing asymptotic boundary conditions, where it is found \cite{Almheiri:2014cka,Almheiri:2019psf} $c=2\pi T_{0}=1/R$, with $T_{0}$ being the temperature of the ``eternal black hole". 

We consider the classic example of a CFT in vacuum restricted to a ball of radius $R$ on the boundary of $\text{AdS}_{2}$  such that the bulk Ryu-Takayanagi surface $z^{2}=R^{2}$ is a bifurcate  Killing horizon generated by $\xi$:
\beq \xi^{a}=-\frac{2\pi}{R}tz\partial_{z}^{a}+\frac{\pi}{R}(R^{2}-z^{2}-t^{2})\partial_{t}^{a}\;. \label{ckvrt}\eeq

 We already have the necessary ingredients to compute the horizon entropy using the Wald formula
\beq S_{\text{Wald}}=-2\pi\frac{\partial\mathcal{L}}{\partial R_{abcd}}\varepsilon_{ab}\varepsilon_{cd}\biggr|_{\text{horizon}}\;.\eeq
Here the ``horizon" is a single point, hence no integral. Then, using $\frac{\partial R}{\partial R_{abcd}}=\frac{1}{2}(g^{ac}g^{bd}-g^{ad}g^{bc})$ together with (\ref{JTlag}) we find
\beq S_{\text{Wald}}=-2\pi\frac{1}{32\pi G_{N}}(\phi_{0}+\phi_{\mathcal{H}})(g^{ac}g^{bd}-g^{bc}g^{ad})\varepsilon_{ab}\varepsilon_{cd}=\frac{1}{4G_{N}}(\phi_{0}+\phi_{\mathcal{H}})\;,\eeq
where $\phi_{\mathcal{H}}$ is the value of the dilaton at the horizon. This entropy is understood as the semi-classical entropy of the (two-sided) ``black hole" and matches Euclidean path integral calculations \cite{Harlow:2018tqv}.


\hspace{2mm}

\subsection{Extended First Law of Entanglement}

Let's now briefly outline the Iyer-Wald formalism extended to include varying coupling constants, as established in \cite{Caceres:2016xjz}. Recall that we define a $(d+1)$ dimensional diffeomorphism invariant theory of gravity coupled to matter, whose Lagrangian is expressed as a $(d+1)$-form
\beq \mathbf{L}(g,\phi,\lambda_{i},\Phi)=\mathcal{L}\varepsilon\;,\eeq
where $g$ is the metric, $\Phi$ any matter fields living on the background and $\lambda_{i}$ are the couplings of the theory and the $(d+1)$-dimensional volume element $\varepsilon$ is given by 
\beq \varepsilon=\sqrt{-g}dt\wedge dx^{1}\wedge...\wedge dx^{d}\;.\eeq
A total variation of the Lagrangian generically takes the form
\beq \delta \mathbf{L} = E^{g}\delta g +E^{\Phi}\delta \Phi+d\Theta(g,\delta g)+\sum_{i}E^{\lambda_{i}}\delta\lambda_{i}\;,\eeq
where $E^{g}$ is the gravitational field equations, $E^{\Phi}$ the Euler-Lagrange equations for the matter content, with $\Theta$ a boundary term obtained when the action is varied, often called the symplectic potential, and where
\beq E^{\lambda_{i}}=\frac{\partial\mathcal{L}}{\partial\lambda_{i}}\varepsilon\;.\eeq

In the Iyer-Wald formalism the first law of extended black hole thermodynamics is derived by varying the Lagrangian $\mathbf{L}$ in two ways: (i) with respect to a variation generated by a vector field $\xi$, and (ii) an arbitrary variation with respect to the bulk fields and couplings. The Noether current $J$ associated with coordinate transformation generated by $\xi$ is given by 
\beq J=\Theta(g,\Phi,\delta_{\xi}g,\delta_{\xi}\Phi)-\xi\cdot \mathbf{L}.\eeq
The dot product is given in the following sense. For an $n$-form $F$
\beq F=\frac{1}{n!}F_{a_{1}a_{2}...a_{n}}dx^{a_{1}}\wedge dx^{a_{2}}\wedge...\wedge dx^{n}\;,\eeq
 we have 
\beq \xi\cdot F=\frac{1}{(n-1)!}\xi^{b}F_{ba_{2}...a_{n}}dx^{a_{2}}\wedge...\wedge dx^{a_{n}}\;.\eeq
Using the equations of motion, we have that on-shell $dJ=0$, such that $J$ is expressed locally as the exterior derivative of a $(d-2)$-form $Q$, the Noether charge, such that $J=dQ$. 

When $\xi$ is a Killing vector, an arbitrary variation of $J$ leads to \cite{Caceres:2016xjz}
\beq d(\delta Q-\xi\cdot\Theta)+\sum_{i}\xi\cdot E^{\lambda_{i}}\delta\lambda_{i}=0\;.\label{maineqn}\eeq
Integrating this over a codimension-1 hypersurface $\Sigma$ and using Stokes' theorem, we arrive to 
\beq \sum_{i}\int_{\Sigma}\xi\cdot E^{\lambda_{i}}\delta\lambda_{i}+\int_{\partial\Sigma}\chi=0\;,\label{firstlawIWext}\eeq
where we have defined
\beq \chi=\delta Q-\xi\cdot\Theta\;.\eeq
We integrate the spatial slice $\Sigma$ between the bifurcate Killing horizon $\mathcal{H}$ and the surface at infinity. When $\mathcal{H}$ is a black hole horizon, (\ref{firstlawIWext}) leads to the extended first law of black hole thermodynamics, where the integral of $\chi$ over the boundary of the bifurcation surface $\partial\Sigma_{\mathcal{H}}$ gives the variation of the $T\delta S$, while the integral of $\chi$ off at infinitym $\partial\Sigma_{\infty}$ gives us the variation in the ADM mass $\delta M$. The integral of the coupling variation over $\Sigma$ leads to $V\delta P$ contribution. 

The Iyer-Wald formalism is well-defined for theories of gravity in $1+1$ dimensions, where now the co-dimension 2 ``surface" $\partial\Sigma$ is a point. In the entanglement set-up, moreover, the integration of $\chi$ over $\partial\Sigma_{\mathcal{H}}$ gives the entropy dual to the CFT boundary entanglement entropy, the integral over $\partial\Sigma_{\infty}$ gives the variation of the modular Hamiltonian, and the $\delta\lambda_{i}$ leads to the extension. The boundary interpretation of the extended bulk first law is the extended first law of entanglement, where the extension is proportional to the variation of the generalized central charge. We  will see in the case of JT gravity that we still have an extended first law due to the variation of the coupling constants of the theory, but it is does not organize itself in terms of a generalized central charge $a_{1}^{\ast}$. 

Let's now write down some explicit expressions needed to compute the extended first law of entanglement for JT gravity. The symplectic current $\Theta$ and Noether charge are given by, respectively \cite{Bueno:2016ypa}
\beq \Theta=\varepsilon_{a}(2P^{abcd}\nabla_{d}\delta g_{bc}-2\nabla_{d}P^{abcd}\delta g_{bc})\;,\eeq
\beq Q=\varepsilon_{ab}(-P^{abcd}\nabla_{c}\xi_{d}-2\xi_{c}\nabla_{d}P^{abcd})\;,\eeq
where $P^{abcd}=\frac{\partial \mathcal{L}}{\partial R_{abcd}}$. Specifically, for the case of JT gravity (\ref{JTact})
\beq P^{abcd}_{JT}=\frac{(\phi_{0}+\phi)}{32\pi G_{N}}(g^{ac}g^{bd}-g^{ad}g^{bc})\;.\label{PJT}\eeq

Moreover, in our conventions, we have a co-dimension 1 volume element, the $d$-form
\beq \varepsilon_{a}=\frac{1}{d!}\epsilon_{ab_{2}...b_{d+1}}dx^{b_{2}}\wedge...\wedge dx^{b_{d+1}}\;,\eeq
and a co-dimension 2 volume $(d-1)$-form
\beq \varepsilon_{ab}=\frac{1}{(d-1)!}\epsilon_{abc_{3}...c_{d+1}}dx^{c_{3}}\wedge...\wedge dx^{c_{d+1}}\;.\eeq
Here $\epsilon$ is the Levi-civita tensor with the sign convention $\epsilon_{tzx^{1}...x^{d-1}}=+\sqrt{-g}$ (in Poincar\'e coordinates). For us, $d=1$ and we have 
\beq 
\begin{split}
&\varepsilon=\sqrt{-g}dt\wedge dz\\
&\varepsilon_{a}=\epsilon_{ab}dx^{b}\\
&\varepsilon_{ab}=\epsilon_{ab}\;.
\end{split}
\eeq

Using (\ref{PJT}) the Noether charge $Q$ and symplectic potential are easily worked out to be
\beq Q=-\frac{1}{16\pi G_{N}}\left[(\phi_{0}+\phi)\nabla^{a}\xi^{b}+2\xi^{a}(\nabla^{b}\phi)\right]\varepsilon_{ab}\;.\label{QJT}\eeq
and 
\beq \Theta=\frac{g^{ac}g^{bd}}{16\pi G_{N}}\left[(\phi_{0}+\phi)(\nabla_{b}\delta g_{cd}-\nabla_{c}\delta g_{bd})-((\nabla_{b}\phi)\delta g_{cd}-(\nabla_{c}\phi)\delta g_{bd})\right]\varepsilon_{a}\;.\eeq

We can simplify the potential $\Theta$ a bit more. As noted in appendix C of \cite{Caceres:2016xjz}, the quantity 
\beq g^{ac}g^{bd}(\nabla_{b}\delta g_{cd}-\nabla_{c}\delta g_{bd})=0\;\eeq
in the Poincar\'e patch. Similarly, expanding out everything using the Christoffel symbols, it is straightforward to show that 
\beq g^{ac}g^{bd}\nabla_{b}\delta g_{cd}=0\;.\eeq
Therefore, our symplectic potential reduces to 
\beq \Theta=-\frac{g^{ac}g^{bd}}{16\pi G_{N}}[(\nabla_{b}\phi)\delta g_{cd}-(\nabla_{c}\phi)\delta g_{bd}]\varepsilon_{a}\label{ThetaJT}\;.\eeq
This is different from Einstein gravity, where $\Theta=0$. Note that $\Theta$ will only be non-zero for $\delta L$ coupling variations to the metric (since the metric itself does not depend on $G_{N}$ or $\phi_{0}$). Explicitly, for $\delta L$ variations
\beq
\begin{split}
\Theta_{\delta L}&=\frac{2z^{2}\delta L}{16\pi G_{N}L^{3}}\left[(\nabla_{z}\phi)\varepsilon_{z}+(\nabla_{t}\phi)\varepsilon_{t}\right]\;.
\end{split}
\eeq
Here $\varepsilon_{z}=\epsilon_{zt}dt=-\sqrt{-g}dt$, and $\varepsilon_{t}=\epsilon_{tz}dz=\sqrt{-g}dz$. Put another way,
\beq \Theta=\Theta^{a}\varepsilon_{a}\;,\quad \Theta^{a}_{\delta L}=\frac{2z^{2}\delta L}{16\pi G_{N}L^{3}}\left[(\nabla_{z}\phi)\delta^{az}+(\nabla_{t}\phi)\delta^{at}\right]\;.\label{varLtheta}\eeq

Let's also write the Noether charge $Q$ more explicitly. Restricting to the $t=0$ surface, we have that the first term in $Q$ is
\beq
\begin{split}
 -\frac{1}{16\pi G_{N}}(\phi_{0}+\phi)\nabla^{a}\xi^{b}\epsilon_{ab}&=-\frac{1}{16\pi G_{N}}(\phi_{0}+\phi)\frac{2z^{2}}{L^{2}}\left(\frac{2\pi z}{R}+\frac{\xi^{t}(t=0)}{z}\right)\varepsilon_{tz}\\
&=-\frac{(\phi_{0}+\phi)}{8RG_{N}}\left(z+\frac{R^{2}}{z}\right)
\end{split}
\;.\label{this}\eeq
We also have the contribution to $Q$:
\beq 
\begin{split}
-\frac{2}{16\pi G_{N}}\xi^{a}(\nabla^{b}\phi)\varepsilon_{ab}&=-\frac{2}{16\pi G_{N}}\frac{\pi}{R}(R^{2}-z^{2})g^{zz}(\nabla_{z}\phi)\epsilon_{tz}\\
&=-\frac{1}{8G_{N}R}(R^{2}-z^{2})(\nabla_{z}\phi)\;,
\end{split}
\eeq
where we used that $\xi^{z}(t=0)=0$. Combined, 
\beq Q=-\frac{1}{8G_{N}R}\left[(\phi_{0}+\phi)\left(z+\frac{R^{2}}{z}\right)+(R^{2}-z^{2})(\nabla_{z}\phi)\right]\;.\label{Qinpoinc}\eeq
Notice that $Q$ does not explicitly depend on the coupling $L$.

We now compute the bulk extended first law using (\ref{firstlawIWext}). The gravitational couplings of JT gravity are $\{\lambda_{i}\}=\{\phi_{0},L,G_{N}\}$, and so 
\beq
\begin{split}
 \sum_{i}\xi\cdot E^{\lambda_{i}}\delta\lambda_{i}&=\sum_{i}\frac{\partial\mathcal{L}}{\partial\lambda_{i}}\delta\lambda_{i}\xi\cdot \varepsilon\\
&=\left(\frac{R}{16\pi G_{N}}\delta\phi_{0}-\frac{4}{16\pi G_{N}L^{3}}\phi\delta L-\frac{\mathcal{L}_{JT}}{G_{N}}\delta G_{N}\right)\xi\cdot\varepsilon\;.
\end{split}
\eeq
At $t=0$, defining our constant time slice $\Sigma$,
\beq \xi\cdot\varepsilon=\xi^{t}\varepsilon_{t}=\frac{\pi}{R}(R^{2}-z^{2})\sqrt{-g}dz\;,\eeq
where we used $\xi^{t}(t=0)=\frac{\pi}{R}(R^{2}-z^{2})$. 

Then, for example, the $\phi_{0}$ variation contribution gives us
\beq
\begin{split}
\frac{\delta\phi_{0}}{16\pi G_{N}}\int_{\Sigma}R\xi\cdot \varepsilon&=\frac{\delta\phi_{0}}{16\pi G_{N}}\int_{\Sigma}\left(-\frac{2}{L^{2}}\right)\frac{\pi}{R}(R^{2}-z^{2})\frac{L^{2}}{z^{2}}dz\\
&=-\frac{\delta\phi_{0}}{8G_{N}R}\int_{\Sigma}\frac{(R^{2}-z^{2})}{z^{2}}dz\\
&=\frac{\delta\phi_{0}}{8G_{N}R}\left[\frac{R^{2}}{z}+z\right]\biggr|_{\epsilon}^{R}\\
&=\frac{\delta\phi_{0}}{4G_{N}}-\frac{\delta\phi_{0}}{8G_{N}}\left(\frac{R}{\epsilon}+\frac{\epsilon}{R}\right)\;.
\end{split}
\label{phi0var}\eeq
We see that we have a $1/\epsilon$ divergence in the limit $\epsilon\to0$. As we will show momentarily, this divergence is cancelled from $\chi$. 

Moreover, using $\phi=\phi_{\mathcal{H}}(\frac{1}{z}+\frac{z}{R^{2}}-\frac{t^{2}}{R^{2}z})|_{t=0}$, we have 
\beq
\begin{split}
-\frac{4\delta L}{16\pi G_{N}L^{3}}\int_{\Sigma}\phi\,\xi\cdot\varepsilon=-\frac{\delta L}{8RG_{N}L}\phi_{\mathcal{H}}\left(\frac{R^{2}}{\epsilon^{2}}+\frac{\epsilon^{2}}{R^{2}}-2\right)\;.
\end{split}
\label{Lvar}\eeq
We again see a divergence in coming from $\epsilon\to0$ limit. The only way for this divergence to be cancelled is via $\chi$, which we move to now. 

We now need to compute
\beq \int_{\Sigma}\chi=-\int_{\Sigma_{\infty}}\chi+\int_{\partial\Sigma_{\mathcal{H}}}\chi\;,\quad \chi=\delta Q-\xi\cdot\Theta\;.\eeq
The `$\delta$' in front of $Q$ and the one appearing in $\Theta$ is a general place holder for the variation with respect to the couplings $\lambda_{i}$ and the metric $g$. The variation with respect to the metric is guaranteed to give us the usual bulk first law of bifurcate Killing horizons, relating the variation of the horizon entropy to the ADM energy, so we won't review it here. We are instead interested in the variations with respect to the couplings $\lambda_{i}$. We therefore split $\chi$ into contributions from each coupling variation
\beq 
\begin{split}
&\chi^{(\delta\phi_{0})}=\delta_{\phi_{0}}Q-\xi\cdot\Theta_{\delta\phi_{0}}\;,\\
&\chi^{(\delta L)}=\delta_{L}Q-\xi\cdot\Theta_{\delta L}\;,\\
&\chi^{(\delta G_{N})}=\delta_{G_{N}}Q-\xi\cdot\Theta_{\delta G_{N}}\;.
\end{split}
\eeq

Starting with the first line, note that $\Theta_{\delta_{0}}=0$ since the metric does not change under variations of $\phi_{0}$. Therefore, 
\beq \chi^{(\delta\phi_{0})}=\delta_{\phi_{0}}Q\;,\eeq
and applying the Noether charge (\ref{Qinpoinc})
\beq \chi^{(\delta_{\phi_{0}})}\biggr|_{\partial\Sigma_{\infty}}=-\frac{\delta\phi_{0}}{8\pi G_{N}}\left(\frac{\pi z}{R}+\frac{\pi R}{z}\right)_{z=\epsilon}=-\frac{\delta\phi_{0}}{8G_{N}}\left(\frac{\epsilon}{R}+\frac{R}{\epsilon}\right)\;.\label{chiphi0var}\eeq

Combining (\ref{phi0var}), (\ref{chiphi0var}) with the extended first law (\ref{firstlawIWext}) we find
\beq 
\begin{split}
0&=\frac{\delta\phi_{0}}{16\pi G_{N}}\int_{\Sigma}R\xi\cdot\varepsilon-\chi^{(\delta_{\phi_{0}})}\biggr|_{\partial\Sigma_{\infty}}+\int_{\partial\Sigma_{\mathcal{H}}}\chi\\
&\Rightarrow \delta_{\phi_{0}}S_{EE}=\frac{\delta\phi_{0}}{4G_{N}}
\end{split}
\label{varyphi0extlaw}\eeq
where $\int_{\partial\Sigma_{\mathcal{H}}}\chi$ always yields the variation of the Wald entropy (which in the boundary limit is the entanglement entropy $S_{EE}$). We see that the $\epsilon\to0$ divergence in (\ref{phi0var}) was precisely cancelled by the divergence in (\ref{chiphi0var}). 

Let's now move onto the variation with respect to $L$. Unlike the case for Einstein gravity in $d\geq2$, we see from (\ref{Qinpoinc}) that 
\beq \delta_{L}Q=0\;.\eeq
We do, however, have a contribution coming from $\xi\cdot\Theta_{\delta L}$. Using (\ref{varLtheta}), and $\phi(z,t=0)=\phi_{\mathcal{H}}\left(\frac{1}{z}+\frac{z}{R^{2}}\right)$, we have
\beq 
\begin{split}
\xi\cdot \Theta_{\delta L}&=\frac{2z^{2}\delta L}{16\pi G_{N}L^{3}}\frac{\pi}{R}(R^{2}-z^{2})(\nabla_{z}\phi)\sqrt{-g}\\
&=-\frac{\phi_{\mathcal{H}}\delta L}{8G_{N}LR}\left[\frac{R^{2}}{z^{2}}+\frac{z^{2}}{R^{2}}-2\right]\;,
\end{split}
\eeq
such that 
\beq \chi^{(\delta L)}\biggr|_{\partial\Sigma_{\infty}}=-\xi\cdot\Theta_{\delta L}=\frac{\phi_{\mathcal{H}}\delta L}{8G_{N}LR}\left[\frac{R^{2}}{\epsilon^{2}}+\frac{\epsilon^{2}}{R^{2}}-2\right]\;.\label{chivarL}\eeq
Note that at the horizon, $z=R$, $\chi^{(\delta L)}=0$. 

Combining (\ref{Lvar}) and (\ref{chivarL}), and substituting them into (\ref{firstlawIWext}), we find that 
\beq \delta_{L} S_{EE}=0\;.\eeq
The divergences coming from \ref{Lvar}) are exactly cancelled from those in (\ref{chivarL}). 

Finally, it is straightforward to show that varying with respect to $G_{N}$ leads to 
\beq \delta _{G_{N}}S_{EE}=-\frac{1}{4G_{N}^{2}}(\phi_{0}+\phi_{h})\delta G_{N}=-\frac{S_{EE}}{G_{N}}\delta G_{N}\;\label{varyGextfirstlaw}\eeq
where one uses $\delta_{G_{N}}Q=-\frac{\delta G_{N}}{G_{N}}Q$ and $\Theta_{\delta G_{N}}=0$.

Putting together the $\phi_{0}$ and $G_{N}$ variation contributions to the extended first law, (\ref{varyphi0extlaw}) and (\ref{varyGextfirstlaw}), respectively, and adding them to the metric variation, we have arrive to the extended first law of entanglement for JT gravity:
\beq 
\begin{split}
\delta\langle H_{\text{Ball}}\rangle&=\delta S_{EE}+\frac{\delta\phi_{0}}{4G_{N}}-\frac{S_{EE}}{G_{N}}\delta G_{N}\\
&=\delta S_{EE}+\frac{1}{\phi_{0}}\left(S_{EE}-\frac{\phi_{\mathcal{H}}}{4G_{N}}\right)\delta\phi_{0}-\frac{S_{EE}}{G_{N}}\delta G_{N}\;.
\end{split}
\eeq
We observe that our extended first law of entanglement for JT gravity does \emph{not} take the usual form $\delta S_{EE}-\frac{S_{EE}}{a_{d}^{\ast}}\delta a_{d}^{\ast}=\delta\langle H\rangle$. This is because we seemingly cannot write $S_{EE}\propto a_{1}^{\ast}$.

\newpage

\bibliography{referencesGR}

\providecommand{\href}[2]{#2}\begingroup\raggedright\begin{thebibliography}{100}

\bibitem{Bekenstein72-1}
J.~D. Bekenstein, \emph{{Black holes and the second law}},
  \href{http://dx.doi.org/10.1007/BF02757029}{\emph{Lett. Nuovo Cim.} {\bf 4}
  (1972) 737--740}.

\bibitem{Hawking74-1}
S.~W. Hawking, \emph{{Black hole explosions}},
  \href{http://dx.doi.org/10.1038/248030a0}{\emph{Nature} {\bf 248} (1974)
  30--31}.

\bibitem{Susskind:1994vu}
L.~Susskind, \emph{{The World as a hologram}},
  \href{http://dx.doi.org/10.1063/1.531249}{\emph{J. Math. Phys.} {\bf 36}
  (1995) 6377--6396}, [\href{https://arxiv.org/abs/hep-th/9409089}{{\tt
  hep-th/9409089}}].

\bibitem{Jacobson:1995ab}
T.~Jacobson, \emph{{Thermodynamics of space-time: The Einstein equation of
  state}}, \href{http://dx.doi.org/10.1103/PhysRevLett.75.1260}{\emph{Phys.
  Rev. Lett.} {\bf 75} (1995) 1260--1263},
  [\href{https://arxiv.org/abs/gr-qc/9504004}{{\tt gr-qc/9504004}}].

\bibitem{Parikh:2017aas}
M.~Parikh and A.~Svesko, \emph{{Einstein’s equations from the stretched
  future light cone}},
  \href{http://dx.doi.org/10.1103/PhysRevD.98.026018}{\emph{Phys. Rev.} {\bf
  D98} (2018) 026018}, [\href{https://arxiv.org/abs/1712.08475}{{\tt
  1712.08475}}].

\bibitem{Parikh14-1}
M.~Parikh and J.~P. van~der Schaar, \emph{{Derivation of the Null Energy
  Condition}}, \href{http://dx.doi.org/10.1103/PhysRevD.91.084002}{\emph{Phys.
  Rev.} {\bf D91} (2015) 084002}, [\href{https://arxiv.org/abs/1406.5163}{{\tt
  1406.5163}}].

\bibitem{Parikh:2015ret}
M.~Parikh and A.~Svesko, \emph{{Thermodynamic Origin of the Null Energy
  Condition}}, \href{http://dx.doi.org/10.1103/PhysRevD.95.104002}{\emph{Phys.
  Rev.} {\bf D95} (2017) 104002}, [\href{https://arxiv.org/abs/1511.06460}{{\tt
  1511.06460}}].

\bibitem{Parikh:2016lys}
M.~Parikh and A.~Svesko, \emph{{Logarithmic corrections to gravitational
  entropy and the null energy condition}},
  \href{http://dx.doi.org/10.1016/j.physletb.2016.07.071}{\emph{Phys. Lett.}
  {\bf B761} (2016) 16--19}, [\href{https://arxiv.org/abs/1612.06949}{{\tt
  1612.06949}}].

\bibitem{Strominger96-1}
A.~Strominger and C.~Vafa, \emph{{Microscopic origin of the Bekenstein-Hawking
  entropy}}, \href{http://dx.doi.org/10.1016/0370-2693(96)00345-0}{\emph{Phys.
  Lett.} {\bf B379} (1996) 99--104},
  [\href{https://arxiv.org/abs/hep-th/9601029}{{\tt hep-th/9601029}}].

\bibitem{Rovelli96-1}
C.~Rovelli, \emph{{Black hole entropy from loop quantum gravity}},
  \href{http://dx.doi.org/10.1103/PhysRevLett.77.3288}{\emph{Phys. Rev. Lett.}
  {\bf 77} (1996) 3288--3291}, [\href{https://arxiv.org/abs/gr-qc/9603063}{{\tt
  gr-qc/9603063}}].

\bibitem{Bombelli:1986rw}
L.~Bombelli, R.~K. Koul, J.~Lee and R.~D. Sorkin, \emph{{A Quantum Source of
  Entropy for Black Holes}},
  \href{http://dx.doi.org/10.1103/PhysRevD.34.373}{\emph{Phys. Rev.} {\bf D34}
  (1986) 373--383}.

\bibitem{Srednicki:1993im}
M.~Srednicki, \emph{{Entropy and area}},
  \href{http://dx.doi.org/10.1103/PhysRevLett.71.666}{\emph{Phys. Rev. Lett.}
  {\bf 71} (1993) 666--669}, [\href{https://arxiv.org/abs/hep-th/9303048}{{\tt
  hep-th/9303048}}].

\bibitem{Maldacena98-1}
J.~M. Maldacena, \emph{{The Large N limit of superconformal field theories and
  supergravity}}, \href{http://dx.doi.org/10.1063/1.59653}{\emph{AIP Conf.
  Proc.} {\bf 484} (1999) 51--63}.

\bibitem{Ryu06-1}
S.~Ryu and T.~Takayanagi, \emph{{Holographic derivation of entanglement entropy
  from AdS/CFT}},
  \href{http://dx.doi.org/10.1103/PhysRevLett.96.181602}{\emph{Phys. Rev.
  Lett.} {\bf 96} (2006) 181602},
  [\href{https://arxiv.org/abs/hep-th/0603001}{{\tt hep-th/0603001}}].

\bibitem{Hubeny:2007xt}
V.~E. Hubeny, M.~Rangamani and T.~Takayanagi, \emph{{A Covariant holographic
  entanglement entropy proposal}},
  \href{http://dx.doi.org/10.1088/1126-6708/2007/07/062}{\emph{JHEP} {\bf 07}
  (2007) 062}, [\href{https://arxiv.org/abs/0705.0016}{{\tt 0705.0016}}].

\bibitem{Lewkowycz:2013nqa}
A.~Lewkowycz and J.~Maldacena, \emph{{Generalized gravitational entropy}},
  \href{http://dx.doi.org/10.1007/JHEP08(2013)090}{\emph{JHEP} {\bf 08} (2013)
  090}, [\href{https://arxiv.org/abs/1304.4926}{{\tt 1304.4926}}].

\bibitem{Faulkner13-1}
T.~Faulkner, A.~Lewkowycz and J.~Maldacena, \emph{{Quantum corrections to
  holographic entanglement entropy}},
  \href{http://dx.doi.org/10.1007/JHEP11(2013)074}{\emph{JHEP} {\bf 11} (2013)
  074}, [\href{https://arxiv.org/abs/1307.2892}{{\tt 1307.2892}}].

\bibitem{Dong:2013qoa}
X.~Dong, \emph{{Holographic Entanglement Entropy for General Higher Derivative
  Gravity}}, \href{http://dx.doi.org/10.1007/JHEP01(2014)044}{\emph{JHEP} {\bf
  01} (2014) 044}, [\href{https://arxiv.org/abs/1310.5713}{{\tt 1310.5713}}].

\bibitem{Casini:2011kv}
H.~Casini, M.~Huerta and R.~C. Myers, \emph{{Towards a derivation of
  holographic entanglement entropy}},
  \href{http://dx.doi.org/10.1007/JHEP05(2011)036}{\emph{JHEP} {\bf 05} (2011)
  036}, [\href{https://arxiv.org/abs/1102.0440}{{\tt 1102.0440}}].

\bibitem{VanRaamsdonk10-1}
M.~Van~Raamsdonk, \emph{{Building up spacetime with quantum entanglement}},
  \href{http://dx.doi.org/10.1007/s10714-010-1034-0,
  10.1142/S0218271810018529}{\emph{Gen. Rel. Grav.} {\bf 42} (2010)
  2323--2329}, [\href{https://arxiv.org/abs/1005.3035}{{\tt 1005.3035}}].

\bibitem{Bianchi12-1}
E.~Bianchi, \emph{{Entropy of Non-Extremal Black Holes from Loop Gravity}},
  \href{https://arxiv.org/abs/1204.5122}{{\tt 1204.5122}}.

\bibitem{Blanco:2013joa}
D.~D. Blanco, H.~Casini, L.-Y. Hung and R.~C. Myers, \emph{{Relative Entropy
  and Holography}},
  \href{http://dx.doi.org/10.1007/JHEP08(2013)060}{\emph{JHEP} {\bf 08} (2013)
  060}, [\href{https://arxiv.org/abs/1305.3182}{{\tt 1305.3182}}].

\bibitem{Wong:2013gua}
G.~Wong, I.~Klich, L.~A. Pando~Zayas and D.~Vaman, \emph{{Entanglement
  Temperature and Entanglement Entropy of Excited States}},
  \href{http://dx.doi.org/10.1007/JHEP12(2013)020}{\emph{JHEP} {\bf 12} (2013)
  020}, [\href{https://arxiv.org/abs/1305.3291}{{\tt 1305.3291}}].

\bibitem{Lashkari13-1}
N.~Lashkari, M.~B. McDermott and M.~Van~Raamsdonk, \emph{{Gravitational
  dynamics from entanglement 'thermodynamics'}},
  \href{http://dx.doi.org/10.1007/JHEP04(2014)195}{\emph{JHEP} {\bf 04} (2014)
  195}, [\href{https://arxiv.org/abs/1308.3716}{{\tt 1308.3716}}].

\bibitem{Faulkner13-2}
T.~Faulkner, M.~Guica, T.~Hartman, R.~C. Myers and M.~Van~Raamsdonk,
  \emph{{Gravitation from Entanglement in Holographic CFTs}},
  \href{http://dx.doi.org/10.1007/JHEP03(2014)051}{\emph{JHEP} {\bf 03} (2014)
  051}, [\href{https://arxiv.org/abs/1312.7856}{{\tt 1312.7856}}].

\bibitem{Faulkner:2017tkh}
T.~Faulkner, F.~M. Haehl, E.~Hijano, O.~Parrikar, C.~Rabideau and
  M.~Van~Raamsdonk, \emph{{Nonlinear Gravity from Entanglement in Conformal
  Field Theories}},
  \href{http://dx.doi.org/10.1007/JHEP08(2017)057}{\emph{JHEP} {\bf 08} (2017)
  057}, [\href{https://arxiv.org/abs/1705.03026}{{\tt 1705.03026}}].

\bibitem{Haehl:2017sot}
F.~M. Haehl, E.~Hijano, O.~Parrikar and C.~Rabideau, \emph{{Higher Curvature
  Gravity from Entanglement in Conformal Field Theories}},
  \href{http://dx.doi.org/10.1103/PhysRevLett.120.201602}{\emph{Phys. Rev.
  Lett.} {\bf 120} (2018) 201602},
  [\href{https://arxiv.org/abs/1712.06620}{{\tt 1712.06620}}].

\bibitem{Jacobson16-1}
T.~Jacobson, \emph{{Entanglement Equilibrium and the Einstein Equation}},
  \href{http://dx.doi.org/10.1103/PhysRevLett.116.201101}{\emph{Phys. Rev.
  Lett.} {\bf 116} (2016) 201101},
  [\href{https://arxiv.org/abs/1505.04753}{{\tt 1505.04753}}].

\bibitem{Susskind:1994sm}
L.~Susskind and J.~Uglum, \emph{{Black hole entropy in canonical quantum
  gravity and superstring theory}},
  \href{http://dx.doi.org/10.1103/PhysRevD.50.2700}{\emph{Phys. Rev.} {\bf D50}
  (1994) 2700--2711}, [\href{https://arxiv.org/abs/hep-th/9401070}{{\tt
  hep-th/9401070}}].

\bibitem{Solodukhin:2011gn}
S.~N. Solodukhin, \emph{{Entanglement entropy of black holes}},
  \href{http://dx.doi.org/10.12942/lrr-2011-8}{\emph{Living Rev. Rel.} {\bf 14}
  (2011) 8}, [\href{https://arxiv.org/abs/1104.3712}{{\tt 1104.3712}}].

\bibitem{Bueno16-1}
P.~Bueno, V.~S. Min, A.~J. Speranza and M.~R. Visser, \emph{{Entanglement
  equilibrium for higher order gravity}},
  \href{https://arxiv.org/abs/1612.04374}{{\tt 1612.04374}}.

\bibitem{Strominger:1997eq}
A.~Strominger, \emph{{Black hole entropy from near horizon microstates}},
  \href{http://dx.doi.org/10.1088/1126-6708/1998/02/009}{\emph{JHEP} {\bf 02}
  (1998) 009}, [\href{https://arxiv.org/abs/hep-th/9712251}{{\tt
  hep-th/9712251}}].

\bibitem{Brown:1986nw}
J.~D. Brown and M.~Henneaux, \emph{{Central Charges in the Canonical
  Realization of Asymptotic Symmetries: An Example from Three-Dimensional
  Gravity}}, \href{http://dx.doi.org/10.1007/BF01211590}{\emph{Commun. Math.
  Phys.} {\bf 104} (1986) 207--226}.

\bibitem{Kubiznak:2016qmn}
D.~Kubiznak, R.~B. Mann and M.~Teo, \emph{{Black hole chemistry: thermodynamics
  with Lambda}}, \href{http://dx.doi.org/10.1088/1361-6382/aa5c69}{\emph{Class.
  Quant. Grav.} {\bf 34} (2017) 063001},
  [\href{https://arxiv.org/abs/1608.06147}{{\tt 1608.06147}}].

\bibitem{Caldarelli:1999xj}
M.~M. Caldarelli, G.~Cognola and D.~Klemm, \emph{{Thermodynamics of
  Kerr-Newman-AdS black holes and conformal field theories}},
  \href{http://dx.doi.org/10.1088/0264-9381/17/2/310}{\emph{Class. Quant.
  Grav.} {\bf 17} (2000) 399--420},
  [\href{https://arxiv.org/abs/hep-th/9908022}{{\tt hep-th/9908022}}].

\bibitem{Sekiwa:2006qj}
Y.~Sekiwa, \emph{{Thermodynamics of de Sitter black holes: Thermal cosmological
  constant}}, \href{http://dx.doi.org/10.1103/PhysRevD.73.084009}{\emph{Phys.
  Rev.} {\bf D73} (2006) 084009},
  [\href{https://arxiv.org/abs/hep-th/0602269}{{\tt hep-th/0602269}}].

\bibitem{Kastor:2009wy}
D.~Kastor, S.~Ray and J.~Traschen, \emph{{Enthalpy and the Mechanics of AdS
  Black Holes}},
  \href{http://dx.doi.org/10.1088/0264-9381/26/19/195011}{\emph{Class. Quant.
  Grav.} {\bf 26} (2009) 195011}, [\href{https://arxiv.org/abs/0904.2765}{{\tt
  0904.2765}}].

\bibitem{Cvetic:2010jb}
M.~Cvetic, G.~Gibbons, D.~Kubiznak and C.~Pope, \emph{{Black Hole Enthalpy and
  an Entropy Inequality for the Thermodynamic Volume}},
  \href{http://dx.doi.org/10.1103/PhysRevD.84.024037}{\emph{Phys.Rev.} {\bf
  D84} (2011) 024037}, [\href{https://arxiv.org/abs/1012.2888}{{\tt
  1012.2888}}].

\bibitem{Kastor:2014dra}
D.~Kastor, S.~Ray and J.~Traschen, \emph{{Chemical Potential in the First Law
  for Holographic Entanglement Entropy}},
  \href{http://dx.doi.org/10.1007/JHEP11(2014)120}{\emph{JHEP} {\bf 11} (2014)
  120}, [\href{https://arxiv.org/abs/1409.3521}{{\tt 1409.3521}}].

\bibitem{Johnson:2014yja}
C.~V. Johnson, \emph{{Holographic Heat Engines}},
  \href{http://dx.doi.org/10.1088/0264-9381/31/20/205002}{\emph{Class. Quant.
  Grav.} {\bf 31} (2014) 205002}, [\href{https://arxiv.org/abs/1404.5982}{{\tt
  1404.5982}}].

\bibitem{Johnson:2018amj}
C.~V. Johnson and F.~Rosso, \emph{{Holographic Heat Engines, Entanglement
  Entropy, and Renormalization Group Flow}},
  \href{http://dx.doi.org/10.1088/1361-6382/aaf1f1}{\emph{Class. Quant. Grav.}
  {\bf 36} (2019) 015019}, [\href{https://arxiv.org/abs/1806.05170}{{\tt
  1806.05170}}].

\bibitem{Johnson:2019wcq}
C.~V. Johnson, V.~L. Martin and A.~Svesko, \emph{{Microscopic description of
  thermodynamic volume in extended black hole thermodynamics}},
  \href{http://dx.doi.org/10.1103/PhysRevD.101.086006}{\emph{Phys. Rev. D} {\bf
  101} (2020) 086006}, [\href{https://arxiv.org/abs/1911.05286}{{\tt
  1911.05286}}].

\bibitem{Parikh:2018anm}
M.~Parikh, S.~Sarkar and A.~Svesko, \emph{{Local first law of gravity}},
  \href{http://dx.doi.org/10.1103/PhysRevD.101.104043}{\emph{Phys. Rev. D} {\bf
  101} (2020) 104043}, [\href{https://arxiv.org/abs/1801.07306}{{\tt
  1801.07306}}].

\bibitem{Svesko:2018qim}
A.~Svesko, \emph{{Equilibrium to Einstein: Entanglement, Thermodynamics, and
  Gravity}}, \href{http://dx.doi.org/10.1103/PhysRevD.99.086006}{\emph{Phys.
  Rev. D} {\bf 99} (2019) 086006},
  [\href{https://arxiv.org/abs/1810.12236}{{\tt 1810.12236}}].

\bibitem{Rosso:2020zkk}
F.~Rosso and A.~Svesko, \emph{{Novel aspects of the extended first law of
  entanglement}}, \href{http://dx.doi.org/10.1007/JHEP08(2020)008}{\emph{JHEP}
  {\bf 08} (2020) 008}, [\href{https://arxiv.org/abs/2003.10462}{{\tt
  2003.10462}}].

\bibitem{Sakharov:1967pk}
A.~D. Sakharov, \emph{{Vacuum quantum fluctuations in curved space and the
  theory of gravitation}}, {\emph{Sov. Phys. Dokl.} {\bf 12} (1968)
  1040--1041}.

\bibitem{Visser:2002ew}
M.~Visser, \emph{{Sakharov's induced gravity: A Modern perspective}},
  \href{http://dx.doi.org/10.1142/S0217732302006886}{\emph{Mod. Phys. Lett.}
  {\bf A17} (2002) 977--992}, [\href{https://arxiv.org/abs/gr-qc/0204062}{{\tt
  gr-qc/0204062}}].

\bibitem{Vassilevich:2003xt}
D.~V. Vassilevich, \emph{{Heat kernel expansion: User's manual}},
  \href{http://dx.doi.org/10.1016/j.physrep.2003.09.002}{\emph{Phys. Rept.}
  {\bf 388} (2003) 279--360}, [\href{https://arxiv.org/abs/hep-th/0306138}{{\tt
  hep-th/0306138}}].

\bibitem{Frolov:1996aj}
V.~P. Frolov, D.~V. Fursaev and A.~I. Zelnikov, \emph{{Statistical origin of
  black hole entropy in induced gravity}},
  \href{http://dx.doi.org/10.1016/S0550-3213(96)00678-5}{\emph{Nucl. Phys.}
  {\bf B486} (1997) 339--352},
  [\href{https://arxiv.org/abs/hep-th/9607104}{{\tt hep-th/9607104}}].

\bibitem{Frolov:1996qh}
V.~P. Frolov, D.~V. Fursaev and A.~I. Zelnikov, \emph{{Black hole statistical
  mechanics and induced gravity}},
  \href{http://dx.doi.org/10.1016/S0920-5632(97)00373-3}{\emph{Nucl. Phys.
  Proc. Suppl.} {\bf 57} (1997) 192--196}.

\bibitem{Frolov:1997xd}
V.~P. Frolov and D.~V. Fursaev, \emph{{Plenty of nothing: Black hole entropy in
  induced gravity}},  \href{https://arxiv.org/abs/hep-th/9705207}{{\tt
  hep-th/9705207}}.

\bibitem{Frolov:1997up}
V.~P. Frolov and D.~V. Fursaev, \emph{{Mechanism of generation of black hole
  entropy in Sakharov's induced gravity}},
  \href{http://dx.doi.org/10.1103/PhysRevD.56.2212}{\emph{Phys. Rev.} {\bf D56}
  (1997) 2212--2225}, [\href{https://arxiv.org/abs/hep-th/9703178}{{\tt
  hep-th/9703178}}].

\bibitem{Eling:2006aw}
C.~Eling, R.~Guedens and T.~Jacobson, \emph{{Non-equilibrium thermodynamics of
  spacetime}},
  \href{http://dx.doi.org/10.1103/PhysRevLett.96.121301}{\emph{Phys. Rev.
  Lett.} {\bf 96} (2006) 121301},
  [\href{https://arxiv.org/abs/gr-qc/0602001}{{\tt gr-qc/0602001}}].

\bibitem{Padmanabhan:2009ry}
T.~Padmanabhan, \emph{{Entropy density of spacetime and thermodynamic
  interpretation of field equations of gravity in any diffeomorphism invariant
  theory}},  \href{https://arxiv.org/abs/0903.1254}{{\tt 0903.1254}}.

\bibitem{Parikh:2009qs}
M.~K. Parikh and S.~Sarkar, \emph{{Beyond the Einstein Equation of State: Wald
  Entropy and Thermodynamical Gravity}},
  \href{https://arxiv.org/abs/0903.1176}{{\tt 0903.1176}}.

\bibitem{Brustein:2009hy}
R.~Brustein and M.~Hadad, \emph{{The Einstein equations for generalized
  theories of gravity and the thermodynamic relation delta Q = T delta S are
  equivalent}}, \href{http://dx.doi.org/10.1103/PhysRevLett.105.239902,
  10.1103/PhysRevLett.103.101301}{\emph{Phys. Rev. Lett.} {\bf 103} (2009)
  101301}, [\href{https://arxiv.org/abs/0903.0823}{{\tt 0903.0823}}].

\bibitem{Guedens:2011dy}
R.~Guedens, T.~Jacobson and S.~Sarkar, \emph{{Horizon entropy and higher
  curvature equations of state}},
  \href{http://dx.doi.org/10.1103/PhysRevD.85.064017}{\emph{Phys. Rev.} {\bf
  D85} (2012) 064017}, [\href{https://arxiv.org/abs/1112.6215}{{\tt
  1112.6215}}].

\bibitem{Carroll16-1}
S.~M. Carroll and G.~N. Remmen, \emph{{What is the Entropy in Entropic
  Gravity?}}, \href{http://dx.doi.org/10.1103/PhysRevD.93.124052}{\emph{Phys.
  Rev.} {\bf D93} (2016) 124052}, [\href{https://arxiv.org/abs/1601.07558}{{\tt
  1601.07558}}].

\bibitem{Verlinde:2010hp}
E.~P. Verlinde, \emph{{On the Origin of Gravity and the Laws of Newton}},
  \href{http://dx.doi.org/10.1007/JHEP04(2011)029}{\emph{JHEP} {\bf 04} (2011)
  029}, [\href{https://arxiv.org/abs/1001.0785}{{\tt 1001.0785}}].

\bibitem{Easson:2010av}
D.~A. Easson, P.~H. Frampton and G.~F. Smoot, \emph{{Entropic Accelerating
  Universe}},
  \href{http://dx.doi.org/10.1016/j.physletb.2010.12.025}{\emph{Phys. Lett.}
  {\bf B696} (2011) 273--277}, [\href{https://arxiv.org/abs/1002.4278}{{\tt
  1002.4278}}].

\bibitem{Verlinde:2016toy}
E.~P. Verlinde, \emph{{Emergent Gravity and the Dark Universe}},
  \href{http://dx.doi.org/10.21468/SciPostPhys.2.3.016}{\emph{SciPost Phys.}
  {\bf 2} (2017) 016}, [\href{https://arxiv.org/abs/1611.02269}{{\tt
  1611.02269}}].

\bibitem{Visser:2011jp}
M.~Visser, \emph{{Conservative entropic forces}},
  \href{http://dx.doi.org/10.1007/JHEP10(2011)140}{\emph{JHEP} {\bf 10} (2011)
  140}, [\href{https://arxiv.org/abs/1108.5240}{{\tt 1108.5240}}].

\bibitem{Brouwer:2016dvq}
M.~M. Brouwer et~al., \emph{{First test of Verlinde's theory of Emergent
  Gravity using Weak Gravitational Lensing measurements}},
  \href{http://dx.doi.org/10.1093/mnras/stw3192}{\emph{Mon. Not. Roy. Astron.
  Soc.} {\bf 466} (2017) 2547--2559},
  [\href{https://arxiv.org/abs/1612.03034}{{\tt 1612.03034}}].

\bibitem{Pardo:2017jun}
K.~Pardo, \emph{{Testing Emergent Gravity with Isolated Dwarf Galaxies}},
  \href{https://arxiv.org/abs/1706.00785}{{\tt 1706.00785}}.

\bibitem{Bianchi:2012ev}
E.~Bianchi and R.~C. Myers, \emph{{On the Architecture of Spacetime Geometry}},
  \href{http://dx.doi.org/10.1088/0264-9381/31/21/214002}{\emph{Class. Quant.
  Grav.} {\bf 31} (2014) 214002}, [\href{https://arxiv.org/abs/1212.5183}{{\tt
  1212.5183}}].

\bibitem{Callan:1994py}
C.~G. Callan, Jr. and F.~Wilczek, \emph{{On geometric entropy}},
  \href{http://dx.doi.org/10.1016/0370-2693(94)91007-3}{\emph{Phys. Lett.} {\bf
  B333} (1994) 55--61}, [\href{https://arxiv.org/abs/hep-th/9401072}{{\tt
  hep-th/9401072}}].

\bibitem{Jacobson:1994iw}
T.~Jacobson, \emph{{Black hole entropy and induced gravity}},
  \href{https://arxiv.org/abs/gr-qc/9404039}{{\tt gr-qc/9404039}}.

\bibitem{Ryu06-2}
S.~Ryu and T.~Takayanagi, \emph{{Aspects of Holographic Entanglement Entropy}},
  \href{http://dx.doi.org/10.1088/1126-6708/2006/08/045}{\emph{JHEP} {\bf 08}
  (2006) 045}, [\href{https://arxiv.org/abs/hep-th/0605073}{{\tt
  hep-th/0605073}}].

\bibitem{Jacobson:2018ahi}
T.~Jacobson and M.~Visser, \emph{{Gravitational Thermodynamics of Causal
  Diamonds in (A)dS}},  \href{https://arxiv.org/abs/1812.01596}{{\tt
  1812.01596}}.

\bibitem{Maldacena:2013xja}
J.~Maldacena and L.~Susskind, \emph{{Cool horizons for entangled black holes}},
  \href{http://dx.doi.org/10.1002/prop.201300020}{\emph{Fortsch. Phys.} {\bf
  61} (2013) 781--811}, [\href{https://arxiv.org/abs/1306.0533}{{\tt
  1306.0533}}].

\bibitem{Hayden:2016cfa}
P.~Hayden, S.~Nezami, X.-L. Qi, N.~Thomas, M.~Walter and Z.~Yang,
  \emph{{Holographic duality from random tensor networks}},
  \href{http://dx.doi.org/10.1007/JHEP11(2016)009}{\emph{JHEP} {\bf 11} (2016)
  009}, [\href{https://arxiv.org/abs/1601.01694}{{\tt 1601.01694}}].

\bibitem{Almheiri:2014lwa}
A.~Almheiri, X.~Dong and D.~Harlow, \emph{{Bulk Locality and Quantum Error
  Correction in AdS/CFT}},
  \href{http://dx.doi.org/10.1007/JHEP04(2015)163}{\emph{JHEP} {\bf 04} (2015)
  163}, [\href{https://arxiv.org/abs/1411.7041}{{\tt 1411.7041}}].

\bibitem{Pastawski:2015qua}
F.~Pastawski, B.~Yoshida, D.~Harlow and J.~Preskill, \emph{{Holographic quantum
  error-correcting codes: Toy models for the bulk/boundary correspondence}},
  \href{http://dx.doi.org/10.1007/JHEP06(2015)149}{\emph{JHEP} {\bf 06} (2015)
  149}, [\href{https://arxiv.org/abs/1503.06237}{{\tt 1503.06237}}].

\bibitem{Harlow:2016vwg}
D.~Harlow, \emph{{The Ryu–Takayanagi Formula from Quantum Error Correction}},
  \href{http://dx.doi.org/10.1007/s00220-017-2904-z}{\emph{Commun. Math. Phys.}
  {\bf 354} (2017) 865--912}, [\href{https://arxiv.org/abs/1607.03901}{{\tt
  1607.03901}}].

\bibitem{Freedman:2016zud}
M.~Freedman and M.~Headrick, \emph{{Bit threads and holographic entanglement}},
  \href{http://dx.doi.org/10.1007/s00220-016-2796-3}{\emph{Commun. Math. Phys.}
  {\bf 352} (2017) 407--438}, [\href{https://arxiv.org/abs/1604.00354}{{\tt
  1604.00354}}].

\bibitem{Bardeen73-1}
J.~M. Bardeen, B.~Carter and S.~W. Hawking, \emph{{The Four laws of black hole
  mechanics}}, \href{http://dx.doi.org/10.1007/BF01645742}{\emph{Commun. Math.
  Phys.} {\bf 31} (1973) 161--170}.

\bibitem{Hawking70-1}
S.~W. Hawking and R.~Penrose, \emph{{The Singularities of gravitational
  collapse and cosmology}},
  \href{http://dx.doi.org/10.1098/rspa.1970.0021}{\emph{Proc. Roy. Soc. Lond.}
  {\bf A314} (1970) 529--548}.

\bibitem{molinaparis99-1}
C.~Molina-Paris and M.~Visser, \emph{{Minimal conditions for the creation of a
  Friedman-Robertson-Walker universe from a 'bounce'}},
  \href{http://dx.doi.org/10.1016/S0370-2693(99)00469-4}{\emph{Phys. Lett.}
  {\bf B455} (1999) 90--95}, [\href{https://arxiv.org/abs/gr-qc/9810023}{{\tt
  gr-qc/9810023}}].

\bibitem{Parikh15-2}
M.~K. Parikh, \emph{{No Open or Flat Bouncing Cosmologies in Einstein
  Gravity}}, \href{http://dx.doi.org/10.1007/JHEP10(2015)089}{\emph{JHEP} {\bf
  10} (2015) 089}, [\href{https://arxiv.org/abs/1501.04606}{{\tt 1501.04606}}].

\bibitem{Hawking92-1}
S.~W. Hawking, \emph{{The chronology protection conjecture}},  in \emph{{Recent
  developments in theoretical and experimental general relativity, gravitation
  and relativistic field theories. Proceedings, 6th Marcel Grossmann Meeting,
  Kyoto, Japan, June 23-29, 1991. Pts. A, B}}, 1991.

\bibitem{Farhi87-1}
E.~Farhi and A.~H. Guth, \emph{{An Obstacle to Creating a Universe in the
  Laboratory}},
  \href{http://dx.doi.org/10.1016/0370-2693(87)90429-1}{\emph{Phys. Lett.} {\bf
  B183} (1987) 149}.

\bibitem{Morris88-2}
M.~S. Morris, K.~S. Thorne and U.~Yurtsever, \emph{{Wormholes, Time Machines,
  and the Weak Energy Condition}},
  \href{http://dx.doi.org/10.1103/PhysRevLett.61.1446}{\emph{Phys. Rev. Lett.}
  {\bf 61} (1988) 1446--1449}.

\bibitem{Freedman99-1}
D.~Z. Freedman, S.~S. Gubser, K.~Pilch and N.~P. Warner, \emph{{Renormalization
  group flows from holography supersymmetry and a c theorem}}, {\emph{Adv.
  Theor. Math. Phys.} {\bf 3} (1999) 363--417},
  [\href{https://arxiv.org/abs/hep-th/9904017}{{\tt hep-th/9904017}}].

\bibitem{Barcelo02-1}
C.~Barcelo and M.~Visser, \emph{{Twilight for the energy conditions?}},
  \href{http://dx.doi.org/10.1142/S0218271802002888}{\emph{Int. J. Mod. Phys.}
  {\bf D11} (2002) 1553--1560},
  [\href{https://arxiv.org/abs/gr-qc/0205066}{{\tt gr-qc/0205066}}].

\bibitem{Rubakov14-1}
V.~A. Rubakov, \emph{{The Null Energy Condition and its violation}},
  \href{http://dx.doi.org/10.3367/UFNe.0184.201402b.0137}{\emph{Phys. Usp.}
  {\bf 57} (2014) 128--142}, [\href{https://arxiv.org/abs/1401.4024}{{\tt
  1401.4024}}].

\bibitem{Chatterjee:2012zh}
S.~Chatterjee, D.~A. Easson and M.~Parikh, \emph{{Energy conditions in the
  Jordan frame}},
  \href{http://dx.doi.org/10.1088/0264-9381/30/23/235031}{\emph{Class. Quant.
  Grav.} {\bf 30} (2013) 235031}, [\href{https://arxiv.org/abs/1212.6430}{{\tt
  1212.6430}}].

\bibitem{Wall:2009wi}
A.~C. Wall, \emph{{Proving the Achronal Averaged Null Energy Condition from the
  Generalized Second Law}},
  \href{http://dx.doi.org/10.1103/PhysRevD.81.024038}{\emph{Phys. Rev.} {\bf
  D81} (2010) 024038}, [\href{https://arxiv.org/abs/0910.5751}{{\tt
  0910.5751}}].

\bibitem{Lashkari:2014kda}
N.~Lashkari, C.~Rabideau, P.~Sabella-Garnier and M.~Van~Raamsdonk,
  \emph{{Inviolable energy conditions from entanglement inequalities}},
  \href{http://dx.doi.org/10.1007/JHEP06(2015)067}{\emph{JHEP} {\bf 06} (2015)
  067}, [\href{https://arxiv.org/abs/1412.3514}{{\tt 1412.3514}}].

\bibitem{Kontou:2015yha}
E.-A. Kontou and K.~D. Olum, \emph{{Proof of the averaged null energy condition
  in a classical curved spacetime using a null-projected quantum inequality}},
  \href{http://dx.doi.org/10.1103/PhysRevD.92.124009}{\emph{Phys. Rev.} {\bf
  D92} (2015) 124009}, [\href{https://arxiv.org/abs/1507.00297}{{\tt
  1507.00297}}].

\bibitem{Bousso:2015wca}
R.~Bousso, Z.~Fisher, J.~Koeller, S.~Leichenauer and A.~C. Wall, \emph{{Proof
  of the Quantum Null Energy Condition}},
  \href{http://dx.doi.org/10.1103/PhysRevD.93.024017}{\emph{Phys. Rev.} {\bf
  D93} (2016) 024017}, [\href{https://arxiv.org/abs/1509.02542}{{\tt
  1509.02542}}].

\bibitem{Faulkner:2016mzt}
T.~Faulkner, R.~G. Leigh, O.~Parrikar and H.~Wang, \emph{{Modular Hamiltonians
  for Deformed Half-Spaces and the Averaged Null Energy Condition}},
  \href{http://dx.doi.org/10.1007/JHEP09(2016)038}{\emph{JHEP} {\bf 09} (2016)
  038}, [\href{https://arxiv.org/abs/1605.08072}{{\tt 1605.08072}}].

\bibitem{Hartman:2016lgu}
T.~Hartman, S.~Kundu and A.~Tajdini, \emph{{Averaged Null Energy Condition from
  Causality}}, \href{http://dx.doi.org/10.1007/JHEP07(2017)066}{\emph{JHEP}
  {\bf 07} (2017) 066}, [\href{https://arxiv.org/abs/1610.05308}{{\tt
  1610.05308}}].

\bibitem{Falkovich:2004}
G.~Falkovich and A.~Fouxon, \emph{{Entropy production and extraction in
  dynamical systems and turbulence}}, {\emph{New J. Phys} {\bf 06} (2004) 50}.

\bibitem{Onsager:1931}
L.~Onsager, \emph{{Reciprocal Relations in Irreversible Processes. I.}},
  {\emph{Phys. Rev.} {\bf 37} (1931) 405}.

\bibitem{Wald:1993nt}
R.~M. Wald, \emph{{Black hole entropy is the Noether charge}},
  \href{http://dx.doi.org/10.1103/PhysRevD.48.R3427}{\emph{Phys. Rev.} {\bf
  D48} (1993) R3427--R3431}, [\href{https://arxiv.org/abs/gr-qc/9307038}{{\tt
  gr-qc/9307038}}].

\bibitem{Birrell82-1}
N.~D. Birrell and P.~C.~W. Davies, \emph{{Quantum Fields in Curved Space}}.
\newblock Cambridge Monographs on Mathematical Physics. Cambridge Univ. Press,
  Cambridge, UK, 1984,
  \href{http://dx.doi.org/10.1017/CBO9780511622632}{10.1017/CBO9780511622632}.

\bibitem{Kaul00-1}
R.~K. Kaul and P.~Majumdar, \emph{{Logarithmic correction to the
  Bekenstein-Hawking entropy}},
  \href{http://dx.doi.org/10.1103/PhysRevLett.84.5255}{\emph{Phys. Rev. Lett.}
  {\bf 84} (2000) 5255--5257}, [\href{https://arxiv.org/abs/gr-qc/0002040}{{\tt
  gr-qc/0002040}}].

\bibitem{Graham07-1}
N.~Graham and K.~D. Olum, \emph{{Achronal averaged null energy condition}},
  \href{http://dx.doi.org/10.1103/PhysRevD.76.064001}{\emph{Phys. Rev.} {\bf
  D76} (2007) 064001}, [\href{https://arxiv.org/abs/0705.3193}{{\tt
  0705.3193}}].

\bibitem{Carlip:2000nv}
S.~Carlip, \emph{{Logarithmic corrections to black hole entropy from the Cardy
  formula}}, \href{http://dx.doi.org/10.1088/0264-9381/17/20/302}{\emph{Class.
  Quant. Grav.} {\bf 17} (2000) 4175--4186},
  [\href{https://arxiv.org/abs/gr-qc/0005017}{{\tt gr-qc/0005017}}].

\bibitem{Govindarajan:2001ee}
T.~R. Govindarajan, R.~K. Kaul and V.~Suneeta, \emph{{Logarithmic correction to
  the Bekenstein-Hawking entropy of the BTZ black hole}},
  \href{http://dx.doi.org/10.1088/0264-9381/18/15/303}{\emph{Class. Quant.
  Grav.} {\bf 18} (2001) 2877--2886},
  [\href{https://arxiv.org/abs/gr-qc/0104010}{{\tt gr-qc/0104010}}].

\bibitem{Fursaev95-1}
D.~V. Fursaev, \emph{{Temperature and entropy of a quantum black hole and
  conformal anomaly}},
  \href{http://dx.doi.org/10.1103/PhysRevD.51.5352}{\emph{Phys. Rev.} {\bf D51}
  (1995) 5352--5355}, [\href{https://arxiv.org/abs/hep-th/9412161}{{\tt
  hep-th/9412161}}].

\bibitem{Mann:1997hm}
R.~B. Mann and S.~N. Solodukhin, \emph{{Universality of quantum entropy for
  extreme black holes}},
  \href{http://dx.doi.org/10.1016/S0550-3213(98)00094-7}{\emph{Nucl. Phys.}
  {\bf B523} (1998) 293--307},
  [\href{https://arxiv.org/abs/hep-th/9709064}{{\tt hep-th/9709064}}].

\bibitem{Cai:2009ua}
R.-G. Cai, L.-M. Cao and N.~Ohta, \emph{{Black Holes in Gravity with Conformal
  Anomaly and Logarithmic Term in Black Hole Entropy}},
  \href{http://dx.doi.org/10.1007/JHEP04(2010)082}{\emph{JHEP} {\bf 04} (2010)
  082}, [\href{https://arxiv.org/abs/0911.4379}{{\tt 0911.4379}}].

\bibitem{Aros10-1}
R.~Aros, D.~E. Diaz and A.~Montecinos, \emph{{Logarithmic correction to BH
  entropy as Noether charge}},
  \href{http://dx.doi.org/10.1007/JHEP07(2010)012}{\emph{JHEP} {\bf 07} (2010)
  012}, [\href{https://arxiv.org/abs/1003.1083}{{\tt 1003.1083}}].

\bibitem{Medved:2004eh}
A.~J.~M. Medved, \emph{{A Comment on black hole entropy or does nature abhor a
  logarithm?}},
  \href{http://dx.doi.org/10.1088/0264-9381/22/1/009}{\emph{Class. Quant.
  Grav.} {\bf 22} (2005) 133--142},
  [\href{https://arxiv.org/abs/gr-qc/0406044}{{\tt gr-qc/0406044}}].

\bibitem{Page05-1}
D.~N. Page, \emph{{Hawking radiation and black hole thermodynamics}},
  \href{http://dx.doi.org/10.1088/1367-2630/7/1/203}{\emph{New J. Phys.} {\bf
  7} (2005) 203}, [\href{https://arxiv.org/abs/hep-th/0409024}{{\tt
  hep-th/0409024}}].

\bibitem{Parikh00-1}
M.~K. Parikh and F.~Wilczek, \emph{{Hawking radiation as tunneling}},
  \href{http://dx.doi.org/10.1103/PhysRevLett.85.5042}{\emph{Phys. Rev. Lett.}
  {\bf 85} (2000) 5042--5045},
  [\href{https://arxiv.org/abs/hep-th/9907001}{{\tt hep-th/9907001}}].

\bibitem{Parikh04-1}
M.~K. Parikh, \emph{{A Secret tunnel through the horizon}},
  \href{http://dx.doi.org/10.1142/S0218271804006498}{\emph{Int. J. Mod. Phys.}
  {\bf D13} (2004) 2351--2354},
  [\href{https://arxiv.org/abs/hep-th/0405160}{{\tt hep-th/0405160}}].

\bibitem{Eling06-1}
C.~Eling, R.~Guedens and T.~Jacobson, \emph{{Non-equilibrium thermodynamics of
  spacetime}},
  \href{http://dx.doi.org/10.1103/PhysRevLett.96.121301}{\emph{Phys. Rev.
  Lett.} {\bf 96} (2006) 121301},
  [\href{https://arxiv.org/abs/gr-qc/0602001}{{\tt gr-qc/0602001}}].

\bibitem{Dey:2016zka}
R.~Dey, S.~Liberati and A.~Mohd, \emph{{Higher derivative gravity: field
  equation as the equation of state}},
  \href{http://dx.doi.org/10.1103/PhysRevD.94.044013}{\emph{Phys. Rev. D} {\bf
  94} (2016) 044013}, [\href{https://arxiv.org/abs/1605.04789}{{\tt
  1605.04789}}].

\bibitem{Price86-1}
R.~H. Price and K.~S. Thorne, \emph{{Membrane Viewpoint on Black Holes:
  Properties and Evolution of the Stretched Horizon}},
  \href{http://dx.doi.org/10.1103/PhysRevD.33.915}{\emph{Phys. Rev.} {\bf D33}
  (1986) 915--941}.

\bibitem{Barbado:2012fy}
L.~C. Barbado and M.~Visser, \emph{{Unruh-DeWitt detector event rate for
  trajectories with time-dependent acceleration}},
  \href{http://dx.doi.org/10.1103/PhysRevD.86.084011}{\emph{Phys. Rev.} {\bf
  D86} (2012) 084011}, [\href{https://arxiv.org/abs/1207.5525}{{\tt
  1207.5525}}].

\bibitem{DeLorenzo:2017tgx}
T.~De~Lorenzo and A.~Perez, \emph{{Light Cone Thermodynamics}},
  \href{http://dx.doi.org/10.1103/PhysRevD.97.044052}{\emph{Phys. Rev.} {\bf
  D97} (2018) 044052}, [\href{https://arxiv.org/abs/1707.00479}{{\tt
  1707.00479}}].

\bibitem{Parikh:1997ma}
M.~Parikh and F.~Wilczek, \emph{{An Action for black hole membranes}},
  \href{http://dx.doi.org/10.1103/PhysRevD.58.064011}{\emph{Phys. Rev.} {\bf
  D58} (1998) 064011}, [\href{https://arxiv.org/abs/gr-qc/9712077}{{\tt
  gr-qc/9712077}}].

\bibitem{Guedens:2012sz}
R.~Guedens, \emph{{Locally inertial null normal coordinates}},
  \href{http://dx.doi.org/10.1088/0264-9381/29/14/145002}{\emph{Class. Quant.
  Grav.} {\bf 29} (2012) 145002}, [\href{https://arxiv.org/abs/1201.0542}{{\tt
  1201.0542}}].

\bibitem{Padmanabhan:2007en}
T.~Padmanabhan and A.~Paranjape, \emph{{Entropy of null surfaces and dynamics
  of spacetime}},
  \href{http://dx.doi.org/10.1103/PhysRevD.75.064004}{\emph{Phys. Rev.} {\bf
  D75} (2007) 064004}, [\href{https://arxiv.org/abs/gr-qc/0701003}{{\tt
  gr-qc/0701003}}].

\bibitem{Bekenstein73-1}
J.~D. Bekenstein, \emph{{Black holes and entropy}},
  \href{http://dx.doi.org/10.1103/PhysRevD.7.2333}{\emph{Phys. Rev.} {\bf D7}
  (1973) 2333--2346}.

\bibitem{Gao:2001ut}
S.~Gao and R.~M. Wald, \emph{{The 'Physical process' version of the first law
  and the generalized second law for charged and rotating black holes}},
  \href{http://dx.doi.org/10.1103/PhysRevD.64.084020}{\emph{Phys. Rev.} {\bf
  D64} (2001) 084020}, [\href{https://arxiv.org/abs/gr-qc/0106071}{{\tt
  gr-qc/0106071}}].

\bibitem{Jacobson:2003wv}
T.~Jacobson and R.~Parentani, \emph{{Horizon entropy}},
  \href{http://dx.doi.org/10.1023/A:1023785123428}{\emph{Found. Phys.} {\bf 33}
  (2003) 323--348}, [\href{https://arxiv.org/abs/gr-qc/0302099}{{\tt
  gr-qc/0302099}}].

\bibitem{Parikh:2005qs}
M.~K. Parikh, \emph{{The Volume of black holes}},
  \href{http://dx.doi.org/10.1103/PhysRevD.73.124021}{\emph{Phys. Rev.} {\bf
  D73} (2006) 124021}, [\href{https://arxiv.org/abs/hep-th/0508108}{{\tt
  hep-th/0508108}}].

\bibitem{Dolan:2011xt}
B.~P. Dolan, \emph{{Pressure and volume in the first law of black hole
  thermodynamics}},
  \href{http://dx.doi.org/10.1088/0264-9381/28/23/235017}{\emph{Class. Quant.
  Grav.} {\bf 28} (2011) 235017}, [\href{https://arxiv.org/abs/1106.6260}{{\tt
  1106.6260}}].

\bibitem{Dolan:2012jh}
B.~P. Dolan, \emph{{Where Is the PdV in the First Law of Black Hole
  Thermodynamics?}},  pp.~291--316.
\newblock INTECH, 2012.
\newblock \href{https://arxiv.org/abs/1209.1272}{{\tt 1209.1272}}.
\newblock \href{http://dx.doi.org/10.5772/52455}{DOI}.

\bibitem{McGrath:2012db}
P.~L. McGrath, R.~J. Epp and R.~B. Mann, \emph{{Quasilocal Conservation Laws:
  Why We Need Them}},
  \href{http://dx.doi.org/10.1088/0264-9381/29/21/215012}{\emph{Class. Quant.
  Grav.} {\bf 29} (2012) 215012}, [\href{https://arxiv.org/abs/1206.6512}{{\tt
  1206.6512}}].

\bibitem{Nielsen:2017hxt}
A.~B. Nielsen and A.~A. Shoom, \emph{{Conformal Killing horizons and their
  thermodynamics}},
  \href{http://dx.doi.org/10.1088/1361-6382/aab505}{\emph{Class. Quant. Grav.}
  {\bf 35} (2018) 105008}, [\href{https://arxiv.org/abs/1708.08015}{{\tt
  1708.08015}}].

\bibitem{Smarr:1972kt}
L.~Smarr, \emph{{Mass formula for Kerr black holes}},
  \href{http://dx.doi.org/10.1103/PhysRevLett.30.521,
  10.1103/PhysRevLett.30.71}{\emph{Phys. Rev. Lett.} {\bf 30} (1973) 71--73}.

\bibitem{Piazza:2010hz}
F.~Piazza, \emph{{Gauss-Codazzi thermodynamics on the timelike screen}},
  \href{http://dx.doi.org/10.1103/PhysRevD.82.084004}{\emph{Phys. Rev.} {\bf
  D82} (2010) 084004}, [\href{https://arxiv.org/abs/1005.5151}{{\tt
  1005.5151}}].

\bibitem{DeLorenzo:2018ghq}
T.~De~Lorenzo and A.~Perez, \emph{{Light Cone Black Holes}},
  \href{http://dx.doi.org/10.1103/PhysRevD.99.065009}{\emph{Phys. Rev.} {\bf
  D99} (2019) 065009}, [\href{https://arxiv.org/abs/1811.03667}{{\tt
  1811.03667}}].

\bibitem{Jacobson:2012yt}
T.~Jacobson, \emph{{Gravitation and vacuum entanglement entropy}},
  \href{http://dx.doi.org/10.1142/S0218271812420060}{\emph{Int. J. Mod. Phys.}
  {\bf D21} (2012) 1242006}, [\href{https://arxiv.org/abs/1204.6349}{{\tt
  1204.6349}}].

\bibitem{Iyer:1994ys}
V.~Iyer and R.~M. Wald, \emph{{Some properties of Noether charge and a proposal
  for dynamical black hole entropy}},
  \href{http://dx.doi.org/10.1103/PhysRevD.50.846}{\emph{Phys. Rev.} {\bf D50}
  (1994) 846--864}, [\href{https://arxiv.org/abs/gr-qc/9403028}{{\tt
  gr-qc/9403028}}].

\bibitem{Jacobson:1993vj}
T.~Jacobson, G.~Kang and R.~C. Myers, \emph{{On black hole entropy}},
  \href{http://dx.doi.org/10.1103/PhysRevD.49.6587}{\emph{Phys. Rev.} {\bf D49}
  (1994) 6587--6598}, [\href{https://arxiv.org/abs/gr-qc/9312023}{{\tt
  gr-qc/9312023}}].

\bibitem{Carroll:2016lku}
S.~M. Carroll and G.~N. Remmen, \emph{{What is the Entropy in Entropic
  Gravity?}}, \href{http://dx.doi.org/10.1103/PhysRevD.93.124052}{\emph{Phys.
  Rev.} {\bf D93} (2016) 124052}, [\href{https://arxiv.org/abs/1601.07558}{{\tt
  1601.07558}}].

\bibitem{Balasubramanian13-1}
V.~Balasubramanian, B.~Czech, B.~D. Chowdhury and J.~de~Boer, \emph{{The
  entropy of a hole in spacetime}},
  \href{http://dx.doi.org/10.1007/JHEP10(2013)220}{\emph{JHEP} {\bf 10} (2013)
  220}, [\href{https://arxiv.org/abs/1305.0856}{{\tt 1305.0856}}].

\bibitem{Bisognano:1976za}
J.~J. Bisognano and E.~H. Wichmann, \emph{{On the Duality Condition for Quantum
  Fields}}, \href{http://dx.doi.org/10.1063/1.522898}{\emph{J. Math. Phys.}
  {\bf 17} (1976) 303--321}.

\bibitem{Casini:2016rwj}
H.~Casini, D.~A. Galante and R.~C. Myers, \emph{{Comments on Jacobson’s
  “entanglement equilibrium and the Einstein equation”}},
  \href{http://dx.doi.org/10.1007/JHEP03(2016)194}{\emph{JHEP} {\bf 03} (2016)
  194}, [\href{https://arxiv.org/abs/1601.00528}{{\tt 1601.00528}}].

\bibitem{Balasubramanian:2013lsa}
V.~Balasubramanian, B.~D. Chowdhury, B.~Czech, J.~de~Boer and M.~P. Heller,
  \emph{{Bulk curves from boundary data in holography}},
  \href{http://dx.doi.org/10.1103/PhysRevD.89.086004}{\emph{Phys. Rev.} {\bf
  D89} (2014) 086004}, [\href{https://arxiv.org/abs/1310.4204}{{\tt
  1310.4204}}].

\bibitem{Lewkowycz:2018sgn}
A.~Lewkowycz and O.~Parrikar, \emph{{The holographic shape of entanglement and
  Einstein’s equations}},
  \href{http://dx.doi.org/10.1007/JHEP05(2018)147}{\emph{JHEP} {\bf 05} (2018)
  147}, [\href{https://arxiv.org/abs/1802.10103}{{\tt 1802.10103}}].

\bibitem{Fursaev:2013fta}
D.~V. Fursaev, A.~Patrushev and S.~N. Solodukhin, \emph{{Distributional
  Geometry of Squashed Cones}},
  \href{http://dx.doi.org/10.1103/PhysRevD.88.044054}{\emph{Phys. Rev.} {\bf
  D88} (2013) 044054}, [\href{https://arxiv.org/abs/1306.4000}{{\tt
  1306.4000}}].

\bibitem{Czech:2014wka}
B.~Czech, X.~Dong and J.~Sully, \emph{{Holographic Reconstruction of General
  Bulk Surfaces}}, \href{http://dx.doi.org/10.1007/JHEP11(2014)015}{\emph{JHEP}
  {\bf 11} (2014) 015}, [\href{https://arxiv.org/abs/1406.4889}{{\tt
  1406.4889}}].

\bibitem{Headrick:2014eia}
M.~Headrick, R.~C. Myers and J.~Wien, \emph{{Holographic Holes and Differential
  Entropy}}, \href{http://dx.doi.org/10.1007/JHEP10(2014)149}{\emph{JHEP} {\bf
  10} (2014) 149}, [\href{https://arxiv.org/abs/1408.4770}{{\tt 1408.4770}}].

\bibitem{Hawking75-1}
S.~W. Hawking, \emph{{Particle Creation by Black Holes}},  in \emph{{1st Oxford
  Conference on Quantum Gravity Chilton, England, February 15-16, 1974}},
  pp.~219--267, 1975.

\bibitem{Johnson:2014xza}
C.~V. Johnson, \emph{{Thermodynamic Volumes for AdS-Taub-NUT and
  AdS-Taub-Bolt}},
  \href{http://dx.doi.org/10.1088/0264-9381/31/23/235003}{\emph{Class. Quant.
  Grav.} {\bf 31} (2014) 235003}, [\href{https://arxiv.org/abs/1405.5941}{{\tt
  1405.5941}}].

\bibitem{Banados:1992wn}
M.~Banados, C.~Teitelboim and J.~Zanelli, \emph{{The Black hole in
  three-dimensional space-time}},
  \href{http://dx.doi.org/10.1103/PhysRevLett.69.1849}{\emph{Phys. Rev. Lett.}
  {\bf 69} (1992) 1849--1851},
  [\href{https://arxiv.org/abs/hep-th/9204099}{{\tt hep-th/9204099}}].

\bibitem{Caceres:2015vsa}
E.~Caceres, P.~H. Nguyen and J.~F. Pedraza, \emph{{Holographic entanglement
  entropy and the extended phase structure of STU black holes}},
  \href{http://dx.doi.org/10.1007/JHEP09(2015)184}{\emph{JHEP} {\bf 09} (2015)
  184}, [\href{https://arxiv.org/abs/1507.06069}{{\tt 1507.06069}}].

\bibitem{Hennigar:2014cfa}
R.~A. Hennigar, D.~Kubiznak and R.~B. Mann, \emph{{Entropy Inequality
  Violations from Ultraspinning Black Holes}},
  \href{http://dx.doi.org/10.1103/PhysRevLett.115.031101}{\emph{Phys. Rev.
  Lett.} {\bf 115} (2015) 031101}, [\href{https://arxiv.org/abs/1411.4309}{{\tt
  1411.4309}}].

\bibitem{Hennigar:2015cja}
R.~A. Hennigar, D.~Kubiznak, R.~B. Mann and N.~Musoke, \emph{{Ultraspinning
  limits and super-entropic black holes}},
  \href{http://dx.doi.org/10.1007/JHEP06(2015)096}{\emph{JHEP} {\bf 06} (2015)
  096}, [\href{https://arxiv.org/abs/1504.07529}{{\tt 1504.07529}}].

\bibitem{Martinez:1999qi}
C.~Martinez, C.~Teitelboim and J.~Zanelli, \emph{{Charged rotating black hole
  in three space-time dimensions}},
  \href{http://dx.doi.org/10.1103/PhysRevD.61.104013}{\emph{Phys. Rev.} {\bf
  D61} (2000) 104013}, [\href{https://arxiv.org/abs/hep-th/9912259}{{\tt
  hep-th/9912259}}].

\bibitem{Feng:2017jub}
X.-H. Feng, H.-S. Liu, W.-T. Lu and H.~Lu, \emph{{Horndeski Gravity and the
  Violation of Reverse Isoperimetric Inequality}},
  \href{http://dx.doi.org/10.1140/epjc/s10052-017-5356-x}{\emph{Eur. Phys. J.}
  {\bf C77} (2017) 790}, [\href{https://arxiv.org/abs/1705.08970}{{\tt
  1705.08970}}].

\bibitem{Chamblin:1998pz}
A.~Chamblin, R.~Emparan, C.~V. Johnson and R.~C. Myers, \emph{{Large N phases,
  gravitational instantons and the nuts and bolts of AdS holography}},
  \href{http://dx.doi.org/10.1103/PhysRevD.59.064010}{\emph{Phys.Rev.} {\bf
  D59} (1999) 064010}, [\href{https://arxiv.org/abs/hep-th/9808177}{{\tt
  hep-th/9808177}}].

\bibitem{Emparan:1999pm}
R.~Emparan, C.~V. Johnson and R.~C. Myers, \emph{{Surface terms as counterterms
  in the AdS / CFT correspondence}},
  \href{http://dx.doi.org/10.1103/PhysRevD.60.104001}{\emph{Phys. Rev.} {\bf
  D60} (1999) 104001}, [\href{https://arxiv.org/abs/hep-th/9903238}{{\tt
  hep-th/9903238}}].

\bibitem{Mann:1999pc}
R.~B. Mann, \emph{{Misner string entropy}},
  \href{http://dx.doi.org/10.1103/PhysRevD.60.104047}{\emph{Phys.Rev.} {\bf
  D60} (1999) 104047}, [\href{https://arxiv.org/abs/hep-th/9903229}{{\tt
  hep-th/9903229}}].

\bibitem{Johnson:2019mdp}
C.~V. Johnson, \emph{{Instability of Super-Entropic Black Holes in Extended
  Thermodynamics}},  \href{https://arxiv.org/abs/1906.00993}{{\tt 1906.00993}}.

\bibitem{Maldacena:1997re}
J.~M. Maldacena, \emph{The large n limit of superconformal field theories and
  supergravity}, {\emph{Adv. Theor. Math. Phys.} {\bf 2} (1998) 231--252},
  [\href{https://arxiv.org/abs/hep-th/9711200}{{\tt hep-th/9711200}}].

\bibitem{Gubser:1998bc}
S.~S. Gubser, I.~R. Klebanov and A.~M. Polyakov, \emph{{Gauge theory
  correlators from noncritical string theory}},
  \href{http://dx.doi.org/10.1016/S0370-2693(98)00377-3}{\emph{Phys. Lett.}
  {\bf B428} (1998) 105--114},
  [\href{https://arxiv.org/abs/hep-th/9802109}{{\tt hep-th/9802109}}].

\bibitem{Witten:1998qj}
E.~Witten, \emph{{Anti-de Sitter space and holography}},
  \href{http://dx.doi.org/10.4310/ATMP.1998.v2.n2.a2}{\emph{Adv. Theor. Math.
  Phys.} {\bf 2} (1998) 253--291},
  [\href{https://arxiv.org/abs/hep-th/9802150}{{\tt hep-th/9802150}}].

\bibitem{Witten:1998zw}
E.~Witten, \emph{Anti-de sitter space, thermal phase transition, and
  confinement in gauge theories}, {\emph{Adv. Theor. Math. Phys.} {\bf 2}
  (1998) 505--532}, [\href{https://arxiv.org/abs/hep-th/9803131}{{\tt
  hep-th/9803131}}].

\bibitem{Dolan:2013dga}
B.~P. Dolan, \emph{{The compressibility of rotating black holes in
  $D$-dimensions}},
  \href{http://dx.doi.org/10.1088/0264-9381/31/3/035022}{\emph{Class.Quant.Grav.}
  {\bf 31} (2014) 035022}, [\href{https://arxiv.org/abs/1308.5403}{{\tt
  1308.5403}}].

\bibitem{Dolan:2014cja}
B.~P. Dolan, \emph{{Bose condensation and branes}},
  \href{http://dx.doi.org/10.1007/JHEP10(2014)179}{\emph{JHEP} {\bf 10} (2014)
  179}, [\href{https://arxiv.org/abs/1406.7267}{{\tt 1406.7267}}].

\bibitem{Couch:2016exn}
J.~Couch, W.~Fischler and P.~H. Nguyen, \emph{{Noether charge, black hole
  volume, and complexity}},
  \href{http://dx.doi.org/10.1007/JHEP03(2017)119}{\emph{JHEP} {\bf 03} (2017)
  119}, [\href{https://arxiv.org/abs/1610.02038}{{\tt 1610.02038}}].

\bibitem{Johnson:2019vqf}
C.~V. Johnson, \emph{{Specific Heats and Schottky Peaks for Black Holes in
  Extended Thermodynamics}},  \href{https://arxiv.org/abs/1905.00539}{{\tt
  1905.00539}}.

\bibitem{Johnson:2019olt}
C.~V. Johnson, \emph{{Holographic Heat Engines as Quantum Heat Engines}},
  \href{https://arxiv.org/abs/1905.09399}{{\tt 1905.09399}}.

\bibitem{Carlip:1994gc}
S.~Carlip and C.~Teitelboim, \emph{{Aspects of black hole quantum mechanics and
  thermodynamics in (2+1)-dimensions}},
  \href{http://dx.doi.org/10.1103/PhysRevD.51.622}{\emph{Phys. Rev.} {\bf D51}
  (1995) 622--631}, [\href{https://arxiv.org/abs/gr-qc/9405070}{{\tt
  gr-qc/9405070}}].

\bibitem{Strominger:1996sh}
A.~Strominger and C.~Vafa, \emph{{Microscopic origin of the Bekenstein-Hawking
  entropy}}, \href{http://dx.doi.org/10.1016/0370-2693(96)00345-0}{\emph{Phys.
  Lett.} {\bf B379} (1996) 99--104},
  [\href{https://arxiv.org/abs/hep-th/9601029}{{\tt hep-th/9601029}}].

\bibitem{Birmingham:1998jt}
D.~Birmingham, I.~Sachs and S.~Sen, \emph{{Entropy of three-dimensional black
  holes in string theory}},
  \href{http://dx.doi.org/10.1016/S0370-2693(98)00236-6}{\emph{Phys. Lett.}
  {\bf B424} (1998) 275--280},
  [\href{https://arxiv.org/abs/hep-th/9801019}{{\tt hep-th/9801019}}].

\bibitem{Cardy:1986ie}
J.~L. Cardy, \emph{{Operator Content of Two-Dimensional Conformally Invariant
  Theories}}, \href{http://dx.doi.org/10.1016/0550-3213(86)90552-3}{\emph{Nucl.
  Phys.} {\bf B270} (1986) 186--204}.

\bibitem{Bloete:1986qm}
H.~W.~J. Bloete, J.~L. Cardy and M.~P. Nightingale, \emph{{Conformal
  Invariance, the Central Charge, and Universal Finite Size Amplitudes at
  Criticality}},
  \href{http://dx.doi.org/10.1103/PhysRevLett.56.742}{\emph{Phys. Rev. Lett.}
  {\bf 56} (1986) 742--745}.

\bibitem{Carlip:1991zk}
S.~Carlip and J.~Gegenberg, \emph{{Gravitating topological matter in
  (2+1)-dimensions}},
  \href{http://dx.doi.org/10.1103/PhysRevD.44.424}{\emph{Phys. Rev.} {\bf D44}
  (1991) 424--432}.

\bibitem{Carlip:1994hq}
S.~Carlip, J.~Gegenberg and R.~B. Mann, \emph{{Black holes in three-dimensional
  topological gravity}},
  \href{http://dx.doi.org/10.1103/PhysRevD.51.6854}{\emph{Phys. Rev.} {\bf D51}
  (1995) 6854--6859}, [\href{https://arxiv.org/abs/gr-qc/9410021}{{\tt
  gr-qc/9410021}}].

\bibitem{Townsend:2013ela}
P.~K. Townsend and B.~Zhang, \emph{{Thermodynamics of “Exotic”
  Bañados-Teitelboim-Zanelli Black Holes}},
  \href{http://dx.doi.org/10.1103/PhysRevLett.110.241302}{\emph{Phys. Rev.
  Lett.} {\bf 110} (2013) 241302}, [\href{https://arxiv.org/abs/1302.3874}{{\tt
  1302.3874}}].

\bibitem{Cong:2019bud}
W.~Cong and R.~B. Mann, \emph{{Thermodynamic Instabilities of Generalized
  Exotic BTZ Black Holes}},  \href{https://arxiv.org/abs/1908.01254}{{\tt
  1908.01254}}.

\bibitem{Carlip:1998qw}
S.~Carlip, \emph{{What we don't know about BTZ black hole entropy}},
  \href{http://dx.doi.org/10.1088/0264-9381/15/11/020}{\emph{Class. Quant.
  Grav.} {\bf 15} (1998) 3609--3625},
  [\href{https://arxiv.org/abs/hep-th/9806026}{{\tt hep-th/9806026}}].

\bibitem{Banados:1992gq}
M.~Banados, M.~Henneaux, C.~Teitelboim and J.~Zanelli, \emph{{Geometry of the
  (2+1) black hole}}, \href{http://dx.doi.org/10.1103/PhysRevD.48.1506,
  10.1103/PhysRevD.88.069902}{\emph{Phys. Rev.} {\bf D48} (1993) 1506--1525},
  [\href{https://arxiv.org/abs/gr-qc/9302012}{{\tt gr-qc/9302012}}].

\bibitem{Gibbons:1976ue}
G.~W. Gibbons and S.~W. Hawking, \emph{{Action Integrals and Partition
  Functions in Quantum Gravity}},
  \href{http://dx.doi.org/10.1103/PhysRevD.15.2752}{\emph{Phys. Rev.} {\bf D15}
  (1977) 2752--2756}.

\bibitem{Frassino:2015oca}
A.~M. Frassino, R.~B. Mann and J.~R. Mureika, \emph{{Lower-Dimensional Black
  Hole Chemistry}},
  \href{http://dx.doi.org/10.1103/PhysRevD.92.124069}{\emph{Phys. Rev.} {\bf
  D92} (2015) 124069}, [\href{https://arxiv.org/abs/1509.05481}{{\tt
  1509.05481}}].

\bibitem{Cadoni:2007ck}
M.~Cadoni, M.~Melis and M.~R. Setare, \emph{{Microscopic entropy of the charged
  BTZ black hole}},
  \href{http://dx.doi.org/10.1088/0264-9381/25/19/195022}{\emph{Class. Quant.
  Grav.} {\bf 25} (2008) 195022}, [\href{https://arxiv.org/abs/0710.3009}{{\tt
  0710.3009}}].

\bibitem{Guica:2008mu}
M.~Guica, T.~Hartman, W.~Song and A.~Strominger, \emph{{The Kerr/CFT
  Correspondence}},
  \href{http://dx.doi.org/10.1103/PhysRevD.80.124008}{\emph{Phys. Rev.} {\bf
  D80} (2009) 124008}, [\href{https://arxiv.org/abs/0809.4266}{{\tt
  0809.4266}}].

\bibitem{Sinamuli:2015drn}
M.~Sinamuli and R.~B. Mann, \emph{{Super-Entropic Black Holes and the Kerr-CFT
  Correspondence}},
  \href{http://dx.doi.org/10.1007/JHEP08(2016)148}{\emph{JHEP} {\bf 08} (2016)
  148}, [\href{https://arxiv.org/abs/1512.07597}{{\tt 1512.07597}}].

\bibitem{Lashkari:2013koa}
N.~Lashkari, M.~B. McDermott and M.~Van~Raamsdonk, \emph{{Gravitational
  dynamics from entanglement 'thermodynamics'}},
  \href{http://dx.doi.org/10.1007/JHEP04(2014)195}{\emph{JHEP} {\bf 04} (2014)
  195}, [\href{https://arxiv.org/abs/1308.3716}{{\tt 1308.3716}}].

\bibitem{Faulkner:2013ica}
T.~Faulkner, M.~Guica, T.~Hartman, R.~C. Myers and M.~Van~Raamsdonk,
  \emph{{Gravitation from Entanglement in Holographic CFTs}},
  \href{http://dx.doi.org/10.1007/JHEP03(2014)051}{\emph{JHEP} {\bf 03} (2014)
  051}, [\href{https://arxiv.org/abs/1312.7856}{{\tt 1312.7856}}].

\bibitem{Dolan:2010ha}
B.~P. Dolan, \emph{{The cosmological constant and the black hole equation of
  state}},
  \href{http://dx.doi.org/10.1088/0264-9381/28/12/125020}{\emph{Class.Quant.Grav.}
  {\bf 28} (2011) 125020}, [\href{https://arxiv.org/abs/1008.5023}{{\tt
  1008.5023}}].

\bibitem{Pufu:2016zxm}
S.~S. Pufu, \emph{{The F-Theorem and F-Maximization}},
  \href{http://dx.doi.org/10.1088/1751-8121/aa6765}{\emph{J. Phys.} {\bf A50}
  (2017) 443008}, [\href{https://arxiv.org/abs/1608.02960}{{\tt 1608.02960}}].

\bibitem{Casini:2017vbe}
H.~Casini, E.~Testé and G.~Torroba, \emph{{Markov Property of the Conformal
  Field Theory Vacuum and the a Theorem}},
  \href{http://dx.doi.org/10.1103/PhysRevLett.118.261602}{\emph{Phys. Rev.
  Lett.} {\bf 118} (2017) 261602},
  [\href{https://arxiv.org/abs/1704.01870}{{\tt 1704.01870}}].

\bibitem{Myers:2010tj}
R.~C. Myers and A.~Sinha, \emph{{Holographic c-theorems in arbitrary
  dimensions}}, \href{http://dx.doi.org/10.1007/JHEP01(2011)125}{\emph{JHEP}
  {\bf 01} (2011) 125}, [\href{https://arxiv.org/abs/1011.5819}{{\tt
  1011.5819}}].

\bibitem{Kastor:2016bph}
D.~Kastor, S.~Ray and J.~Traschen, \emph{{Extended First Law for Entanglement
  Entropy in Lovelock Gravity}},
  \href{http://dx.doi.org/10.3390/e18060212}{\emph{Entropy} {\bf 18} (2016)
  212}, [\href{https://arxiv.org/abs/1604.04468}{{\tt 1604.04468}}].

\bibitem{Caceres:2016xjz}
E.~Caceres, P.~H. Nguyen and J.~F. Pedraza, \emph{{Holographic entanglement
  chemistry}}, \href{http://dx.doi.org/10.1103/PhysRevD.95.106015}{\emph{Phys.
  Rev.} {\bf D95} (2017) 106015}, [\href{https://arxiv.org/abs/1605.00595}{{\tt
  1605.00595}}].

\bibitem{Lan:2017xcl}
S.-Q. Lan, G.-Q. Li, J.-X. Mo and X.-B. Xu, \emph{{On the first law of
  entanglement for Quasi-Topological gravity}},
  \href{http://dx.doi.org/10.1007/s10714-018-2426-9}{\emph{Gen. Rel. Grav.}
  {\bf 50} (2018) 106}, [\href{https://arxiv.org/abs/1710.01553}{{\tt
  1710.01553}}].

\bibitem{Teitelboim:1983ux}
C.~Teitelboim, \emph{{Gravitation and Hamiltonian Structure in Two Space-Time
  Dimensions}},
  \href{http://dx.doi.org/10.1016/0370-2693(83)90012-6}{\emph{Phys. Lett.} {\bf
  126B} (1983) 41--45}.

\bibitem{Jackiw:1984je}
R.~Jackiw, \emph{{Lower Dimensional Gravity}},
  \href{http://dx.doi.org/10.1016/0550-3213(85)90448-1}{\emph{Nucl. Phys.} {\bf
  B252} (1985) 343--356}.

\bibitem{Ryu:2006bv}
S.~Ryu and T.~Takayanagi, \emph{{Holographic derivation of entanglement entropy
  from AdS/CFT}},
  \href{http://dx.doi.org/10.1103/PhysRevLett.96.181602}{\emph{Phys.Rev.Lett.}
  {\bf 96} (2006) 181602}, [\href{https://arxiv.org/abs/hep-th/0603001}{{\tt
  hep-th/0603001}}].

\bibitem{Hung:2011xb}
L.-Y. Hung, R.~C. Myers and M.~Smolkin, \emph{{On Holographic Entanglement
  Entropy and Higher Curvature Gravity}},
  \href{http://dx.doi.org/10.1007/JHEP04(2011)025}{\emph{JHEP} {\bf 04} (2011)
  025}, [\href{https://arxiv.org/abs/1101.5813}{{\tt 1101.5813}}].

\bibitem{Rosso:2019lsm}
F.~Rosso, \emph{{Localized thermal states and negative energy}},
  \href{http://dx.doi.org/10.1007/JHEP10(2019)246}{\emph{JHEP} {\bf 10} (2019)
  246}, [\href{https://arxiv.org/abs/1907.07699}{{\tt 1907.07699}}].

\bibitem{Emparan:1999gf}
R.~Emparan, \emph{{AdS / CFT duals of topological black holes and the entropy
  of zero energy states}},
  \href{http://dx.doi.org/10.1088/1126-6708/1999/06/036}{\emph{JHEP} {\bf 06}
  (1999) 036}, [\href{https://arxiv.org/abs/hep-th/9906040}{{\tt
  hep-th/9906040}}].

\bibitem{Unruh76-1}
W.~G. Unruh, \emph{{Notes on black hole evaporation}},
  \href{http://dx.doi.org/10.1103/PhysRevD.14.870}{\emph{Phys. Rev.} {\bf D14}
  (1976) 870}.

\bibitem{Strobl:1999wv}
T.~Strobl, \emph{{Gravity in two space-time dimensions}}.
\newblock PhD thesis, Aachen, Tech. Hochsch., 1999.
\newblock \href{https://arxiv.org/abs/hep-th/0011240}{{\tt hep-th/0011240}}.

\bibitem{Strominger:1998yg}
A.~Strominger, \emph{{AdS(2) quantum gravity and string theory}},
  \href{http://dx.doi.org/10.1088/1126-6708/1999/01/007}{\emph{JHEP} {\bf 01}
  (1999) 007}, [\href{https://arxiv.org/abs/hep-th/9809027}{{\tt
  hep-th/9809027}}].

\bibitem{Cadoni:1999ja}
M.~Cadoni and S.~Mignemi, \emph{{Asymptotic symmetries of AdS(2) and conformal
  group in d = 1}},
  \href{http://dx.doi.org/10.1016/S0550-3213(99)00398-3}{\emph{Nucl. Phys.}
  {\bf B557} (1999) 165--180},
  [\href{https://arxiv.org/abs/hep-th/9902040}{{\tt hep-th/9902040}}].

\bibitem{Hartman:2008dq}
T.~Hartman and A.~Strominger, \emph{{Central Charge for AdS(2) Quantum
  Gravity}}, \href{http://dx.doi.org/10.1088/1126-6708/2009/04/026}{\emph{JHEP}
  {\bf 04} (2009) 026}, [\href{https://arxiv.org/abs/0803.3621}{{\tt
  0803.3621}}].

\bibitem{Castro:2008ms}
A.~Castro, D.~Grumiller, F.~Larsen and R.~McNees, \emph{{Holographic
  Description of AdS(2) Black Holes}},
  \href{http://dx.doi.org/10.1088/1126-6708/2008/11/052}{\emph{JHEP} {\bf 11}
  (2008) 052}, [\href{https://arxiv.org/abs/0809.4264}{{\tt 0809.4264}}].

\bibitem{Alishahiha:2008tv}
M.~Alishahiha and F.~Ardalan, \emph{{Central Charge for 2D Gravity on AdS(2)
  and AdS(2)/CFT(1) Correspondence}},
  \href{http://dx.doi.org/10.1088/1126-6708/2008/08/079}{\emph{JHEP} {\bf 08}
  (2008) 079}, [\href{https://arxiv.org/abs/0805.1861}{{\tt 0805.1861}}].

\bibitem{Cvetic:2016eiv}
M.~Cvetič and I.~Papadimitriou, \emph{{AdS$_{2}$ holographic dictionary}},
  \href{http://dx.doi.org/10.1007/JHEP12(2016)008,
  10.1007/JHEP01(2017)120}{\emph{JHEP} {\bf 12} (2016) 008},
  [\href{https://arxiv.org/abs/1608.07018}{{\tt 1608.07018}}].

\bibitem{Saad:2019lba}
P.~Saad, S.~H. Shenker and D.~Stanford, \emph{{JT gravity as a matrix
  integral}},  \href{https://arxiv.org/abs/1903.11115}{{\tt 1903.11115}}.

\bibitem{Mann:1992ar}
R.~B. Mann and S.~F. Ross, \emph{{The D ---> 2 limit of general relativity}},
  \href{http://dx.doi.org/10.1088/0264-9381/10/7/015}{\emph{Class. Quant.
  Grav.} {\bf 10} (1993) 1405--1408},
  [\href{https://arxiv.org/abs/gr-qc/9208004}{{\tt gr-qc/9208004}}].

\bibitem{Faraoni:1999hp}
V.~Faraoni and E.~Gunzig, \emph{{Einstein frame or Jordan frame?}},
  \href{http://dx.doi.org/10.1023/A:1026645510351}{\emph{Int. J. Theor. Phys.}
  {\bf 38} (1999) 217--225},
  [\href{https://arxiv.org/abs/astro-ph/9910176}{{\tt astro-ph/9910176}}].

\bibitem{Postma:2014vaa}
M.~Postma and M.~Volponi, \emph{{Equivalence of the Einstein and Jordan
  frames}}, \href{http://dx.doi.org/10.1103/PhysRevD.90.103516}{\emph{Phys.
  Rev.} {\bf D90} (2014) 103516}, [\href{https://arxiv.org/abs/1407.6874}{{\tt
  1407.6874}}].

\bibitem{Maldacena:2016upp}
J.~Maldacena, D.~Stanford and Z.~Yang, \emph{{Conformal symmetry and its
  breaking in two dimensional Nearly Anti-de-Sitter space}},
  \href{http://dx.doi.org/10.1093/ptep/ptw124}{\emph{PTEP} {\bf 2016} (2016)
  12C104}, [\href{https://arxiv.org/abs/1606.01857}{{\tt 1606.01857}}].

\bibitem{Harlow:2018tqv}
D.~Harlow and D.~Jafferis, \emph{{The Factorization Problem in
  Jackiw-Teitelboim Gravity}},
  \href{http://dx.doi.org/10.1007/JHEP02(2020)177}{\emph{JHEP} {\bf 02} (2020)
  177}, [\href{https://arxiv.org/abs/1804.01081}{{\tt 1804.01081}}].

\bibitem{Bergshoeff:2009hq}
E.~A. Bergshoeff, O.~Hohm and P.~K. Townsend, \emph{{Massive Gravity in Three
  Dimensions}},
  \href{http://dx.doi.org/10.1103/PhysRevLett.102.201301}{\emph{Phys. Rev.
  Lett.} {\bf 102} (2009) 201301}, [\href{https://arxiv.org/abs/0901.1766}{{\tt
  0901.1766}}].

\bibitem{Bergshoeff:2009aq}
E.~A. Bergshoeff, O.~Hohm and P.~K. Townsend, \emph{{More on Massive 3D
  Gravity}}, \href{http://dx.doi.org/10.1103/PhysRevD.79.124042}{\emph{Phys.
  Rev.} {\bf D79} (2009) 124042}, [\href{https://arxiv.org/abs/0905.1259}{{\tt
  0905.1259}}].

\bibitem{Sinha:2010ai}
A.~Sinha, \emph{{On the new massive gravity and AdS/CFT}},
  \href{http://dx.doi.org/10.1007/JHEP06(2010)061}{\emph{JHEP} {\bf 06} (2010)
  061}, [\href{https://arxiv.org/abs/1003.0683}{{\tt 1003.0683}}].

\bibitem{Marolf:2016dob}
D.~Marolf and A.~C. Wall, \emph{{State-Dependent Divergences in the
  Entanglement Entropy}},
  \href{http://dx.doi.org/10.1007/JHEP10(2016)109}{\emph{JHEP} {\bf 10} (2016)
  109}, [\href{https://arxiv.org/abs/1607.01246}{{\tt 1607.01246}}].

\bibitem{Casini:2015woa}
H.~Casini, M.~Huerta, R.~C. Myers and A.~Yale, \emph{{Mutual information and
  the F-theorem}}, \href{http://dx.doi.org/10.1007/JHEP10(2015)003}{\emph{JHEP}
  {\bf 10} (2015) 003}, [\href{https://arxiv.org/abs/1506.06195}{{\tt
  1506.06195}}].

\bibitem{Ryu:2006ef}
S.~Ryu and T.~Takayanagi, \emph{{Aspects of holographic entanglement entropy}},
  {\emph{JHEP} {\bf 08} (2006) 045},
  [\href{https://arxiv.org/abs/hep-th/0605073}{{\tt hep-th/0605073}}].

\bibitem{Hartman:2013mia}
T.~Hartman, \emph{{Entanglement Entropy at Large Central Charge}},
  \href{https://arxiv.org/abs/1303.6955}{{\tt 1303.6955}}.

\bibitem{Faulkner:2013yia}
T.~Faulkner, \emph{{The Entanglement Renyi Entropies of Disjoint Intervals in
  AdS/CFT}},  \href{https://arxiv.org/abs/1303.7221}{{\tt 1303.7221}}.

\bibitem{Calabrese:2004eu}
P.~Calabrese and J.~L. Cardy, \emph{{Entanglement entropy and quantum field
  theory}}, \href{http://dx.doi.org/10.1088/1742-5468/2004/06/P06002}{\emph{J.
  Stat. Mech.} {\bf 0406} (2004) P06002},
  [\href{https://arxiv.org/abs/hep-th/0405152}{{\tt hep-th/0405152}}].

\bibitem{Cardy:2016fqc}
J.~Cardy and E.~Tonni, \emph{{Entanglement hamiltonians in two-dimensional
  conformal field theory}},
  \href{http://dx.doi.org/10.1088/1742-5468/2016/12/123103}{\emph{J. Stat.
  Mech.} {\bf 1612} (2016) 123103},
  [\href{https://arxiv.org/abs/1608.01283}{{\tt 1608.01283}}].

\bibitem{Solodukhin:2008dh}
S.~N. Solodukhin, \emph{{Entanglement entropy, conformal invariance and
  extrinsic geometry}},
  \href{http://dx.doi.org/10.1016/j.physletb.2008.05.071}{\emph{Phys. Lett.}
  {\bf B665} (2008) 305--309}, [\href{https://arxiv.org/abs/0802.3117}{{\tt
  0802.3117}}].

\bibitem{Maldacena:2016hyu}
J.~Maldacena and D.~Stanford, \emph{{Remarks on the Sachdev-Ye-Kitaev model}},
  \href{http://dx.doi.org/10.1103/PhysRevD.94.106002}{\emph{Phys. Rev.} {\bf
  D94} (2016) 106002}, [\href{https://arxiv.org/abs/1604.07818}{{\tt
  1604.07818}}].

\bibitem{Jensen:2016pah}
K.~Jensen, \emph{{Chaos in AdS$_2$ Holography}},
  \href{http://dx.doi.org/10.1103/PhysRevLett.117.111601}{\emph{Phys. Rev.
  Lett.} {\bf 117} (2016) 111601},
  [\href{https://arxiv.org/abs/1605.06098}{{\tt 1605.06098}}].

\bibitem{Faulkner:2013ana}
T.~Faulkner, A.~Lewkowycz and J.~Maldacena, \emph{{Quantum corrections to
  holographic entanglement entropy}},
  \href{http://dx.doi.org/10.1007/JHEP11(2013)074}{\emph{JHEP} {\bf 11} (2013)
  074}, [\href{https://arxiv.org/abs/1307.2892}{{\tt 1307.2892}}].

\bibitem{Jafferis:2019wkd}
D.~L. Jafferis and D.~K. Kolchmeyer, \emph{{Entanglement Entropy in
  Jackiw-Teitelboim Gravity}},  \href{https://arxiv.org/abs/1911.10663}{{\tt
  1911.10663}}.

\bibitem{Penington:2019npb}
G.~Penington, \emph{{Entanglement Wedge Reconstruction and the Information
  Paradox}}, \href{http://dx.doi.org/10.1007/JHEP09(2020)002}{\emph{JHEP} {\bf
  09} (2020) 002}, [\href{https://arxiv.org/abs/1905.08255}{{\tt 1905.08255}}].

\bibitem{Carroll04-1}
S.~M. Carroll, \emph{{Spacetime and geometry: An introduction to general
  relativity}}.
\newblock 2004.

\bibitem{Kallosh92-1}
R.~Kallosh, A.~D. Linde, T.~Ortin, A.~W. Peet and A.~Van~Proeyen,
  \emph{{Supersymmetry as a cosmic censor}},
  \href{http://dx.doi.org/10.1103/PhysRevD.46.5278}{\emph{Phys. Rev.} {\bf D46}
  (1992) 5278--5302}, [\href{https://arxiv.org/abs/hep-th/9205027}{{\tt
  hep-th/9205027}}].

\bibitem{Bekenstein74-1}
J.~D. Bekenstein, \emph{{Generalized second law of thermodynamics in black hole
  physics}}, \href{http://dx.doi.org/10.1103/PhysRevD.9.3292}{\emph{Phys. Rev.}
  {\bf D9} (1974) 3292--3300}.

\bibitem{Poisson04-1}
E.~Poisson, \emph{A Relativist's Toolkit: The Mathematics of Black-hole
  mechanics}.
\newblock Cambridge University Press, New York, 2004.

\bibitem{Misner73-1}
C.~W. Misner, K.~S. Thorne and J.~A. Wheeler, \emph{{Gravitation}}.
\newblock W. H. Freeman, San Francisco, 1973.

\bibitem{Hawking71-1}
S.~W. Hawking, \emph{{Gravitational radiation from colliding black holes}},
  \href{http://dx.doi.org/10.1103/PhysRevLett.26.1344}{\emph{Phys. Rev. Lett.}
  {\bf 26} (1971) 1344--1346}.

\bibitem{Sudarsky:1992ty}
D.~Sudarsky and R.~M. Wald, \emph{{Extrema of mass, stationarity, and
  staticity, and solutions to the Einstein Yang-Mills equations}},
  \href{http://dx.doi.org/10.1103/PhysRevD.46.1453}{\emph{Phys. Rev.} {\bf D46}
  (1992) 1453--1474}.

\bibitem{Traschen:2001pb}
J.~H. Traschen and D.~Fox, \emph{{Tension perturbations of black brane
  space-times}},
  \href{http://dx.doi.org/10.1088/0264-9381/21/1/021}{\emph{Class. Quant.
  Grav.} {\bf 21} (2004) 289--306},
  [\href{https://arxiv.org/abs/gr-qc/0103106}{{\tt gr-qc/0103106}}].

\bibitem{Henneaux:1984ji}
M.~Henneaux and C.~Teitelboim, \emph{{The Cosmological Constant as a Canonical
  Variable}},
  \href{http://dx.doi.org/10.1016/0370-2693(84)91493-X}{\emph{Phys.Lett.} {\bf
  B143} (1984) 415--420}.

\bibitem{Teitelboim:1985dp}
C.~Teitelboim, \emph{{The Cosmological Constant as a Thermodynamic Black Hole
  Parameter}},
  \href{http://dx.doi.org/10.1016/0370-2693(85)91186-4}{\emph{Phys.Lett.} {\bf
  B158} (1985) 293--297}.

\bibitem{Henneaux:1989zc}
M.~Henneaux and C.~Teitelboim, \emph{{The Cosmological Constant and General
  Covariance}},
  \href{http://dx.doi.org/10.1016/0370-2693(89)91251-3}{\emph{Phys.Lett.} {\bf
  B222} (1989) 195--199}.

\bibitem{Kaloper:2013zca}
N.~Kaloper and A.~Padilla, \emph{{Sequestering the Standard Model Vacuum
  Energy}}, \href{http://dx.doi.org/10.1103/PhysRevLett.112.091304}{\emph{Phys.
  Rev. Lett.} {\bf 112} (2014) 091304},
  [\href{https://arxiv.org/abs/1309.6562}{{\tt 1309.6562}}].

\bibitem{Kaloper:2015jra}
N.~Kaloper, A.~Padilla, D.~Stefanyszyn and G.~Zahariade, \emph{{Manifestly
  Local Theory of Vacuum Energy Sequestering}},
  \href{http://dx.doi.org/10.1103/PhysRevLett.116.051302}{\emph{Phys. Rev.
  Lett.} {\bf 116} (2016) 051302},
  [\href{https://arxiv.org/abs/1505.01492}{{\tt 1505.01492}}].

\bibitem{Svesko:2018cbo}
A.~Svesko and G.~Zahariade, \emph{{On the Constraint Structure of Vacuum Energy
  Sequestering}},
  \href{http://dx.doi.org/10.1088/1475-7516/2019/12/033}{\emph{JCAP} {\bf 1912}
  (2019) 033}, [\href{https://arxiv.org/abs/1812.11625}{{\tt 1812.11625}}].

\bibitem{Banerjee:2010qc}
S.~Banerjee, R.~K. Gupta and A.~Sen, \emph{{Logarithmic Corrections to Extremal
  Black Hole Entropy from Quantum Entropy Function}},
  \href{http://dx.doi.org/10.1007/JHEP03(2011)147}{\emph{JHEP} {\bf 03} (2011)
  147}, [\href{https://arxiv.org/abs/1005.3044}{{\tt 1005.3044}}].

\bibitem{Denef:2009kn}
F.~Denef, S.~A. Hartnoll and S.~Sachdev, \emph{{Black hole determinants and
  quasinormal modes}},
  \href{http://dx.doi.org/10.1088/0264-9381/27/12/125001}{\emph{Class. Quant.
  Grav.} {\bf 27} (2010) 125001}, [\href{https://arxiv.org/abs/0908.2657}{{\tt
  0908.2657}}].

\bibitem{David:2009xg}
J.~R. David, M.~R. Gaberdiel and R.~Gopakumar, \emph{{The Heat Kernel on AdS(3)
  and its Applications}},
  \href{http://dx.doi.org/10.1007/JHEP04(2010)125}{\emph{JHEP} {\bf 04} (2010)
  125}, [\href{https://arxiv.org/abs/0911.5085}{{\tt 0911.5085}}].

\bibitem{Keeler:2018lza}
C.~Keeler, V.~L. Martin and A.~Svesko, \emph{{Connecting quasinormal modes and
  heat kernels in 1-loop determinants}},
  \href{http://dx.doi.org/10.21468/SciPostPhys.8.2.017}{\emph{SciPost Phys.}
  {\bf 8} (2020) 017}, [\href{https://arxiv.org/abs/1811.08433}{{\tt
  1811.08433}}].

\bibitem{Keeler:2019wsx}
C.~Keeler, V.~L. Martin and A.~Svesko, \emph{{BTZ one-loop determinants via the
  Selberg zeta function for general spin}},
  \href{http://dx.doi.org/10.1007/JHEP10(2020)138}{\emph{JHEP} {\bf 10} (2020)
  138}, [\href{https://arxiv.org/abs/1910.07607}{{\tt 1910.07607}}].

\bibitem{Martin:2019flv}
V.~L. Martin and A.~Svesko, \emph{{Normal modes in thermal AdS via the Selberg
  zeta function}},
  \href{http://dx.doi.org/10.21468/SciPostPhys.9.1.009}{\emph{SciPost Phys.}
  {\bf 9} (2020) 009}, [\href{https://arxiv.org/abs/1910.11913}{{\tt
  1910.11913}}].

\bibitem{Martin:2020api}
V.~L. Martin and A.~Svesko, \emph{{Higher spin partition functions via the
  quasinormal mode method in de Sitter quantum gravity}},
  \href{http://dx.doi.org/10.21468/SciPostPhys.9.3.039}{\emph{SciPost Phys.}
  {\bf 9} (2020) 039}, [\href{https://arxiv.org/abs/2004.00128}{{\tt
  2004.00128}}].

\bibitem{Padmanabhan07-1}
T.~Padmanabhan and A.~Paranjape, \emph{{Entropy of null surfaces and dynamics
  of spacetime}},
  \href{http://dx.doi.org/10.1103/PhysRevD.75.064004}{\emph{Phys. Rev.} {\bf
  D75} (2007) 064004}, [\href{https://arxiv.org/abs/gr-qc/0701003}{{\tt
  gr-qc/0701003}}].

\bibitem{Vollick07-1}
D.~N. Vollick, \emph{{Noether Charge and Black Hole Entropy in Modified
  Theories of Gravity}},
  \href{http://dx.doi.org/10.1103/PhysRevD.76.124001}{\emph{Phys. Rev.} {\bf
  D76} (2007) 124001}, [\href{https://arxiv.org/abs/0710.1859}{{\tt
  0710.1859}}].

\bibitem{Lovelock71-1}
D.~Lovelock, \emph{{The Einstein tensor and its generalizations}},
  \href{http://dx.doi.org/10.1063/1.1665613}{\emph{J. Math. Phys.} {\bf 12}
  (1971) 498--501}.

\bibitem{Glavan:2019inb}
D.~Glavan and C.~Lin, \emph{{Einstein-Gauss-Bonnet Gravity in Four-Dimensional
  Spacetime}},
  \href{http://dx.doi.org/10.1103/PhysRevLett.124.081301}{\emph{Phys. Rev.
  Lett.} {\bf 124} (2020) 081301},
  [\href{https://arxiv.org/abs/1905.03601}{{\tt 1905.03601}}].

\bibitem{Lu:2020iav}
H.~Lu and Y.~Pang, \emph{{Horndeski gravity as $D \rightarrow 4$ limit of
  Gauss-Bonnet}},
  \href{http://dx.doi.org/10.1016/j.physletb.2020.135717}{\emph{Phys. Lett. B}
  {\bf 809} (2020) 135717}, [\href{https://arxiv.org/abs/2003.11552}{{\tt
  2003.11552}}].

\bibitem{Hennigar:2020lsl}
R.~A. Hennigar, D.~Kubiz\v{n}\'ak, R.~B. Mann and C.~Pollack, \emph{{On taking
  the $D \rightarrow 4$ limit of Gauss-Bonnet gravity: theory and solutions}},
  \href{http://dx.doi.org/10.1007/JHEP07(2020)027}{\emph{JHEP} {\bf 07} (2020)
  027}, [\href{https://arxiv.org/abs/2004.09472}{{\tt 2004.09472}}].

\bibitem{Easson:2020mpq}
D.~A. Easson, T.~Manton and A.~Svesko, \emph{{$D\to4$ Einstein-Gauss-Bonnet
  gravity and beyond}},
  \href{http://dx.doi.org/10.1088/1475-7516/2020/10/026}{\emph{JCAP} {\bf 10}
  (2020) 026}, [\href{https://arxiv.org/abs/2005.12292}{{\tt 2005.12292}}].

\bibitem{Jacobson:1993xs}
T.~Jacobson and R.~C. Myers, \emph{{Black hole entropy and higher curvature
  interactions}},
  \href{http://dx.doi.org/10.1103/PhysRevLett.70.3684}{\emph{Phys. Rev. Lett.}
  {\bf 70} (1993) 3684--3687},
  [\href{https://arxiv.org/abs/hep-th/9305016}{{\tt hep-th/9305016}}].

\bibitem{Maldacena:2001kr}
J.~M. Maldacena, \emph{{Eternal black holes in anti-de Sitter}},
  \href{http://dx.doi.org/10.1088/1126-6708/2003/04/021}{\emph{JHEP} {\bf 04}
  (2003) 021}, [\href{https://arxiv.org/abs/hep-th/0106112}{{\tt
  hep-th/0106112}}].

\bibitem{Headrick:2019eth}
M.~Headrick, \emph{{Lectures on entanglement entropy in field theory and
  holography}},  \href{https://arxiv.org/abs/1907.08126}{{\tt 1907.08126}}.

\bibitem{Casini:2010kt}
H.~Casini and M.~Huerta, \emph{{Entanglement entropy for the n-sphere}},
  \href{http://dx.doi.org/10.1016/j.physletb.2010.09.054}{\emph{Phys. Lett.}
  {\bf B694} (2011) 167--171}, [\href{https://arxiv.org/abs/1007.1813}{{\tt
  1007.1813}}].

\bibitem{Hung:2011nu}
L.-Y. Hung, R.~C. Myers, M.~Smolkin and A.~Yale, \emph{{Holographic
  Calculations of Renyi Entropy}},
  \href{http://dx.doi.org/10.1007/JHEP12(2011)047}{\emph{JHEP} {\bf 12} (2011)
  047}, [\href{https://arxiv.org/abs/1110.1084}{{\tt 1110.1084}}].

\bibitem{Johnson:2018bma}
C.~V. Johnson, \emph{{Physical Generalizations of the Rényi Entropy}},
  \href{https://arxiv.org/abs/1807.09215}{{\tt 1807.09215}}.

\bibitem{Dias:2010ma}
O.~J.~C. Dias, R.~Monteiro, H.~S. Reall and J.~E. Santos, \emph{{A Scalar field
  condensation instability of rotating anti-de Sitter black holes}},
  \href{http://dx.doi.org/10.1007/JHEP11(2010)036}{\emph{JHEP} {\bf 11} (2010)
  036}, [\href{https://arxiv.org/abs/1007.3745}{{\tt 1007.3745}}].

\bibitem{Belin:2013uta}
A.~Belin, L.-Y. Hung, A.~Maloney, S.~Matsuura, R.~C. Myers and T.~Sierens,
  \emph{{Holographic Charged Renyi Entropies}},
  \href{http://dx.doi.org/10.1007/JHEP12(2013)059}{\emph{JHEP} {\bf 12} (2013)
  059}, [\href{https://arxiv.org/abs/1310.4180}{{\tt 1310.4180}}].

\bibitem{Belin:2014mva}
A.~Belin, L.-Y. Hung, A.~Maloney and S.~Matsuura, \emph{{Charged Renyi
  entropies and holographic superconductors}},
  \href{http://dx.doi.org/10.1007/JHEP01(2015)059}{\emph{JHEP} {\bf 01} (2015)
  059}, [\href{https://arxiv.org/abs/1407.5630}{{\tt 1407.5630}}].

\bibitem{Hayden:2011ag}
P.~Hayden, M.~Headrick and A.~Maloney, \emph{{Holographic Mutual Information is
  Monogamous}}, \href{http://dx.doi.org/10.1103/PhysRevD.87.046003}{\emph{Phys.
  Rev. D} {\bf 87} (2013) 046003}, [\href{https://arxiv.org/abs/1107.2940}{{\tt
  1107.2940}}].

\bibitem{Headrick:2010zt}
M.~Headrick, \emph{{Entanglement Renyi entropies in holographic theories}},
  \href{http://dx.doi.org/10.1103/PhysRevD.82.126010}{\emph{Phys.Rev.} {\bf
  D82} (2010) 126010}, [\href{https://arxiv.org/abs/1006.0047}{{\tt
  1006.0047}}].

\bibitem{Karch:2015rpa}
A.~Karch and B.~Robinson, \emph{{Holographic Black Hole Chemistry}},
  \href{http://dx.doi.org/10.1007/JHEP12(2015)073}{\emph{JHEP} {\bf 12} (2015)
  073}, [\href{https://arxiv.org/abs/1510.02472}{{\tt 1510.02472}}].

\bibitem{Kothawala:2010bf}
D.~Kothawala, \emph{{The thermodynamic structure of Einstein tensor}},
  \href{http://dx.doi.org/10.1103/PhysRevD.83.024026}{\emph{Phys. Rev.} {\bf
  D83} (2011) 024026}, [\href{https://arxiv.org/abs/1010.2207}{{\tt
  1010.2207}}].

\bibitem{Calabrese:2009qy}
P.~Calabrese and J.~Cardy, \emph{{Entanglement entropy and conformal field
  theory}}, \href{http://dx.doi.org/10.1088/1751-8113/42/50/504005}{\emph{J.
  Phys.} {\bf A42} (2009) 504005}, [\href{https://arxiv.org/abs/0905.4013}{{\tt
  0905.4013}}].

\bibitem{Iyer:1995kg}
V.~Iyer and R.~M. Wald, \emph{{A Comparison of Noether charge and Euclidean
  methods for computing the entropy of stationary black holes}},
  \href{http://dx.doi.org/10.1103/PhysRevD.52.4430}{\emph{Phys. Rev.} {\bf D52}
  (1995) 4430--4439}, [\href{https://arxiv.org/abs/gr-qc/9503052}{{\tt
  gr-qc/9503052}}].

\bibitem{Almheiri:2014cka}
A.~Almheiri and J.~Polchinski, \emph{{Models of AdS$_{2}$ backreaction and
  holography}}, \href{http://dx.doi.org/10.1007/JHEP11(2015)014}{\emph{JHEP}
  {\bf 11} (2015) 014}, [\href{https://arxiv.org/abs/1402.6334}{{\tt
  1402.6334}}].

\bibitem{Almheiri:2019psf}
A.~Almheiri, N.~Engelhardt, D.~Marolf and H.~Maxfield, \emph{{The entropy of
  bulk quantum fields and the entanglement wedge of an evaporating black
  hole}}, \href{http://dx.doi.org/10.1007/JHEP12(2019)063}{\emph{JHEP} {\bf 12}
  (2019) 063}, [\href{https://arxiv.org/abs/1905.08762}{{\tt 1905.08762}}].

\bibitem{Bueno:2016ypa}
P.~Bueno, P.~A. Cano, V.~S. Min and M.~R. Visser, \emph{{Aspects of general
  higher-order gravities}},
  \href{http://dx.doi.org/10.1103/PhysRevD.95.044010}{\emph{Phys. Rev. D} {\bf
  95} (2017) 044010}, [\href{https://arxiv.org/abs/1610.08519}{{\tt
  1610.08519}}].

\end{thebibliography}\endgroup

\end{document}